\documentclass[12pt]{elsart1p}

\usepackage{color}
\usepackage{amsmath}
\usepackage{amssymb}
\usepackage{sidecap}
\usepackage{graphicx}
\usepackage{pdfpages}
\usepackage{ulem}
\usepackage{cite}
\usepackage{float}
\usepackage{newclude}

\newcommand{\MD}[1]{{\color{black} #1}}

\begin{document}

\begin{frontmatter}
\title{Electrons as probes of dynamics in molecules and clusters : a contribution from Time Dependent Density Functional Theory}
\author{P. Wopperer$^{a,b}$, P.~M.~Dinh\corauthref{cor}$^{a,b}$}
\author{, P.-G.~Reinhard$^c$, and E.~Suraud$^{a,b,d}$}

\corauth[cor]{Corresponding author\\{\it Email-address}~:
  dinh@irsamc.ups-tlse.fr} 

\address{$^a$CNRS, LPT (IRSAMC)\\
  118 route de Narbonne F-31062 Toulouse C\'edex, France}
\address{$^b$Universit\'e de Toulouse, UPS,  Laboratoire de Physique
  Th\'{e}orique (IRSAMC)\\
  118 route de Narbonne F-31062 Toulouse C\'edex, France}
\address{$^c$Institut f{\"u}r Theoretische Physik, Universit{\"a}t
  Erlangen,\\
  Staudtstrasse 7 D-91058 Erlangen, Germany}
\address{$^d$Physics Department, University at Buffalo, The State University New York, Buffalo, NY 14260, USA}

\begin{abstract}
There are various ways to analyze the dynamical response of clusters
and molecules to electromagnetic perturbations. Particularly rich
information can be obtained from measuring the properties of electrons
emitted in the course of the excitation dynamics. Such an analysis of
electron signals covers observables such as total ionization, Photo-Electron
Spectra (PES), Photoelectron Angular Distributions (PAD), and ideally
combined PES/PAD. It has a long history in molecular physics and was
increasingly used in cluster physics as well.  Recent progress in the
design of new light sources (high intensity, high frequency, ultra
short pulses) opens new possibilities for measurements and thus has
renewed the interest on these observables, especially for the analysis of 
various dynamical scenarios, well beyond a simple access to electronic density of states.
This, in turn, has
motivated many theoretical investigations of the dynamics of electronic
emission for molecules and clusters up to such a complex and
interesting system as C$_{60}$.  A theoretical tool of choice is here Time-Dependent 
Density Functional Theory (TDDFT) propagated in real time
and on a spatial grid,
and augmented by a Self-Interaction Correction (SIC).  This provides a
pertinent, robust, and efficient description of electronic emission
including the detailed pattern of PES and PAD.  A direct comparison
between experiments and well founded elaborate microscopic theories is
thus readily possible, at variance with more demanding observables
such as for example fragmentation or dissociation cross sections.

The purpose of this paper is to describe the theoretical tools
developed on the basis of real-time and real-space TDDFT and to address in a realistic
manner the analysis of electronic emission following irradiation of
clusters and molecules by various laser pulses.  After a general
introduction, we shall present in a second part the available
experimental results motivating such studies, starting from the
simplest total ionization signals to the more elaborate PES and PAD,
possibly combining them and/or resolving them in time. This
experimental discussion will be complemented in a third part by a presentation of
available theoretical tools focusing on TDDFT and detailing the
methods used to address ionization observables.  We shall also discuss
the shortcomings of standard versions of TDDFT, especially what
concerns the SIC problem, and show how to improve 
formally and practically the theory on that aspect. A long fourth part will be devoted to
representative  results. We shall illustrate the use of total ionization
in pump and probe scenarios with fs lasers for tracking  ionic dynamics in
clusters. More challenging from the experimental point of view is 
pump and probe setups using attosecond pulses.
The effort there is more on the capability to
define proper signals to be measured/computed at such a short time
scale. TDDFT analysis provides here a valuable tool in the search
for the most efficient observables. 
PES and PAD will allow one to address more directly electronic
dynamics itself by means of fs or ns laser pulses. 
We shall in particular discuss the impact of the dynamical regime in PES and PAD.  
We shall end this fourth part by addressing the role of temperature in
PES and PAD. When possible, the results will be
directly compared to experiments. 
The fifth part of the paper will be
devoted to future directions of investigations. From the rich choice
of developments, we shall in particular address two aspects. We shall start to
discuss the information content of energy/angular spectra of emitted electrons
in case of excitation by swift and highly
charged ions rather than lasers. The
second issue concerns the account of dissipative effects in
TDDFT to be able to consider longer laser pulses where the
competition between direct electron emission and thermalization is
known to play a role as, e.g., in experiments with C$_{60}$.  Although
such questions have been superficially addressed in the simple case of
alkaline clusters by means of semi-classical methods, no satisfying quantum
formulation, compulsory for most realistic systems, is yet
available. First encouraging results will be
presented on that occasion. We shall finally give a short conclusion.
\end{abstract}

\begin{keyword}
Time-Dependent Density Functional Theory 
\sep
Electronic observables
\sep
Ionization
\sep
Lasers
\sep
Charged projectiles
\sep
Photo-Electron Spectrum
\sep
Photoelectron Angular Distribution
\sep
Orientation averaging
\sep 
Self-interaction correction
\sep
Time-resolved observables
\sep
Temperature effects
\sep
Dissipation effects

\PACS
34.10.+x\sep 34.35.+a\sep 34.50.-s\sep 34.50.Gb\sep 36.40.-c\sep 61.46.Bc
\end{keyword}
\end{frontmatter} 

\tableofcontents

\section{General introduction and physical context}
\label{sec:intro}

%
%
Irradiation of matter constitutes a key tool in physics, chemistry,
and biology, for analyzing structural and dynamical properties of
atoms, molecules, clusters and bulk material. Lasers offer here an
especially flexible and powerful instrument which has been
widely exploited, especially during the last decades with the enormous
technological progress reached in 
the manipulation of laser light \cite{Kel03,Rul05,Pas08}.
We dispose now of a broad choice of laser intensities,
  frequencies, pulse lengths, and pulse shapes.  Collisions with
charged projectiles \cite{Ish13} are also used as sources of short electromagnetic
pulses. However, they often require access to dedicated
  facilities.

Radiation damage is the other side of irradiation studies and it is of
high current interest, for example in connection with biological
tissues ("human-controlled" as in a medical context or "natural" when
referring to earth or space radiations) \cite{Gar12}. There are also other
interesting domains of application. A typical example is the case of
the irradiation of materials (especially insulators) with applications
to nuclear waste management. The field is rather unexplored from the
microscopic dynamical point of view and any possibility of treating
irradiation scenarios on large systems would be here of invaluable
help \cite{Rei16}.  In both above examples, though, the lack of
understanding of microscopic mechanisms calls for dedicated studies on
prototype, finite systems. Let us cite as an example the detailed
studies of irradiation of molecules of biological interest coated by
a finite and well known number of water molecules \cite{Liu06}. The
study of the irradiation of finite molecular systems and clusters is
thus not only of interest for basic science but also for a wide
  range of practical applications.

In all cases, the immediate electronic response of the irradiated
system plays a key role as the doorway to all subsequent dynamical
scenarios. A basic feature is here the optical response corresponding
to electronic oscillations \cite{Mie08,Kre93}. It delivers
 a first overview of the coupling
between irradiation and matter in a large variety of dynamical
situations, from gentle to strong perturbations 
\cite{Bra93,Hee93,Rei03a,Saa06aR}.
Optical response related to photo-absorption is the leading
  signal in the case of gentle perturbations. It has been explored in
  great detail for a large variety of electronic systems, from bulk
down to atoms. For the case of stronger perturbations, further
response channels, especially ionization, become highly
relevant \cite{Cal00,Rei03a,Fen10}. Still, the optical response spectrum, which characterizes the
structural coupling of the system to light, provides a highly valuable
information on any ensuing response mechanism, especially on
ionization pattern.  A typical example here is the case of resonant
ionization occurring when the laser frequency comes close to an
eigenfrequency of the system \cite{Cal00}.

Equally important in energetic irradiation processes is electron
transport, particularly electron emission. As typical examples, one
can cite the many studies on irradiation of clusters by short and
intense laser pulses \cite{Fen10}, providing invaluable information
especially through energy (Photo-Electron Spectra, PES \cite{Cam00})
and, more recently, angle-resolved \cite{Bar09} distributions of
emitted electrons (Photoelectron Angular Distributions, PAD).  Electron
  emission can also change the resonant ionization conditions in the
  course of time evolution which, in turn, influences back again the
optical response, making the whole scenario extremely rich \cite{Fen10}.  
Secondary
electrons in DNA damage \cite{Bou00} also provide a remarkable example
where a microscopic understanding of irradiation damage in biological
systems will only be achieved when including such complex non-linear
electronic effects. A deeper understanding of the underlying
  mechanisms is highly desirable, as this example is of great practical
  interest, especially in relation to oncology \cite{Gar12}.

The analysis and understanding of electronic emission from a finite
system is thus a key issue in a wide range of physical, chemical and
biological processes. Electrons are usually the first constituents to
respond to an electromagnetic pulse. Strong excitations lead to
immediate ionization of the system, often with dramatic long-time
effects as, e.g., dissociation or 
Coulomb explosion \cite{Saa06aR}. It implies
electronic transport and possible indirect effects on neighboring
species. A typical example of indirect effects is provided by
Dissociative Electron Attachment (DEA) 
in biological systems \cite{Gar12} where
electrons emitted somewhere else are attached to a target
biological molecule which, in turn, leads to the break up of the
latter. Emitted electrons may also provide valuable insight into
reaction pathways when properly tracked. Typical examples are
here PES and PAD. Moreover, Time-Resolved (TR) PES and PAD have been
recorded in molecules \cite{Sei02} and more recently in clusters, 
{see e.g.~\cite{Roh10}.}
Electrons are thus leading players at all
stages of an excitation of a system subject to an external
electromagnetic perturbation (i.e. an irradiation). They are the
  first to respond at short time scales and distribute then the
excitation more or less quickly to other degrees of freedom. They are
finally useful probes along the whole dynamical process, especially
when emitted from the system and properly recorded.

Analyzing the characteristics of emission properties of clusters and
molecules is thus at the core of the understanding of irradiation
processes. The numerous new experimental developments in analysis of
electronic emission (PES, PAD) now allow an ever improving detailed
access to electron dynamics in irradiated species. In turn, a
theoretical description of these highly involved dynamical scenarios
calls for dedicated modeling. It is the aim of this paper to provide
an overview of the theoretical description of observables from
electron emission on the basis of the well established theoretical
framework of Time-Dependent Density
 Functional Theory (TDDFT) \cite{Mar12}.  This
will be done with a view on applications, as far as possible in direct
relation to ongoing experiments. Before going into the details, we
will in this introductory section briefly remind the reader the
typical systems (and associated scales) that we aim at describing. It
is also of relevance to address here basic properties of laser pulses
as presently accessible experimentally.

\subsection{On the typical systems considered in this paper}

In order to provide a basis for the forthcoming discussions, we shortly present here
a few typical systems we shall consider in the
following. This will be the occasion to remind typical scales
associated to these systems, especially in terms of times and
energies.

\begin{figure}[htbp]
\centerline{\includegraphics[width=0.92\linewidth,angle=0]{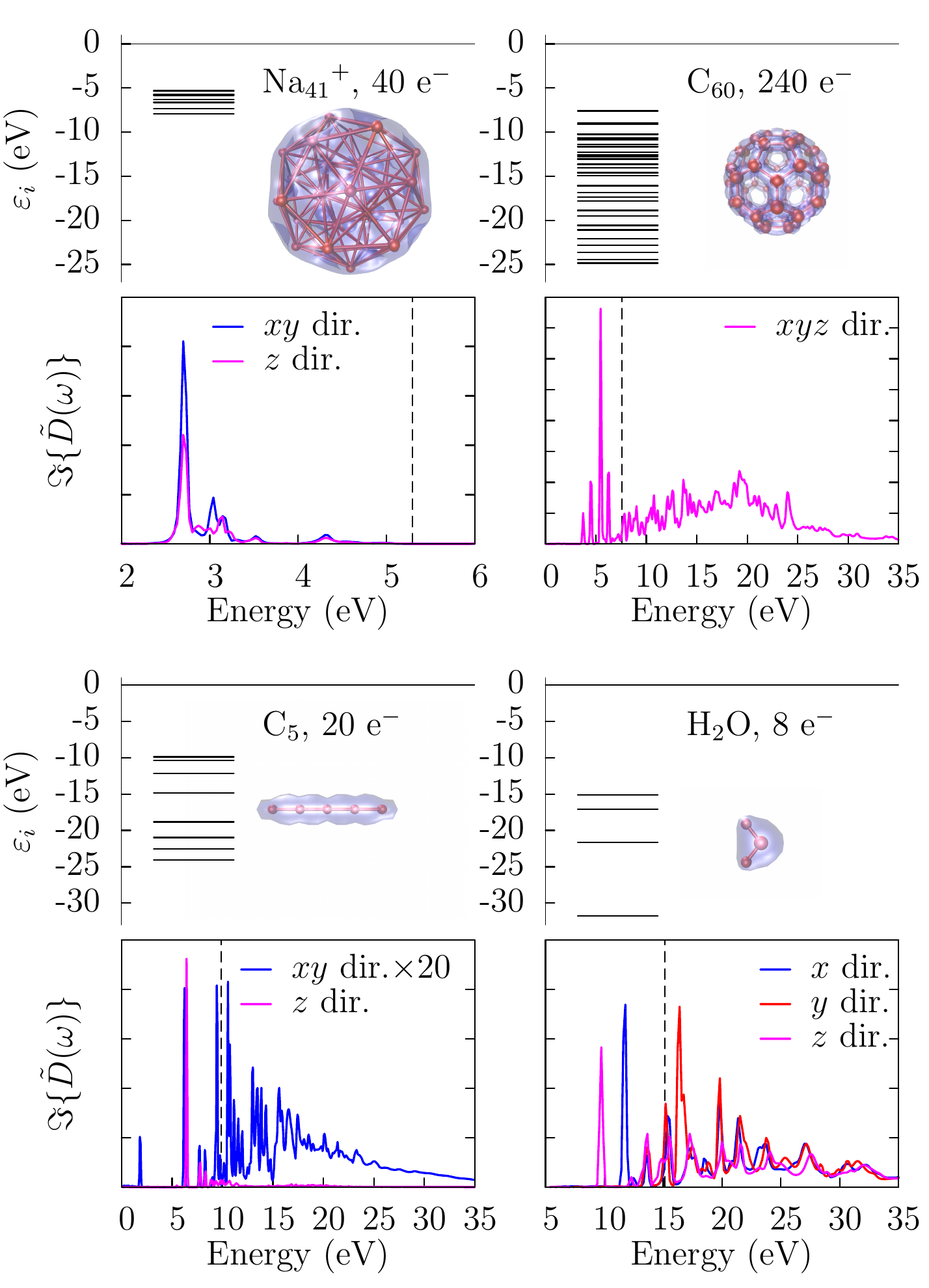}}
\caption{Four typical examples of molecules and clusters explored
  theoretically in this review, namely the metal cluster ${{\rm
      Na}_{41}}^+$ (upper left), the Buckminster fullerene
  C$_{60}$ (upper right), the carbon chain C$_5$ (lower left), and the 
covalent molecule H$_2$O (lower right). For each panel: 
Top row~: ionic structures (all plotted at the
  same scale) and single particle energies of the valence electrons
  whose number is indicated; Bottom row~: corresponding optical
  response. For C$_5$, the transverse spectrum is multiplied by 20 to
  ease the comparison with the longitudinal response. The vertical
  dashes indicate the position of the ionization potential in each
  case.
\label{fig:systems}}
\end{figure}

Fig.~\ref{fig:systems} provides four examples of systems computed
with the tools described in Secs.~\ref{sec:theo}, \ref{sec:sic_stat} and \ref{sec:sic_optresp}. 
They cover several different binding types and properties.
The figure shows single-particle (s.p.) energies, optical response and
ionic structure.  The
four presented systems are ${{\rm Na}_{41}}^+$ as an example of a simple
metal cluster, C$_{60}$ for its outstanding properties and many
applications, H$_2$O as a prototype of a covalent molecule, and C$_5$
as a simple carbon chain, which displays interesting optical
properties. Let us analyze each system separately to
extract typical properties.

We start with the ${{\rm Na}_{41}}^+$ cluster (upper left block in
Fig.~\ref{fig:systems}) which is a medium size alkaline cluster. It
contains 40 valence electrons forming an electronic shell
closure. This leads to a particularly abundant/stable species. The
s.p.  energies span an energy range of order 2.6~eV and the Ionization
Potential (IP) is of order 5.3~eV. Such values are typical of
alkaline clusters. The optical response displays a pronounced
collective, essentially single, peak around 2.6~eV. This is called
  the Mie surface plasmon and it is a typical mode for simple metal
clusters. In larger clusters, the density of s.p. states grows, which
leads to more Landau fragmentation and  somewhat broadens the plasmon
peak. The Mie plasmon frequency is related to a typical time scale of
order 1.5 fs, again a characteristic time scale for simple
metals. Ionic time scales (not shown in Fig.~\ref{fig:systems})
  are more sensitive to the actual material due to the largely
  differing atomic weight.  In Na clusters, vibrational modes
typically lie in the 10 meV range and are associated to ionic motion
in the 100 fs range.

The second example is C$_{60}$ (upper right block) with 240 valence electrons (4 per C atom). The
s.p. energies now ranges in a span of about 17.3~eV, much wider than
in Na. The IP is 7.6~eV. The deeper binding and broader span of
energies is typical of carbon, and more generally of organic systems
with a covalent binding. Due to the high symmetry close to sphericity,
the optical response exhibits the same behavior in all spatial
directions. It has, however, a more complex structure than in ${{\rm
    Na}_{41}}^+$. One can identify two prominent features, a strong
resonance peak just below the IP and a much broadened peak centered
around 20 eV. The latter part of the optical response lies well above
the IP, whence its highly fragmented structure. It is considered to
  represent the Mie surface plasmon in C$_{60}$.  The energies are
higher and thus the associated time scales much smaller than in Na,
typically well sub-fs.  Ionic vibration energies typically  lie in the 
40-200 meV  range with associated time scales of order 20-100 fs.

The case of the small carbon chain C$_5$ (lower left block) is 
complementing C$_{60}$ in the sense that it has the
  same binding type, but a different geometry and thus different
  optical response. The s.p. energies span of the 20 valence
electrons is of order 14 eV, and the IP of order 9.9~eV. These values
are of the same order of magnitude for larger chains.  According to
the linear geometry of the chain, the optical response shows a
dominant resonance peak along the longitudinal direction at a frequency
of 6.4~eV. 
The transverse modes are suppressed by at least one order
of magnitude (mind that transverse strengths have been multiplied by a
factor of 20 to allow a better graphical comparison with the
longitudinal modes) and are significantly fragmented. There are three
main peaks~: one at the same energy as the longitudinal plasmon peak,
and two other ones at higher energies near the IP energy. The all
dominant longitudinal mode lies well below the IP, a feature common to
all carbon chains. Associated time scales are now typically ranging
from sub-fs to fs.  Ionic vibration energies lie again in the 0.15~eV
range with associated time scales of order 27.6~fs.

We finally discuss the case of the prototypical water molecule H$_2$O
(lower right block) which has 8 active
valence electrons in our calculations (6 for O and 1 per each H). The
s.p. energy span is now with 16~eV even larger than in
C$_{60}$, in spite of the much smaller number of electrons. The IP is
of order 15.1 eV. Such large IP's are typical of covalent systems of 
small to moderate size. The optical response, as well, is typical of covalent
molecules with its highly fragmented structure above the IP, and some
isolated low energy peaks below the IP.  Associated time scales lie well
below fs. Ionic vibrations are more energetic than in other systems
because of the especially light H species and the strong covalent
binding between H and O. The O-H ionic vibration energy is about
0.5~eV with associated period of 8.3~fs.

All in all, the four above examples point out the diversity and richness of
the various systems nowadays accessible to both experimental and
theoretical investigations. The various cases also show that the range of
energy and time scales to be investigated is rather large from
attosecond to several fs for electrons, and from fs to ps for ions. In
addition, the optical spectra exhibit different pattern. Specific 
for metals is the especially well marked Mie surface plasmon
with simple scaling properties with size \cite{Bra93}.  The
case of covalent systems is more involved with basically no simple scaling
properties, but nevertheless some generic trends. Optical spectra
are generally much more fragmented below and even more so above IP.
Pure carbon systems contain besides covalent binding a fraction
of metallic binding which produces also plasmon structures amongst the
highly fragmented spectrum.

Optical response is the key to understanding the coupling of the
system to laser light, at least in the frequency-dominated regime (see
Secs.~\ref{sec:lasers} and \ref{sec:mecha}). This will constitute a
mostly used tool of investigation of dynamical scenarios in the
following discussions.  Before introducing the actual observables
which can be attained that way, we will briefly discuss present days
capabilities of lasers and the description of the electromagnetic fields
they deliver. This aspect is addressed in the following
Sec.~\ref{sec:lasers}.

\subsection{On excitation mechanisms}
\label{sec:lasers}

Cluster dynamics requires excitation of the cluster formerly
  resting in its ground state. In this paper, we will exclusively addess
   excitation by electromagnetic fields, predominantly by
  laser pulses and in a few cases by short pulses from collisions with
  highly charged ions. The corresponding excitation mechanisms are
  shortly explained in this section. Thereby, we focus on laser
  properties and finally address ion collisions in a short paragrph.

\subsubsection{Laser pulse characteristics}
\label{sec:las_pulse}

Laser science has experienced impressive progress during the last few
decades \cite{Pas08}. The versatility of laser pulses has increased remarkably, thus
allowing one to shape a wide range of dynamical scenarios in the course of
irradiation processes. We briefly remind here key quantities of the
laser pulses we are going to use in the following. Throughout this
paper, we shall work in the dipole approximation which requires that
the irradiated system is much smaller than the laser wavelength
$\lambda= 2\pi c/\omega_\mathrm{las}$. In practice, the dipole
approximation is well justified in the optical domain ($\lambda \sim
\mu$m) for systems of nm size. It may become questionable for XUV
photons and very large clusters in which field variations {\it inside}
the system itself should be accounted for. But we shall not consider
such cases here. In the non-relativistic regime, linearly polarized
laser pulses acting on atoms, molecules or clusters can then be
described as a homogeneous time-dependent electric field of the form
\begin{equation}
  \mathbf{E}(t) 
  = 
  \mathbf{e}_\mathrm{pol} \, E_0 \, f(t) \, \sin(\omega_\mathrm{las}t 
  + 
  \varphi) 
  \quad.
\label{eq:Elaser}
\end{equation}
In this expression, $\mathbf{e}_\mathrm{pol}$ denotes the (linear)
polarization, $E_0$ is the peak field strength, $\omega_\mathrm{las}$
is the photon frequency, and $\varphi(t)$ is some phase shift, usually
assumed to be zero. Finally $f(t)$ is the pulse envelop. The
  peak laser intensity is $I_0 = c \, \varepsilon_0\, E_0^2/2$ ($c$
being velocity of light in vacuum) usually expressed in W/cm$^2$.
The net yield in a laser pulse is often characterized by the
  fluence $\mathcal{F}=\int \textrm dt I(t)\approx I_0T_\mathrm{FWHM}$, 
{where the latter time
$T_{\rm FWHM}$ stands for the Full Width at Half Maximum of the pulse}. This
  allows one to compare the energy impact of laser pulses with different
  durations.
  
For the sake of simplicity, we keep in the present discussion a fixed
value of $\omega_\mathrm{las}$ but the latter quantity can also be
made time-dependent ("chirped") which can induce interesting effects
\cite{Pas08}. One may also render the phase $\varphi$
  time-dependent, which could produce interesting phenomena.  We shall
  not discuss these aspects here. The laser polarization is usually
taken linear but there also exists experiments/calculations using
circularly polarized light \cite{Pas08}. Again, we shall not
discuss much such cases in the following and thus recur to a linear
polarization for the present discussion.

The laser pulse envelop can be varied in a large range. Most flexible,
and most widely used, are optical lasers with pulse lengths from
nano-seconds down to atto-seconds \cite{Bra00aR,Kra09}.  Free Electron
Lasers (FEL) \cite{Fel05,Pas08} are yet on their way to comparable flexibility, with
present pulse lengths down to 20 fs. It should also be noted that the
actual shape of $f(t)$ is not exactly known experimentally.  In many
situations, the actual pulse of interest is built upon a (hopefully
harmless) background of a long, low intensity, pulse. Moreover,
  the peak intensity has a spatial variation decreasing towards the
  edges of the pulse.  This has to be kept in mind when assigning the
observed signal to the laser pulse characteristics.  Ignoring
background, experimental short laser pulses have a pulse profile of
Gaussian type.  The theoretical situation is simpler as the pulse
profile can be exactly specified.  The Gaussian profile is
theoretically not welcome since it never fully vanishes and requires
unnecessarily long computation times to cover the pulse sufficiently
well.  Therefore, we mostly use for computations a $\sin^2$
pulse~:
\begin{equation}
  f(t) 
  = 
  \sin^2 \left(\frac{t\pi}{T_\mathrm{pulse}} \right) 
  \, \theta(t) \, \theta(T_\mathrm{pulse}-t) 
  \quad. 
\label{eq:cos2}
\end{equation}
where $\theta$ stands here for the Heaviside function.  This
  pulse is limited to a finite time interval
  $t\in[0,T_\mathrm{pulse}]$, but soft enough to deliver a clean
  frequency spectrum.  It can be simply characterized by its FWHM which
is in this case $T_\mathrm{FWHM}=T_\mathrm{pulse}/2$.  
Note that the FWHM of the intensity $I(t)$, which is proportional to the 
square of the field $\mathbf E(t)$, is rather $T_{\rm pulse}/3$. 
The pulse maximum occurs at $t=T_\mathrm{pulse}/2$.  Note that the sin$^2$
profile is written here for the laser field amplitude, which means that
the time profile of the intensity time has a sin$^4$ shape.
Thus far, we have discussed simple one-peak pulses.  More
  flexibility is conceivable. The next important tool are dual pulses
as used in pump-and-probe experiments in which the laser irradiation
is performed in two steps. We shall illustrate such cases at several
places below.

In practice, the effect of the laser field will be accounted for in
our calculations as an external potential $U_{\rm ext}(\mathbf{r},t)$
which delivers a time-dependent perturbation. In the long wavelength
limit, the electric field is homogeneous and delivers the
potential~:
\begin{equation}
  U_{\rm ext}(\mathbf{r},t)
  =
  -e\, E_0\, f(t)\, \sin(\omega_\mathrm{las}t)\, 
  \mathbf{e}_\mathrm{pol}\cdot\mathbf{r} 
  \quad,
\label{eq:lasfield-x}
\end{equation}
where $f(t)$ is the time profile usually taken according to
Eq.~(\ref{eq:cos2}). This is, in fact, the laser field in space gauge.
Equivalently, one can use the velocity gauge for which the laser
field is described by the interaction operator~:
\begin{equation}
  \hat{U}_{\rm ext}^{(v)}
  =
  -\frac{e}{c} \, E_0 \, F(t)\, \mathbf{e}_\mathrm{pol}\cdot\hat{\mathbf{p}}
  \quad.
\label{eq:lasfield-v}
\end{equation}
The rules of gauge transformation relate time profiles and wave functions by~:
\begin{eqnarray}
  F(t)
  &=&
  \displaystyle \int^t_{-\infty} \mathrm dt'\,f(t') \, \sin(\omega_\mathrm{las}t')
  \quad,
\label{eq:gaugeF}
\\
  \varphi_i^{(v)}(\mathbf{r},t)  
  &=&
  \varphi_i^{\mbox{}}(\mathbf{r},t)  
  \exp{\left[\mathrm{i}\, E_0\, F(t)\, \mathbf{e}_\mathrm{pol}\cdot\mathbf{r}\right]}
  \quad.
\label{eq:gaugepsi}
\end{eqnarray}
Both gauges are fully equivalent. Which one is to be preferred 
is a matter of the actual
numerical scheme. Most observables are
not even sensitive to gauge. An exception is the evaluation of
photoelectron spectra where the phase of the wave function plays a
role. In this case, one has to consider gauges carefully. This will be
addressed in more detail in Sec.~\ref{sec:observ_theo}.

\subsubsection{Varying laser characteristics}
\label{sec:las_charac}

\begin{figure}[htbp]
\centerline{\includegraphics[width=0.9\linewidth]{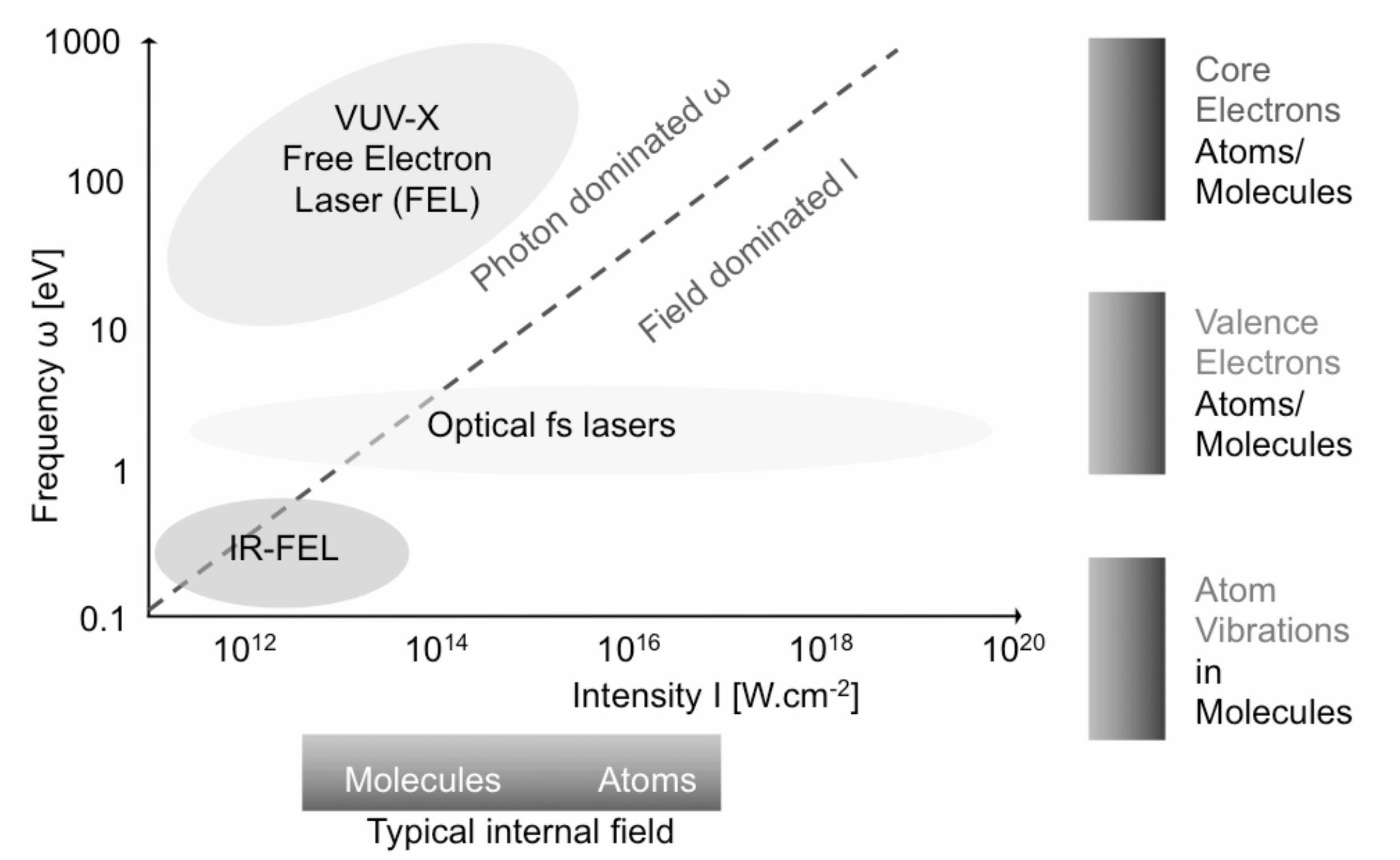}}
\caption{Schematic representation of various dynamical regimes as a
  function of laser intensity $I$ and photon frequency $\omega$. The
  dashed diagonal line represents frequency-intensity combinations
  with constant Keldysh parameter $\gamma=1$, see text for
  details. This line characterizes the transition from
  photon-dominated to field-dominated regime for an assumed IP of a
  few eV. The blocks to the right side indicate typical frequency
  ranges as labeled. The block below the plot indicates typical atomic
  field strengths related to  given laser intensities.
\label{fig:laser-regimes}
}
\end{figure}

As pointed out above, all laser parameters can be tuned in rather
large ranges. The point is illustrated in Fig.~\ref{fig:laser-regimes}
which displays typical regions of interest in the intensity-frequency
plane.
One can notice the enormous large intensity range of optical
lasers. But the range of available conditions is also dramatically
extended by FEL, which exist for photons in the IR, VUV and X-ray
regime. Also indicated outside the axes are corresponding regions of
relevance in atoms and molecules in terms of energy/frequency and
field/intensity. The lowest frequencies in the deep IR are associated
with molecular vibrations, while the range around visible light
belongs to the dynamics of valence electrons and core electrons move
at much higher frequencies in X-ray regime. The gray box below the
plot indicates typical atomic and molecular field strengths in terms
of an equivalent laser intensity.

Laser characteristics have to be considered in relation to the
electronic response. This is usually quantified via the ponderomotive
potential $U_p$ and the associated Keldysh parameter $\gamma$. $U_p$
represents the electron kinetic energy (averaged over one photon cycle)
of a freely oscillating electron (pure quiver motion, no drift
velocity) in a laser field. At peak laser intensity, it
reads~:
\begin{eqnarray}
  U_p
  =
  \frac{e^2\, E_0^2}{4\, m_{\rm el}\, \omega_{\rm las}^2} 
  =
  9.33\times10^{-14}{\rm eV}\times\,
  I_0[{\rm W/cm^2}]\,(\lambda_\mathrm{las}[{\rm \mu m}])^2
  \quad ,
\label{eq_U_pond}
\end{eqnarray}
where $\lambda_\mathrm{las}$ is the photon wavelength. The other
aspect concerns the electronic binding in the system, which can be
quantified by the ionization potential (IP) with associated energy
$E_\mathrm{IP}$. What counts is the relation between $E_\mathrm{IP}$
an $U_p$, quantified by the Keldysh parameter \cite{Kel64}~:
\begin{equation}
  \gamma
  =
  \sqrt{\frac{E_\mathrm{IP}}{2U_p}}
  =
  \sqrt{\frac{2\, E_\mathrm{IP}\, \omega_\mathrm{las}^2}{I_0}}
  \quad.
  \label{eq:Keldysh}  
\end{equation}
The value $\gamma = 1$ (see Fig.~\ref{fig:laser-regimes} in the case
$E_\mathrm{IP}=1$ eV) separates two regimes.  For $\gamma\ll 1$,
direct ionization (over barrier or tunneling) prevails. This regime is
dominated by laser intensity and not so much by laser frequency
(field-dominated regime). For $\gamma\gg 1$, emission proceeds through
multi-photon ionization in a regime of weak perturbations. There, the
results sensitively depend on laser frequency (photon- or frequency-dominated regime).

\subsubsection{Not on lasers: collisions with fast ions}
\label{sec:ioncoll}

There is an alternative excitation mechanism by collisions with charged
projectiles.  We shall also marginally consider a few examples of
collisions with fast ions and thus comment briefly about this tool
here.  Experiments with charged, fast ions often require access to
large scale facilities. Thus there are much less experiments with
irradiation by charged projectiles than by the more easily accessible
and versatile lasers. Although collisions with charged particles also
provide a strong electromagnetic perturbation (often in form of a
short pulse as soon as the projectile velocity is large enough), the
characteristics of the perturbing field are significantly different
from those delivered by a laser pulse. While lasers provide (up to
details) an electromagnetic field with a well defined frequency band
(basically the laser frequency), collisions with charged projectiles
deliver a perturbation covering a very broad band of frequencies, the
broader the shorter the pulse. This delivers useful, complementing
information to that attained from lasers. It is important to note that
collisions with charged projectiles also concern a wide range of
potential applications of irradiation dynamics, especially in relation
to radiation damage and applications thereof.  The present review
concentrates on laser excitations. Nevertheless, we shall discuss a few
cases with high energy projectiles. For them, the delivered
electromagnetic perturbation can be modeled as an instantaneous boost
($\propto \delta(t)$) at the initial time of the simulation. This is
the way we shall treat this case in the following (see in particular
Sec.~\ref{sec:projectile}).


\section{From integrated to detailed observables}
\label{sec:observ}
%
%

Electronic emission can be analyzed at various levels of sophistication, starting from
fully integrated quantities (total ionization) down to energy-resolved (Photo-Electron
Spectra, PES) and angle-resolved (Photo-Angular Distribution, PAD) quantities. Time is
also a key quantity as ionization signals can be followed in time, leading to
Time-Resolved (TR) results. We briefly describe in this section the various types of
observables experimentally accessible, starting from the simplest one, that is the total
ionization, to the most elaborate ones (TR-PES and PAD). In terms of cross sections, this
means that we go from integrated ones to single-differential and even
double-differential ones, all possibly time-resolved. Before discussing these various
observables, we briefly introduce key mechanisms of ionization, again focusing the
discussion on laser induced ionization.

\subsection{Ionization mechanisms}
\label{sec:mecha}

\begin{figure}[htbp]
\centering\resizebox{1.\columnwidth}{!}{\includegraphics{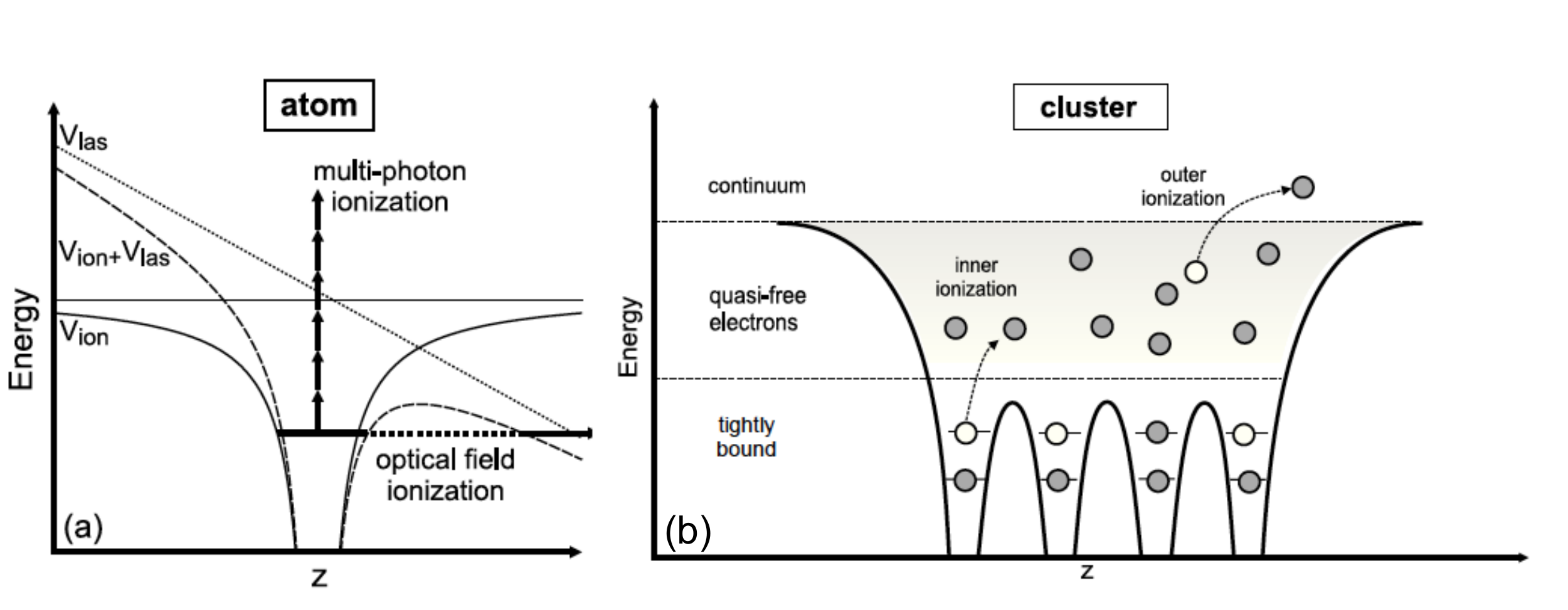}}
\caption{\label{fig:ioni_mecha} Schematic view of ionization mechanisms in atoms and
clusters. Occupied electron states are indicated by horizontal bars. (a)~: multiphoton and optical field
ionization in an atom, for which are drawn the potentials of the unperturbed ion $V_{\rm ion}$ (full line), of the
laser $V_{\rm las}$ (dots), and of the sum of both (dashes).
(b)~: inner and outer ionizations in a cluster, with the effective electron potential (without laser)
 shown as a solid line. Adapted from \cite{Fen10}.}
\end{figure}

Basic ionization mechanisms are illustrated in Fig.~\ref{fig:ioni_mecha}.
We start from the simplest case of an atom (left panel in Fig.~\ref{fig:ioni_mecha}) to
introduce two basic ionization mechanisms. The first one corresponds to a vertical
excitation of a bound electron by absorption of one or several ($\nu$) photons
(Multi-Photon Ionization or MPI). This mechanisms may spread over several laser cycles and
prevails in weak and moderate fields, usually quoted perturbative regime. It is associated
to large values of the Keldysh parameter ($\gamma \gg 1$). MPI can promote electrons far
above threshold into the continuum and then, it also stands for Above Threshold Ionization
(ATI). It is a typical mechanism underlying PES and PAD measurements in the perturbative
regime (see Sec.~\ref{sec:pes_pad}), providing mostly structural information. The second
mechanism illustrated in the case of atoms is known as Optical Field Ionization (OFI) in
which the laser acts as a quasi stationary field. For sufficiently large fields, bound
electron can tunnel through the barrier, which means that both barrier height and width
(thus tunnel characteristic time) allow ionization. This typically corresponds to moderate
values of the Keldysh parameter ($\gamma \lesssim 1$). The limiting case corresponds to
full barrier suppression which can be associated to a critical laser intensity in atoms
and which reasonably matches ion appearance intensities in atomic gases \cite{Aug89}.

The cases of molecules and clusters mix the above considerations with structural
properties of the considered systems. For example, ionization barriers are influenced by
neighbouring ions. A typical example is the case of strong field ionization of diatomic
molecules \cite{Sei95,Zuo95} in which an appropriate internuclear separation leads to
lowering or suppression of inner and outer potential barriers, thus leading to enhanced
ionization. The effect was also studied in small clusters \cite{Ven01,Sie02}. In the case
of large clusters, one should also mention the separation between inner and outer
ionization~\cite{Las99} (see right panel in Fig.~\ref{fig:ioni_mecha}), especially
important in the case of strong fields. Inner ionization leads to the formation of a set
of quasi free electrons constituting sort of a metallic phase. A final excitation may
promote them to the continuum for final escape and then will appear as the total ionization of
the system. In most of the cases, we shall discuss in the following we shall not consider
strong enough fields to use this concept further.
On the other hand, we shall deal
with situations where another key ingredient, already mentioned previously, enters the
picture. It concerns the optical response of the irradiated species. Indeed the optical
response provides the eigenfrequencies with which a given system does couple to light. It
is thus crucial to integrate it in the discussion of ionization mechanisms, especially in
the case of metal clusters in which the plasmon plays a leading role.

\begin{figure}[htbp]
\centering{\includegraphics[width=0.65\linewidth]{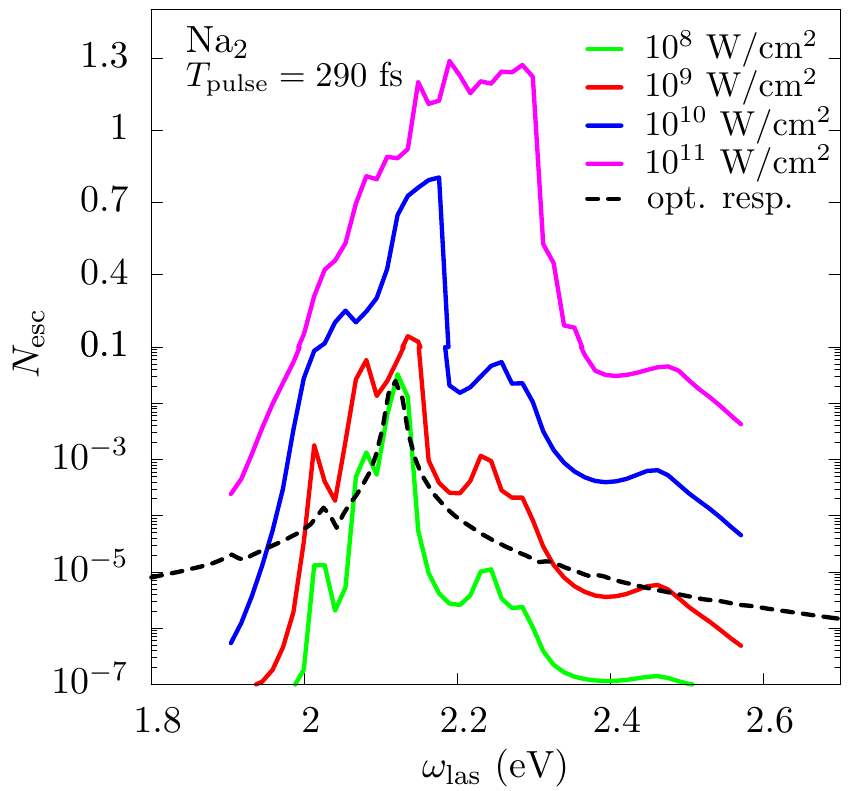}}
\caption{Total ionization $N_{\rm esc}$ of Na$_2$ after irradiation by laser pulses
polarized along the cluster axis, with duration of 290~fs, as a function of laser
frequency $\omega_{\rm las}$, for four different intensities as indicated. Dashed
curve~: optical response (power spectrum) of Na$_2$. Mind that for $N_{\rm esc}<0.1$, 
the vertical axis is in logarithmic scale, while a linear scale is used above 0.1.
\label{fig:ioniz_plasmon} 
}
\end{figure}

The point is illustrated in Fig.~\ref{fig:ioniz_plasmon} in the case of a small size
sodium cluster Na$_2$ in which the notion of collective plasmon is hard to disentangle
from that of a molecular dipole transition. Na$_2$ possesses two valence electrons.
The mechanism actually remains the same and is thus illustrative of the role of the
optical response. We shall consider plasmon effects on some examples later on (see in
particular Sec.~\ref{sec:basic_ioniz} and Fig.~\ref{fig:nesc_om}). The dashed curve shows
the optical response of the system with a well identified peak at 2.12~eV. The full curves
display the total ionization as a function of laser frequency for a set various laser
intensities between $10^8$ and 10$^{11}$ W/cm$^2$. 
One clearly observes that the ionization signal directly follows the optical
response~: attaching a resonance peak leads to enhanced ionization. The effect is
especially visible at low intensity and vanishes with increasing intensities. We gradually
leave the photon-dominated regime (low intensity) to reach the field-dominated one. In
terms of the Keldysh parameter $\gamma$, it decreases. In the present test case, $\gamma$
takes values typically between 60 and 250 in the resonance region at low intensity, and
reaches values between 2 and 10 in the high intensity case. The role of resonance peaks is
thus crucial here and it should be noted that it does not reduce to the linear regime of
excitation. The total ionization may reach rather large values (more than half of the
available valence electrons) with increasing laser intensity, and still, the resonance
enhancement remains very clear. This indicates that it will have to be considered whatever
the dynamical regime in the following, especially in the case of metals. Although the
basic enhancement mechanisms remain similar in non-metallic systems (see
Sec.~\ref{sec:basic_ioniz}), resonances are usually less collective and more narrow so
that their impact is somewhat different. Still, in many systems such as for example
C$_{60}$, one observes a wide bunch of resonances above continuum threshold which very
clearly play a key role in the dynamics.

\subsection{Total ionization}
\label{sec:nesc}

Total ionization is the simplest ionization signal one can measure. Still, it already
brings interesting information, although not highly detailed, on irradiation mechanisms,
as we just discussed in the previous section. We here illustrate the point on two examples
taken from rather original scenarios. The first case results from an irradiation with
extremely large frequencies obtained from a FEL, and the second one
directly addresses the dynamical evolution of the system in a time-resolved experiment.

\begin{figure}[htbp]
\centering{\includegraphics[width=0.5\linewidth]{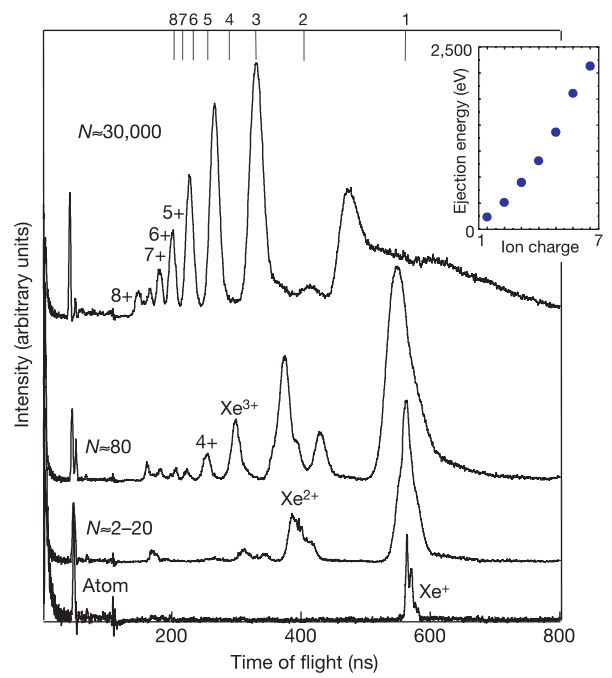}}
\caption{
Time-of-flight mass spectra of ionization products of Xe atoms (bottom curve) and clusters
of various sizes $N$ as indicated. Irradiation was performed by a free-electron laser of
wavelength of 98 nm and an average intensity of $2\times 10^{13}$ W/cm$^2$. The line
splitting of the atomic spectrum (bottom curve) is due to different isotopes. Inset~: ion
kinetic energies as a function of ion charge, in the case of 1500-atom clusters.
From~\cite{Wab02}.
\label{fig:examp_FEL}
}
\end{figure}

Fig.~\ref{fig:examp_FEL} shows time-of-flight (TOF) spectra of Xe clusters irradiated by a
FEL of frequency around 12 eV \cite{Wab02}. 
The TOF gives access to the various charge
states attained after irradiation by a laser of intensity $2\times 10^{13}$ W/cm$^2$ and
pulse length of 100 fs. The striking point of the figure is the differences observed
between the various cluster sizes in terms of attained charge states. While the atomic
gas, under the present laser conditions, only allows to access singly charged cations,
increasing cluster size allows to progressively reach larger and larger charge states,
clearly up to $8+$ in the largest system of about 3,000 atoms. The case very nicely
illustrates the well known difference between energy absorption by single atoms and
clusters, as discussed on many occasions in the past (see for example \cite{Fen10} and
references therein). The mass peaks are rather broad. They are furthermore displaced with
respect to the calculated flight times indicated by thin vertical lines (corresponding to
the different charge states) in the top of the figure. This is an indication that ions
have high kinetic energies. Not surprisingly, one can also note that the higher the charge
state, the higher the ejection energy (see inset in Fig.~\ref{fig:examp_FEL}) and the
larger the above mentioned peak displacements.

Analysis of total electron emission (or alternatively of charge state of ionized clusters)
also gives information on the dynamics of the charging process. One of the most striking
early example of such an analysis can be found in a series of experiments led by the
Rostock group on large size Pb clusters \cite{Koe99,Sch99,Doe00}. These experiments have
shown strong enhancement of cluster ionization for optimal pulse durations. More
specifically, one observed Pb ions with very large state states, much larger than those
attained in an atomic gas. Moreover, the attained charge state $q$ strongly depends on the
pulse duration. The shortest and most intense pulses of duration 150 fs yield ions up to
charge state $q = 20$. When increasing pulse duration, both the maximum
charge state and the signal intensity do grow towards a maximum attained for an optimal
pulse width of 800 fs. Charge states up to $q = 28$ can then be identified. For longer
pulses, both maximum $q$ and signal decrease again. Although other mechanisms can be
envisioned, the efficient charging for a certain pulse duration was in most cases
attributed to resonant heating (plasmon-enhanced ionization)
\cite{Sur00,Saa03,Doe05b,Saa06aR}.

\begin{figure}[htbp]
\centering{\includegraphics[width=0.7\linewidth]{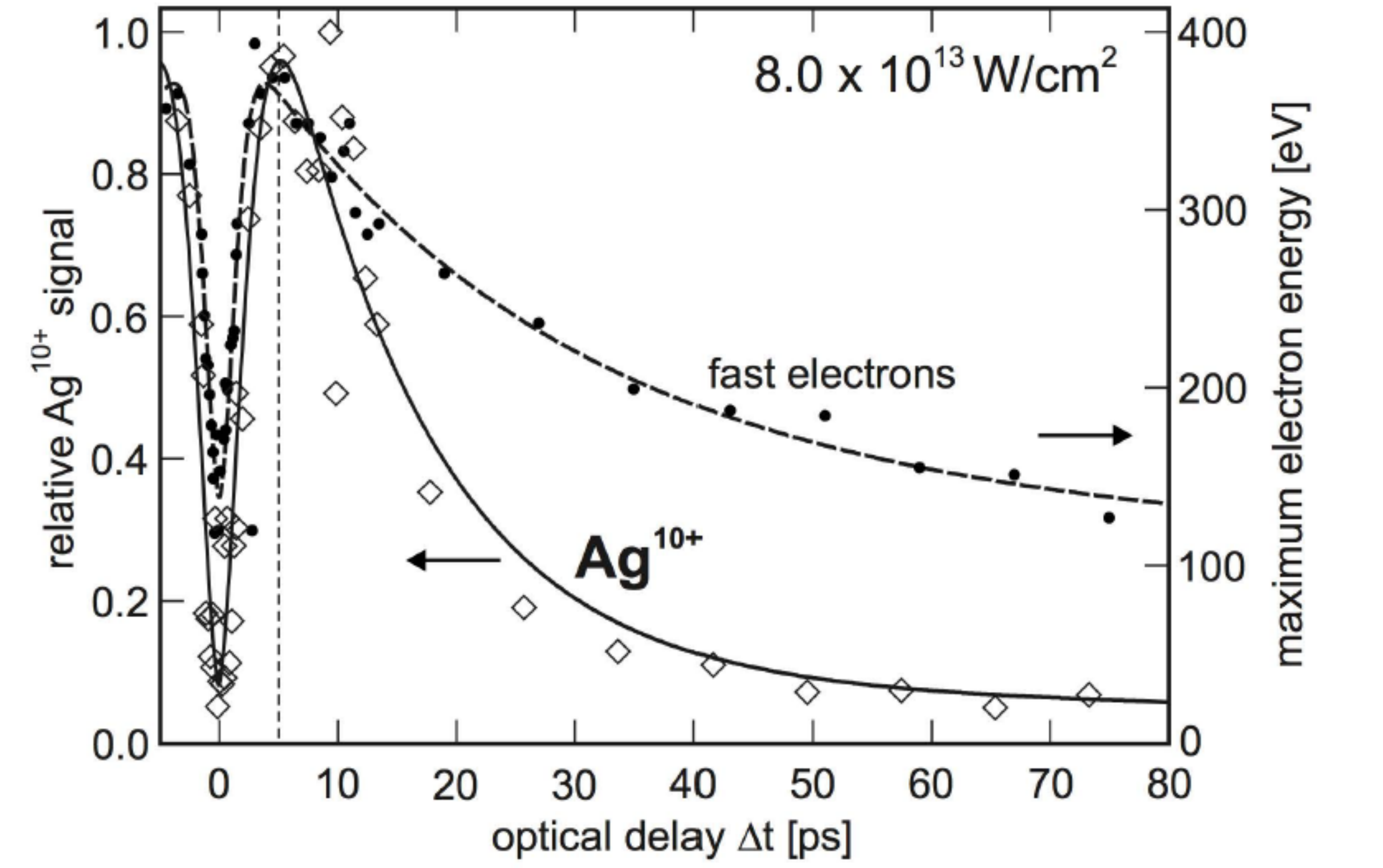}}
\caption{
Ionic charge state Ag$^{10+}$ yield (diamonds, left axis) in relation to the maximum
kinetic energy of the emitted electrons (dots, right axis) following a laser excitation of
Ag clusters of about 20\,000 atoms with dual 100 fs laser pulses of intensity of $8 \times
10^{13}$~W/cm$^2$ and wavelength of 800~nm. From \cite{Doe06}.
\label{fig:pulse_length}
}
\end{figure}

The above case of ionization enhancement was attained with a single laser pulse and only
provides a rather indirect indication on the ionization mechanism. More detailed
investigations were led with dual (or pump-and-probe) pulses, especially in the case of Ag
clusters of about 20\,000 atoms. An example of such experiments is shown in
Fig.~\ref{fig:pulse_length} where one focuses on the yield of Ag$^{10+}$ and the maximum
of the emitted electrons as a function of pulse separation.
One observes a strong variation of ionic signal as a function of pulse separation (delay
between pump and probe) with a clear maximum around 5 ps \cite{Doe06}. Such a behaviour
indicates that cluster activation and enhanced ionization can be clearly disentangled,
which is also found in numerical simulations \cite{Sie05,Doe06,Bor07}. This again provides
an interesting insight the dynamical evolution of the system. There are even clear
indications that a sequence of two pulses might constitute an optimal pulse profile for the
production of very high charge ions \cite{Zam04}, provided a proper tuning of pulse
parameters. And pushing again the argument, one can even envision a route for targeted
control of the cluster dynamics \cite{Doe05b}. Finally a word on the maximal electron
energy shown in Fig.~\ref{fig:pulse_length} is in place. The coincidence of high
ionization yield and maximal electron energy again points out the leading role of
collective excitations, and this in both channels. This is compatible with other
observations \cite{Sha96,Spr03}.

\subsection{Energy- and angular-resolved ionization}
\label{sec:pes_pad}

The next step in the analysis of electronic emission consists in 
characterizing the properties of the emitted electrons in terms of 
kinetic energy and angular distribution. This leads to Photo-Electron Spectra (PES) 
for the energy analysis and Photo-Angular Distribution (PAD) for angular signals. 
The terms come from laser irradiation but the signals themselves can as well 
be recorded in any ionization scenario, for example from collisions with highly 
charged ions (see Sec.~\ref{sec:projectile}). PES and PAD signals turn out 
to provide extremely rich information, both from a structural and from a dynamical 
viewpoint. We briefly discuss their properties in this section and illustrate them on a few
examples covering several dynamical situations. 

\begin{figure}[htbp]
\centerline{
\includegraphics[width=0.8\linewidth]{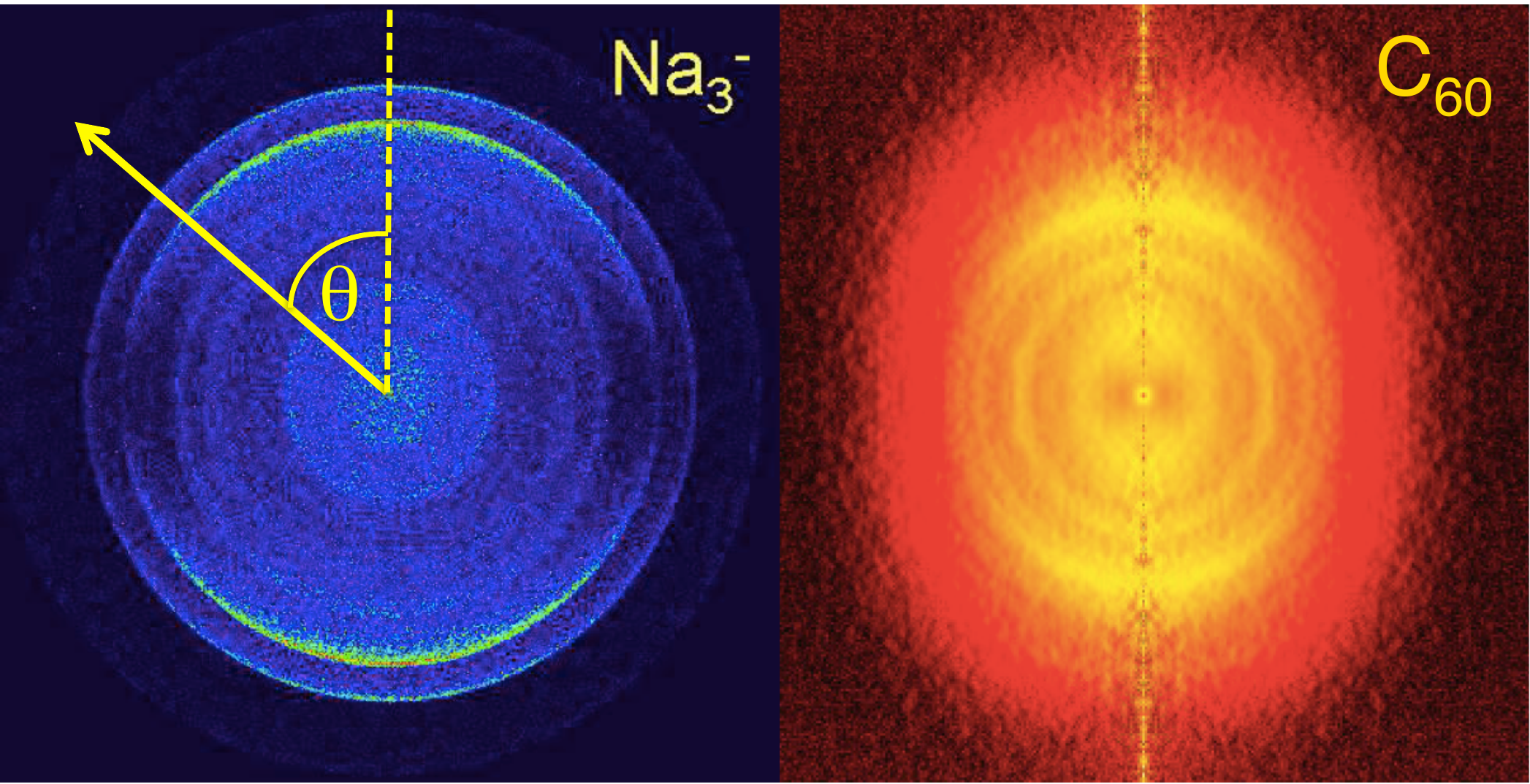}
}
\caption{Left~: Raw velocity map image of ${\mathrm{Na}_3}^-$ irradiated by a laser pulse
of frequency of 4.02~eV and polarized vertically~\cite{Kos07a}. The length of the arrow
stands for the velocity of the photoelectron, and $\theta$ its angle to the laser
polarization axis. Right~: Inverted momentum map image of C$_{60}$ irradiated by a laser
pulse of FWHM of $150\pm 5$~fs, intensity of $1.25\times 10^{13}$~W/cm$^2$, frequency of
3.68~eV, and polarization along the vertical axis~\cite{Kje10}.}
\label{fig:exampl_vmi}
\end{figure}

Kinetic energies of emitted electrons can be measured in several ways. TOF devices provide
here a versatile tool, but this time applied to the electrons themselves (while they are
traditionally used for ions). Because of the well defined mass to charge ratio for an
electron, the arrival delay directly maps the electron kinetic energy, provided a
carefully guiding of the electron flow e.g., by a magnetic mirror \cite{Hee93}. Another
very interesting technique is provided by photo-imaging spectroscopy, also known as
Velocity Map Imaging ({VMI}). This technique is more and more routinely used and provides
a remarkable tool of investigation. It is based on a static electrical field which allows
one to map the distribution of electron velocities onto definite positions on a detection
screen \cite{Bor96}. This is a polar representation of a velocity-resolved (or a momentum
one) and angular-resolved photoelectron spectrum. Two experimental examples are presented
in Fig.~\ref{fig:exampl_vmi}, one in the metal cluster ${\mathrm{Na}_3}^-$ in the
monophoton regime~\cite{Kos07a}, and another one in C$_{60}$ in the multiphoton
regime~\cite{Kje10}.
In these two examples, the vertical direction stands for the laser polarization axis. If
one draws an arrow from the origin of the circle, its length represents the norm of the
velocity (or the momentum), while the angle $\theta$ to the vertical direction is the
angle of the photoelectron with respect to the laser polarization axis, and the lighter
the extremum of the arrow, the higher the yield at this point. The observed circles
correspond to peaks in the PES (not shown here). 
 left panel of Fig.~\ref{fig:exampl_vmi} is a raw
image, while the right one is obtained after some inversion analysis. An approach such as
VMI allows a simultaneous determination of PES together with PAD, which is extremely
interesting. From the thus combined PES/PAD distribution (double differential, energy- and
angle-resolved, cross section) it is then easy to recover PES or PAD separately by proper
energy or angular integration. Still, mostly because of signal intensity, the double
differential cross section can rarely be used as a whole. Therefore, energy or angular
integration usually allow a simpler access to the data. It is also simpler and usually
more quantitative to compare theory to experiments in simpler representations, where
rather than the double differential cross section $\textrm d\sigma/\textrm dE d\Omega$
PES/PAD, one considers singly differential PES $\textrm d\sigma/\textrm dE$ or PAD
$\textrm d\sigma/\textrm d\Omega$ cross sections. We shall thus
explore now in more detail integrated PES and PAD.

\subsubsection{Photoelectron spectroscopy}
\label{sec:PES}

A PES typically results from a multiphoton ionization (MPI) mechanism (see
Sec.~\ref{sec:mecha}). Electrons absorb a certain number of photons to reach the continuum
and be emitted. They can absorb more than the number of photons
required to reach the
continuum threshold, which leads to copies of the signal (although much reduced in
intensity). The kinetic energies of the emitted electrons are then directly related to the
single electron energies $\varepsilon_i$ of the initially occupied electron states $i$
inside the cluster through the simple relation~:
\begin{equation}
  \varepsilon_{\rm kin}
  =
  \varepsilon_i
+
  \nu \hbar\omega_{\rm las} \quad,
\label{eq:ekin1}
\end{equation}
where $\nu$ is the number of photons involved in the process. 
 
\begin{figure}[htbp]
{\includegraphics[width=\linewidth,angle=0]{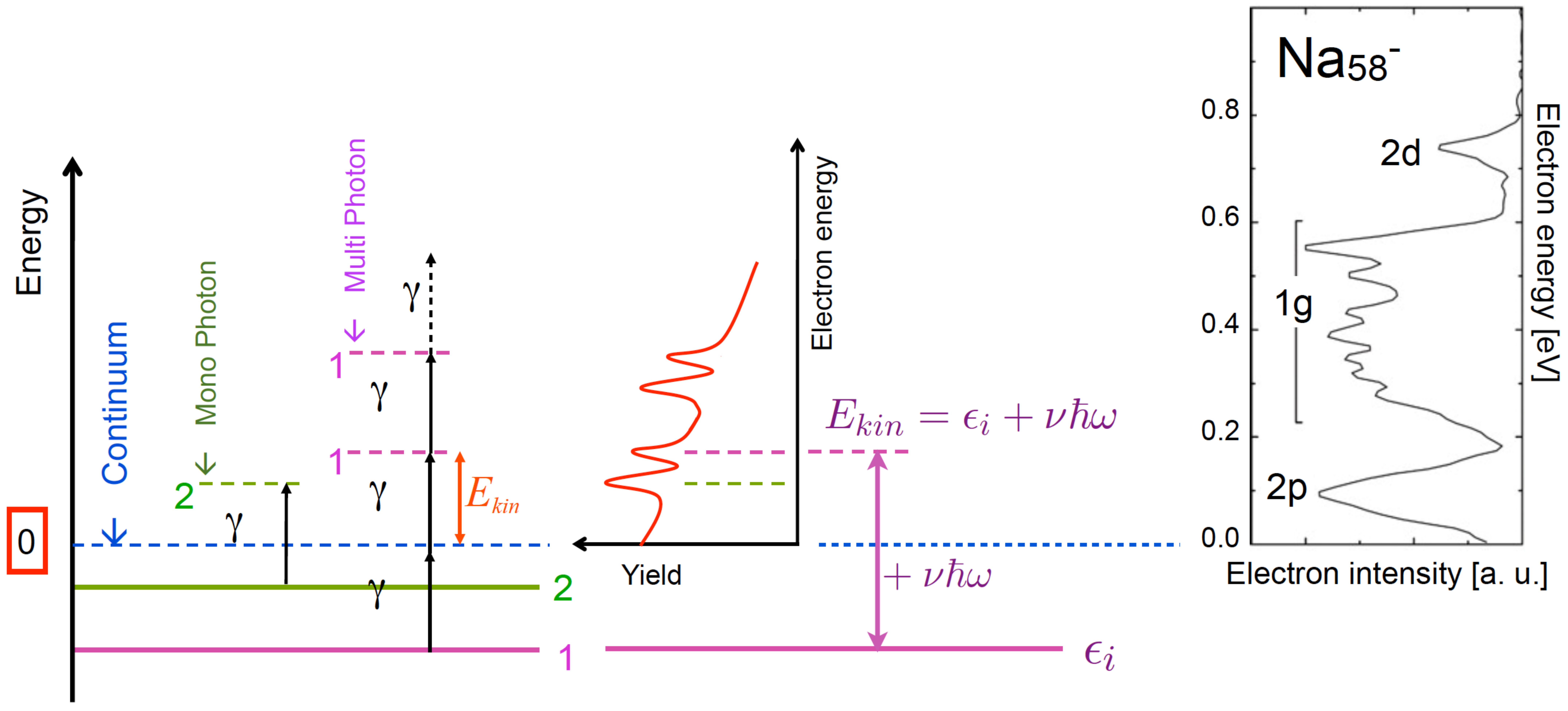}}
\caption{\label{fig:pes_mecha} Left~: schematic view of photoelectron spectroscopy (PES),
including multiphoton ionization scenario for the most bound state. The sample system has
two single electron states, $\epsilon_1<\epsilon_2<0$. The emission threshold is taken as
the reference of zero energy, here the ionization potential IP$=-\epsilon_2$. The measured
kinetic energies of emitted electrons are then recorded from threshold on upwards
involving a varying number $\nu$ of photons which produces successive copies of the single
electron energies, separated by the laser frequency. Right~: experimental example of PES
measurement for the ${{\rm Na}_{58}}^-$ cluster (monophoton regime), obtained 
with photons of energy 4.02 eV~\cite{Kos07a}.}
\end{figure}

Fig.~\ref{fig:pes_mecha} illustrates the principle of PES in terms of a scheme (left
part), both in the case of mono- and multi-photons.
A hypothetical system with two accessible valence states is considered (levels 1 and 2)
whose electrons can reach the continuum via 1 or 2 photon absorption. In the multiphoton
case, the resulting PES displays copies of the original PES, separated by the laser
frequency. The PES is furthermore illustrated on an experimental example (right part) from
${{\rm Na}_{58}}^-$, in the monophoton case. The case of anionic clusters is emblematic of
one-photon PES. Indeed, in such clusters, valence electron states are little bound so that
they can easily be turned to continuum electrons according to Eq.~(\ref{eq:ekin1}) with
one photon in the visible. These measurements basically provide a structural information
on the system. In the present case, the PES exhibits well resolved peaks associated to the
single electron states, as indicated in standard spectroscopic notation. One can note that
the degeneracy of the $1g$ state is split into a series of sub-peaks because of symmetry
breaking of the ionic configuration. Such one-photon measurements on anionic clusters were
thus already performed in the early 1990's \cite{Che90}. More recent measurements nowadays
allow one to access PES for neutral or even cationic clusters, as those from neutral
fullerenes \cite{Cam00,Kje10} and from positively charged metal clusters \cite{Wri02}.

MPI, as already indicated in Eq. (\ref{eq:ekin1}) with $\nu>1$, is also possible thanks to
the high coherence of laser pulses. The impact on PES will be discussed at length in
Secs.~\ref{sec:mono&multi} and \ref{sec:pes_I}. Let us however give a few words here.
For moderate laser intensities, the MPI maps in the PES further copies of the occupied
electron spectrum with increasing kinetic energy, each copy separated by $\hbar
\omega_{\rm las}$. For larger intensities, the regular pattern of copies of the single
electron spectrum is blurred because of large ionization affecting the spectrum itself. At
even higher intensities the signal mostly becomes exponential with basically no structure
left \cite{Fen10,Poh04a}.

\begin{figure}[htbp]
\centering{\includegraphics[width=0.55\linewidth]{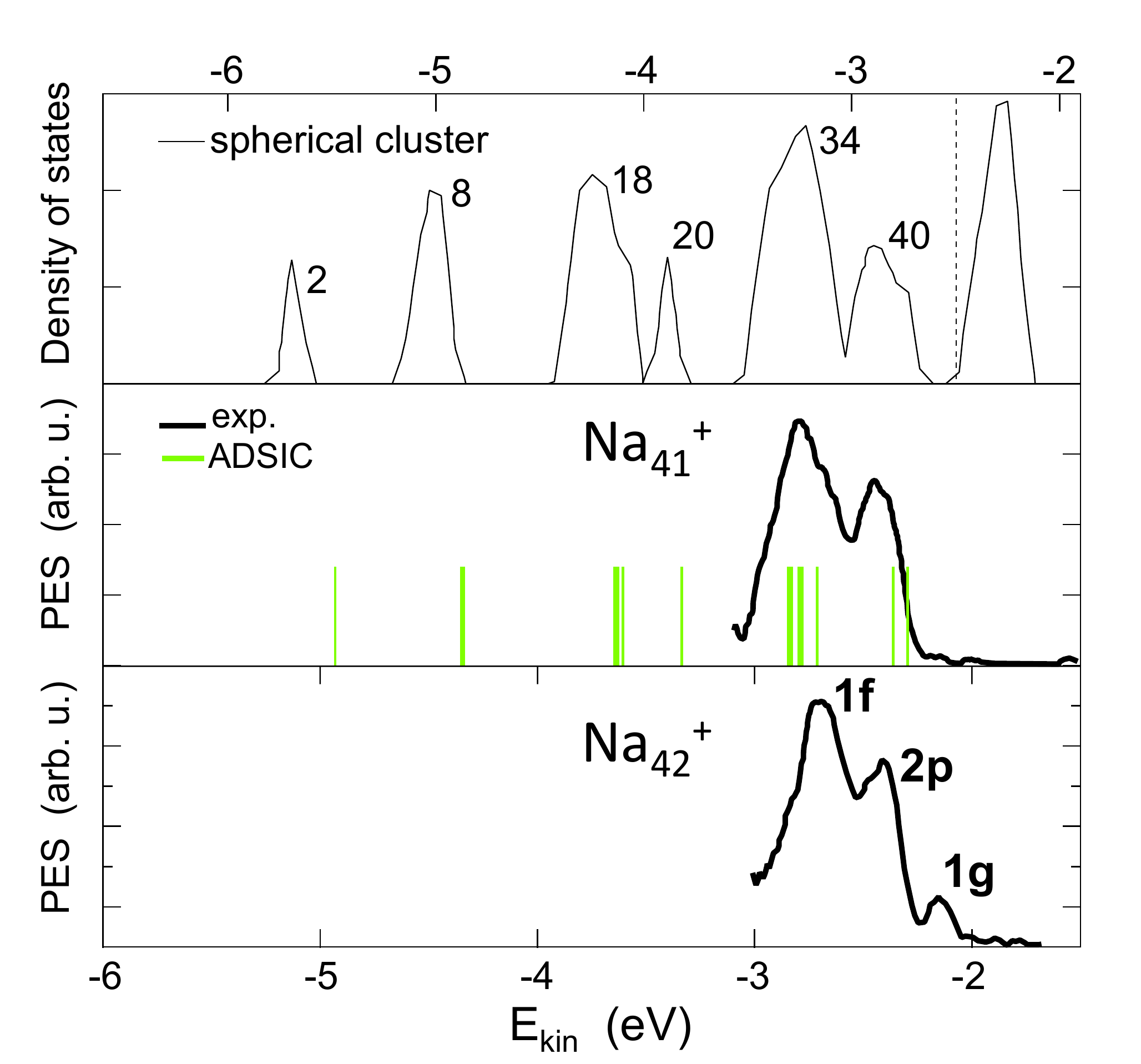}}
\caption{
Lower and middle panels: Experimental photoelectron spectra for ${{\rm Na}_{41}}^+$
(bottom) and ${{\rm Na}_{42}}^+$ (middle) obtained 
by  irradiation  from an ArF$^-$
excimer laser  of frequency 
$\hbar\omega_\mathrm{las}$ = 6.42 eV \cite{Wri02}.
Vertical lines in the middle panel~: static Kohn-Sham single particle energies of ${{\rm
Na}_{41}}^+$ calculated in ADSIC. Upper panel: Kohn-Sham density of states for a spherical
neutral Na$_{40}$ calculated in LDA \cite{Ryt98}. 
\label{fig:issendorf}
}\end{figure}

Fig.~\ref{fig:issendorf} shows a typical example of a PES measurement, in this case performed 
on cationic species.  
The chosen material is sodium in which it is well known that electronic shell closure
leads to especially stable configurations \cite{Bra93}. In turn, the PES is expected to
display the corresponding shell structure. The figure focuses on the region of 40
electrons (which corresponds to a shell closure). For comparison, the expected shell
sequence, as computed in the Clemenger-Nilsson approach \cite{Hee93}, is indicated in the
upper panel. Note that in that case only the two least bound shells altogether containing
20 electrons were measured. The figure exhibits several interesting features. First the
comparison between ${{\rm Na}_{41}}^+$ and ${{\rm Na}_{42}}^+$ (which contains 41 electrons) very clearly
points out the shell closure at 40 electrons with the appearance of one single electron in
the 1g level around 5.2~eV. This also complies with the expected level sequence displayed in the upper
panel. Finally, for the sake of completeness, we have also indicated the results of a DFT
calculation performed with the ADSIC correction (see section \ref{sec:sic_theo} for
details). The agreement obtained without any adjustment is remarkable. It should
nevertheless be noted that in that case the PES mostly provides a structural information
on the system by giving access to the sequence of energies of occupied single electron
levels. As we shall see below, PES, especially in the MPI regime, can also provide
valuable information on the dynamics of the electron cloud.

\begin{figure}[htbp]
\centerline{\includegraphics[width=0.8\linewidth]{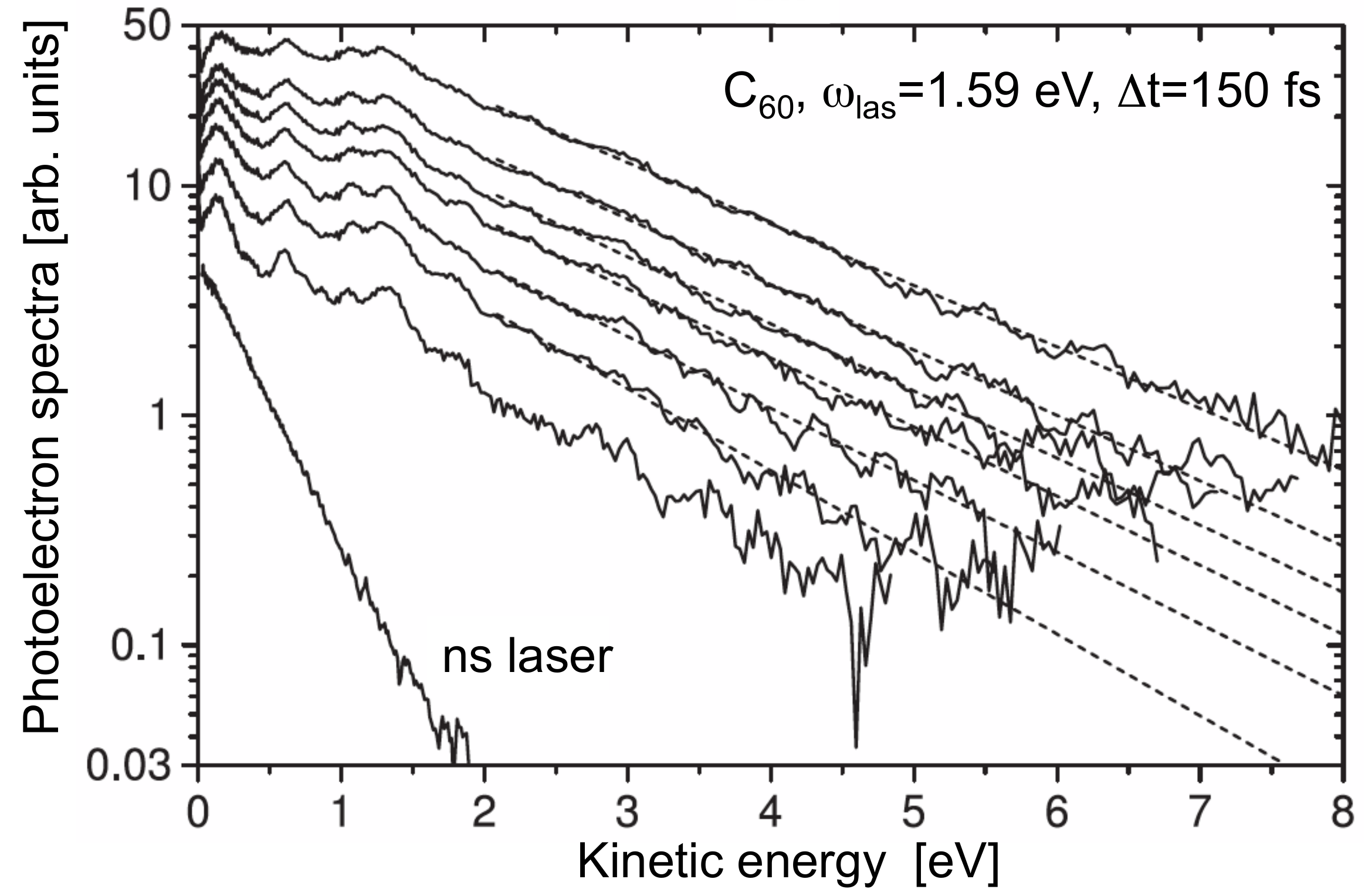}}
\caption{Photoelectron spectra of C$_{60}$ irradiated by a laser of frequency of 1.59 eV, pulse duration of 150 fs,
and various fluences of (from top to bottom) 2.19, 1.84, 1.70, 1.56, 1.42, 1.27, and
1.13 J/cm$^2$. Adapted from~\cite{Kje10}.
\label{fig:campbell1}
}
\end{figure}

As a first example of study of electron dynamics by PES, an example on which we shall come
back later (see section \ref{sec:dissipe}), we consider the case of C$_{60}$ irradiated by
laser pulses of various fluences, but fixed pulse duration of 150 fs \cite{Kje10}, see
Fig.~\ref{fig:campbell1}.
At variance with the spectroscopic character of the PES in Fig.~\ref{fig:issendorf}, the
PES presented here display an almost monotonous exponential shape with little structures
on top. The latter structures are interpreted as signals from single-photon ionization of
Rydberg states \cite{Cam00}. The exponential slope is explained as reflecting thermal
electron emission \cite{Han03a}. In this picture, the energy deposited by the laser is
concerted into thermal electron energy. Concluding on the nature of the energy conversion
on the single basis of the PES is nevertheless a bit questionable as exponential PES are
also naturally obtained by considering higher and higher MPI processes \cite{Poh04a}. On
the other hand, the experiments of \cite{Kje10} also measured the PAD of emitted electrons
and clearly identified a strong isotropic component which might indeed be associated to
thermal emission. The interpretation of \cite{Kje10} is thus certainly to be considered
very seriously. We shall come back on that point in Sec.~\ref{sec:dissipe} when
discussing effects of dissipation on electronic observables in more detail. At present
stage, it is sufficient to conclude that PES clearly opens the door to the analysis of
electron dynamics. And that PAD offers for sure an invaluable complement to such studies
(see next section \ref{sec:PADetal}).

\begin{figure}[htbp]
\centering{
\includegraphics[width=0.8\linewidth]{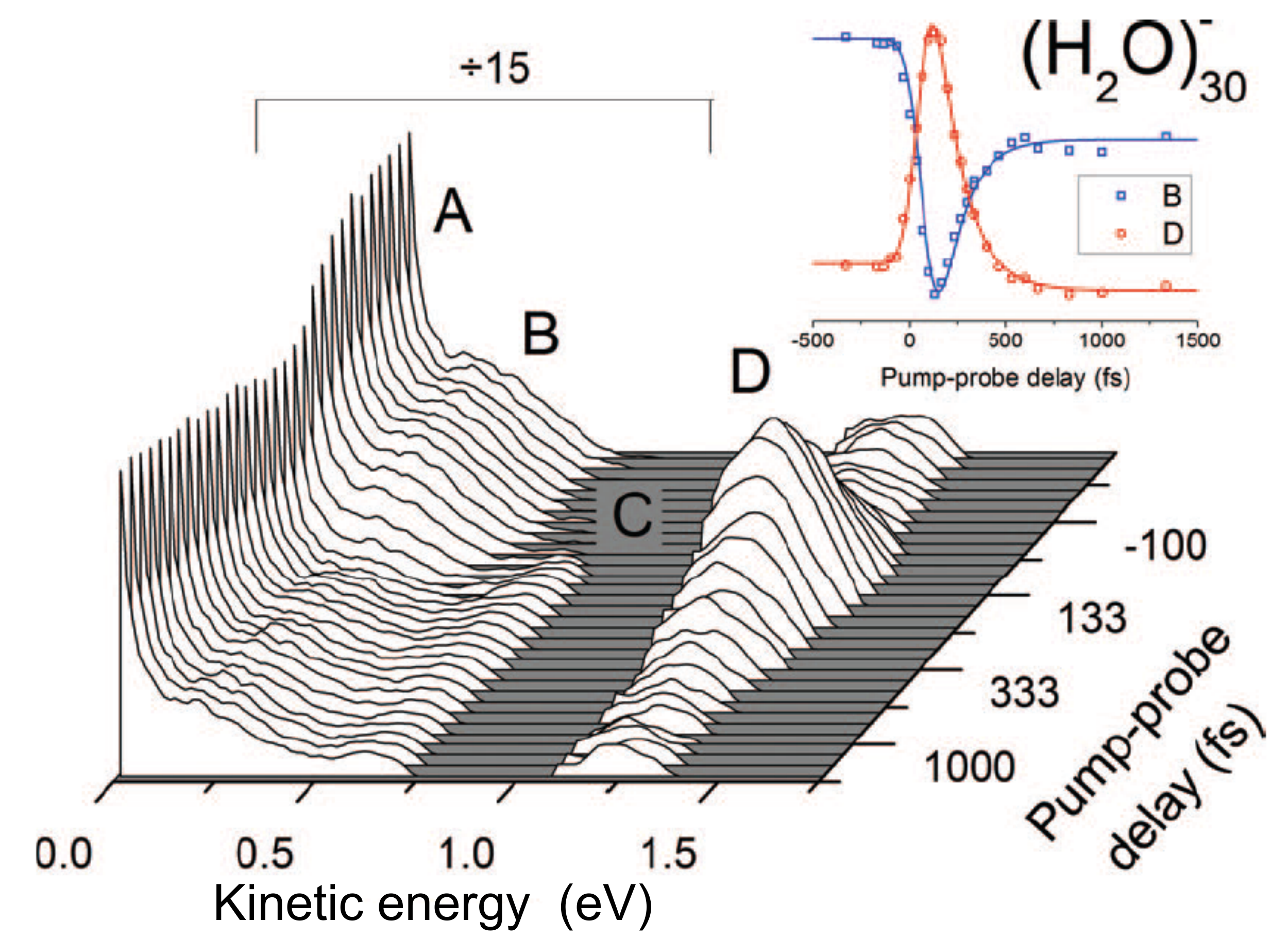}
}
\caption{
Time-resolved photoelectron spectra of (H$_2$O)$_{30}^-$ irradiated by a pump of 1.00 eV
and a probe of 1.57 eV. The spectrum below 1 eV has been multiplied by 15. See text for
the explanation of the features A, B, C and D. Inset: integrated intensity of features B
and D as a function of pump-probe delay.
From~\cite{You12}.
\label{fig:trpes}
}\end{figure}

Finally, and before discussing PAD 
we would like to discuss another possible application of 
the PES now involving rather long  time scales.  
Fig.~\ref{fig:trpes} shows an example of a time-resolved photoelectron spectrum (TRPES)  
measured in (H$_2$O)$_{30}^-$ \cite{You12}. 
The irradiation process is performed with a pump-and-probe setup of laser frequencies of 1
and 1.57 eV respectively, and similar intensities (50-100 $\mu$J per pulse). The PES
exhibit a clear dependence on the delay between the pump and the probe. The four major
structures are indicated by capital letters. The low energy
structure (A) is associated to excited-state autodetachment, while direct probe detachment
from the ground state (B) is observed around 0.25 eV. Structure around 0.6 eV (C) is
attributed to resonant two-photon detachment from the pump , and finally transient
excited-state signal (pump-probe, D) appears in the 1.00--1.50 eV kinetic energy range.
Integrating the intensities of these structures provides the associated population
dynamics, which is indicated in the inset of Fig.~\ref{fig:trpes} for structures (B) and
(D). Both exhibit a similar decay time. To summarize, the above result clearly shows that
a TRPES provides an extremely rich tool of investigation of details of electron dynamics.
Such measurements, possibly complemented by theoretical investigations, should thus help
to reveal crucial information on irradiation scenarios. Even more so PAD bring an
invaluable complement to PES, as we shall see in the next section.

\subsubsection{Photo-Angular Distributions (PAD) and PES/PAD}
\label{sec:PADetal}

\begin{figure}[htbp]
\centerline{\includegraphics[width=\linewidth]{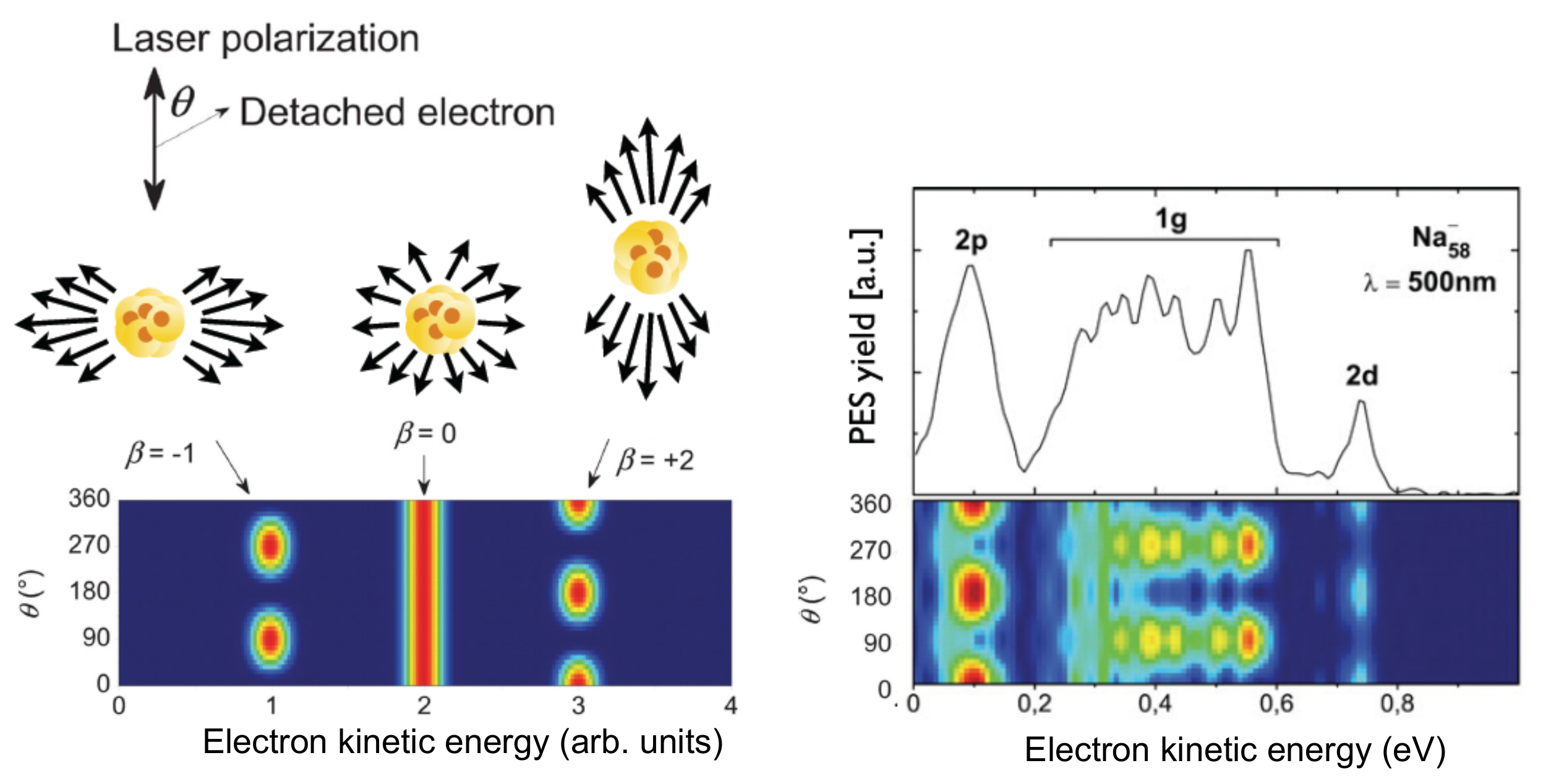}}
\caption{Top left~: schematic view of various types of photoemission distributions 
(from left to right~: oblate, isotropic, and prolate). Bottom left~: ideal photoangular 
distributions corresponding to each case, with the respective  value of the anisotropy parameter $\beta$.
Right~: experimental PES (top) and PES/PAD (bottom) of ${{\rm Na}_{58}}^-$ cluster, irradiated
by linearly polarized pulses from a dye laser
(pulse width of about 10 ns, peak intensity below  10$^5$ W/cm$^2$ and 
photon energy 4.02 eV).
Adapted from~\cite{Bar09}.
\label{fig:examp_pad}
}
\end{figure}

Photo-Angular Distributions bring an invaluable complement to Photo-Electon Spectra. 
An example of PAD is shown in Fig.~\ref{fig:examp_pad}, adapted from \cite{Bar09}.
The PAD are plotted as a function of electron kinetic energies, so that they in fact
represent a combined PES/PAD. The mere PAD can then be obtained by integrating over
kinetic energies. It should be immediately noted that the notion of PAD requires a proper
definition of a reference frame. The reference direction is given by the laser
polarisation axis and angular distributions are thus measured with respect to this axis.
But it should also be noted that, in the gas phase, the actual orientation of clusters or
molecules with respect to this polarisation axis is unknown so that one has at best access
to an orientation averaged (of the molecule with respect to the laser polarisation)
signal. This in particular reduces the angular distribution to a dependence on the angle
between the laser polarization axis and the detection angle, because of angular averaging
around the polarization axis. For then, in the case of single photon absorption, the cross
section takes the simple form :
\begin{equation}
\frac{\textrm d\sigma}{\textrm d\Omega} \propto 1 + \beta_2 P_2\left( \cos \theta \right) \quad,
\label{eq:beta2}
\end{equation}
where $\theta$ is the direction of the emitted electron measured with respect to the laser
polarisation, $P_2$ is the second order Legendre parameter and $\beta_2$ is known as the
anisotropy parameter. In the simple case of one photon processes, the angular distribution
is thus fully characterised by the anisotropy parameter $\beta_2$ which takes values
between -1 and 2. Three values of $\beta_2$ are thus special, as illustrated in the left
part of Fig.~\ref{fig:examp_pad} : $\beta_2$=2 corresponds to a prolate-like form of the
(orientation averaged) emission cloud along the laser polarisation, so that signal will
gather around 0$^\circ$ and 180$^\circ$; $\beta_2$=-1 corresponds to a purely transverse
emission, oblate-like shape, with signal gathering around 90$^\circ$ and 270$^\circ$;
finally $\beta_2$=0 corresponds to a fully isotropic emission.

A realistic measurement is shown in the right part of Fig.~\ref{fig:examp_pad} to
complement the schematic part. The measurement has again been performed in ${{\rm
Na}_{58}}^-$, thus complementing the PES example of Fig.~\ref{fig:pes_mecha}. We
nevertheless indicate the latter PES for completeness and to ease the explanation of the
features of the combined PES/PAD. As already noted in the discussion of
Fig.~\ref{fig:pes_mecha}, the case demonstrates a clear dependence of the photoemission on
the nature of the electronic wave functions (indicated with spectroscopic notations in the
figure). Comparing the PES (upper right panel) to the PES/PAD (lower right panel), one can
see that the $2p$ and $2d$ electrons are emitted parallel to the laser polarization. On
the contrary, the emission from all the $1g$ states occurs preferentially aligned in the
transverse direction. This demonstrates that PAD certainly adds further useful information
on the spatial structure of the emitting states.

\begin{figure}[htbp]
\centerline{\includegraphics[width=0.8\linewidth]{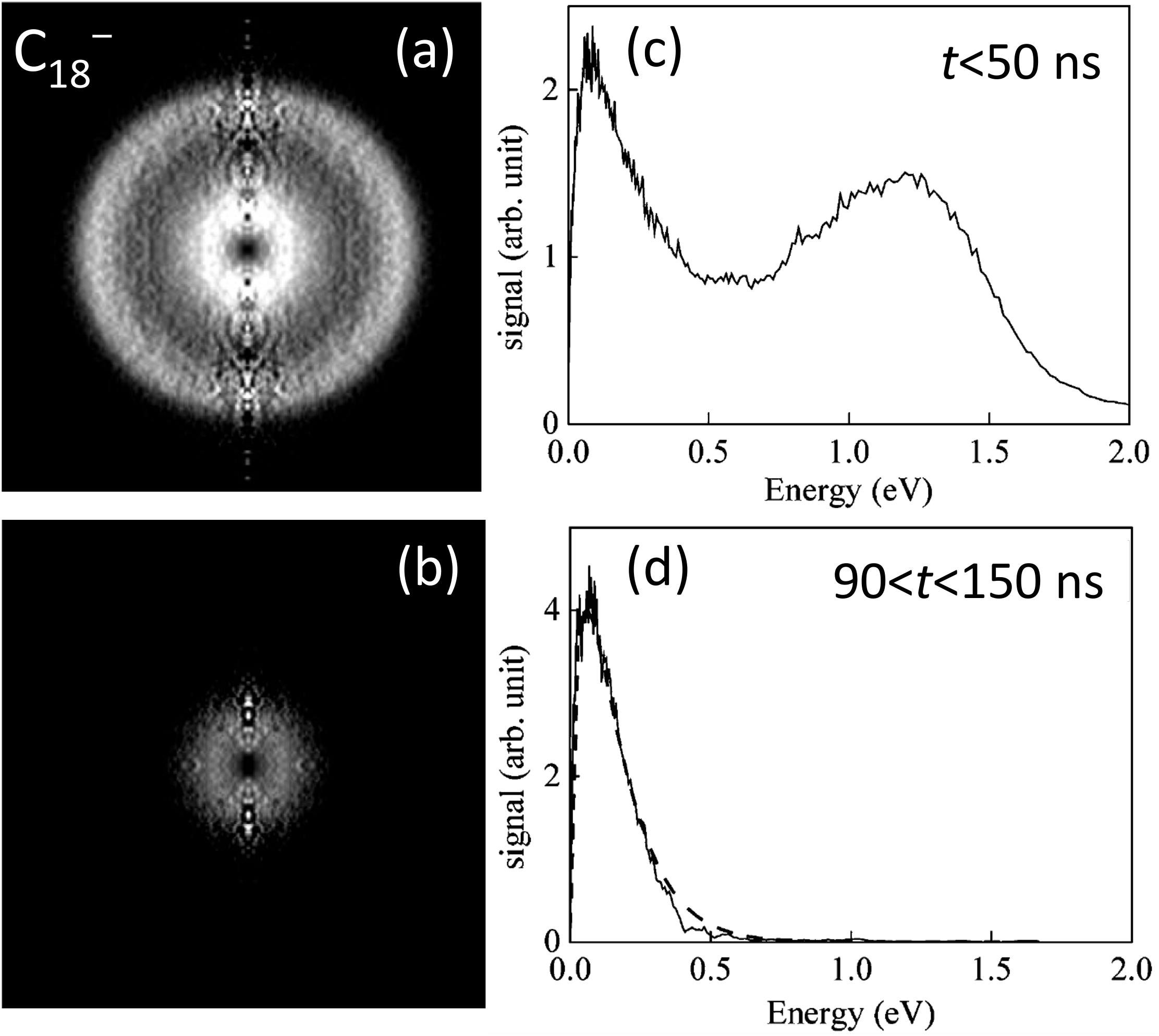}}
\caption{
Velocity map images (left panels) of ${{\rm C}_{18}}^-$, and corresponding photoelectron spectra (right panels),
after irradiation by laser pulses with frequency of 4.025~eV and duration of a few ns. Top row~: yields
accumulated for $t<50$~ns. Bottow row~: yields accumulated for $90<t<150$~ns.
Adapted from~\cite{Wil04}. 
\label{fig:lepine2004}
}
\end{figure}

Another example of PES/PAD, this time in the standard VMI presentation, is shown in
Fig.~\ref{fig:lepine2004} in the case of ${{\rm C}_{18}}^-$~\cite{Wil04}.
In this representation, the VMI provides a polar image of the directions (angle) and
kinetic energies (radius) of the emitted electrons, again with a well defined reference
axis provided by the laser polarization. The example of Fig.~\ref{fig:lepine2004} is
furthermore time-resolved, or at least allows one to separate well separated time scales
of emission. Irradiation is performed with photons of frequency 4.025 eV and pulse
durations in the ns range. Panel (a) provides an image of photoelectrons emitted during
the first 50 ns after excitation, and the corresponding PES is plotted in panel (c). The
PES exhibit two maxima, also visible as rings in panel (a), one at high energy and one at
low energy. The high energy signal is associated to direct emission from the
photo-excitation itself. The low energy component is attributed to thermionic emission in
which the original laser energy has been partly equipartitioned between vibronic degrees
of freedom of the cluster, prior to electron emission. The time scale and the typical
energies associated to thermoionic emission are thus much larger than the ones associated
to direct emission. In the present experiment the typical time scales of thermoionic
emission lie in the tens to hundreds of ns, which in that case can be identified
experimentally. The scenario is confirmed in panels (b) and (d) which present the VMI and
the PES recorded in a late time window, that is between 90 and 150 ns. The PES is now
fully concentrated at low energies, with no sign of a high energy, direct emission,
component. This confirms the thermoionic nature of this late, low energy, emission.
Therefore, even at a coarse time level, such an analysis exemplifies the capabilities of
PES and PES/PAD to analyse electron dynamics in detail. We shall come back on those
aspects later,  see in particular Secs. \ref{sec:pes_dyn}, \ref{sec:pad_results} and \ref{sec:temper}.

\section{Theoretical approaches}
\label{sec:theo}

Many-particle systems such as molecules or clusters, are highly
correlated, and exact calculations of their properties are extremely
involved, mostly beyond feasibility for finite systems without major
symmetries.  The main issue concerns here the treatment of
electrons. Except for some specific cases, ions can be treated as
classical particles. This will always be the case in the following. To
deal with the electronic problem, a variety of approaches has been
developed, each one being a compromise between precision and
expense. In this section, we present the most widely used schemes,
paying a particular attention to density-functional theory (DFT) which
is one of the most efficient tools in cluster dynamics.

\begin{figure}[htbp]
\centerline{\includegraphics[width=\linewidth]{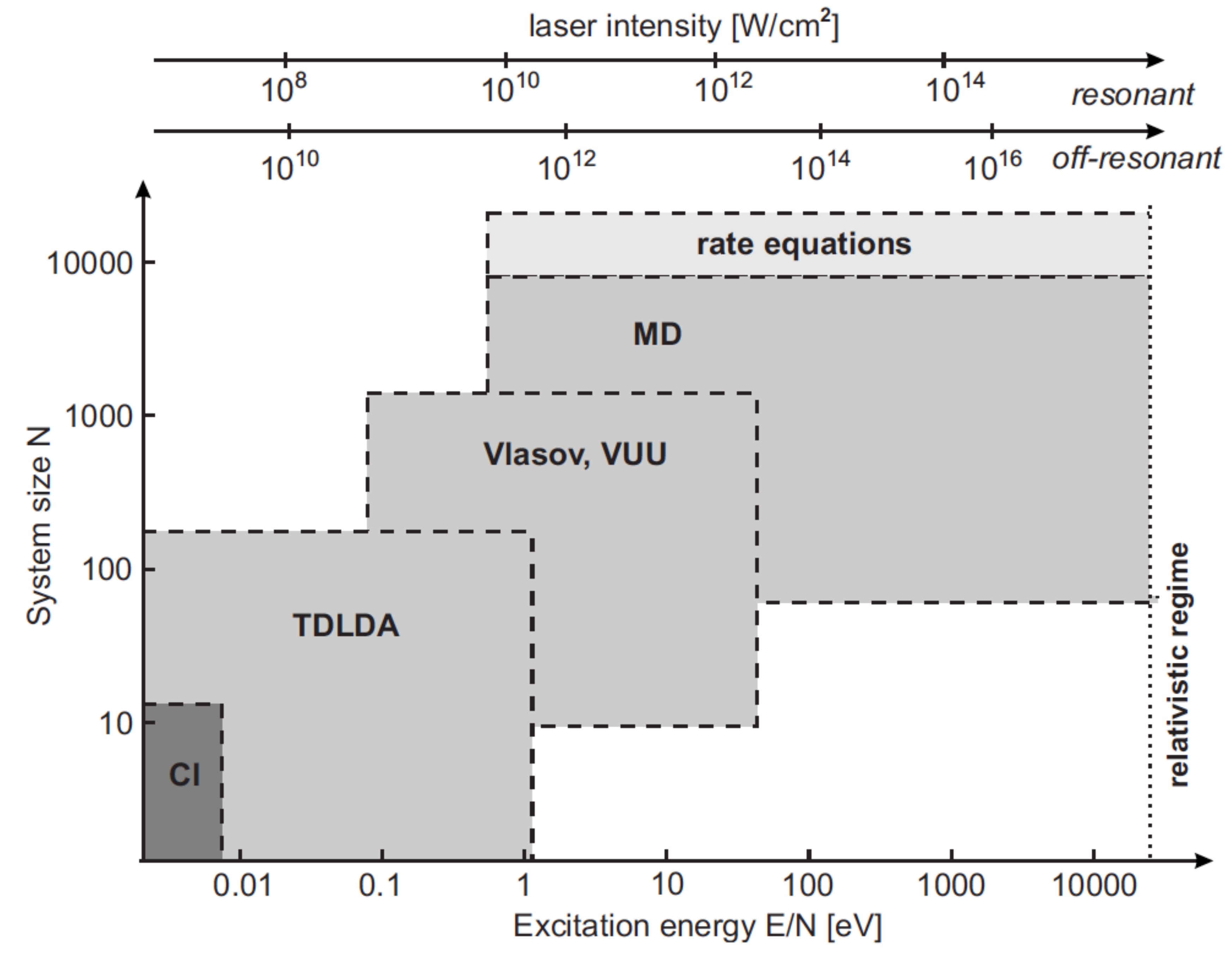}}
\caption{
Schematic view of applicability of different approaches (see text for details) in a landscape of
system size versus excitation energy per atom. 
The  excitation energy can be
loosely related to typical laser intensities in the optical range. This is
indicated by the intensity scales on top, which are, however, also strongly
dependent on the response of the particular system, i.e. resonant or
non-resonant.
\label{fig:theo_schemes}
}
\end{figure}

Before going into details of the theoretical treatment, we
schematically summarize in Fig.~\ref{fig:theo_schemes} the most
widely used theoretical approaches and sketch the regimes of their
applicability in a plane of excitation energy and particle number.
The boundaries of the regimes are to be understood as very soft with
large zones of overlap between the models because the choice of a
method also depends on several aspects (e.g., demand on precision,
material, time span of simulation).
The most elaborate models are the ``ab-initio'' methods which deal in
a systematic manner with a Hamiltonian as exact as possible. The
simplest example is the Hartree-Fock approximation which, however,
misses the crucial electronic correlations. A typical example of the
more elaborate approaches is the Configuration Interaction (CI) method
which relies on an expansion of the exact many-body wave function into
a superposition of Slater states \cite{Yam94,Kra07a}. The limitations
for CI (and other {\it ab-initio} methods) are purely a matter of
practicability. The limitation is nevertheless even more severe for
dynamical applications of such theories, which are thus presently
restricted to rather small system sizes and small excitation
energies. The range of applicability will slowly grow with the
steadily increasing computer power.
Density functional theory (DFT) describes a system effectively in
terms of a set of single-electron states (see Sec.~\ref{sec:tddft}). 
It is limited in system size for practical reasons
and in excitation energy for physical reasons, because of the missing
dynamical correlations from electron-electron
collisions. Nevertheless, DFT and even more so its time dependent
extension TDDFT (especially when realized in full real time)
nevertheless provide a most robust and versatile tool in the field.
A semi-classical mean-field description is provided by the Vlasov
equation originally designed for plasma physics \cite{Vla50}. This
approach ignores quantum effects such as shell structure or tunneling
and thus becomes questionable at low energies. It is furthermore
reasonably tuned to metal electrons because of their ressemblance to a
Fermi gas, but more difficult to apply in other materials, especially
in covalent bound systems. On the other hand, the semi-classical
treatment allows one to include dynamical correlations due to
electron-electron collisions, leading to the Vlasov-Uehling-Uhlenbeck
(VUU) approach \cite{Ber88,Dom00c,Fen04}, which extends the
applicability to larger energies than those allowed by TDLDA.
Even higher excitations and system sizes are the realm of electronic
Molecular Dynamics approaches and rate equations which, however, are
even more limited than VUU for low energies and small systems
\cite{Sie06}. The upper limit in energy is given by the onset of the
relativistic regime, where retardation effects within the coupling
begin to severely influence the dynamics.

In the following, we shall use real-time TDDFT as the basic
theory to describe ionization dynamics. We shall occasionally use VUU
in order to discuss electronic temperature effects, as observed in
some experiments.  We thus briefly describe in this section basics of
TDDFT and practical implementations thereof. We discuss in some detail
the self-interaction correction strategy to be developed to properly
account for ionization in a dynamical way within standard
approximations of DFT. We next present in detail the tools developed
to access PES and PAD in TDDFT.  We in particular discuss the
demanding inclusion of orientation effects of the irradiated clusters
and molecules with respect to laser polarization.  We finally remind
basics of VUU for completeness.

\subsection{Basic formalism}

\subsubsection{Handling of the ionic background}
\label{sec:ions}

The interaction between the ions in a cluster and the electrons is
usually described by pseudopotentials. This allows one to eliminate the
inert, deep lying electron states around each ion and to concentrate on 
the relevant valence electrons. For a detailed discussion of
pseudopotentials see, e.g., \cite{Sza85}. We go here a pragmatic way
and take published pseudopotentials. For simple metals, we consider
the soft, local pseudopotentials of \cite{Kue99}. In more general
cases, we employ mostly the local and non-local pseudopotentials in
separable form as introduced in \cite{Goe96}. 
More precisely, for each type of atom with $Z$ valence
electrons, we use the pseudopotential $V_\mathrm{PsP}$ of the following form~:
\begin{subequations}
\begin{eqnarray}
\label{eq:PsP} 
  V_\mathrm{PsP}(\mathbf r) \, \varphi_j (\mathbf r) 
  &=& 
  V_\mathrm{loc}(r) \, \varphi_j(\mathbf r)
  +
  \int \textrm d^3 \mathbf r'\, V_\mathrm{nloc} (\mathbf r,\mathbf r')\,
  \varphi_j(\mathbf r') 
  \;,
\\
  V_\mathrm{loc}(r) 
  &=& 
  - \frac{Z}{r} \, {\rm erf}\left( x/\sqrt{2} \right) 
  + \exp \left( - x^2/2 \right) \left[ C_1 + C_2\, x^2 \right], 
 \;
  x=\frac{r}{r_\mathrm{loc}}
  \;,
\\
  V_\mathrm{nloc} (\mathbf r,\mathbf r') 
  &=& 
  p(r)\, h_0\, p(r')
  \;,
\\
  p(r) 
  &=&
  \frac{ \sqrt{2} } { {r_\mathrm{nloc}}^{3/2} \sqrt{\Gamma\left(3/2\right)}}  
  \exp\left( -\frac{r^2}{2 {r_\mathrm{nloc}}^2}\right)
  \;.
\end{eqnarray}
\end{subequations}
Here $\varphi_j$ denotes the wave function of state $j$, ${\rm erf}$
the error function, $\Gamma$ the Gamma function, and
$x=r/r_\mathrm{loc}$. The refitted parameters are $C_1$, $C_2$,
$r_\mathrm{loc}=r_\mathrm{loc}$, and $h_0$.  The standard parameters
are given in \cite{Goe96}.  However, for the results presented in
Section~\ref{sec:results}, we use often refitted parameters which
employ larger radii $r_\mathrm{loc}$ and $r_\mathrm{nloc}$, thus
softer pseudopotentials for more robust numerical handling, see
Section~\ref{sec:num}.

There also exists a commonly used alternative to pseudopotentials for
metallic systems. In particular, simple metals have valence electrons
with long mean free path throughout. The fine details of the ionic
background can thus be seen by the electrons only in an average
manner. This motivates the jellium approximation in which the ionic
background is smeared out to a constant positive background
charge. This is a standard approach in bulk metals \cite{Ash76}. It
has been generalized to finite clusters. In its simplest form, one
carves from the bulk a finite, homogeneously and positively charged
sphere of radius $R=r\, N_\mathrm{ion}^{1/3}$, whose total ion charge
reproduces the given ionic charge $eN_\mathrm{ion}$.
A more flexible approach is achieved when allowing for a finite
surface width, yielding the soft jellium model
\begin{equation}
  \rho_\mathrm{jel}(\mathbf{r})
  =
  \frac{3}{4\pi r_s^3}
  \left[1+
     \exp\left(\frac{|\mathbf{r}|-R_\mathrm{jel}}
                    {\sigma_\mathrm{jel}}\right)
  \right]^{-1}
  \quad,
\label{eq:softJ}
\end{equation}
with $R_\mathrm{jel}$ being defined by normalization to the total particle
number $ \int \textrm d^3\mathbf{r}\, \rho_\mathrm{jel} =
N_\mathrm{ion}.$ The central density reproduces the bulk density
$\rho_0={3}/({4\pi r_s^3})$. The parameter $\sigma_\mathrm{jel}$
accounts for a smooth surface transition from $\rho_0$ to zero. The
surface width (transition from 90\% to 10\% bulk density) is about
$4\sigma_\mathrm{jel}$.  The model can be extended to also describe
deformations which can have a considerable influence in metal cluster
spectroscopy depending on the system \cite{Eka91,Mon95b}.


\subsubsection{Density Functional Theory and its Time-Dependent version}
\label{sec:tddft}

The goal of DFT is to develop self-consistent equations which employ
effective potentials for the contributions from exchange and
correlation. These potentials are to be expressed in terms of the
total local electron density $\rho(\mathbf{r})$ of the system. The
success of DFT depends on a diligent choice of these effective
potentials. For the brief review of DFT, we take here a practitioners
approach and discuss the Kohn-Sham (KS) scheme from a given energy
functional. We do not address the theoretical foundations of DFT in
terms of the much celebrated Hohenberg-Kohn theorem~\cite{Hoh64} and
Kohn-Sham formalism~\cite{Koh65}. The many aspects of foundation and
derivation can be found, e.g., in \cite{Par89,Dre90,Gro96}.

\paragraph{The energy functional}

The starting point is an energy functional for the total electronic
energy $E_\mathrm{total,el}$. In the Kohn-Sham (KS) scheme, one
represents the $N$ (valence) electrons, by $N$ non-interacting
Kohn-Sham (KS) orbitals (or s.p. states) $\varphi_i(\mathbf r)$,
$i\in\{1,\ldots,N\}$. The total energy is then separated into kinetic
energy (which then takes a simple form) and interaction energy
(associated to the above mentioned effective pseudopotentials). The
total electronic density is expressed from the KS orbitals as~:
\begin{equation}
\label{eq:rho}
\rho(\mathbf r) = \sum_{i=1}^N \big| \varphi_i(\mathbf r) \big|^2 \quad.
\end{equation}
Note that DFT schemes allow one to treat spin-up and spin-down density
separately. For simplicity of presentation, we drop the spin
dependence in the following. The total electronic energy is then
composed as
\begin{subequations}
\label{eq:DFT_Etot}
\begin{eqnarray}
\label{eq:Etotal}
  E_\mathrm{total,el}[\rho]
  &=&
  E_\mathrm{kin}(\{\varphi_i\})
  +
  E_\mathrm{H} [ \rho ]
  +
  E_\mathrm{xc} [ \rho ]
  +
  E_\mathrm{coupl}
  +
  E_\mathrm{ext} \quad,
\\
\label{eq:DFT_kin}
   E_\mathrm{kin} \left( \{ \varphi_i \} \right) 
  &=& 
  - \frac{\hbar^2 }{2m} \ \int \textrm d^3\mathbf r 
   \sum_{i=1}^N \varphi_i^*(\mathbf r) \ \nabla^2 \ \varphi_i(\mathbf r) 
   \quad,
\\
\label{eq:hartree_nrj}
   E_\mathrm{H}[\rho] 
   &=& 
  \frac{e^2}{2} \iint \textrm d^3 \mathbf r \ \textrm d^3 \mathbf r' \ 
  \frac{\rho(\mathbf r) \rho(\mathbf r')}{|\mathbf r - \mathbf r'|} 
  = 
  \frac{1}{2} \int \textrm d^3 \mathbf r \ 
  \rho(\mathbf r) \ U_\mathrm{H}[\rho] 
  \quad,
\\
\label{eq:V_coupl}
  E_\mathrm{coupl}
  &=&
  \int \textrm d^3 \mathbf{r}\,\sum_{i=1}^N
  \varphi_i^*(\mathbf r)\ \hat{V}_\mathrm{coupl}\ \varphi_i(\mathbf r)
  \quad, 
\\
\label{eq:Eext}
  E_\mathrm{ext}
  &=&
  \int \textrm d^3\mathbf{r}\ \rho(\mathbf{r})\  U_\mathrm{ext}(\mathbf{r}) 
  \quad.
\end{eqnarray}
\end{subequations}
The kinetic energy is a functional of the s.p. orbitals $\varphi_i$
which serves to maintain the quantum shell structure in the KS
calculations. The non-trivial correlation part of the exact kinetic
energy is summarized in the interaction energy. The interacting term
is mapped to the density functionals $E_\mathrm{H}[\rho] +
E_\mathrm{xc}[\rho]$. The first term $E_\mathrm{H}$ is the standard
(direct) Coulomb Hartree energy, which naturally is a functional of
$\rho$. We have introduced here the notation $U_\mathrm{H}$ for the
corresponding Hartree potential. Conceptually simple are
$E_\mathrm{coupl}$ from the coupling to the ions
($\hat{V}_\mathrm{coupl}$ is the potential operator built from the
pseudopotentials) and the energy $E_\mathrm{ext}$ modeling an external
electromagnetic field $U_\mathrm{ext}(\mathbf{r})$. Both these
contributions couple to single electrons and are naturally well
represented by an independent particle picture in terms of
$\varphi_i$.

Finally, there is the exchange-correlation energy $E_\mathrm{xc}$
which accumulates all pieces of the exact energy not yet accounted
for. This is the most problematic part in the scheme, since its
functional expression is not exactly known. Many approximations
thereof do exist, among which the simplest and most robust one is the
Local Density Approximation (LDA). The construction of LDA is
simple. One computes the ground state of the homogeneous electron gas
as exactly as possible and obtains the exchange-correlation energy per
volume $E_\mathrm{xc}/V=\rho\,\epsilon_\mathrm{xc}(\rho)$. Here this
energy is still a function of the (homogeneous) density $\rho$. The
crucial point is to allow now for an inhomogeneous, and time-dependent
if needed, density $\rho(\mathbf{r},t)$ in that expression. It amounts
to considering the energy as composed piecewise from an infinite
electron gas of densities $\rho(\mathbf{r},t)$ which is a bold
approximation. Nonetheless, LDA provides a robust description for a
wide variety of systems. There is an enormous body of literature
pondering successes and failures, for a more detailed discussion, see
e.g.  \cite{Dre90}. Note that a functional depending on
$\rho(\mathbf{r},t)$ employs the instantaneous density and thus
excludes any memory effect. This time-dependent generalization is
often called adiabatic LDA (ALDA). Again, we use in the following the
generic notation LDA.

The validity of LDA depends very much on the system under
consideration. One of the major problems is the self-interaction
error~: the single particle state $\varphi_i$ is included in the
density $\rho$, and thus contributes to the mean-field Hamiltonian
$\hat{h}_\mathrm{KS}$ (see Eq.~(\ref{eq:hamKS}) below) which acts on
$\varphi_i$. This yields a wrong asymptotics for the Coulomb mean
field. For example, for a neutral system described in LDA, it decays
exponentially at large distances instead of $\propto e^2/r$ as it
should. An attempt to reduce the self-interaction error is the
generalized-gradient approximation (GGA) which augments LDA with an
additional dependence on $\nabla\rho$ \cite{Bec88,Per96}. GGA yields a
significant improvement in the computation of atomic and molecular
binding. For example, it lifts the description of dissociation
energies to a quantitative level. However, GGA does not fully remove
the self-interaction error. Thus there are various attempts for further
improvement as, e.g., adding kinetic terms to DFT
\cite{Per99}. Another line of development is to explicitly implement a
Self-Interaction Correction (SIC). This helps to deliver correct
ionization properties, which is crucial in describing PES and PAD
dynamically. We will therefore discuss this approach in more detail in
Sec.~\ref{sec:sic_theo}.

As indicated above, time-dependent DFT, effectively using ALDA, makes
also an adiabatic approximation. In order to account for dynamical
effects, a Current DFT (CDFT) has been developed, which is based on
LDA augmented by a dependence on electronic currents evaluated in the
linear response \cite{Dha87a,Vig96,Ull12}. The response kernels in the
extended functional include memory effects and allow one to describe
relaxation \cite{Ago06a}. CDFT is rather involved and thus there exist
so far only applications to symmetry restricted systems as, e.g., in
solids\cite{Mai03a}. The underlying linear response modeling makes
CDFT an extension for low excitation energies and/or amplitudes. Real
electron-electron collisions become important for more energetic
processes. These are often treated by a quantum generalization of the
Boltzmann collision term. We will address this extension of DFT in
Secs.~\ref{sec:temperature_theo} and \ref{sec:qdissip}.

\paragraph{The Kohn-Sham equations}

The stationary KS equations are derived by variation of the total
energy with respect to the s.p. wave functions $\varphi_i^*$,
yielding~:
\begin{subequations}
\label{eq:KS}
\begin{eqnarray}
  \hat{h}_\mathrm{KS}[\rho]\ \varphi_i (\mathbf r) 
  &=&
  \varepsilon_i \ \varphi_i(\mathbf r) 
  \quad,
\label{eq:statKS}
\\
  \hat{h}_\mathrm{KS} [\rho]
  &=&  
  -\frac{\hbar^2 \nabla^2}{2m}
  +
    U_\mathrm{KS} [\rho] 
    +
    \hat{V}_\mathrm{coupl}
    +
    U_\mathrm{ext}  
  \quad,
\label{eq:hamKS}
\\
  U_\mathrm{KS} [\rho]
  &=&  
    U_\mathrm{H} [\rho] 
    +
    U_\mathrm{xc} [\rho] 
   \quad.
\label{eq:UKS}
\end{eqnarray}
\end{subequations}
The local and density-dependent Kohn-Sham potential
$U_\mathrm{KS}$ consists in the direct Coulomb term
$U_\mathrm{H}$ and the exchange-correlation potential, which is a
standard functional derivative
$U_\mathrm{xc}=\delta{E}_\mathrm{xc}/\delta \rho$. Coupling
potentials to ions and to the external field are trivially given.

The time-dependent KS equations analogously read~:
\begin{equation}
  {\rm i} \hbar \ \partial_t\varphi_i(\mathbf r,t) = \hat{h}_\mathrm{KS}[\rho] \,
\varphi_i (\mathbf r,t) \quad,
\label{eq:TDKS}
\end{equation}
where $\hat{h}_\mathrm{KS}$ is composed in the same manner as above,
provided that one replaces $\rho(\mathbf{r})$ by
$\rho(\mathbf{r},t)$. This assumes an instantaneous adjustment of the
total electronic density, although memory effects can play in some
cases an important role, especially in $E_\mathrm{xc}$ \cite{Gro96}.

The stationary KS equations (\ref{eq:statKS}) pose an eigenvalue
problem. They provide the electronic ground state of a system. This is
a highly non-linear problem due to the self-consistent feedback of the
local density in the KS hamiltonian. It is usually solved by iterative
techniques~\cite{Cal00}. The time-dependent KS equations imply an
initial value problem. The natural starting point is the ground state
obtained from the stationary KS equations. The time-dependent KS
system can then be solved by standard methods of first order
differential equations~\cite{Cal00}.  We finally remind that we wrote
spinless KS equations. One can easily include the electron spin in
Eqs.~(\ref{eq:KS}). We refer the reader to~\cite{Par89,Dre90,Gro96} for
more details.

\subsubsection{A few words on numerical implementation}
\label{sec:num}

A representation of the s.p. wave functions
$\varphi_\alpha(\mathbf{r},t)$ and the fields $\rho(\mathbf{r},t)$ and
$U_\mathrm{KS}(\mathbf{r},t)$ on a coordinate-space grid is
strongly recommended if one aims at computing electronic emission
properties. Conceptually straightforward is a Cartesian 3D grid with
equally spaced mesh points. This leaves two choices for the
description of the kinetic energy, that is finite difference schemes
\cite{Bon78a,Ter96b,Cas06,And12} or the Fourier definition exploiting
the extremely efficient fast Fourier transformation (FFT)
\cite{Cus76a,Cal95a,Cal00}. The Coulomb problem is solved either by
iterative methods (e.g., successive over-relaxed iterations) in
connection with finite-difference schemes or by Fourier techniques in
case of FFT. In the latter case, one can produce an exact solution on
the grid by using a double grid (in each direction) \cite{Eas79} or,
somewhat faster, by eliminating the long-range terms by a separate
analytical treatment \cite{Lau94}. A fast scheme for the static
solution is provided by the damped gradient iteration
\cite{Rei82}. The scheme for time evolution depends on the
representation of the kinetic energy. For finite-difference schemes,
one typically uses a second order predictor-corrector with a Taylor
expansion of the KS time-evolution operator while for FFT schemes, the
time-splitting (also called $\hat{T}$-$\hat{V}$ splitting) technique
is preferable in connection with the FFT representation
\cite{Fei82}. An extensive comparative study of the various griding
and iteration techniques can be found in \cite{Blu92}.

Let us end with a few words in the case of symmetric systems. For
instance, the jellium model allows a description in higher symmetry,
that is a representation on an axial 2D grid \cite{Dav81a}. In the
case of explicit ions in simple metal clusters, they can be described
by soft, local pseudopotentials. This allows an averaging over axial
angle, leading to the cylindrically averaged pseudopotential scheme
(CAPS) \cite{Mon94a} which is extremely efficient and thus has been
used in many explorative studies. The CAPS allows one to treat
explicit ionic structure in full 3D with pseudopotentials while
keeping electrons with cylindrical symmetry \cite{Cal00}. This turns
out to be a very good approximation for metals and it is even exact
for linear molecules such as carbon chains \cite{Ber02}. It has even
allowed to step forth to rather complex systems as, e.g., embedded
clusters \cite{Din07b,Din09aR}. The Fourier representation of kinetic
energy cannot be applied in this geometry. Finite differences are the
method of choice in axial grids. The static solutions use the same
iterative schemes as in 3D. A particularly suitable time-stepping
scheme for axial 2D is the Peaceman-Rachford step, which is a
separable version of the well known Crank-Nicholson step
\cite{Var62,Pre92}.

\subsection{The self-interaction problem in DFT and TDDFT}
\label{sec:sic_theo}

As outlined above, LDA is plagued by the self-interaction error which
is particularly harmful for ionization properties. The safest way to
deal with that is to introduce an explicit Self-Interaction Correction
(SIC). A conceptually simple and robust SIC was introduced by
J. Perdew and A. Zunger in which all single-particle self-interactions
are subtracted from the DFT energy \cite{Per81}~:
\begin{equation}
  E_\mathrm{SIC}[\rho]
  =
  E_\mathrm{LDA}[\rho]
  -
  \sum_{\alpha=1}^N
  E_\mathrm{LDA}[\rho_\alpha]
  \quad,\quad
  \rho_\alpha(\mathbf{r})
  =
  \left|\psi_\alpha(\mathbf{r})\right|^2
  \quad,
\label{eq:PZSIC_Etot}
\end{equation}
where $E_\mathrm{LDA}=E_\mathrm{H}+E_\mathrm{xc}$ is the LDA
functional for the Coulomb-Hartree term as well as exchange and
correlations. Note that we have changed here our standard notation for
the single electron KS orbitals from $\varphi_i$ to
$\psi_\alpha$. This is done on purpose as will become clear below in
Eq.~(\ref{eq:ut}). The self-interaction corrected KS equations are
then derived again by variation. A problem is that the emerging SIC-KS
hamiltonian $\hat{h}_\mathrm{SIC}$ then becomes state-dependent
because
$\delta{E}_\mathrm{LDA}[\rho_\alpha]/\delta\psi_\beta^*\propto\delta_{\alpha\beta}$
is state selective. We formulate this in terms of a projector and
obtain~:
\begin{eqnarray}
  \hat{h}_\mathrm{SIC}
  &=&
  \hat{h}_\mathrm{LDA}
  -
  \sum_\alpha U_{\alpha}|\psi_\alpha\rangle\langle\psi_\alpha|
  \quad,\quad
  U_{\alpha}
  =
  \frac{\delta{E}_\mathrm{LDA}[\rho_\alpha]}{\delta\rho_\alpha}
  \quad.
\label{eq:SICmf}
\end{eqnarray}
It becomes apparent that the state dependence leads to a non-Hermitian
SIC hamiltonian.  This leads to a violation of orthonormality of the
$\psi_\alpha$. To restore it, we have to add a constraint
$\sum_{\alpha\beta}\lambda_{\alpha\beta}\langle\psi_\beta|\psi_\alpha\rangle$
to the SIC energy with the (hermitian) Lagrange multiplier
$\lambda_{\alpha\beta}$. The SIC mean-field equations thus become~:
\begin{equation*}
  \hat{h}_\mathrm{SIC} |\psi_\alpha \rangle
  =
  \sum_\beta \lambda_{\beta \alpha}|\psi_\beta \rangle
\end{equation*}
for the static case and 
\begin{equation*}
\left ( \hat{h}_\mathrm{SIC} - \mathrm{i} \hbar \partial_t \right )
|\psi_\alpha \rangle
  =
  \sum_\beta \lambda_{\beta \alpha}|\psi_\beta \rangle
\end{equation*}
for the dynamic case. In both cases, these equations have to be
complemented by the "symmetry condition"
\begin{equation}
  \langle\psi_\beta|U_\mathrm{\beta}-U_{\rm\alpha}|\psi_\alpha\rangle=0
\label{eq:sym}
\end{equation}
which is the crucial ingredient in the scheme stemming from the
orthonormality constraint \cite{Mes09,Ped84}. These SIC equations are
hard to solve directly in the static case and near to impossible in
dynamics. The key to success is to introduce a second s.p. basis set
$\{\varphi_l,l=1, \ldots, N\}$ which is connected to the set
$\{\psi_\alpha,\alpha=1, \ldots, N\}$ by a unitary transformation
\cite{Mes08a}
\begin{equation}
  |\varphi_j \rangle
  =
  \sum_{\alpha=1}^N u_{j\alpha} \, | \psi_\alpha \rangle
\label{eq:ut}
\end{equation}
tuned to diagonalize the matrix of $\lambda_{\beta\alpha}$.
This simplifies the static and dynamic SIC-KS equations to
\begin{equation}
  \hat{h}_\mathrm{SIC} |\varphi_j\rangle
  =
  \varepsilon_j|\varphi_j\rangle
  \;,\;
  \left ( \hat{h}_\mathrm{SIC} - \mathrm{i} \hbar \partial_t \right )
 |\varphi_j\rangle
  =
  0
  \;.
\label{eq:SICeqs}
\end{equation}
Eqs. (\ref{eq:SICmf}--\ref{eq:SICeqs}) formulate the SIC problem in
the ``2setSIC'' scheme. They are solved by interlaced iterations. One
performs a step of static or dynamics mean-field problem
(\ref{eq:SICeqs}) and then adjusts the unitary transformation
(\ref{eq:ut}) to accomodate the symmetry condition (\ref{eq:sym}). A
detailed representation of the scheme can be found in
\cite{Mes08a,Mes09}.

\begin{figure}[htbp]
\centerline{\includegraphics[width=0.7\linewidth]{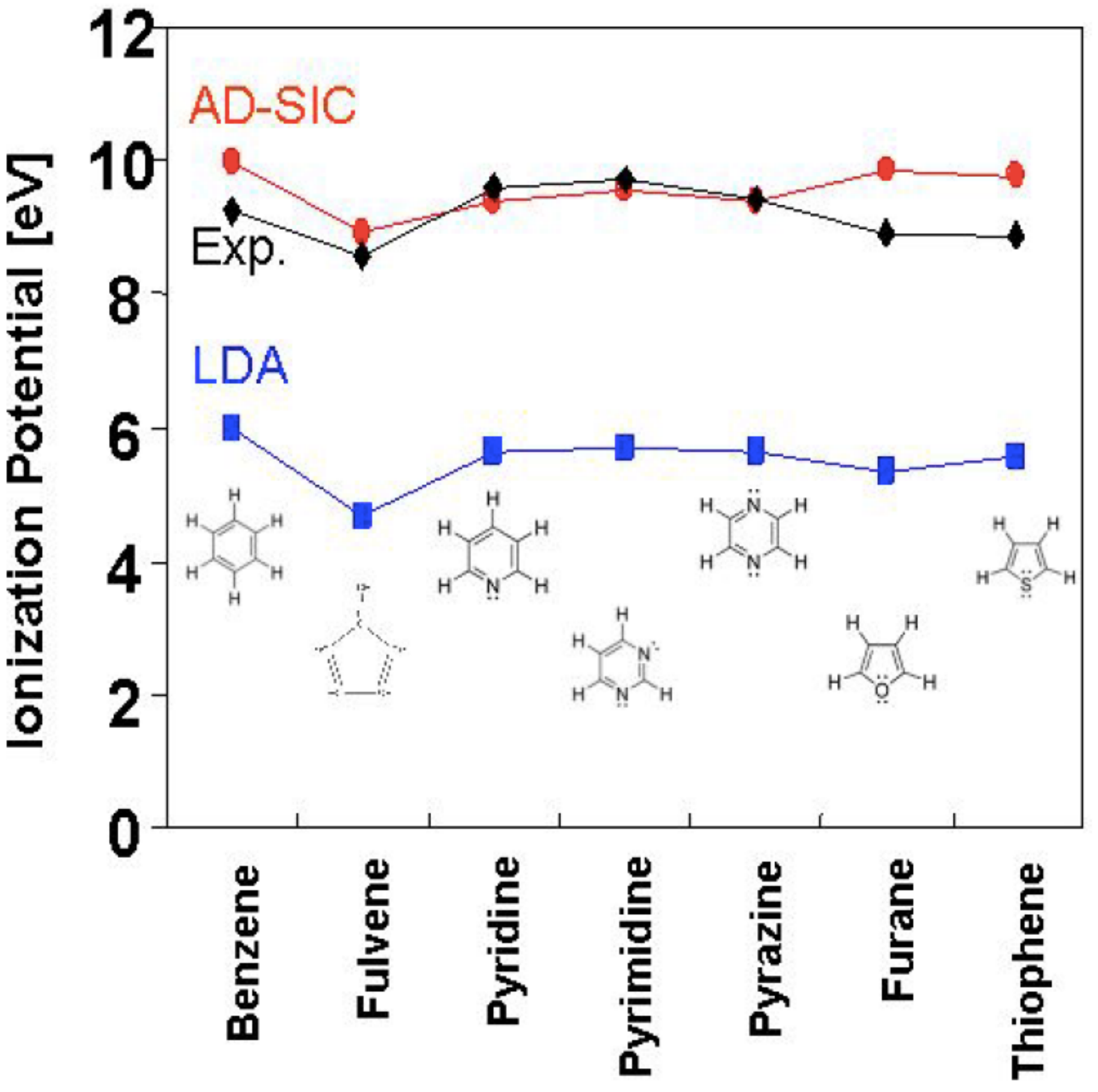}}
\caption{Ionization potentials calculated from the energy of the
  HOMO, for a selection of conjugated molecules. Compared are results
  from LDA and ADSIC with experimental data. Adapted
  from~\cite{Cio05}. 
\label{fig:adsic_mol}
}
\end{figure}

Although we dispose with 2setSIC of a powerful technique to solve the
static and dynamical equations with SIC, it remains a tedious
task. There are several interesting simplifications around. With the
help of the optimized effective potential method (OEP) \cite{Kue08},
one has developed implementations of SIC in terms of state-independent
local potentials, which lead to the Krieger-Li-Iafrate (KLI) method
\cite{Kri92a} and, one step simpler, to the Slater approximation to
SIC \cite{Sha53} (for a detailed discussion in connection with
clusters, see \cite{Leg02}). Metal clusters are special in the sense that their
valence electrons have all very similar spatial extension and stay
close in energy (see for instance the s.p. energies of
${\mathrm{Na}_{41}}^+$ in Fig.~\ref{fig:systems}).  This allows one to
replace the detailed $N$ s.p. densities $\rho_\alpha$ in SIC by a
single averaged representative $\bar\rho_{1e}=\rho/N$ which then
defines the energy functional for Average-Density SIC (ADSIC) as~:
\begin{equation}
  E_\mathrm{ADSIC}(\rho)
  =
  E_\mathrm{LDA}^{\mbox{}}(\rho)
  -
  N\, E_\mathrm{LDA}^{\mbox{}}(\rho/N)
  \;.
\end{equation}
ADSIC is simple, robust, and reliable. It provides the correct
asymptotics of the KS field while it is formally as simple to handle
as LDA. The correct asymptotics and simplicity renders ADSIC very
useful in calculating electron emission and its observables. Most
examples in this article were computed with ADSIC. Although motivated
by metal electrons, ADSIC also performs surprisingly well for covalent
molecules, see \cite{Cio05,Klu13} and Fig.~\ref{fig:adsic_mol}.
Within ADSIC, the concept of s.p. densities is not needed anymore such
that ADSIC is also applicable to semi-classical schemes
\cite{Leg02}. In fact, it was first proposed by Fermi \cite{Fer34} in
a semi-classical context.

Fig.~\ref{fig:adsic_mol} demonstrates the effect of ADSIC for a couple
of basic organic molecules. Ionization potentials (IP) presented in
this figure have been computed using the energy of the Highest
Occupied Molecular Orbital (HOMO) obtained from the ground state
configuration of each system.
The wrong asymptotic Coulomb Kohn-Sham potential of pure LDA leads to
less binding and thus to much reduced IP. The deviation is
uncomfortably large. Correcting the self-interaction error even with
the simple ADSIC suffices to obtain a very satisfying reproduction of
the experimental IP.

\begin{figure}[htbp]
\centerline{\includegraphics[width=0.8\linewidth]{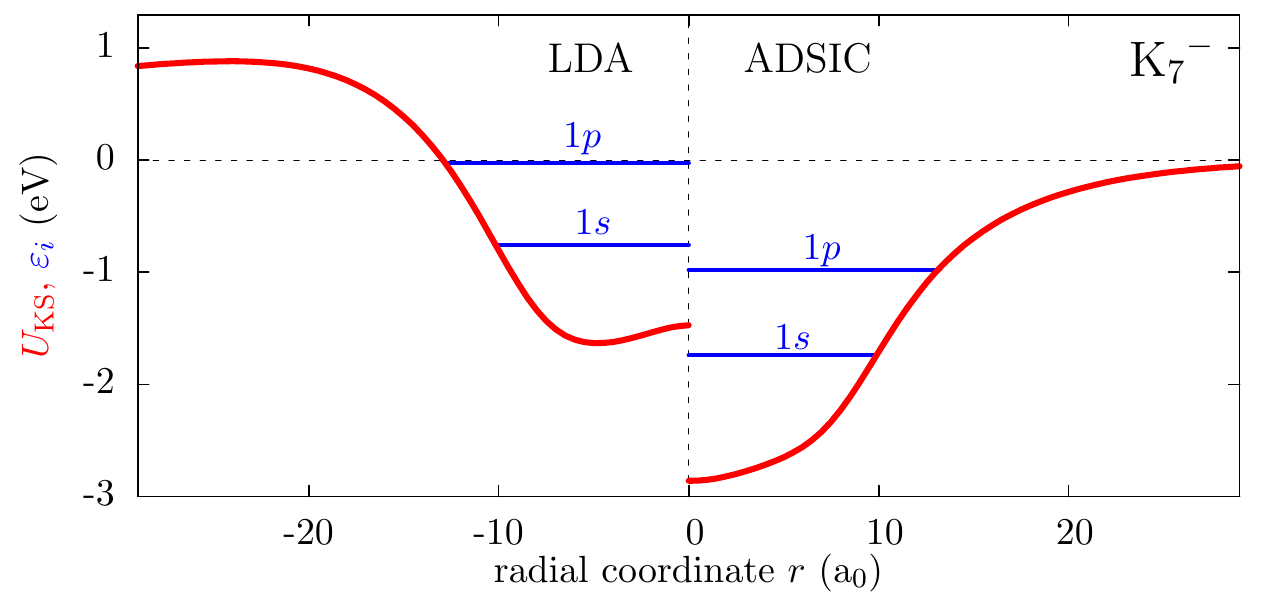}}
\caption{Kohn-Sham potential and single particles energies for ${{\rm K}_7}^{-}$
  described with spherical jellium background. Compared are
  results from LDA (left) with those from ADSIC (right).
\label{fig:K7m-SIC}
}
\end{figure}

Fig.~\ref{fig:K7m-SIC} demonstrates the effect of SIC in more detail.
As test case, we consider the cluster anion ${{\rm K}_7}^-$ where the
failure of LDA is particularly apparent. We use here a spherical soft
jellium ionic background, see Eq.(\ref{eq:softJ}), with a Wigner-Seitz
radius $r_s=5$ a$_0$ and surface parameter $\sigma=1.4$ a$_0$. The
cluster as a whole has a negative charge. Consequently, the total
Coulomb potential as it is used in LDA has an asymptotics
$\propto+{e}^2/r$ and produces a Coulomb barrier between inside and
outside. ADSIC, on the other hand, sees asymptotically the Coulomb
potential of all electrons minus the one which is departing. This is
the potential of a neutral system which converges exponentially to
zero from below. The different asymptotic potentials mostly lead to a
global shift of the s.p. energies, while the energy differences
between the occupied states are less affected. This global shift is
particurlarly disastrous for this anion. The system is hardly bound
with LDA, while ADSIC produces comfortable and realistic binding,
although weak.

\begin{figure}[htbp]
\centerline{\includegraphics[width=0.65\linewidth]{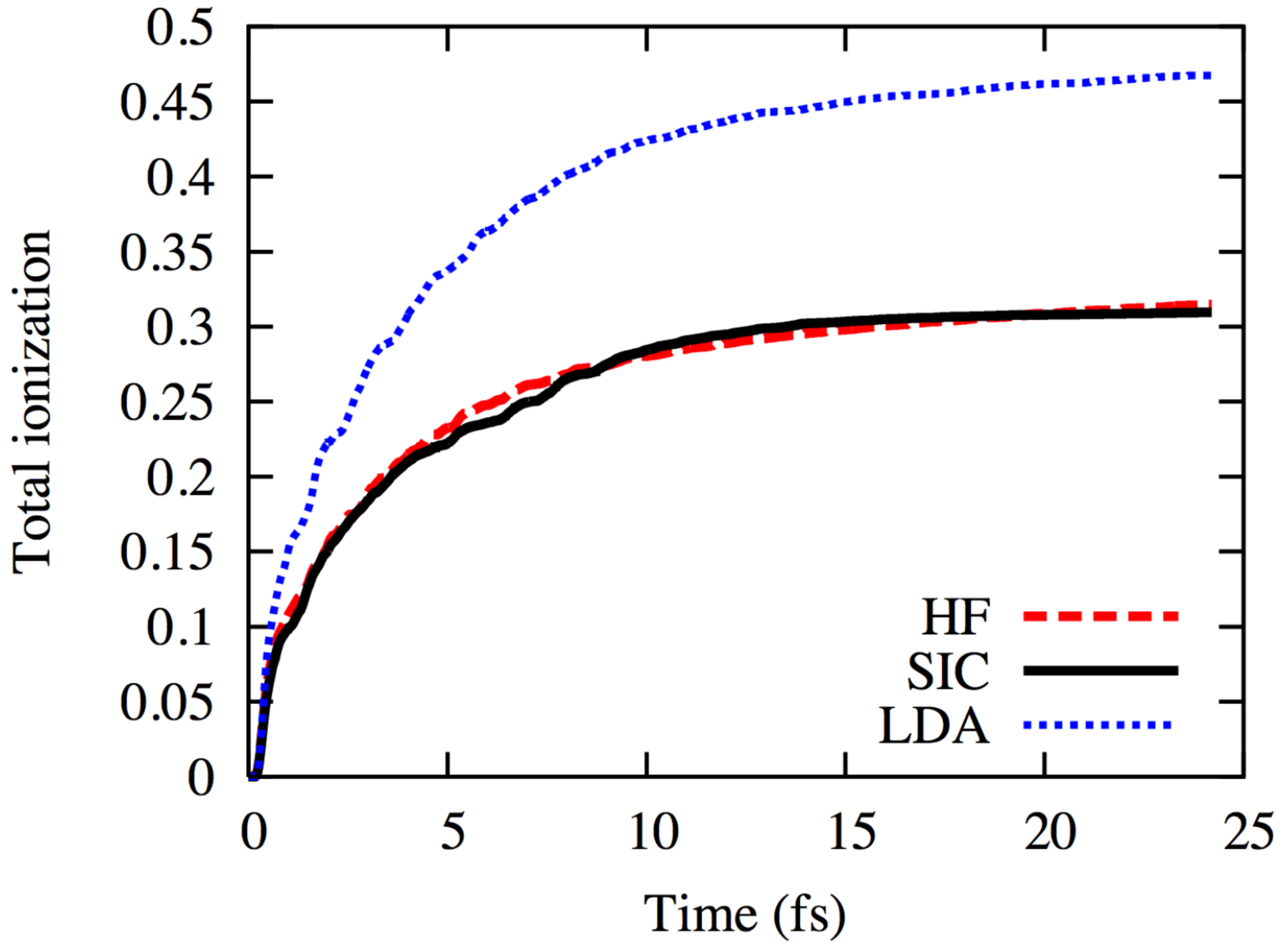}}
\caption{Time evolution of total ionization after an instantaneous boost for a 1D model of H$_2$
in triplet state, calculated in LDA (blue dots), Hartree-Fock (red dashes) and 2setSIC
(black full curve). Adapted from \cite{Mes08a}.
\label{fig:TD_2setSIC_1D}
}
\end{figure}

Finally, we consider an example of time-dependent SIC (TDSIC) solved
in the 2setSIC framework and directly analyze the ionization dynamics
of a molecule. To that end, we use a simple 1D model for a H$_2$ dimer
molecule with a smoothed Coulomb potential \cite{Hen98}. It is well
adapted for a consistent test of SIC \cite{Mes08a}. The two electrons
have aligned spins (triplet state) to make a non-trivial test of
SIC. We work at the level of ``exchange only'', so that the benchmark
becomes time-dependent Hartree-Fock (computed with exact exchange). To
test LDA consistently, a density functional for exchange has been
developed for this 1D model within LDA. This density functional is
also used as a basis for SIC \cite{Mes08a}. As SIC has a large impact
on the IP, we take the time evolution of ionization as a critical
test. A result for an instantaneous boost is shown in
Fig.~\ref{fig:TD_2setSIC_1D}.
The failure of TDLDA (blue dots) for this observable becomes
obvious. The IP is grossly underestimated and consequently, the
ionization is too high. The 2setSIC (full line) cures the problem
almost perfectly, as is visible in the excellent agreement with the
Hartree-Fock calculation (red dashes).

\subsection{Total ionization, PES and PAD in TDDFT}
\label{sec:observ_theo}

In this section, we discuss detailed observables from direct electron
emission. By direct emission, we mean those processes which are caused
without delay by the electronic excitation process. They dominate at
moderate excitations and short laser pulses with duration of some tens
of fs where the competing process, that is thermalization and
subsequent thermal emission, is less important. At this short time
scale, we can also often neglect explicit ionic motion.

\subsubsection{An example of PES and PAD as preview}
\label{sec:example_pespad}

\begin{figure}[htbp]
\centerline{\includegraphics[width=\linewidth]{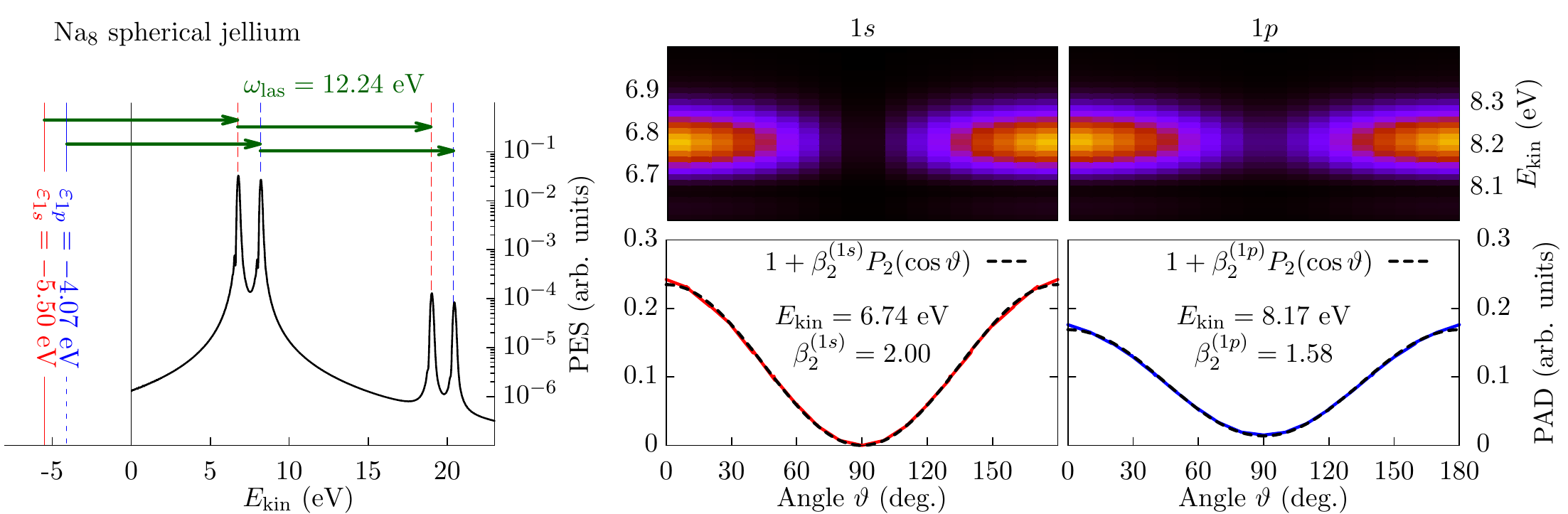}}
\caption{ Photoelectron spectrum (left), photoangular distributions
  (bottom middle and right) and combined PES/PAD (top middle and
  right) from Na$_8$ with spherical jellium background, see
  Eq. (\ref{eq:softJ}), using $r_s=3.65$ a$_0$ and $\sigma=1$ a$_0$
  after excitation with a linearly polarized laser pulse, see
  Eq.~(\ref{eq:Elaser}), of intensity $I=6.9\times10^{13}$W/cm$^2$,
  frequency $\omega_\mathrm{las}=12.24$~eV, and pulse duration
  $T_\mathrm{pulse}=90$ fs. The two occupied s.p. states lie at
  $\varepsilon_{1s}=-5.5$ eV and $\varepsilon_{1p}=-4.07$ eV. The
  total ionization is $N_\mathrm{esc}=0.003$ electron. The angle
  $\vartheta$ is measured with respect to the laser polarization.
\label{fig:example}
}
\end{figure}

As a preview, we show in Fig.~\ref{fig:example} PES and PAD for a
simple example, Na$_8$ with spherical jellium background. This system
has only two occupied levels $1s$ (twice degenerate) and $1p$
(six-fold degenerate) which simplifies the analysis.
The left panel shows the PES, that is the distribution of kinetic
energies of emitted electrons. Left of the vertical axis, the two
originally occupied s.p. states are indicated. One photon adds 0.9 Ry
energy and so places a peak at $-5.5+12.24$ eV, or $-4.07+12.24$ eV
respectively. The energy shift by the photon is indicated by
horizontal arrows. The peaks in the PES directly map the occupied
states. Further 12.24~eV higher, one sees another two peaks. These are
due to two-photon processes moving the electrons
$2\hbar\omega_\mathrm{las}$ up in energy. The two upper right panels
of Fig.~\ref{fig:example} show combined PES/PAD with an energy window
around the first (middle) and second (right) peaks of the PES. The
lower panels show the corresponding PAD for emission from the $1s$
states (middle) and $1p$ states (right). The PAD have a very simple
structure. This test case with closed
electron shells and spherical jellium background is spherical
throughout. The laser defines a preferred direction thus leaving axial
symmetry for the process. Therefore the PAD depend only on the angle
$\vartheta$ relative to the laser polarization axis. Moreover, they 
consist out of a constant
contribution plus a cos$^2$ profile. We will see later on that this is
the only possible structure for PAD from one-photon processes. The
figure also indicates the anisotropy $\beta_2$, as defined later in
Eq. (\ref{eq:anisotropy}), for each case. The PAD from a perfectly
spherical $1s$ state has the maximal possible value $\beta_2=2$, which
corresponds to strong alignment with the laser polarization and
vanishing emission perpendicular to it. The less symmetrical $1p$
states yield a somewhat lower anisotropy. This simple example already
demonstrates the richness of PES and PAD. It also shows that one needs
get acqaueinted with technical details to better understand the
content and behavior of both PES and PAD.

\subsubsection{Absorbing boundary conditions}
\label{sec:abso}

A grid representation naturally leads to reflecting or periodic
boundary conditions.  Reflection emerges for finite difference
schemes.  A representation of the kinetic energy by complex Fourier
transformation is associated with periodic boundary conditions where
flow leaving the box at one side is re-fed at the opposite
side. Both can lead to artifacts if a sizable
fraction of electronic flow hits the boundaries.  There are several
ways to solve the problem. The conceptually simplest approach is to
enhance the size of the numerical box. However, this is not a
realistic option as the expense grows cubically with the box length in
3D and quadratically in 2D. Very recently, a multi-grid method has
been proposed \cite{deGio12} which renders the use of enlarged boxes
feasible (although still at the edge of present days computer
capabilities). Perfect removal of escaping particles is achieved by or
exact boundary conditions \cite{Bou97,Man98a} which, again, are not
yet practicable in 3D calculations. Robust and efficient are absorbing
boundary conditions by an especially tailored imaginary potential
\cite{Nak05a} or by applying a mask function during time evolution
\cite{Kra92a}. The latter technique is particularly easy to implement
and has been widely used in the past. Its robustness and efficiency
allow one to develop advanced analyzing techniques on the grid as,
e.g., the computation of PES and PAD \cite{Rei06f}. A detailed
description and discussion of this approach and its proper choice of
numerical parameters is found in \cite{Rei06c}. In the following, we
will only address the mask technique for absorbing bounds.

\begin{figure}[htbp]
\centerline{\includegraphics[width=0.7\linewidth]{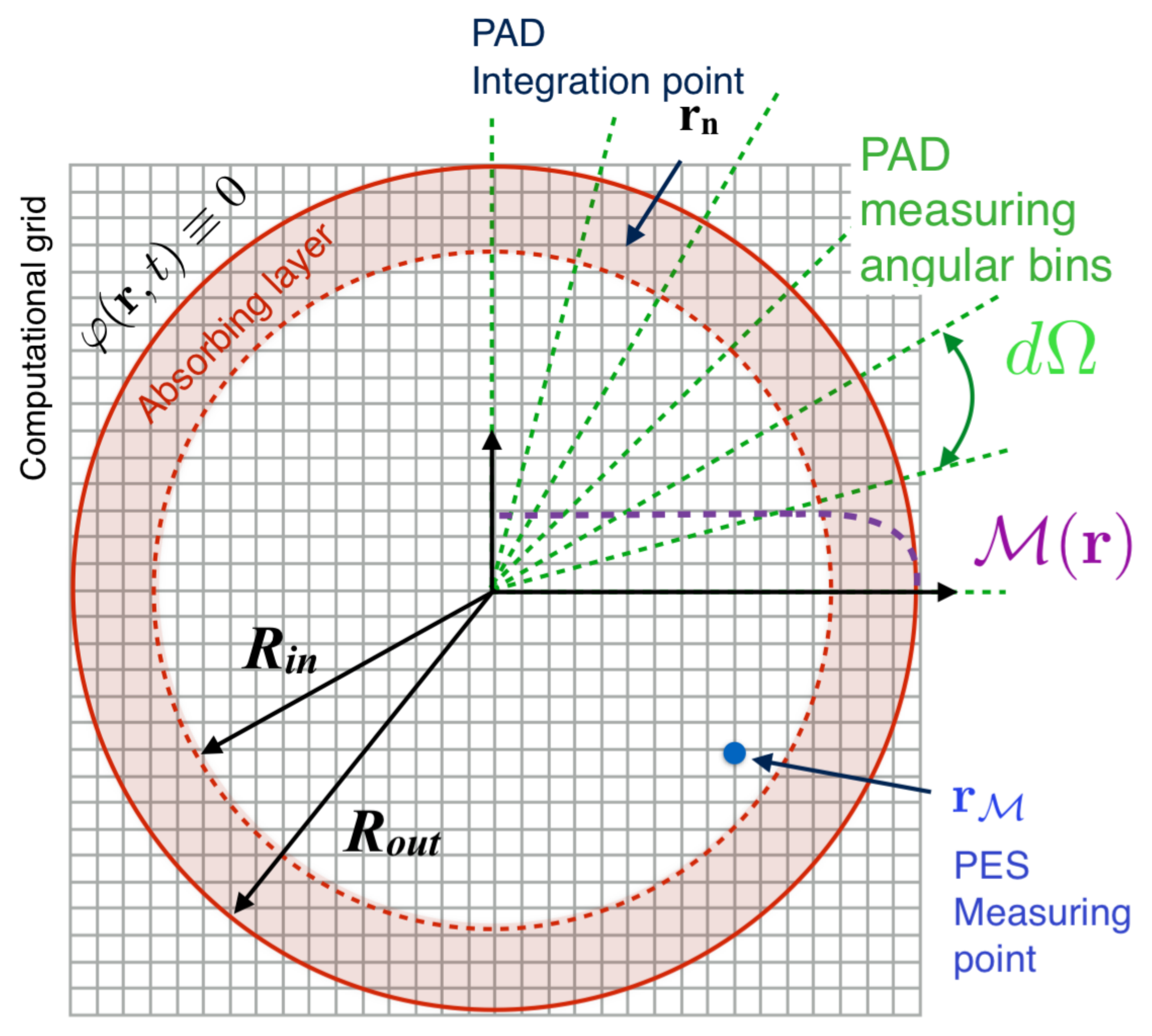}}
\caption{Schematic view of a coordinate-space grid with
absorbing bounds (ring zone), a sampling direction for accumulating PAD, and
measuring points $\mathbf{r}_\mathcal{M}$ for the PES.
 \label{fig:mask}}
\end{figure}

Fig.~\ref{fig:mask} sketches the implementation of absorbing boundary
conditions with computation of PES and PAD on a coordinate space grid.
Proper handling of electron emission requires absorbing boundary
conditions. These are indicated by the ring area in the figure
covering here 3 grid points in each direction (actual calculations
typically use 6--8 points.) The absorption is performed in each time
step as~:
\begin{subequations}
\begin{eqnarray}
  \varphi(\mathbf{r},t)
  &\longrightarrow&
  \tilde\varphi(\mathbf{r},t\!+\!\delta t)
  =
  \hat{\mathcal{U}}_\mathrm{KS}(t\!+\!\delta t,t)\, \varphi(\mathbf{r},t)
  \quad,
\label{eq:KSpart}\\
  \varphi(\mathbf{r},t\!+\!\delta t)
  &=&
  \mathcal{M}(\mathbf{r})\, \tilde\varphi(\mathbf{r},t\!+\!\delta t)
  \quad,
\label{eq:maskact}
\\
  \mathcal{M}(\mathbf r)
  &=&
  \left\{\begin{array}{lll}
  1 & \mbox{for} &|\mathbf{r}|<R_\mathrm{in}
  \quad,
  \\
  \displaystyle
 \cos\left(
  \frac{|\mathbf{r}|-R_\mathrm{in}}{R_\mathrm{out}-R_\mathrm{in}}\frac{\pi}{2}
 \right)^{\gamma_\mathcal{M}}
  &\mbox{for}&
  R_\mathrm{in}<|\mathbf{r}|<R_\mathrm{out}
  \quad,
  \\
  0 
  &\mbox{for}&
  R_\mathrm{out}<|\mathbf{r}|
  \quad.
  \end{array}\right.
\label{eq:mask}
\end{eqnarray}
\end{subequations}
First comes the standard step (\ref{eq:KSpart}) in terms of the TDLDA
(or TDSIC) propagator $\hat{\mathcal{U}}_\mathrm{KS}$, which yields
the intermediate wave function
$\tilde\varphi(\mathbf{r},t\!+\!\delta{t})$. This is followed by the
action in Eq.  (\ref{eq:maskact}) of the mask function $\mathcal{M}$
defined in Eq.(\ref{eq:mask}), which removes gradually any amplitude
towards the bounds. We use here a spherically symmetric mask. The
spherical profile is helpful to minimize griding artifacts when
computing angular distributions \cite{Poh04b}. The absorbing bounds
steadily reduce the norm of the wave functions from the inner mask
radius $R_\mathrm{in}$ to the outer one $R_\mathrm{out}$. The mask
technique is however not perfect. One will always encounter a small
amount of reflected flow, particularly for electrons with low kinetic
energy. One can minimize the back-flow by proper choice of the
exponent $\gamma_\mathcal{M}$ entering the mask profile, see
Eq.~(\ref{eq:mask}). This depends, however, on the actual numerics
(number of absorbing points, size of time step), for a detailed
discussion see \cite{Rei06c}. Typical values of $\gamma_\mathcal{M}$
are of order $1/8$.

\subsubsection{Ionization}
\label{sec:ionization}

The first observable which can be computed using working absorbing
boundaries is the total ionization, i.e. the number of escaping
electrons $N_\mathrm{esc}$. This can be computed simply from the, now
decreasing, single-particle norms as~:
\begin{equation}
  N_\mathrm{esc}(t)
  =
  \sum_{i=1}^N N_{\mathrm{esc},i}(t)
  \quad,\quad
  N_{\mathrm{esc},i}(t)
  =
  1- \langle \varphi_i(t)|\varphi_i(t)\rangle
  \quad.
\label{eq:nesc}
\end{equation}
This shows that we have access to even more than the mere net
ionization. Indeed each $N_{\mathrm{esc},i}$ yields the depletion of
s.p. state $i$ separately. Both, total ionization and detailed level
depletion are very instructive observables, see e.g.
Fig.~\ref{fig:pes-deple_c60}.

\subsubsection{Photoemission angular distributions (PAD)}
\label{sec:pad}

The angular distributions $\textrm d\sigma/\textrm d\Omega
(\vartheta,\phi)$ are evaluated in angular segments labeled by the
azimuthal angle $\vartheta $ and the polar angle $\phi$. The reference
frame for these two angles is usually one axis, called the $z$ axis,
of the Cartesian 3D grid designed to be identical with the laser
polarization axis, for details see Sec.~\ref{sec:orientaver}.
We collect all probability which was removed by the absorption step
(\ref{eq:maskact}) and accumulate it.  A straightforward collection of
grid points in a segment tends to produce noisy results because the
number of grid points per segment fluctuates. We therefore associate
with each grid point a smoothing function $\mathcal{W}(\mathbf{r})$
which distributes the strength over a vicinity of order of grid
spacing.  This suffices to produce acceptable smooth distributions.
The PAD is thus computed as~:
\begin{subequations}
\label{eq:PADfixed}
\begin{eqnarray}
  \mathcal{A}(\vartheta,\phi)
  &=&
  \sum_{i=1}^N\mathcal{A}^{(i)}(\vartheta,\phi)
  \quad,
\\
  \mathcal{A}^{(i)}(\vartheta,\phi)
  &=&
  \sum_{\mathbf{n}\in\mbox{abs.b.c.}}
  \int \textrm dr\,r^2\,\mathcal{W}(r\mathbf{e}_r-\mathbf{r}_\mathbf{n})
   \, n_{\mathrm{esc},i}(\mathbf{r}_\mathbf{n})
  \quad,
\\
  \mathcal{W}(\mathbf{r})
  &=&
  \frac{\mbox{max}(\Delta x-|x|,0)}{\Delta x} \, 
  \frac{\mbox{max}(\Delta y-|y|,0)}{\Delta y} \,
  \frac{\mbox{max}(\Delta z-|z|,0)}{\Delta z}
  \quad,
\\
  n_{\mathrm{esc},i}(\mathbf{r}_\mathbf{n})
  &=&
  \int \textrm dt\,\left|\tilde{\varphi}_i(\mathbf{r}_\mathbf{n},t)\right|^2
  \left[1-\mathcal{M}(\mathbf{r}_\mathbf{n},t)\right]
  \quad,
\end{eqnarray}
\end{subequations}
where $\mathbf{e}_r=\left(\sin\vartheta\cos\phi,
\sin\vartheta\sin\phi,\cos\vartheta\right)$ is the unit vector in the
direction of the wanted angles. The smoothing is done by simple tent
functions which comply with the integration rule used in the
normalization.  The angular segments in Fig.~\ref{fig:mask} try to
symbolize this smoothing which collects (weighted) information in the
vicinity of a ray. The above recipe applies to state specific PAD
$\mathcal{A}_i$ as well as the total PAD $\mathcal{A}$.
Alternatively, one uses a cross-section-like notation $\textrm
d\sigma/\textrm d\Omega$ for PAD. The present choice is more flexible
for the presentation of orientation averaging, see Sec.~\ref{sec:orientaver}.

\subsubsection{Photoemission spectra (PES)}
\label{sec:pes}

The PES can be deduced from the temporal phase oscillations of the
wave functions at measuring points $\mathbf{r}_\mathcal{M}$ close
to the absorbing bounds.  This technique had been introduced in
\cite{Poh00,Poh03a}.  A detailed discussion of the method and
extension to strong laser fields is found in \cite{Din13a}. We
summarize it here briefly.  To explain the computation of PES, we
first confine the considerations to 1D in order to keep things simple
and extend it finally to the general 3D case. The measuring point is
thus denoted for a while $z_\mathcal{M}$.  Starting point is the
solution of the Schr\"odinger equation for {one} electron in a
laser field in velocity-gauge, see Eq. (\ref{eq:lasfield-v}) in
section \ref{sec:las_pulse}. In this gauge, the electronic wave
function $\varphi^{(v)}$ at the sampling point $z_\mathcal{M}$ reads
\begin{subequations}
\label{eq:solve-wfall}
\begin{eqnarray}
  \varphi^{(v)}(z_\mathcal{M},t)
  &=&
  \int\frac{\textrm dk}{\sqrt{2\pi}}\,
  e^{\mathrm{i}k z_\mathcal{M}} \, \widetilde{\varphi}^{(v)}_0(k) \, 
  e^{-\mathrm{i}\omega_kt+\mathrm{i}k\delta q-\mathrm{i}\delta\Omega}
  \quad,
\label{eq:solve-wf}
\\
  \omega_k
  &=&
  \frac{k^2}{2}
  \quad\longleftrightarrow\quad
  k
  =\sqrt{2\omega_k}
  \quad,
\label{eq:solve-k-w}
\\
  \delta q(t)
  &=&
  E_0 \int_{0}^t\mathrm dt'\,F(t')
  \quad,
\label{eq:delq}\\
  \delta\Omega(t)
  &=&
  \frac{E_0^2}{2}\int_{0}^t\mathrm dt'\,F(t')^2
  \quad,
\label{eq:delOmega}
\end{eqnarray}
\end{subequations}
where $F(t)$ is the time integrated laser pulse introduced in
Eq.~(\ref{eq:gaugeF}).  It should also be noted that
Eq.~(\ref{eq:solve-k-w}) exploits the fact that $z_\mathcal{M}$ is
near the absorbing bounds, such that only outgoing waves with $k>0$
pass through this point.  The aim is to deduce the momentum
distribution $\left|\widetilde{\varphi}^{(v)}_0(k)\right|$ from the
sampled $\varphi^{(v)}(z_\mathcal{M},t)$. This is straightforward for
weak laser fields where we can neglect $\delta q$ and $\delta\Omega$.
For then, a time to frequency Fourier transformation of
$\varphi^{(v)}$, namely
$\int\mathrm{dt}\,e^{\mathrm{i}\omega{t}}\varphi^{(v)}(z_\mathcal{M},t)$,
produces in the right-hand side of Eq. (\ref{eq:solve-wf}) a
$\delta(\omega-\omega_k)$ which, in turn, reduces the $k$ integration
to the point $k=\sqrt{2\omega}$ thus delivering the wanted
$\widetilde{\varphi}^{(v)}_0(\sqrt{2\omega})$.

Trying to directly apply this time-frequency transformation when the
field strength is not small runs into trouble due to the non-trivial
time dependencies induced by the factors $\delta q(t)$ and
$\delta\Omega(t)$ in Eqs.~(\ref{eq:solve-wfall}).  A simple solution
is to counter-weight the disturbing phase factors by a
phase-correction factor $e^{\mathrm{i}\Phi}$ before the
transformation. We thus consider
\begin{eqnarray*}
  \int\frac{\textrm dt}{\sqrt{2\pi}}\,e^{\mathrm{i}\omega t
              -\mathrm{i}\sqrt{2\omega}\delta q
              +\mathrm{i}\delta\Omega}\varphi^{(v)}(z_\mathcal{M},t)
  &=&
  \int\frac{\textrm dk}{2\pi}\int \textrm dt\,
  e^{\mathrm{i}k z_\mathcal{M}} \, \widetilde{\varphi}^{(v)}_0(k) \, 
  e^{\mathrm{i}(\omega-\omega_k)t-\mathrm{i}(\sqrt{2\omega}-k)\delta q}
\\
  &\approx&
  \int{\textrm dk}\,\delta(\omega-\omega_k)
  e^{\mathrm{i}k z_\mathcal{M}} \, \widetilde{\varphi}^{(v)}_0(k) 
\\
  &=&
  e^{\mathrm{i}\sqrt{2\omega}z_\mathcal{M}} \,
  \widetilde{\varphi}^{(v)}_0(\sqrt{2\omega}) 
  \quad.
\end{eqnarray*}
We finally evaluate the PES yield $\mathcal{Y}_\mathcal{M}$ at
measuring point $z_\mathcal{M}$ by~:
\begin{equation*}
  \mathcal{Y}_\mathcal{M}(E_\mathrm{kin})
  \propto
  \left|\widetilde{\varphi}^{(v)}_0(\sqrt{2\omega})\right|^2
  =
  \left|
  \int\frac{dt}{\sqrt{2\pi}}\,e^{\mathrm{i}\omega t
              -\mathrm{i}\sqrt{2\omega}\delta q
              +\mathrm{i}\delta\Omega}\varphi^{(v)}(z_\mathcal{M},t)
  \right|^2
  \quad,\quad
  \omega\equiv E_\mathrm{kin}
  \quad.
\end{equation*}
The approximation here consists in assuming that the time integration,
although complicated by the time profile in $\delta q(t)$, will still
deliver $\omega\approx k^2/2$ and thus
$e^{-\mathrm{i}(\sqrt{2\omega}-k)\delta q}\approx 1$. This
approximation is valid anyway for weak fields. It extends the
applicability of the time-frequency transformation to stronger fields.
Fourier transformation of a phase-augmented wave function thus allows
one to deduce the wanted momentum amplitude
$\widetilde{\varphi}^{(v)}_0$ for a wide range of field strengths.
The recipe may fail, however, for very strong fields where the
temporal variation of $(\sqrt{2\omega}-k)\delta q$ dominates over
$(\omega-\omega_k)t$.

For the extension to 3D, we have to take into account that there is a
whole vector of $\mathbf{k}$ rather than just two directions
$k=\pm\sqrt{2\omega}$. We exploit the fact that the 3D analyzing point
$\mathbf{r}_\mathcal{M}$ is close to the absorbing bounds (no
reflection) and sufficiently far from the emitting zone. Thus the
prevailing outgoing momentum $\mathbf{k}$ at this point has the
direction of $\mathbf{r}_\mathcal{M}$, i.e.
\begin{equation}
  \mathbf{e}_k
  =
  \frac{\mathbf{k}}{k}
  =
  \frac{\mathbf{r}_\mathcal{M}}{r_\mathcal{M}}
  =
  \mathbf{e}_\mathcal{M}
  \quad.
\end{equation}
The frequency-momentum relation  Eq. (\ref{eq:solve-k-w}) is then
generalized to $\mathbf{k}=+\mathbf{e}_\mathcal{M}\sqrt{2\omega}$
and the PES yield at $\mathbf{r}_\mathcal{M}$ is identified as
\begin{equation}
  \mathcal{Y}_\mathcal{M}(E_\mathrm{kin})
  \propto
  \left|
  \int\frac{\textrm dt}{\sqrt{2\pi}}\,e^{\mathrm{i}\omega t
              -\mathrm{i}\sqrt{2\omega}\delta q
              +\mathrm{i}\delta\Omega}\varphi^{(v)}(\mathbf r_\mathcal{M},t)
  \right|^2
  \quad.
\label{Eqn:PA3D}
\end{equation}
This is the straightforward 3D generalization of the 1D formula
above.  The phase
$\Phi(t)=-\mathrm{i}\sqrt{2\omega}\delta{q}(t)+\mathrm{i}\delta\Omega(t)$ is
negligible in case of weak fields $E_0\ll\sqrt{\omega/2}$. It extends
the applicability of the method to stronger fields.

A word is in order about the choice of gauge. The above evaluation of
PES is formulated in velocity gauge because the exact solution of the
Schr\"odinger equation in the laser field is much simpler in this
gauge, see Eq. (\ref{eq:solve-wfall}). If one prefers to perform
numerical calculations in space gauge, one just has to apply the
transformations Eq.~(\ref{eq:gaugeF}) and Eq. (\ref{eq:gaugepsi})
before using Eq.~(\ref{Eqn:PA3D}).

\begin{figure}[htbp]
\centerline{\includegraphics[width=0.99\linewidth]{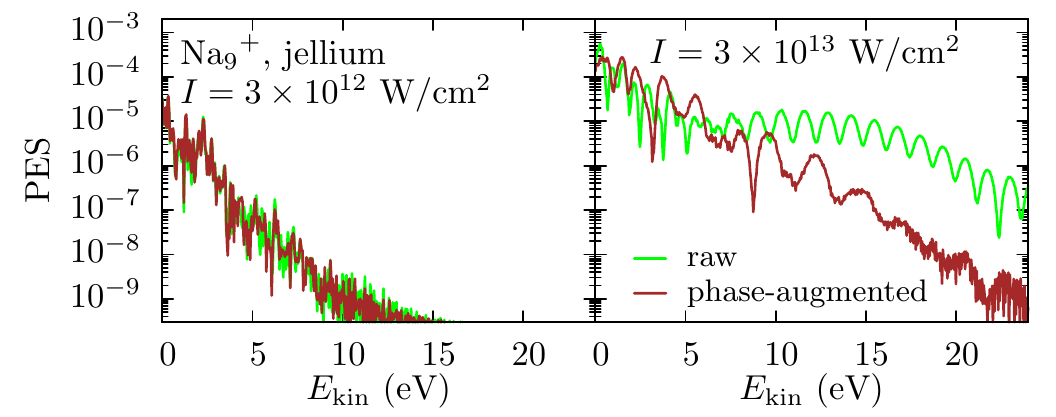}}
\caption{Ionization properties for ${{\rm Na}_9}^+$ with jellium background
    under the influence of a laser pulse having
    frequency $\omega_\mathrm{las}=1.4$~eV, 
    pulse length $T_\mathrm{pulse}=12$ fs, and
    intensity as indicated in each panel. 
    Spherical absorbing bounds were used covering at least 16 grid points.
    The ``phase-augmented" results (brown lines) are obtained from Eq.~(\ref{Eqn:PA3D}),
while the ``raw'' results (light green lines) only use the time-frequency Fourier transform of 
$\varphi^{(v)}(\mathbf r_\mathcal{M},t)$,  see Eq.~(\ref{eq:solve-wf}) in the 1D case.
\label{fig:gauge}
}
\end{figure}

The effect of the phase correction in the recipe Eq.~(\ref{Eqn:PA3D}) is
demonstrated in Fig.~\ref{fig:gauge}. 
The field strength for $I=3\times 10^{12}$W/cm$^2$ is obviously
sufficiently small such that there is practically no effect from the
phase correction. The stronger field with $I=3\times10^{13}$W/cm$^2$
clearly needs the phase correction and including it yields still a
reliable PES.  Results for significantly larger field strengths cannot
be trusted.

\subsubsection{Alternative routes to compute PES/PAD}

\begin{figure}[htbp]
\centerline{\includegraphics[width=\linewidth]{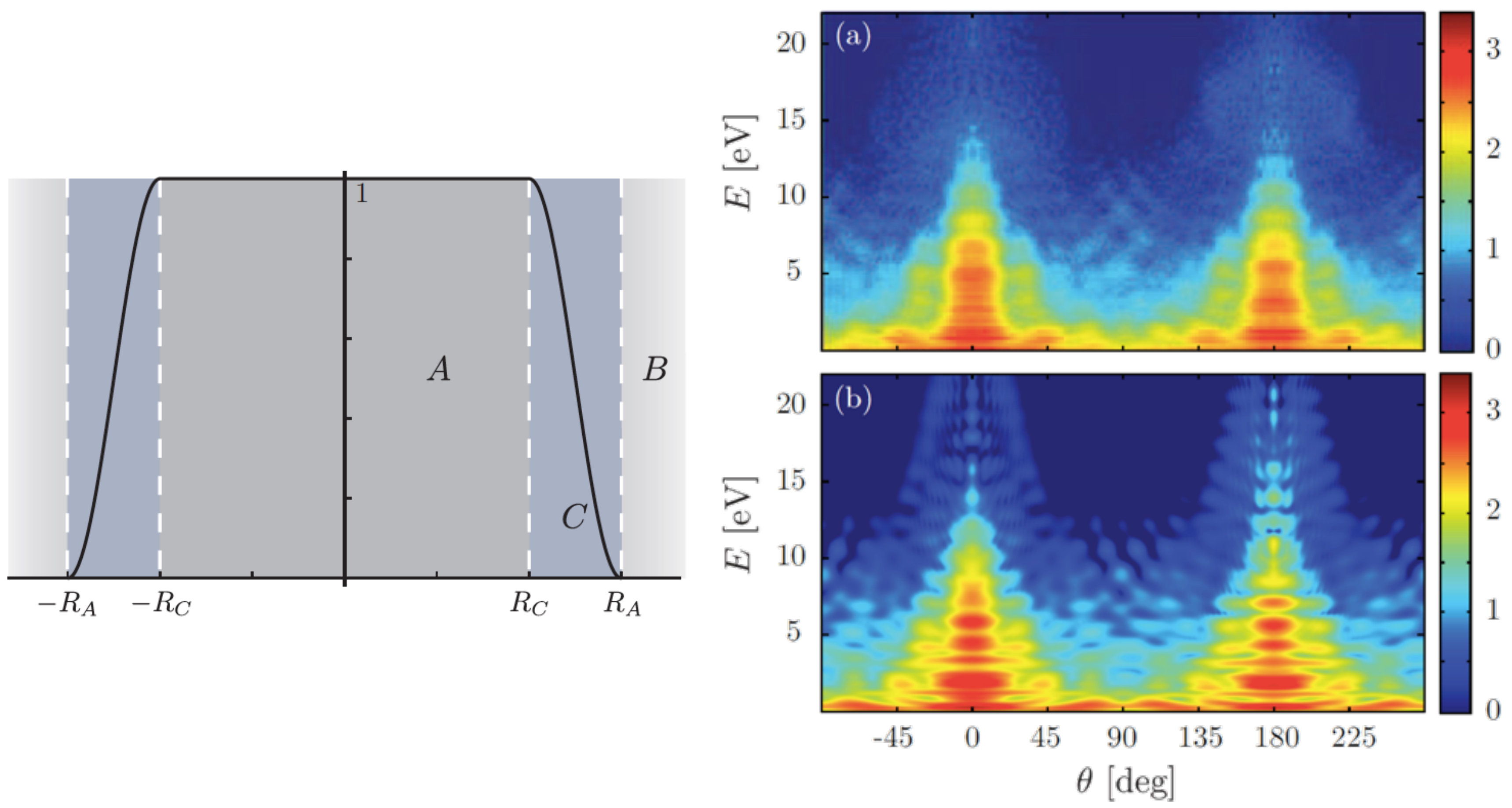}}
\caption{Example of calculated combined PES/PAD.  Left~: 1D view of
  the mask function applied in region $A$ where time-dependent
  Kohn-Sham equations are solved numerically in real-space, while wave
  functions are propagated analytically in region $B$. Region $C$
  serves a a buffer region where wave functions, evaluated in $A$,
  overlap with those evaluated in $B$~\cite{deGio12}.  Right~:
  Experimental (top) combined PES/PAD~\cite{Gaz11} and theoretical
  ones (bottom)~\cite{deGio12} in logarithmic scale for N$_2$
  molecules in a 6 cycle infrared laser pulse ($\lambda_{\rm
    las}=750$~nm, $I = 4.3 \times 10^{13}$ W/cm$^2$). The theoretical
  PES/PAD is averaged over four orientations of the N$_2$ molecule.}
\label{fig:deGio}
\end{figure}

An alternative theoretical approach to evaluate PES/PAD has been
proposed very recently in \cite{deGio12}. The principle is
schematically explained in the left panel of Fig.~\ref{fig:deGio}.
This method is very similar to ours in the sense that in region $A$,
the TDDFT equations are solved in a real-coordinate space grid, with
the application of a mask function $\mathcal{M}(\mathbf r)$ in region $C$. The
new feature comes with region $B$ where Volkov states are analytically
propagated in momentum space to account for the
interaction with the laser field. More precisely, state $i$ is
described by the following wave functions $\varphi_i$~:
\begin{subequations}
\begin{eqnarray}
  \varphi_{A,i}(\mathbf r, t) 
  &=& 
  \eta_{A,i}(\mathbf r, t) + \eta_{B,i}(\mathbf r, t)\quad,
\\
  \tilde{\varphi}_{B,i}(\mathbf p, t) 
  &=& 
  \tilde{\xi}_{A,i}(\mathbf p, t) + \tilde{\xi}_{B,i}(\mathbf p, t) 
  \quad,
\end{eqnarray}
\end{subequations}
where
\begin{subequations}
\begin{eqnarray}
  \eta_{A,i}(\mathbf r, t) 
  &=&
  \mathcal{M}(\mathbf r)\, U(t, t')\, \varphi_{A,i}(\mathbf r, t')
  \quad, 
\label{eq:etaA}\\
  \eta_{B,i}(\mathbf r, t) 
  &=&
  \mathcal{M}(\mathbf r) \int \frac{\textrm d\mathbf p}{(2\pi)^{d/2}} 
  \, e^{i \mathbf p \cdot \mathbf r}
  \, U_V (t, t') \, \tilde{\varphi}_{B,i}(\mathbf p, t) 
  \quad, 
\label{eq:etaB}\\
  \tilde{\eta}_{A,i}(\mathbf p, t) 
  &=& 
  \int \frac{\textrm d\mathbf r}{(2\pi)^{d/2}}, e^{-i \mathbf p \cdot \mathbf r}
  \, \left[1 - \mathcal{M}(\mathbf r)\right] \, U(t, t')\,
  \varphi_{A,i}(r, t') 
  \quad, 
\label{eq:tilde_etaA}\\
  \tilde{\xi}_{B,i}(\mathbf p, t) 
  &=&
  U_V (t, t') \, \tilde{\varphi}_{B,i}(\mathbf p, t') 
  −
  \int \frac{\textrm d\mathbf r}{(2\pi)^{d/2}} \, 
  e^{-i \mathbf p \cdot \mathbf r} \, \eta_{B,i}(\mathbf r, t) 
  \quad. 
\label{eq:tilde_etaB}
\end{eqnarray}
\end{subequations}
In Eqs.~(\ref{eq:etaA}) and (\ref{eq:tilde_etaA}), $U(t,t')$ stands
for the propagator from time $t'$ to $t$ with the full Hamiltonian
including external fields, while in Eqs.~(\ref{eq:tilde_etaA}) and
(\ref{eq:tilde_etaB}), only the laser field enters the Volkov time
propagator $U_V(t,t')$. Finally the momentum distribution of the
photoelectrons is approximated to~:
\begin{equation}
\frac{\textrm d \sigma}{\textrm d\Omega_\textrm p} \simeq \sum_i 
\tilde{\varphi}_{B,i}(\mathbf p, t\rightarrow \infty) \quad . 
\label{eq:deGio_distrib}
\end{equation}
This procedure has been applied to N$_2$ irradiated by a 6 cycle laser
pulse of wavelength 750~nm and intensity of $4.3 \times 10^{13}$
W/cm$^2$, with an averaging over 4 orientations ($0^\circ$,
$30^\circ$, $60^\circ$, $90^\circ$, see right bottom panel of
Fig.~\ref{fig:deGio}. The grid spacing in region $A$ of radius
$R_A=35$~a$_0$ is 0.38~a$_0$, and the buffer region has a radius
$R_C=25$~a$_0$. These results are compared with experimental
measurements~\cite{Gaz11} (see top right panel). The latter show that
photoelectrons are preferentially emitted parallel to the laser
polarization axis (0 and 180$^\circ$). Note that the signal at
$0^\circ$ is slightly different from that at $180^\circ$. This can be
explained by the fact that the laser pulse is so short that the
symmetry along the laser polarization axis is
broken~\cite{deGio12,Mil06}. The comparison with the theoretical
PES/PAD is fairly good, especially at high kinetic energies. More
discrepancies are observed at low kinetic energies, probably because
of the limited size of the numerical box.

\subsection{Orientation Averaging PAD (OAPAD)}
\label{sec:orientaver}

The PAD obtained by Eqs. (\ref{eq:PADfixed}) determine the distribution
for a fixed orientation of the molecule/cluster relative to the laser
polarization axis. However measurements of free clusters are usually
done in the gas phase covering an isotropic distribution of cluster
orientations. There are techniques for aligning molecules by strong
laser pulses, for a review see \cite{Sta03aR} and for proposals using
chains of pulses see \cite{Pab10a,Pab10b,Rei05f}. Nonetheless, these
have not yet been used in connection with measuring PAD on
clusters. Thus we have to perform orientation averaging of the TDDFT
results to establish contact with existing measurements. Efficient
techniques for orientation averaging of clusters have been developed
in \cite{Wop10a,Wop10b}. We summarize them here briefly.

\subsubsection{Direct averaging scheme}
\label{sec:direct_oapad}

\begin{figure}[htbp]
\centerline{\includegraphics[width=0.35\linewidth]{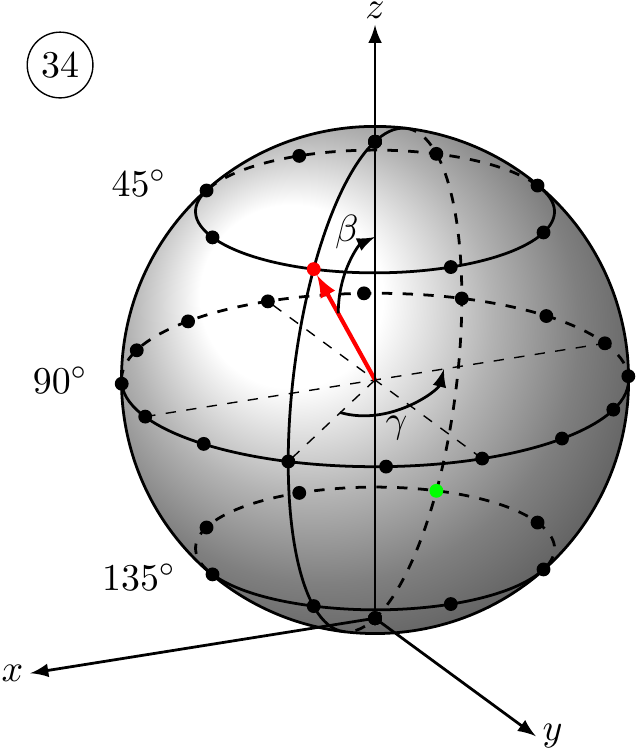}}
\caption{Example of a set of orientation points (black dots)
  used in direct orientation averaging.  The cluster is first rotated
  by $\gamma$ about the $z$-axis (laser polarization) and then by
  $\beta$ about the $y$-axis of the laboratory frame.
\label{fig:orient_av}
}
\end{figure}

The cross-section detailed in Eqs.~(\ref{eq:PADfixed}) stems from a
TDDFT calculation with one fixed configuration in which the cluster
orientation relative to laser polarization is known. What we are
looking for is the average of the cross-section over all possible
cluster orientations with equal weight.  For its evaluation, we
distinguish the laboratory frame and the cluster frame (in which all
quantities are primed). The laboratory frame is defined by the laser
polarization axis, such that the polarization vector points along the
$z$-axis of the 3D Cartesian coordinate system,
i.e. $\mathbf{e}_\mathrm{pol}=\mathbf{e}_z$. The observed emission
angles $(\vartheta,\phi)$ are defined with respect to this laboratory
frame, where $\vartheta$ is the angle with respect to the $z$-axis and
$\phi$ the angle in the $x$-$y$-plane.  A cluster has three principle
axes. The cluster orientation is defined by the Euler angles
$(\alpha,\beta,\gamma)$ of these three axes with respect to the
laboratory frame \cite{Edm57}.  Thus we deal, in fact, with an
ensemble of PAD
$\mathcal{A}^{(i)}(\vartheta,\phi;\alpha,\beta,\gamma)$ of the same
cluster with different orientations.  The orientation averaged
one-photon PAD for emission from the s.p. state $\varphi_i$ then
becomes
\begin{eqnarray}
 \overline{\mathcal{A}^{(i)}}(\vartheta,\phi)
  &=&
  \!\int\! \frac{\rm d\alpha \, \rm d\!\cos\beta \, \mathrm{d}\gamma}{8\pi^2}
  \mathcal{A}^{(i)}(\vartheta,\phi;\alpha,\beta,\gamma)
  \approx
  \!\sum_{m=1}^M\lambda_m \,
  \mathcal{A}^{(i)}(\vartheta,\phi;\alpha_m,\beta_m,\gamma_m)
  \;.
\label{eq:aver-orient}
\end{eqnarray}
where the sum from 1 to $M$ runs over a discrete set of points on
which the integral will be discretized (see below). Since spherical
absorbing boundaries are used, the rotation by $\alpha$ about the
laser axis can be done a posteriori and does not require any
additional TDDFT calculation.  This leaves averaging over $\beta$ and
$\gamma$ which is approximated by a finite-element representation of
the integral, see rightmost part of Eq.~(\ref{eq:aver-orient}). The
chosen values for $\beta_m$ and $\gamma_m$ can be illustrated on a
unit sphere. Figure \ref{fig:orient_av} shows a sampling over 34
orientation points.
The weight factors $\lambda_m$ are determined by cutting the surface
of the unit sphere into segments $\mathcal{S}_m$ around unit direction
$\mathbf{e}_m$ of element $m$. The $\mathcal{S}_m$ is defined as the
collection of points on the unit sphere which are closer to
$\mathbf{e}_m$ than to any other $\mathbf{e}_{m'}$. The summation
weights are then simply the areas of $\mathcal{S}_m$ divided by the
area of the whole sphere,
i.e. $\lambda_m=\mbox{Area}(\mathcal{S}_m)/(4\pi)$.

\subsubsection{OAPAD from one-photon processes in the perturbative regime}
\label{sec:perturbative_oapad}

The direct averaging scheme is conceptually simple and can be applied
in any situation. However, it requires a considerable amount of
reference orientations (segments) to achieve sufficiently reliable
results. A great simplification can be worked out for one-photon
processes in the perturbative regime. In that case, one can deduce,
using perturbation theory formally, that the rotated PAD can be
represented as~:
\begin{eqnarray}
 \mathcal{A}^{(i)}(\vartheta\phi,\alpha\beta\gamma)
 =
 \sum\limits_{\mu\mu',lmm'}
  D_{\mu'0}^{(1)*}(\alpha\beta\gamma)\,
  D_{\mu 0}^{(1)}(\alpha\beta\gamma)\,
  D_{m'm}^{(l)}(\alpha\beta\gamma)   \,
  a_{\mu\mu',lm'}^{(i)} Y_{lm}(\vartheta\phi)
  \;.
\label{eq:Arepresent}
\end{eqnarray}
in terms of a couple of expansion coefficients
$a_{\mu\mu',lm'}^{(i)}$, rotation functions $D_{\mu\mu'}^{(l)}$
\cite{Edm57}, and spherical harmonics. In this form, the integration
over Euler angles $(\alpha,\beta,\gamma)$ can be worked out
analytically (for details, see \cite{Wop10a,Wop10b}), yielding finally
the Orientation Averaged PAD as
\begin{eqnarray}
  \overline{\frac{{\rm d}\sigma_{i}}{{\rm d}\Omega}}
  \equiv
  \overline{\mathcal{A}^{(i)}}(\vartheta\phi)
  &=&
  C_{0}^{(i)}Y_{00}(\vartheta\phi)
  +
  C_{2}^{(i)}Y_{20}(\vartheta\phi)
  \quad,
\label{eq:fincrosssect-i}
\\
  C_{0}^{(i)}
  &=&
  \frac{1}{3}\sum_\mu a_{\mu\mu,00}^{(i)}
  \quad,
\label{eq:fincrosssect-i2}
\\
  C_{2}^{(i)}
  &=&
  \sum_\mu a_{\mu\mu',2\,\mu-\mu'}^{(i)}(-1)^\mu
  \times\left(\begin{array}{ccc}
   1 & 1 & 2 \\ 0 & 0 & 0
  \end{array}\right)
  \left(\begin{array}{ccc}
   1 & 1 & 2 \\ -\mu & \mu' & \mu\!-\!\mu'
  \end{array}\right)
  \quad.
\label{eq:fincrosssect-i3}
\end{eqnarray}
The corresponding total PAD is obtained by summing over the s.p. PAD,
i.e. $\overline{\mathcal{A}}=\sum_i\overline{\mathcal{A}^{(i)}}$. It
remains to determine the expansion coefficients
$a_{\mu\mu',lm'}^{(i)}$. Fortunately, there are only very few, that is
three for $l=0$ and six for $l=2$. They can be determined from the PAD
in only six properly chosen orientations because the different $l$ can
be produced from one PAD, see Eqs.~(\ref{eq:Arepresent}).  The six
reference orientations should be chosen such that
Eqs.~(\ref{eq:Arepresent}) can be solved in a stable manner as a linear
equation for the $a_{\mu\mu',lm'}^{(i)}$. A recommended set is
$\mathbf{e}^{(1)}=\mathbf{e}_x$,
$\mathbf{e}^{(2)}=\mathbf{e}_y$,
$\mathbf{e}^{(3)}=\mathbf{e}_z$,
$\mathbf{e}^{(4)}=(\mathbf{e}_x\!+\!\mathbf{e}_y)/\sqrt{2}$,
$\mathbf{e}^{(5)}=(\mathbf{e}_x\!+\!\mathbf{e}_z)/\sqrt{2}$,
and
$\mathbf{e}^{(6)}=(\mathbf{e}_y\!+\!\mathbf{e}_z)/\sqrt{2}$.

\begin{figure}[htbp]
\centerline{\includegraphics[width=\linewidth]{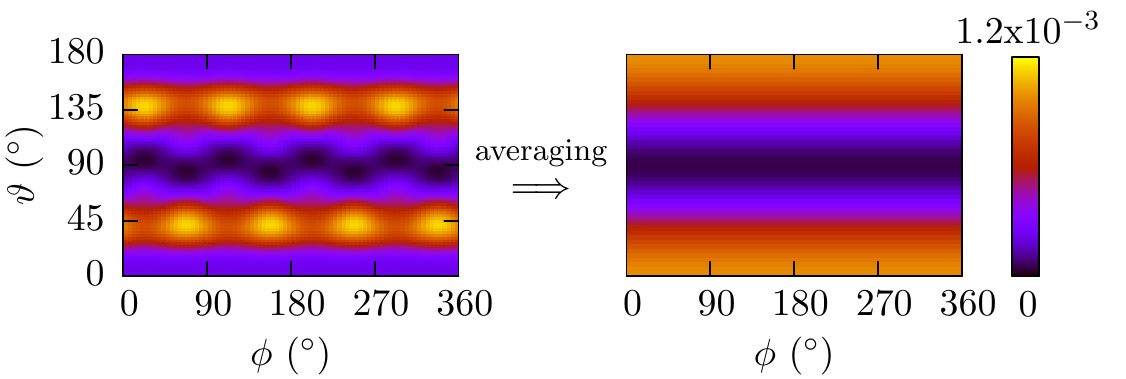}}
\caption{Total PAD $\mathcal{A}(\vartheta,\phi)$ for Na$_8$ with
  explicit ionic background for one fixed orientation (left panel) and
  orientation-averaged (right panel), after laser excitation with
  frequency $\omega_\mathrm{las}=7.5$~eV, intensity
  $I=10^{11}$W/cm$^2$, and pulse length with $T_{\rm pulse}=60$~fs. The emission
  angles $\vartheta$ and $\phi$ are measured with respect to the laser
  polarization.
\label{fig:orient_av2}
}
\end{figure}

The effect of orientation averaging on PAD is demonstrated in Fig.~\ref{fig:orient_av2}.
Test case is Na$_8$ with detailed ionic structure which is, unlike the
jellium case, not rotational invariant. The laser frequency is
sufficiently high such that one-photon emission dominates. The left
panel shows the PAD for a fixed cluster orientation where the cluster
symmetry axis is aligned with the laser polarization axis. One sees a
pronounced pattern, in particular the four-fold structure of the ionic
rings and the rotation of the upper ring by 45$^\circ$ relative to the
lower ring.  The emission maxima are located at $\vartheta\approx
45^\circ$ and $\vartheta\approx 135^\circ$, i.e. sidewards to the
laser polarization.  The orientation averaged PAD becomes independent
of $\phi$, as it should be, and the emission is forward/backward
dominated with maxima at $\vartheta=0$ and
$\vartheta=180^\circ$. Altogether, the effect of orientation averaging
is dramatic. Calculations for a single orientation have thus little
predictive value.

As already pointed out, the OAPAD only
depend on the angle $\vartheta$, and not on $\phi$ anymore.
Moreover, they are symmetric with respect to the transformation
$\vartheta\leftrightarrow-\vartheta$. 
Thus they can be expanded in a standard manner in terms of even Legendre
polynomials $P_l(\cos\vartheta)$ as
\begin{equation}
  \overline{\frac{{\rm d}\sigma}{{\rm d}\Omega}}
  \propto
  1+\beta_2 P_2(\cos\vartheta)
  +\beta_4 P_4(\cos\vartheta)
  +\beta_6 P_6(\cos\vartheta)
  +...
\label{eq:anisotropy}
\end{equation}
Most important is $\beta_2$, called the anisotropy parameter. It is
the only remaining parameter in the perturbative regime where all
$\beta_{>2}=0$. Note that we had already recognized the simple
$1+\beta_2P_2 (\cos \vartheta)$ structure in Fig.~\ref{fig:example}.

\subsection{Thermal effects}
\label{sec:temperature_theo}

The previous discussions were led assuming that electrons remain at
zero temperature. Introducing an ionic temperature in the formalism
raises no difficulty as the ions are treated classically. An ionic
temperature can thus either naturally build up in the course of time,
or be initially introduced by giving the proper velocities to
ions. The system, of course, remains globally microcanonical, and the
temperature is thus subject to the corresponding fluctuations. The
electronic temperature, in turn, is explicitly set to zero by
construction. TDLDA does not allow occupation numbers of KS orbitals
to vary in time, which prevents an account of thermal effects at the
side of the electrons. This is certainly a major formal limitation of
the formalism. We have seen that (possibly sizable) electronic
temperatures have been observed experimentally (see for example Figure
\ref{fig:campbell1}). These effects may play a significant role in the
understanding of the dynamics of the system, and on the analysis of
experimental results. A theoretical account of such thermal effects is
thus to be developed.

A standard path to accommodate thermal effects in dynamical systems is
to recur to kinetic theory.  This theory was originally formulated in
the framework of classical mechanics with the Boltzmann equation as a
prototypical example \cite{Hua63}.  Quantum systems add two more
complications.  The Pauli principle prevents collisions into occupied
states. The uncertainty principle, in turn, raises difficulties when
trying to treat collisions locally (as done in the Boltzmann
equation).  Fully quantal kinetic equations are nevertheless
conceivable but much more involved than the classical ones
\cite{Kad62,Bal75} and very hard to apply in finite quantum systems
with their discrete spectra. In practice, there is thus no practical
quantum theory of such collisional correlations available yet.  In
turn, semi-classical approaches were developed over the years and
allow to cover some situations. The basics is then a version of the
Boltzman equation, adapted to account for Pauli principle and known as
the Boltzmann-Uehling-Ulhenbeck or Vlasov-Uehling-Ulhenbeck (VUU)
equation.  It was formally introduced in the early 1930's
\cite{Ueh33}, then widely used in nuclear dynamics \cite{Ber88,Dur00},
and explored more recently in simple metal clusters
\cite{Dom98b,Dom00a,Fen04}. By construction, the VUU equation is
semi-classical, which means that details of quantum shell effects are
washed out. It is thus applicable, at best, only in highly dissipative
situations where the latter shell effects do indeed disappear, or in
homogeneous fermionic systems such as electrons in solids
\cite{Che05aB} or nucleons in a neutron star \cite{Uec88aB}.  In
finite fermion systems like clusters and molecules, they are thus
bound to high excitation energies, which certainly strongly limits
their range of application.

Another point is worth being mentioned here.  In realistic
calculations, and in spite of the semi-classical treatment, one wants
to recover an acceptable description of ground state properties of the
studied systems. Experience shows that in electronic systems, this is
practically viable only in  metal clusters such as sodium
clusters~\cite{Dom00a,Fen04}. 
This again strongly
restricts the range of applicability of such methods.  In particular,
it does not allow to attack such widely studied systems as C$_{60}$ in
a fully realistic manner.  Still, results obtained in simple metals
are interesting and demonstrate the importance of this inclusion of
collisional correlations. We will thus briefly present VUU and discuss
some of these results here.

There are various ways of introducing VUU but in the case of finite
fermionic systems, the simplest and probably best founded presentation
is to recur to a semi-classical approximation on top of TDLDA. To
perform such a semi-classical limit, we first recast the
time-dependent KS (TDKS) equations in a matrix form, by introducing
the one-body density matrix ${\hat \rho}_\mathrm{KS}$ associated to KS
states, which reads in real space representation~:
\begin{equation}
  {\hat \rho}_\mathrm{KS}({\bf r},{\bf r}') 
  = 
  \sum_{i=1}^N \varphi_i^*({\bf r}') \, \varphi_i({\bf r}) \quad.
\label{eq:roks}
\end{equation}
This one-body density matrix fulfills the TDKS equations which, in a
matrix form, read~:
\begin{equation}
  \mathrm{i}\frac{\partial {\hat \rho}_\mathrm{KS}}{\partial t} 
  =  
  [{\hat h}_\mathrm{KS},{\hat \rho}_\mathrm{KS}] 
  \quad,
\label{eq:rotdks}
\end{equation}
with $\hat{h}_\mathrm{KS}$ given by Eq.~(\ref{eq:hamKS}). The
semi-classical limit can then be attained by performing a Wigner
transform (or, even better, an Husimi transform \cite{Hus40}) of
Eqs. (\ref{eq:roks}) and (\ref{eq:rotdks}). This leads to the
introduction of the phase space distribution $f({\bf r},{\bf p},t)$
which then fulfills, at lowest order in $\hbar$, the Vlasov equation~:
\begin{equation}
  \frac{\partial f}{\partial t} 
  = 
  \{h_\mathrm{KS}({\bf r},{\bf p},t),f({\bf r},{\bf p},t)\} \quad,
\label{eq:vlasov}
\end{equation}
where we have introduced Poisson brackets.  At the LDA level, for
which the xc functional is local in density, and for simple metals,
for which the pseudopotential is local as well, the semi-classical KS
hamiltonian takes the following simple form~:
\begin{equation}
  {h}_\mathrm{KS}({\bf r},{\bf p},t) 
  = 
  \frac{{\bf p}^2}{2m} + U_\mathrm{KS}[\rho] +  V_\mathrm{coupl}({\bf r})
  + U_\mathrm{ext}({\bf r},t) 
  \quad,
\label{eq:schamKS}
\end{equation}
where the local density is now obtained from momentum space
integration of $f({\bf r},{\bf p},t)$:
\begin{equation}
  \rho({\bf r}) 
  = 
  \int \frac{\textrm d{\bf p}}{(2\pi\hbar)^3} f({\bf r},{\bf p},t) \quad.
\label{eq:rovlasov}
\end{equation}

The Vlasov(-LDA) equation can now be complemented by the effect of
"two-body" collisions as usually done in kinetic theory.  This leads
to the Vlasov-Uehling-Ulhenbeck (VUU) equation~:
\begin{equation}
  \frac{\partial f}{\partial t} 
  = 
  \{h_\mathrm{KS}({\bf r},{\bf p},t),f({\bf r},{\bf p},t)\} 
  + 
  I_\mathrm{UU}[{\bf r},{\bf p}]  
  \quad,
\label{eq:vuu}
\end{equation}
obtained as the Vlasov-LDA equation complemented by a UU collision term~:
\begin{equation}
  I_\mathrm{UU}[{\bf r},{\bf p}] 
  = 
  \int \textrm d\Omega \, \textrm d{\bf p}_1 \frac{|{\bf p}-{\bf p}_1|}{m}
  \frac{\textrm d\sigma}{\textrm d\Omega}
  \left[f_{\bf p'}f_{\bf p'_1}(1-{\tilde f}_{\bf p})(1-{\tilde f}_{\bf p_1})-
  f_{\bf p}f_{\bf p_1}(1-{\tilde f}_{\bf p'})(1-{\tilde f}_{\bf p'_1})
  \right]  
  \;.
\label{eq:UU}
\end{equation}
which exhibits a local gain-loss balance for elastic electron
scattering $({\bf p},{\bf p}_1) \leftrightarrow ({\bf p}',{\bf
  p}'_1)$. The associated cross-section for electron-electron
"two-body" collisions is $\textrm d \sigma/\textrm d\Omega$ and
depends on the relative momentum $|{\bf p}-{\bf p}_1|$ and possibly on the
scattering angle. We have furthermore used the shorthand notations
${\tilde f}_{\bf p}= 2\pi \hbar^3 f_{\bf p}/2$ and $f_{\bf p}= f({\bf
  r},{\bf p},t)$, the collision term being local in space and time. To
avoid a double counting with the mean field $h_\mathrm{KS}$ the
differential cross-section can be evaluated with a screened Coulomb
interaction following standard scattering theory
\cite{Dom98b,Dom00a,Fen04,Koe08}.

The Vlasov and VUU equations are best solved using test particle
methods rather than grid methods.  Practically, this amounts to
represent the phase space distribution $f({\bf r},{\bf p},t)$ by a
swarm of numerical classical particles $\{({\bf r}_i,{\bf p}_i), i =1,
...\}$, each one with a given weight $\omega$ and following classical
equations of motion with forces derived from the mean field
hamiltonian $h_\mathrm{KS}({\bf r},{\bf p},t)$ \cite{Rei95a}.
Finally, the coupling to classical ionic motion, similarly as in
quantal TDLDA, is performed, leading altogether to a Vlasov/VUU-LDA-MD
approach. The theory was introduced in cluster physics in the late
1990's \cite{Dom98b} and further used since then \cite{Fen04},
but, as already mentioned, only applied to simple metal clusters.

The analysis of Vlasov or VUU dynamics proceeds in a rather simple
manner as the elementary degrees of freedom are classical test
particles. The total ionization is directly given by the number of test
particles ($\times \omega$) outside a large given box in which the
Coulomb field is computed. It is equivalently obtained as the integral
of the semi-classical density Eq. (\ref{eq:rovlasov}) over this
computational box. The latter density also enters the expression of
the dipole moment (see Eq. (\ref{eq:dipole}) in the next section) to
provide an analysis of optical response.  The distribution of kinetic
energies and the angular distribution are directly extracted from the
velocities of the "emitted" test particles. Note that the
semi-classical nature of the approach makes the PES always
exponentially decreasing, whatever the excitation, at variance with
the quantum case which exhibits the electronic single particles
energies. The quantum/classical comparison is then meaningful only at
sufficiently large excitations for which the quantum PES is also
exponentially decreasing with no more shell structure. The PAD, in
turn, can be compared in a meaningful way whatever the excitation
energy. For a detailed discussion, see Sec.~\ref{sec:dissipe}.


\section{Illustrative results}
\label{sec:results}

\begin{figure}[htbp]
\centerline{\includegraphics[width=0.85\columnwidth]{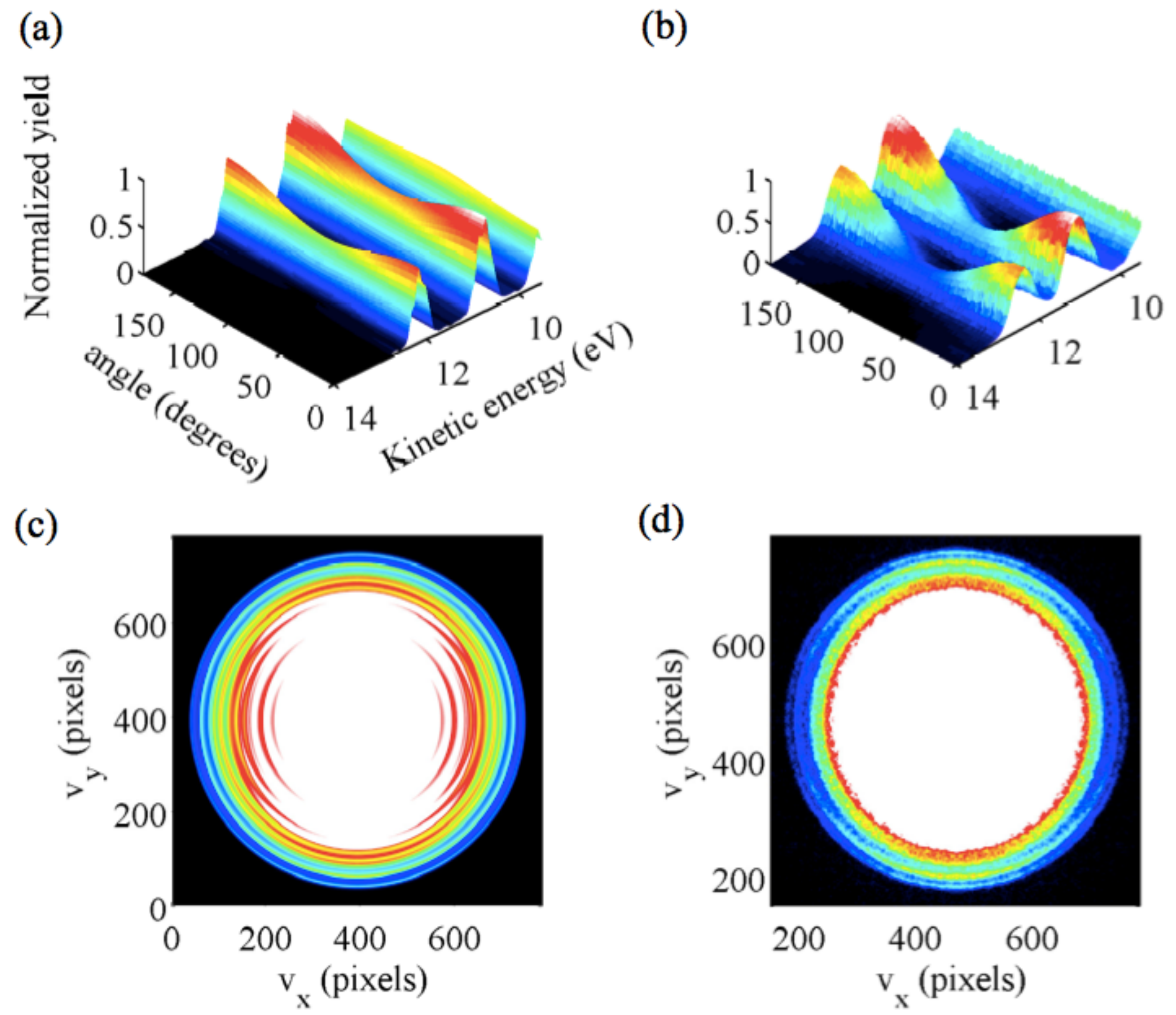}}
\caption{
Top~: Combined PES/PAD of C$_{60}$ obtained at $\omega_{\rm las}=20$~eV given by (a)
TDLDA-ADSIC calculations ($I=7.8\times 10^9$~W/cm$^2$ and duration of 60~fs,{ C$_{60}$
radius of 6.44~a$_0$)} and (b) experimental measurements using synchrotron radiation with
duration of about 1 ps. Bottom~: the corresponding velocity map image of (a) is presented
in panel (c), and that of (b) in panel (d).
}
\label{fig:c60-vmi_xuv}
\end{figure}

The coordinate-space and real-time technique to solve TDDFT as presented in
Sec.~\ref{sec:theo} offers a powerful tool to describe a broad range of scenarios for
the dynamics of clusters and molecules. We present in this section a few illustrative
examples of such theoretical studies thereby concentrating on electron ionization and
related observables. As a starter, Fig.~\ref{fig:c60-vmi_xuv} shows a calculated combined
PES/PAD (left column) for C$_{60}$, compared with an experimental one (right column).
The experimental spectrum was recorded at the Maxlab Synchrotron facility, using an oven
to produce the C$_{60}$ molecular beam~\cite{Epp97}. Experiments and calculations are both
performed at a photon energy of 20~eV. The top panels present the angle- and
energy-resolved distributions of photoelectrons for three selected energies close to the
HOMO level in a way similar to Fig.~\ref{fig:example}, while the bottom panels show the
distribution in a polar representation called velocity map images (VMI), see introductory
discussion in Sec.~\ref{sec:PADetal} and Fig.~\ref{fig:exampl_vmi}. The upper panel shows
nice agreement between experimental and theoretical results. The VMI in the lower panels
also nicely agree for larger energies near the IP (outer part of the circle). But they
differ in the low-energy region because it is filled with electrons from thermal emission
not accounted for in TDDFT, see also Fig.~\ref{fig:pes-deple_c60} later on. However,
although the physical content of such VMI is very rich, one cannot easily read off a
quantitative comparison from such a figure. Therefore, we will in the following sections
preferably discuss integrated VMI~: PES obtained by integrating over the angle, PAD
obtained by integrating over the kinetic energy, and total ionization obtained by
integrating over both, energy and angle.

As already mentioned in our first example in Sec.~\ref{sec:observ_theo} and
Fig.~\ref{fig:example}, the peaks observed in a PES are fingerprints of the single
particle (s.p.) energies of the electrons before they had been ejected by the photon.
However, s.p. energies are usually not well described by LDA. It has to be complemented by
a self-interaction correction (SIC) to attain realistic ionization potentials (IP) and
thus ionization properties (see Sec.~\ref{sec:sic_theo}). Before presenting detailed
results on PAD and PES, we first illustrate in Sec.~\ref{sec:sic_stat} the capabilities of
SIC to properly describe ionization dynamics. We remind again that we describe the
processes fully dynamically (see the methodology of Sec.~\ref{sec:pes}). This means, e.g.,
that discussing PES does not amount to a mere comparison of a computed static s.p.
spectrum with a measured PES. Since our access to PES, PAD, and ionization is fully
dynamical, it is thus applicable to any dynamical regime and free of any adjustable
parameter. This is why having a proper IP is a crucial step in our dynamical description,
especially at low energy where ionization occurs mostly close to threshold, whence the
importance of a SIC.
{Even when dealing with appropriate IP, a word of caution is to be
 added concerning the interpretation of PES as map of s.p. spectra.
 There are cases where one finds slight deviations for deep lying
 s.p. states. These can be understood as correction from final state
 interaction \cite{Mun06b}. They should be considered for a high
 precision analysis of data. We will ignore the effect in the more
 principle considerations of this section.}

This section is thus organized as follows. In Sec.~\ref{sec:sic_results}, we will first
demonstrate the impact of SIC on structural properties as single-electron energies and
optical response. In Sec.~\ref{sec:ioniz}, we will discuss ionization as a signal from
laser-induced dynamics. Then, in Sec.~\ref{sec:pes_dyn}, we will address PES, and in
Sec.~\ref{sec:pad_results} PAD, both in the one-photon and in the multi-photon regime.
{We finally discuss the impact of temperature, either ionic or electronic, on PES and
PAD in Sec.~\ref{sec:temper}.}

\subsection{Impact of the self-interaction correction on electronic emission}
\label{sec:sic_results}

As just mentioned, ionization properties are very sensitive to the
s.p. energies, whence the importance of SIC. We will here
demonstrate the impact of SIC on electronic properties, such as
s.p. spectrum or optical response, and finally discuss a
typical example of PES.

\subsubsection{Ionization potentials and single electron spectra}
\label{sec:sic_stat}

Before starting a discussion of the ionization potential (IP), we have to specify that in
more detail. For a system with $N$ electrons, the IP is defined by the difference
$I_\mathrm{adia}=E(N-1)-E(N)$. This is, in fact, called the ``adiabatic IP'' if the final
values of $E(N-1)$ is taken after waiting for full ionic rearrangement. However, this
adiabatic IP is a rather involved observable as it mixes electronic and ionic properties.
Much easier to handle and to interpret is the ``vertical IP''
$I_\Delta=E_\mathrm{fix}(N-1)-E(N)$ which is obtained from the energy $E_\mathrm{fix}$ of
the ionized system while still maintaining the ions in their original configuration. This
vertical IP is a purely electronic observable which renders it very instructive. In
practice, induced ionization processes (laser pulse, ion collision) are so fast that the
ionic configuration is almost inert during the emission process. This motivates the use of
the vertical IP, henceforth called simply ``the IP''.

As already discussed in Sec.~\ref{sec:sic_theo}, correct s.p. energies are strongly
related to the correct asymptotic behavior of the KS potential, see the example of ${{\rm
K}_7}^-$ in Fig.~\ref{fig:K7m-SIC}. And as just explained above, the IP is defined as
difference of two stationary energies. However, this definition is not well suited to
dynamical calculations where one aims at tracking ionization "on the fly". For then, it
becomes crucial to fulfill Koopmans' theorem \cite{Koo34} which states
that the vertical IP
should be identical with the s.p. energy of the last bound electron (HOMO level), i.e.
$I_\varepsilon=-\varepsilon_\mathrm{HOMO}$. Koopmans' theorem is violated in LDA and usually
recovered when invoking SIC. This rules out LDA for a fully dynamical approach to
ionization and calls for some SIC. It is thus crucial to test the performance of TDDFT in
this respect. As discussed in section \ref{sec:sic_theo}, we have basically a full SIC
(practically implemented via the 2setSIC scheme {and simply denoted in the following by
SIC}) and ADSIC at our disposal. ADSIC is orders of magnitude simpler than SIC and thus
certainly worth being considered very seriously. As a first step, we shall thus check the
capabilities of both approaches with respect to reproducing IP's along the lines of a
large systematic study in \cite{Klu13}.

\begin{figure}[htbp]
\centerline{\includegraphics[width=0.75\linewidth]{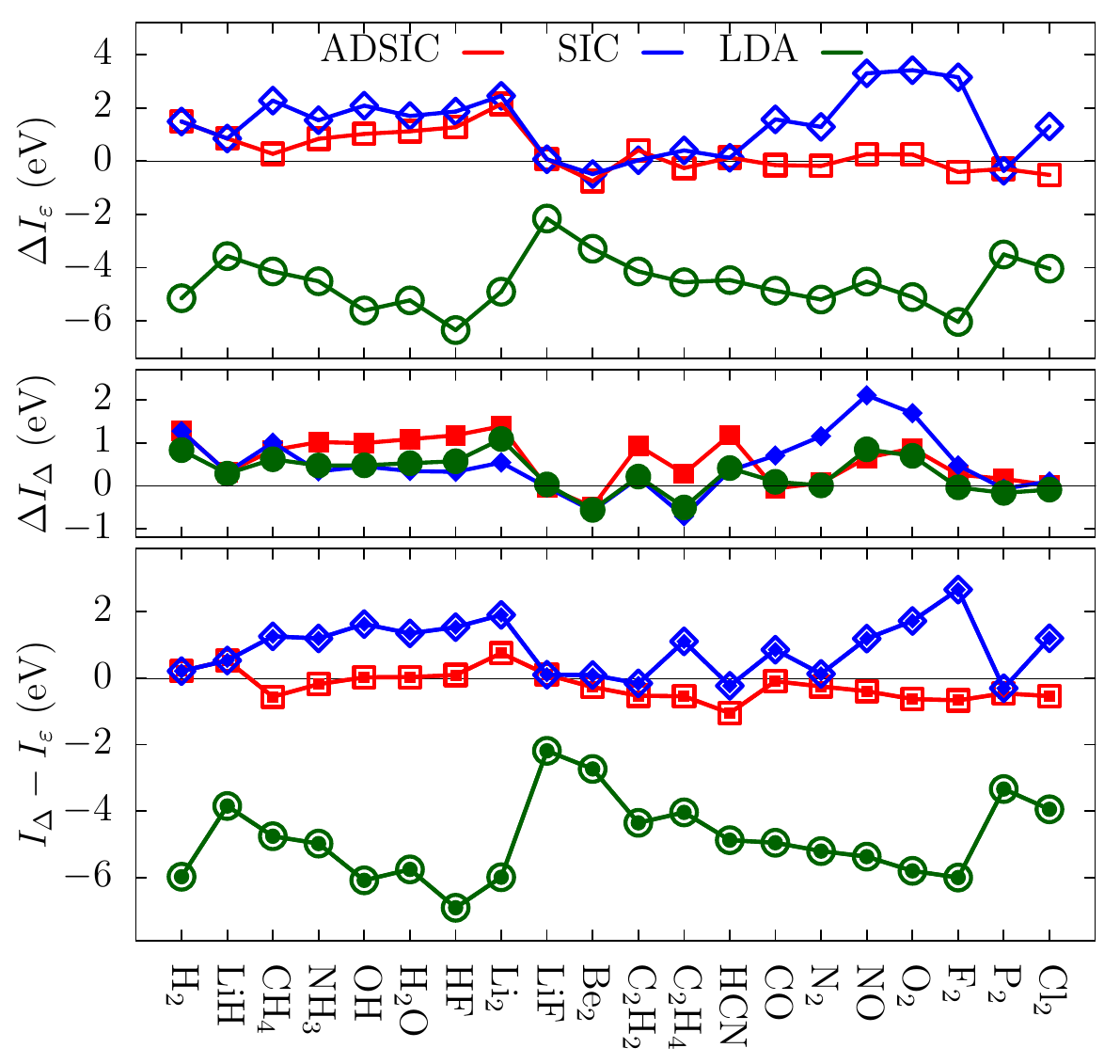}}
\caption{Calculated ionization potentials (IP) of a selection of molecules with different
  bonding types for various level of SIC (see text for details). Top~:
  difference $\Delta I_\varepsilon=I_\varepsilon-I_\mathrm{exp}$ of the IP
  $I_\varepsilon$ deduced from the energy of the HOMO with the experimental
  (vertical) IP $I_\mathrm{exp}$. Middle~: difference $\Delta
  I_\Delta= I_\Delta-I_\mathrm{exp}$ for the IP $I_\Delta$ calculated
  as the difference of binding energies. Lower~: non-Koopmans'
  error $I_\Delta-I_\varepsilon$.  Adapted
  from~\cite{Klu13}.
\label{fig:sic_molecules}
}
\end{figure}

Fig.~\ref{fig:sic_molecules} depicts the difference between the calculated IP and the experimental 
one, for a selection of molecules. 
Both
ways are used for the calculations~: from the HOMO level as $-\varepsilon_\mathrm{HOMO}$ 
(upper panel, open symbols) or
from the difference of binding energies $I_\Delta$ (middle panel and filled symbols). 
The upper panel shows that, as expected, LDA performs badly when one
considers the IP $I_\epsilon$ from the energy of the HOMO. SIC and even
more so ADSIC come much closer to the experimental IP.  The middle
panel compares the IP $I_\Delta$ from energy differences. Here we see
better agreement for all three methods, demonstrating that the
$I_\Delta$ is the more robust definition.  To emphasize the
discrepancy between both estimates of IP's, we plot their difference in
the lower panel of Fig.~\ref{fig:sic_molecules}.  Vanishing
difference signifies fulfillment of Koopmans' theorem.
LDA produces large errors while SIC and
ADSIC do well.  It is a bit of a surprise that the simpler ADSIC often
performs better than the more elaborate SIC.

\begin{figure}[htbp]
\centering{
\includegraphics[width=0.62\linewidth]{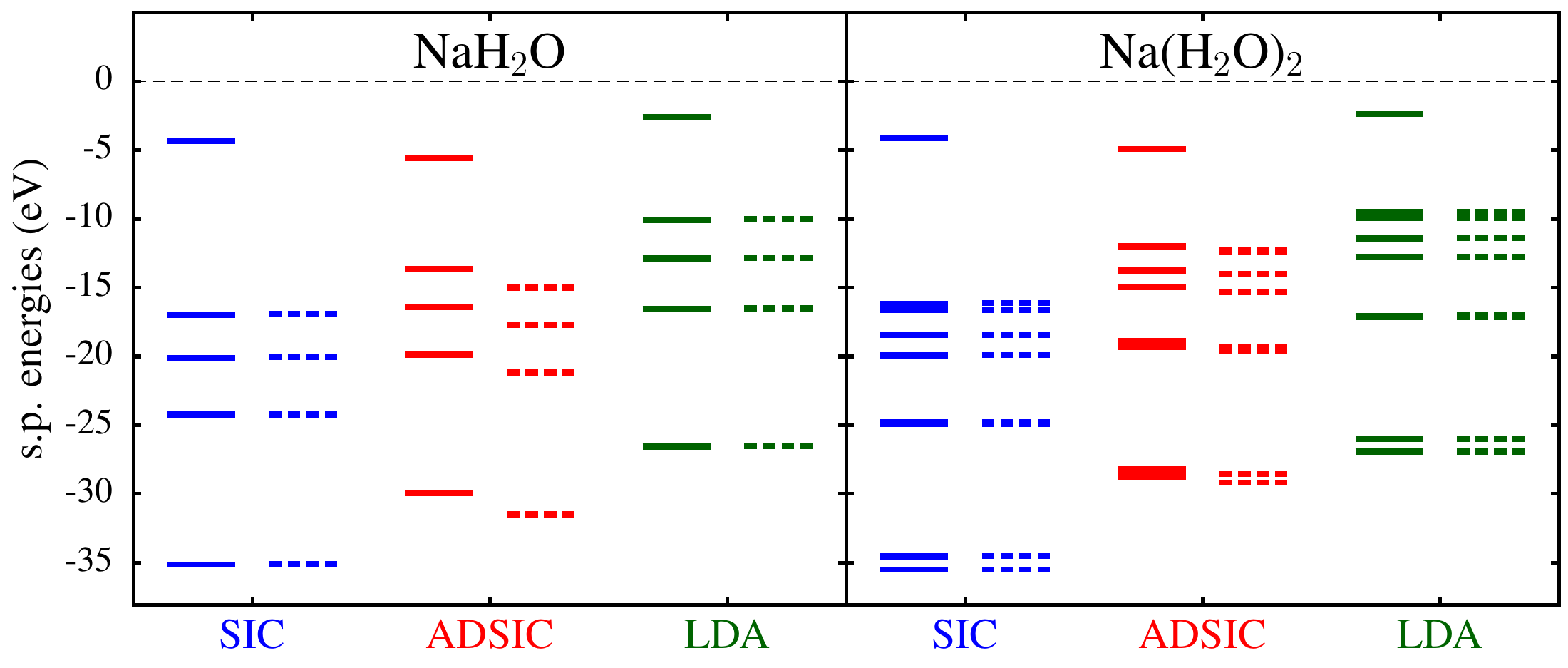}
\includegraphics[width=0.34\linewidth]{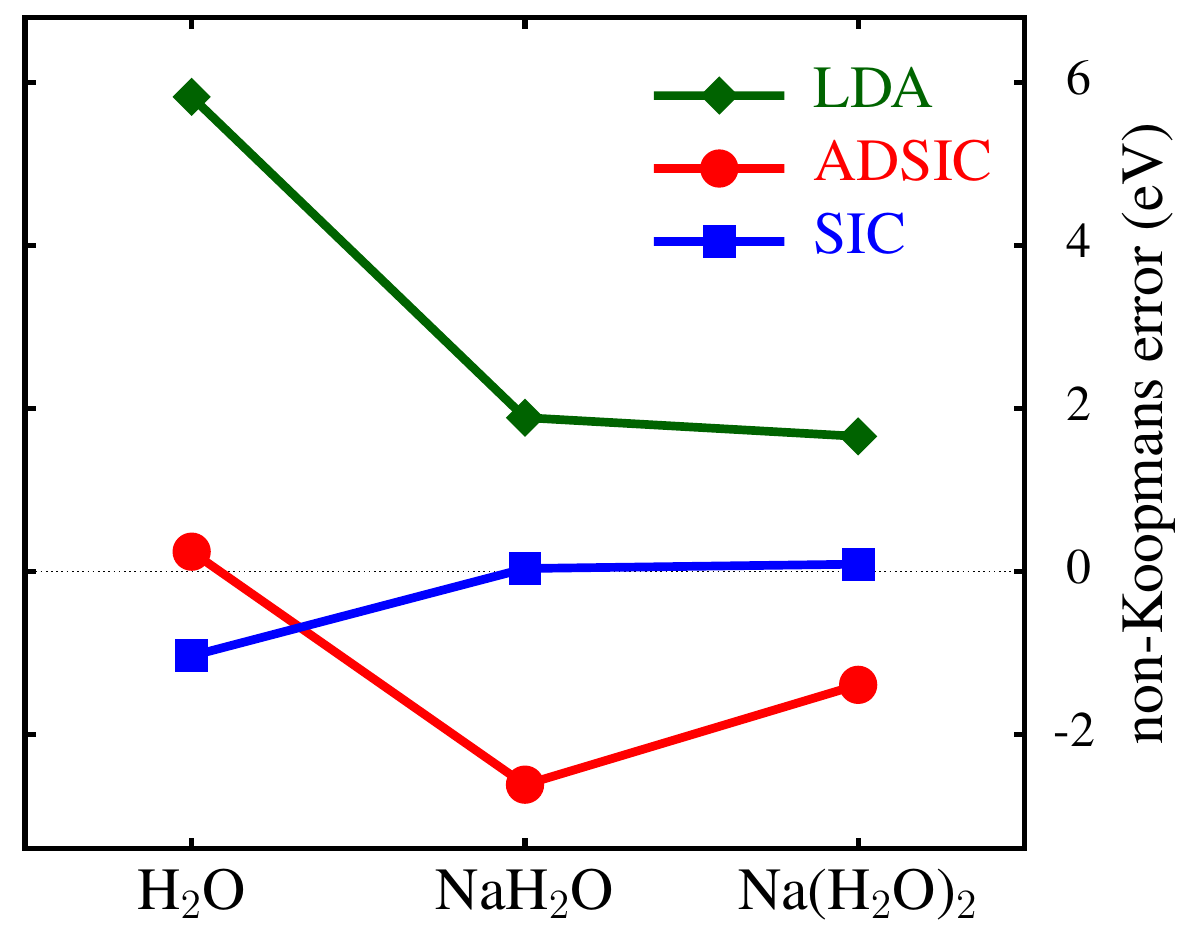}
}
\caption{Single particle energies of NaH$_2$O (left) and
  Na(H$_2$O)$_2$ (middle) in LDA, ADSIC and SIC, for spins up (full lines) and
  down (dashes).  Right~: non-Koopmans' error calculated in various
  levels of DFT as indicated.  From~\cite{Din14}.
\label{fig:nah2o_sic}
}
\end{figure}

By construction, ADSIC is a priori well suited to systems with metallic
binding \cite{Leg02}. It was nevertheless soon realized that it also
performs well in covalent systems \cite{Cio05}. The very systematic
study of \cite{Klu13} led to the unexpected result that for an
enormous range of atoms, molecules, carbon chains, and fullerenes
ADSIC leads very often to smaller non-Koopmans errors than SIC.  It
was also shown that, when comparing theoretical IP's to experimental
ones, again ADSIC was providing the best results for a huge range of
molecules. That however does not mean that ADSIC is the
ultimate solution to the self-interaction problem. We have already
mentioned its intrinsic limitations (essentially due to the fact that
the functional explicitly depends on the number of electrons) in
Sec.~\ref{sec:sic_theo} and we should add here that the scaling
properties of ADSIC with increasing system size are also raising some
problems \cite{Klu13}. Furthermore, there are a few specific cases
which raise difficulties. These are molecules where very different
types of bonding coexist, such as covalent and metallic bonding. The
point is illustrated in Fig.~\ref{fig:nah2o_sic} which presents the
s.p. energies of the two complexes NaH$_2$O and
Na(H$_2$O)$_2$, calculated in LDA, ADSIC and SIC. The
right part of the figure complements the picture by showing the
corresponding non-Koopmans errors.
Not surprisingly the water molecule with one prevailing bonding type
is well described in ADSIC, while SIC yields a small but somewhat
larger non-Koopmans' error. The situation becomes different in the
mixed complexes Na(H$_2$O)$_n$, where both ADSIC and LDA exhibit
similarly large errors (although with opposite signs), while the
Koopmans' theorem is perfectly fulfilled in SIC. Thus ADSIC should not
be used for Na(H$_{2}$O)$_{n}$. This may not be a total surprise if
one reminds that ADSIC applies the same correction to every orbital
while, in such a mix of bonding types, one encounters both highly
delocalized metallic orbitals and much more localized covalent
orbitals.  As visible in the left panel of Fig.~\ref{fig:nah2o_sic},
the HOMO in NaH$_{2}$O at 5.14~eV arises from the Na atom whereas the
deeper orbitals mostly come from the H$_{2}$O molecule, which has an
IP of 12.6 eV (see Fig.~\ref{fig:systems}). 
The electronic density of the HOMO is thus clearly
different from that of the other orbitals, so that self-interaction
for different electronic states becomes very different.

The example of a metal-covalent complex shows that ADSIC is not a safe fire solution to
the self-interaction problem. It is likely to fail whenever the s.p. states cover grossly
different regions of space. This also occurs in the fragmentation of a molecule and in
processes with high ionization. Nonetheless, ADSIC provides very often a remarkably
accurate and simple approximation to SIC. It is always worthchecking whether a given
problem allows one to employ ADSIC. In the following examples, we will often recur to
ADSIC.

\subsubsection{Optical response}
\label{sec:sic_optresp}

One of the most prominent observables of electronic dynamics 
is the optical response measured in
terms of the photo-absorption strength. It gives insight into the spectrum of dipole
transitions and provides useful information on the collective modes and the particle-hole
excitations of the system. Note that the word "optical" is generic in the sense that it
covers also the spectrum outside the optical range of frequencies. Consider, for instance,
covalent systems where the dominant peaks rather lie in the UV domain. Here, we want to
explore the impact of SIC on the optical response.

The optical response can be calculated in various
ways. It is often evaluated by
  computing the response function directly in linearized TDLDA. Having
  a fully fledged TDLDA code at hand, it is technically and
  conceptually simpler to employ spectral analysis for that purpose
  \cite{Cal95a,Yab96,Cal97b}. To that end, we initialize the
  electronic dynamics by applying an instantaneous dipole boost to the
  electronic wave functions. We then record the time-dependent dipole
  momentum~:
\begin{equation}
\label{eq:dipole}
   \mathcal{D}(t) 
   =
   \int \textrm{d}^3 \mathbf r\,
   (\mathbf{r}-\mathbf{R}_\mathrm{cm,ion})\,\rho(\mathbf{r},t) \quad,
\end{equation}
where $\mathbf{R}_\mathrm{cm,ion}$ denotes the center of mass of the
ions.  The dipole strength $S_D(\omega)$ is obtained by
  Fourier transforming
  $\mathcal{D}(t)\longrightarrow\tilde{\mathcal{D}}(\omega)$ yielding
  finally $S_D(\omega)\propto\Im\{\tilde{\mathcal{D}}(\omega)\}$.
{One can alternatively look at the power spectrum $|\tilde{\mathcal D}(\omega)|^2$
which basically contains the same information as the dipole strength.}

\begin{figure}[htbp]
\centerline{\includegraphics[width=0.7\linewidth]{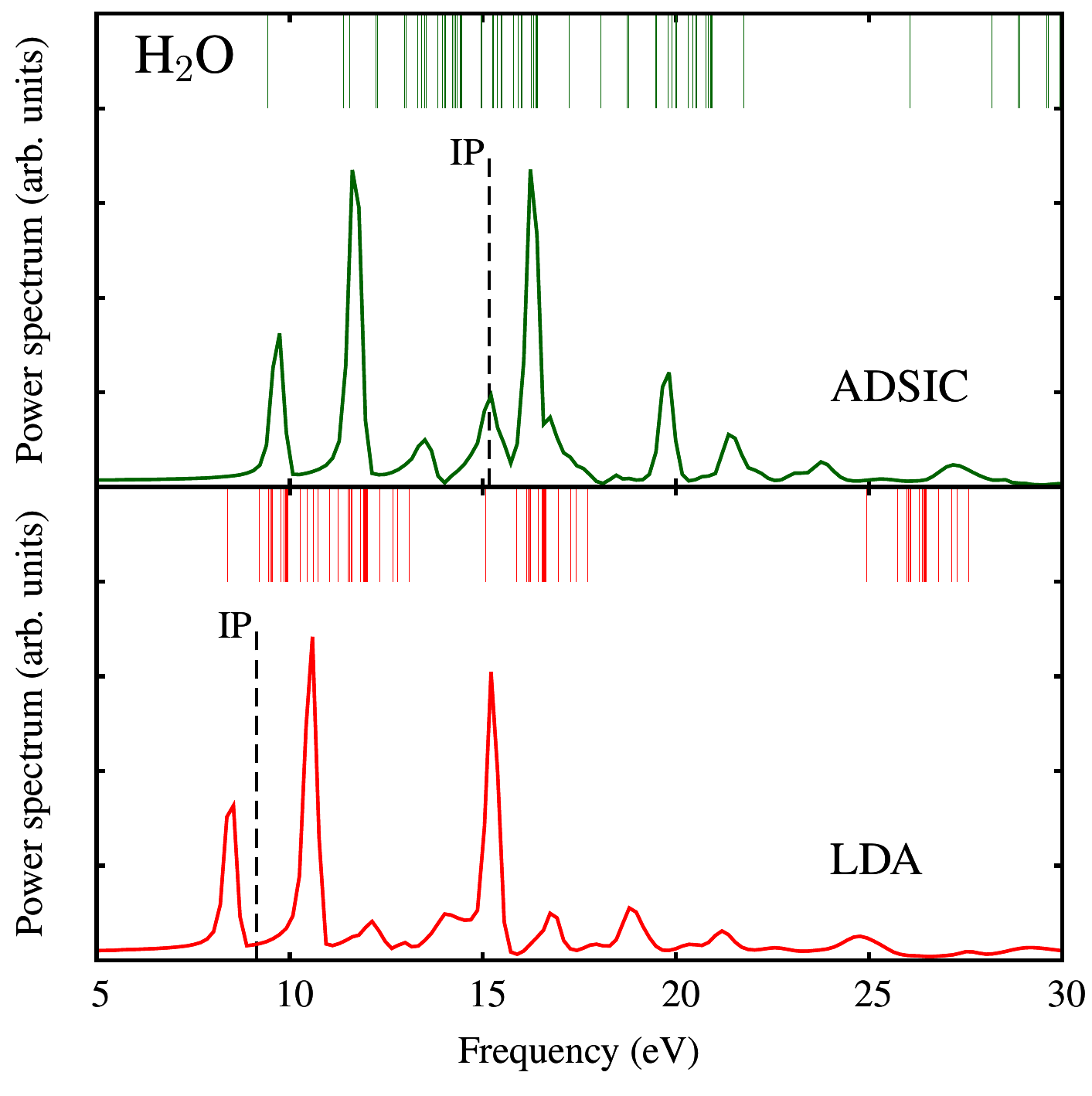}}
\caption{Optical response of H$_2$O calculated in LDA (bottom) and
  ADSIC (top). The dashes indicate the ionization potential (IP) calculated
  as the opposite of the HOMO energy. The vertical full lines stand
  for the possible static dipole transitions.
\label{fig:sic_optresp}
}
\end{figure}

{We already presented in the bottom right part of Fig.~\ref{fig:systems}
the optical response  calculated in ADSIC of H$_2$O in the three spatial directions.}
Fig.~\ref{fig:sic_optresp} now displays the power strength averaged over the three spatial directions,
and compares the calculations done in LDA with those in ADSIC.
Each photo-absorption spectrum is complemented by the sequence of one-particle-one-hole
($1ph$) states with dipole character calculated from the static s.p. energies of occupied
and empty states. First, we notice that the distribution of $1ph$ states in LDA is very
different to that in ADSIC. This reflects the different s.p. spectra of LDA and ADSIC.
ADSIC tends to localize the wave functions more than LDA which results in a more compact
electron cloud (associated to a larger IP) and a wider span of dipole transitions. It is
also interesting to note that most LDA dipole transitions lie in the continuum (IP$\simeq
9.1$~eV). In spite of these large differences in $1ph$ spectra, the dipole spectra look
very similar. The reason is that the recoupling of the pure $1ph$ states to the true
excited states is dominated by the Coulomb Hartree term \cite{Rei92a,Rei96c} which is the
same in LDA and SIC. This defines the overall position of dominant dipole strength which
is more or less robust. The underlying $1ph$ structure has an impact on the detailed
fragmentation pattern which can depend more sensitively on the level of SIC treatment. The
payoff between Coulomb interaction and $1ph$ structure depends, of course, on the system.
Metal clusters have the pronounced Mie plasmon mode which is dominated by the Coulomb
interaction, thus very robust against SIC. On the other hand, systems with fuzzy dipole
spectra are more critical. The example H$_2$O is somehow in between.

\subsubsection{SIC-revisited photoelectron spectra}
\label{sec:sic_dyn}

\begin{figure}[htbp]
\centerline{\includegraphics[width=0.7\linewidth]{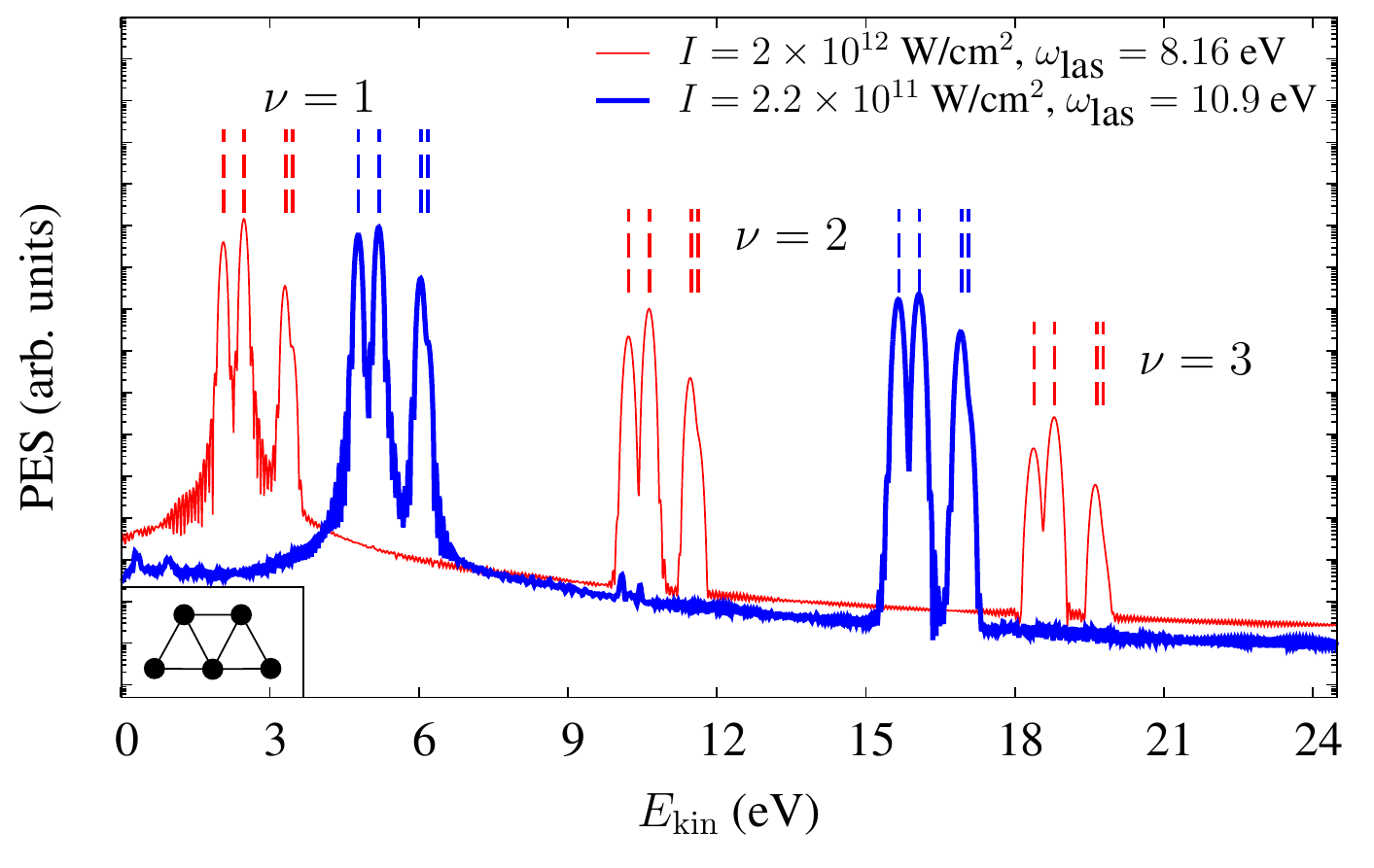}}
\caption{Photoelectron spectra for Na$_5$ (ionic configuration in the
  inset) irradiated by two different laser pulses with pulse duration
  of 60 fs, intensity $I$ and frequency $\omega_\mathrm{las}$ as
  indicated, and polarized along the direction normal to the Na$_5$
  plane. The static Kohn-Sham orbital energies, shifted by
  $n\omega_\mathrm{las}$ where $\nu$ is the number of involved photons,
  see Eq.~(\ref{eq:ekin1}), are also indicated as vertical dashed
  lines in both cases. Adapted from~\cite{Vin13}.
\label{fig:sic_pes}
}
\end{figure}

Since the early days of DFT, the interpretation of Kohn-Sham (KS)
orbitals has been a matter of debate. A direct comparison of
experimental PES spectra to s.p. spectra or even better so
to our dynamically computed PES (see Sec.~\ref{sec:pes}) is an a
posteriori proof of the meaning to be given to KS s.p.
energies, following the basic multiphoton ionization (MPI) relation (\ref{eq:ekin1}).  The
point is rather easy to accept at the LDA or ADSIC level to the extent
that the KS hamiltonian ${\hat h}_{\rm KS}$ (see Eq.(\ref{eq:hamKS}))
is well defined and common to all KS orbitals. The situation is more
involved in the case of SIC where the KS hamiltonian (\ref{eq:SICmf})
becomes state-dependent.  The "2setSIC'' solution scheme
  (\ref{eq:SICmf}--\ref{eq:SICeqs}) allows, nonetheless, to define
  unambiguously s.p. energies. It is thus interesting to
see how SIC performs for computing PES, to see to which extent
the SIC s.p. energies have a similar meaning as in LDA or
ADSIC.  Rather than making a comparison to experiment, which will not
test the internal capabilities of the theory, it is here more
interesting to check the evolution of the PES peaks within varying the
laser frequency and see whether they follow the MPI rule
(\ref{eq:ekin1}). This would give an indication on their
possible interpretation. The point is illustrated for the case of the
planar metal cluster Na$_5$.  Two different laser pulses have been
used with frequencies 8.16 and 10.9~eV.  The laser intensities have
been adjusted in each case to obtain about the same low total
ionization around 0.006, thus well in the perturbative regime
where PES signals are not yet blurred by Coulomb shift (see
  Secs.~\ref{sec:pes_I} and {\ref{sec:TRPES}}).  The laser polarization is taken normal
to the cluster's plane. Fig.~\ref{fig:sic_pes} displays both PES.
For a given laser frequency $\omega_\mathrm{las}$, one clearly
observes copies of the same pattern which are separated by
$\omega_\mathrm{las}$. Each pattern exhibits peaks which are
positioned at values of the kinetic energy $E_\mathrm{kin}$ following
the standard MPI relation Eq.(\ref{eq:ekin1}), i.e.
$\varepsilon_{\mathrm{kin},j}=\varepsilon_j+\nu\,\hbar\omega_\mathrm{las}$.
The $\varepsilon_j$ entering this equation are the eigenvalues of the
stationary equation $h_\mathrm{SIC} |\varphi_j\rangle = \varepsilon_j
|\varphi_j\rangle$, see Eq.~(\ref{eq:SICeqs}), while $\nu$ corresponds
to the number of photons involved. The remarkable fact that the peaks
of the PES, obtained from the calculation of a time propagation of the
2 sets $\{\varphi_j\}$ and $\{\psi_\alpha\}$, coincide with the static
$\varepsilon_j$ validates the interpretation (and the definition) of
these energies as sound s.p. energies. This also supports
the identification of the $\varphi_j$'s as the physical wave functions
of the associated s.p. states whose characteristics are
measurable via the PES.

\subsection{Using ionization as an observable}
\label{sec:ioniz}

In this section, we discuss basic mechanisms in the laser irradiation
of an electronic system leading to significant electronic emission
and their analysis in terms of total ionization as an observable. For
moderate laser intensities, a major issue is the relation of the laser
frequency with the optical response peaks, especially collective ones.
We will present here two generic scenarios for this resonance
effect. We will first explore laser irradiation of a water molecule
demonstrating off- and on-resonant ionization. In
Sec.~\ref{sec:P&P_ion}, we will take advantage of resonant
enhancement of ionization to explore the ionic dynamics via the
use of a pump-and-probe (P\&P) setup. In Sec.~\ref{sec:P&P_ele}, we will
finally consider again a P\&P setup, but this time within
using a train of attosecond pulses.

\subsubsection{Off- and on-resonant ionization}
\label{sec:basic_ioniz}

\begin{figure}[htbp]
\centerline{\includegraphics[width=0.6\linewidth]{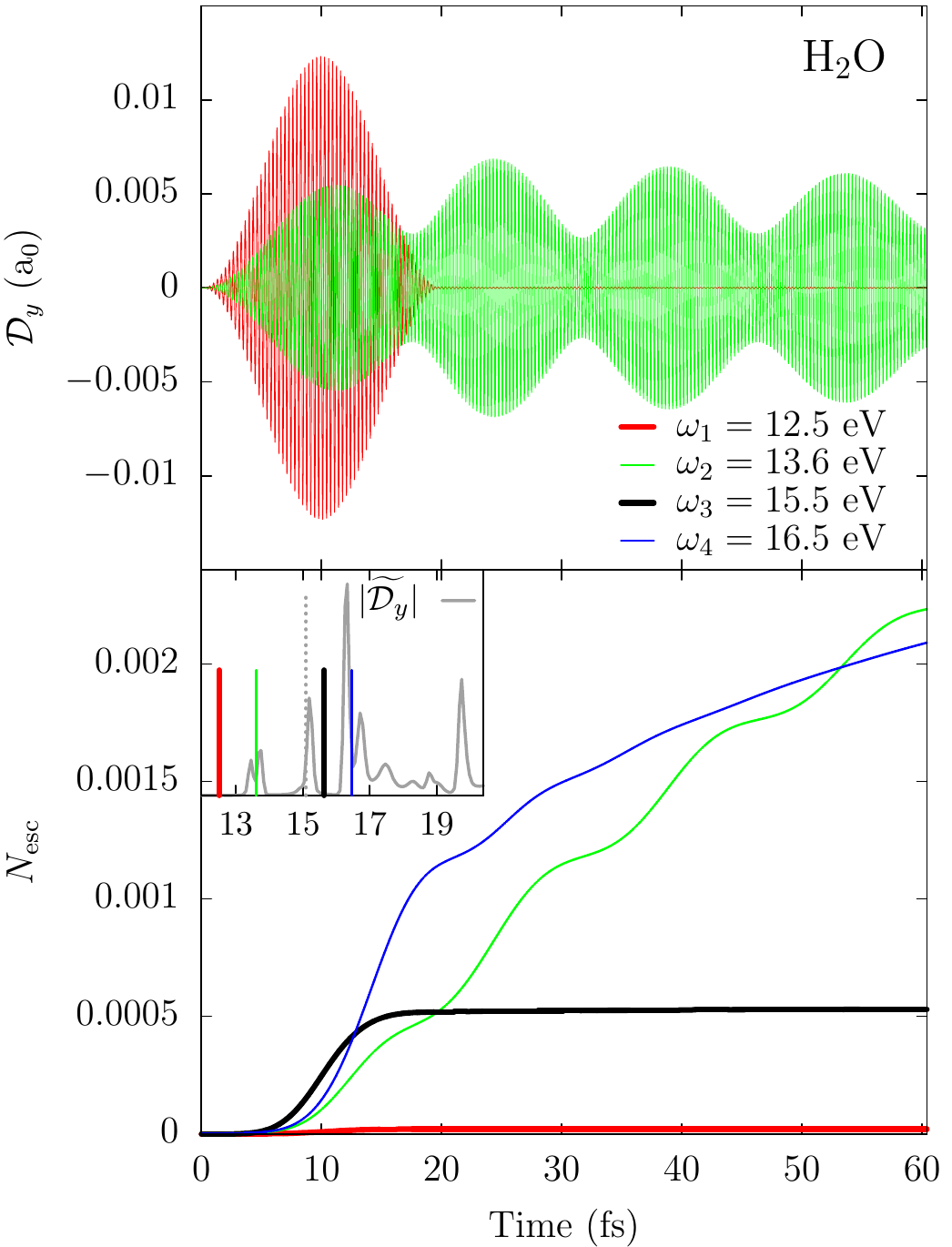}}
\caption{Resonant and off-resonant irradiation of H$_2$O by laser
  pulses polarized along the symmetry axis of H$_2$O, denoted by $y$,
  and of various intensities ($I_1=10^{12}$, $I_2=I_3=10^{11}$,
  $I_4=10^{9}$~W/cm$^2$) and various frequencies ($\omega_1=12.5$,
  $\omega_2=13.6$, $\omega_3=15.5$, $\omega_4=16.5$~eV). 
  Calculations had been done with ADSIC and pseudopotentials
  to leave core electrons inert. Top~: time
  evolution of the electronic dipole in $y$ direction. Bottom~: time
  evolution of the total ionization $N_\mathrm{esc}$. Inset~: optical
  strength of H$_2$O in $y$ direction with horizontal energy axis 
in eV. The vertical full lines
  indicate the chosen laser frequencies, and the dotted one corresponds
  to the IP, here at 15.1~eV.  }
\label{fig:h2o_irrad}
\end{figure}

As already discussed in Sec.~\ref{sec:lasers}, the great
  versatility of lasers through the choice of frequency, intensity,
pulse duration and shape, offers experimentalists and theoreticians a
world of dynamical scenarios.  To gather orientation in this huge
  landscape of options, we first explore the impact of laser
  frequency. To that end, we consider the dynamics of laser excitation 
of a H$_2$O molecule, {with techniques similar to those used in~\cite{Ndo10}}. 
The results are shown in Fig.~\ref{fig:h2o_irrad}.
The duration of the laser pulses is in all cases 20 fs. Four
frequencies have been explored, namely $\omega_1=12.5$,
$\omega_2=13.6$, $\omega_3=15.5$ and $\omega_4=16.5$~eV.  The
corresponding intensities are $I_1=10^{12}$, $I_2=I_3=10^{11}$ and
$I_4=10^{9}$~W/cm$^2$ to keep the maximal dipole amplitude
  similar. The laser polarization is along the symmetry axis of the
water molecule, denoted here by $y$.

Let us start with $\omega_1$ and $\omega_2$. Both frequencies are
below the IP of H$_2$O (15.1~eV).  The optical response of H$_2$O in
$y$ direction is shown in the inset in the bottom panel~: $\omega_2$
lies on a double peak and, as we will see, corresponds to a resonant
frequency, whereas $\omega_1$ does not match any dipole
transition and is thus off-resonant. The top panel of
Fig.~\ref{fig:h2o_irrad} shows the time evolution of the electronic
dipoles. As expected \cite{Cal00}, the red curve for the off-resonant
$\omega_\mathrm{las}=\omega_1$ nicely follows the laser pulse profile
and dies out with the laser signal at 20 fs.  The total ionization
$N_\mathrm{esc}$ shown in the bottom panel (red line) stays very close
to zero for this low frequency case. {The resonant $\omega_{\rm
    las}=\omega_2$ proceeds differently~:} {During the first 15
  fs, the dipole signal (light green curve in top panel) still follows
  the laser profile, but then continues to oscillate with large
  amplitude} long after the laser pulse is switched off. Such
{resonant oscillations come along with} a larger deposit of energy
in the molecule and {thus stronger ionization}.  This is clearly
demonstrated in the bottom panel of Fig.~\ref{fig:h2o_irrad},
where $N_\mathrm{esc}$ (green line) steadily increases with visible
steps perfectly correlated to the maxima in the dipole
oscillations. The total ionization $N_\mathrm{esc}$ in the resonant
case is orders of magnitude larger than in the off-resonant one,
although the laser intensity in the resonant case is 10 times
smaller. This can be understood in terms of the Keldysh parameter
(\ref{eq:Keldysh}) which is, in both cases, much larger than 1. This
indicates that we are in the frequency-dominated regime where such
differences matter, see Sec.~\ref{sec:las_charac}.

\MD{
Note also the initial profile of the ionization is determined by the laser pulse such 
that the maximum slope 
coincides with the laser peak amplitude. But this profile is delayed by the time it takes for the 
escaping electrons to reach the box bounds. The final non-vanishing slope is related
to the non-vanishing dipole oscillations.
}

The next two cases consider an off-resonant frequency ($\omega_3$) and
a resonant one ($\omega_4$), now both above the IP. For the sake of
clarity, the corresponding dipole signals are not displayed.
They show again the same typical pattern of resonant (long
  standing after-oscillations) and off-resonant (signal dies out with
  the laser) response. We only show the associated total
ionization $N_\mathrm{esc}$ in the bottom panel of
Fig.~\ref{fig:h2o_irrad} (black thick line for $\omega_3$ and blue
thin line for $\omega_4$). The off-resonant case 
increases with a slope following the amplitude of the dipole
  oscillations and levels off to a plateau after the laser pulse is
over. The case $\omega_3$ yields much higher ionization than the case
$\omega_1$, even if its intensity $I_3$ is an order of magnitude
smaller than $I_1$. This happens because $\omega_3$ stays above
  the IP and can ionize directly with one-photon processes.  Finally
  we compare the two resonant cases $\omega_2<$IP with $\omega_4>$IP.
Although $I_2/I_4=100$, both $N_{\rm esc}$ are very similar.  Two
  effects cooperate here~: {\it i)} $\omega_4>$IP and {\it ii)}
the strength of the mode excited at $\omega_4$ is at least 3 times
stronger than that at $\omega_2$. This demonstrates, once again,
the importance of the laser frequency in relation to the optical
spectrum in order to drive large ionization.

\begin{figure}[htbp]
\centering{
\includegraphics[width=0.495\linewidth]{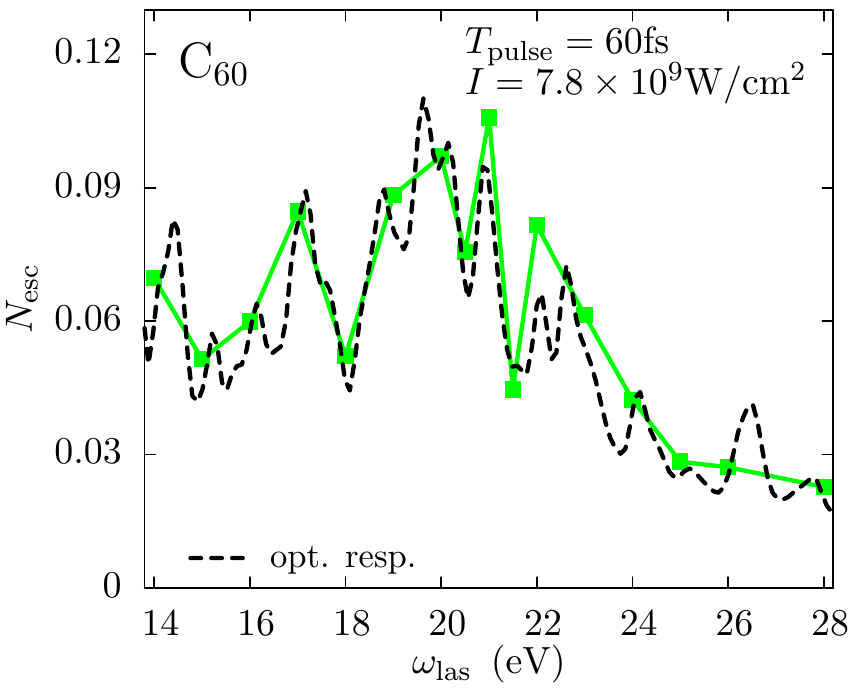}
\includegraphics[width=0.495\linewidth]{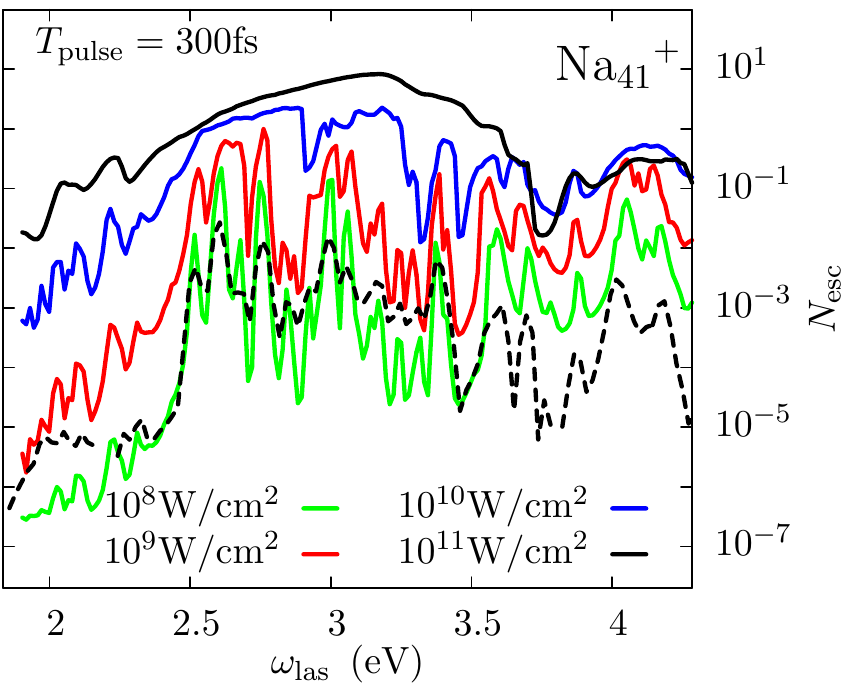}
}
\caption{Total ionization $N_\mathrm{esc}$ as a function of laser frequency $\omega_{\rm las}$. 
Left~: case of C$_{60}$,
{calculated in full 3D with a radius of 6.44 a$_0$,} irradiated by a laser pulse with
$T_\mathrm{pulse} = 60$~fs and $I = 7.8\times 10^9$~W/cm$^2$. 
Right~: case of ${{\rm Na}_{41}}^+$ irradiated by laser pulses with
$T_\mathrm{pulse}=300$~fs and intensities $I$ as indicated {(using
CAPS)}. The black dashed line represents the optical response in both cases.}
\label{fig:nesc_om}
\end{figure}

The point is again illustrated, this time in a more systematic manner,
in Fig.~\ref{fig:nesc_om} which displays the dependence on laser
frequency of the total ionization $N_\mathrm{esc}$ of irradiated
C$_{60}$ (left) and ${{\rm Na}_{41}}^+$ (right).  
The figure is a continuation
to the pedagogical Fig.~\ref{fig:ioniz_plasmon} discussed in Sec.
\ref{sec:mecha}, this time for two more complex systems though.
For C$_{60}$, with the chosen laser parameters ($I = 7.8\times
10^9$~W/cm$^2$ and $\omega_\mathrm{las}=$14--28~eV), we are again in
the frequency-dominated regime.  The ionization
$N_\mathrm{esc}(\omega_\mathrm{las})$ (light green curve) exhibits
strong oscillations with $\omega_\mathrm{las}$ which match remarkably
well the optical response of C$_{60}$ (black dashed line).
This shows that the signal of photoemission
  $N_\mathrm{esc}(\omega_\mathrm{las})$ is close to the signal of
  photo-absorption, at least above the emission threshold.

The situation is similar in ${\mathrm{Na}_{41}}^+$ at the lowest laser
intensity (light green curve in right panel) where $N_\mathrm{esc}$
exhibits strong oscillations with $\omega_\mathrm{las}$, once again
fitting fairly well those of the optical spectrum. However, if the
laser intensity is increased, we progressively leave the
frequency-dominated regime to enter the field- (intensity-)dominated
domain (see Sec.~\ref{sec:las_charac}).  And indeed, the fragmented
structure of $N_{\rm esc}$ steadily broadens to be finally washed out
at the highest intensity (see top black curve). At the same time, the
values of $N_\mathrm{esc}$ for a given $\omega_\mathrm{las}$ also
increase, since there are more and more photons pulling on the valence
electrons of the cluster.

\subsubsection{Pump and probe (P\&P) analysis of ionic dynamics}
\label{sec:P&P_ion}

The emergence of fs lasers allowed the development of time-resolved
studies of molecular reactions through pump-and-probe (P\&P)
experiments.  The typical strategy of such a fs spectroscopy is
simple. An initial short laser pulse (pump) excites the
electronic system which leads to subsequent ionic motion. This motion in
turn changes the electronic response according to the actual ionic
configuration. This change is explored by the response (e.g.,
ionization) of the system to a second laser pulse, the probe,
sent after a certain time delay. Scanning the reaction strength as a
function of delay time allows one to map the time evolution of the
molecular system.  There are, of course, many variants of this generic
strategy according to the variety of molecules and flexibility of
laser pulses.  Altogether, fs spectroscopy has become an extremely
powerful analyzing tool in physics and chemistry, for early reviews, see
\cite{Zew94,Gar95}.

Of course, P\&P analysis is also  an extremely interesting tool in
cluster physics. Very small clusters allow scenarios very similar to those in
simple molecules, see e.g. experiments on trimers
\cite{Lei99,Hei00} and associated theory \cite{Har98b}. Larger
clusters are too complex for the very subtle and detailed pathways
followed in small molecules. One better looks for global properties of
the ionic background as, e.g., radius or deformation, and one needs
prominent signal in the dense electronic spectrum. Metal clusters are
distinguished by the dominant Mie surface plasmon resonance
\cite{Kre93}, whose peak frequency is predominantly determined by
cluster radius and deformation. Thus there are many P\&P
studies on metal clusters, either free, deposited on a surface, or embedded
in a substrate, see e.g. Sec.~5.3.4 of \cite{Rei03a}.

As P\&P experiments on clusters are rather demanding, early studies
achieved a comparable, although coarser, effect by varying the
temporal width of a single laser pulse. For an early example on Pt
clusters, see \cite{Koe99}. We exemplify this type of analysis here for
the case of Ag clusters embedded in a He droplet for better handling
\cite{Doe07a}.
\begin{figure}[htbp]
\centerline{\includegraphics[width=0.6\linewidth]{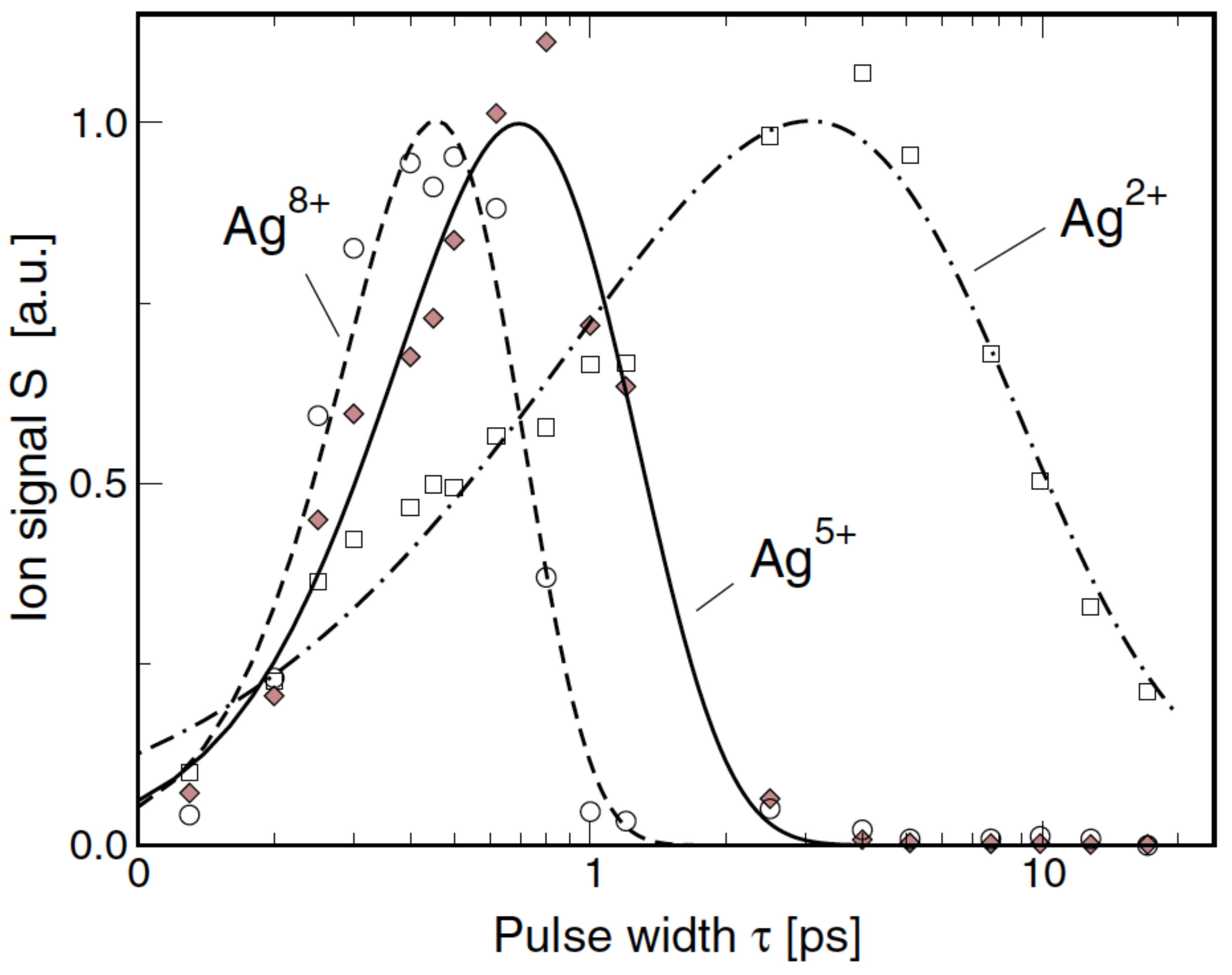}}
\caption{Yields for selected Ag$^{q+}$ ions after irradiation of a Ag
  cluster with a laser pulse of wavelength 800 nm, drawn as a function
  of the width $\tau$ of the the laser pulse. For the shortest pulse
  of 130 fs, the peak intensity is $I_0 = 1.2 \times 10^{14}$
  W/cm$^2$. For other $\tau$, the fluence $\propto \tau\,I_0$ has been kept
  constant.  The data are normalized for better comparison, and the
  fit curves serve as a guide to the eye. From~\cite{Doe07a}.
\label{fig:pulse_length2}
}
\end{figure}
The clusters are irradiated with laser pulses of fluence
$\tau\, I_0=156$ W\,fs/cm$^2$ (where $\tau$ is the pulse width and $I_0$
the peak intensity). This strong pulse leads to a disintegration of
clusters producing all sorts of fragments and highly ionized Ag
atoms. The charge state $q$ of the emerging Ag$^{q+}$ ions is an
indicator for the violence of the reaction and thus for the strength
of the laser-cluster coupling. Fig.~\ref{fig:pulse_length2} shows
the ion yield as a function of pulse width.  All ionization stages $q$
show a strong dependence on $\tau$ with a distinct maximum for a
certain $\tau$.  This optimum pulse width, which was already observed
in \cite{Koe99}, results from an interplay of (ongoing) laser pulse,
ionic expansion, and plasmon frequency. The IR laser pulse first triggers
ionization. The Coulomb pressure thus generated leads to a slow
expansion of the cluster. And the plasmon frequency (originally in the
visible range) decreases with increasing radius, until the laser comes
into resonance with the plasmon with subsequently strong energy
absorption and violent reaction. If the laser pulse is too short, it
is over before resonant conditions are reached. If it is too long, it
becomes too weak (remind the constant fluence thus implying decreasing
intensity with increasing pulse duration) to trigger sufficient
expansion. Such a maximum is seen in Fig.~\ref{fig:pulse_length2}
for each charge state, however at different delay times $\tau$. The
interpretation given in \cite{Doe07a}, furthermore, addresses a subtle
point in laser experiments.  The laser intensity is not constant over
the spatial width of the beam. It decreases when going away from the
focus. Thus clusters outside the focus receive a weaker signal than
those right in the focus of the beam. It is assumed that these lower
charge states are related to lower intensities out of focus. Therefore, the
experiment so to say produces at once results for different laser
fluences.

The rather involved P\&P analysis is easier for clusters in/on a
substrate because this allows a higher density of reactive
centers. Therefore, most P\&P studies on clusters are performed
in/on substrate. The typical setup is that of a chromophore in an
inert substrate. The latter thus serves mainly as a support for the
cluster. The principles and the richness of P\&P analysis remain
unaffected by the inert substrate. There is a couple of measurements
of an electronic property, the electronic relaxation time, for
clusters on surfaces in a variety of material combinations
\cite{Kle97,Kle99,Mer00}. More typical for P\&P analysis is the study
of ionic oscillations which has been performed, e.g., for Ag clusters
embedded in glass matrix \cite{Per00a,Sei00}.  A much more gentle
support is provided by liquid He clusters, which were already used as
useful laboratory for studying molecular properties under well
controlled conditions \cite{Sti01}.  The He environment couples such
softly to any other material that one can consider the embedded system
as being practically free. There are then several instructive P\&P
experiments of Ag clusters in He droplets, e.g.  \cite{Doe05b,Doe05a}
(called ``dual pulse'' experiments in these publications). A detailed
description of the large scale dynamics of Ag clusters is very
expensive. Theoretical investigations are thus often performed for Na
clusters as practicable model systems for metal clusters
\cite{And02,And04,Doe05a}.

\begin{figure}[htbp]
\centerline{\includegraphics[width=0.5\linewidth]{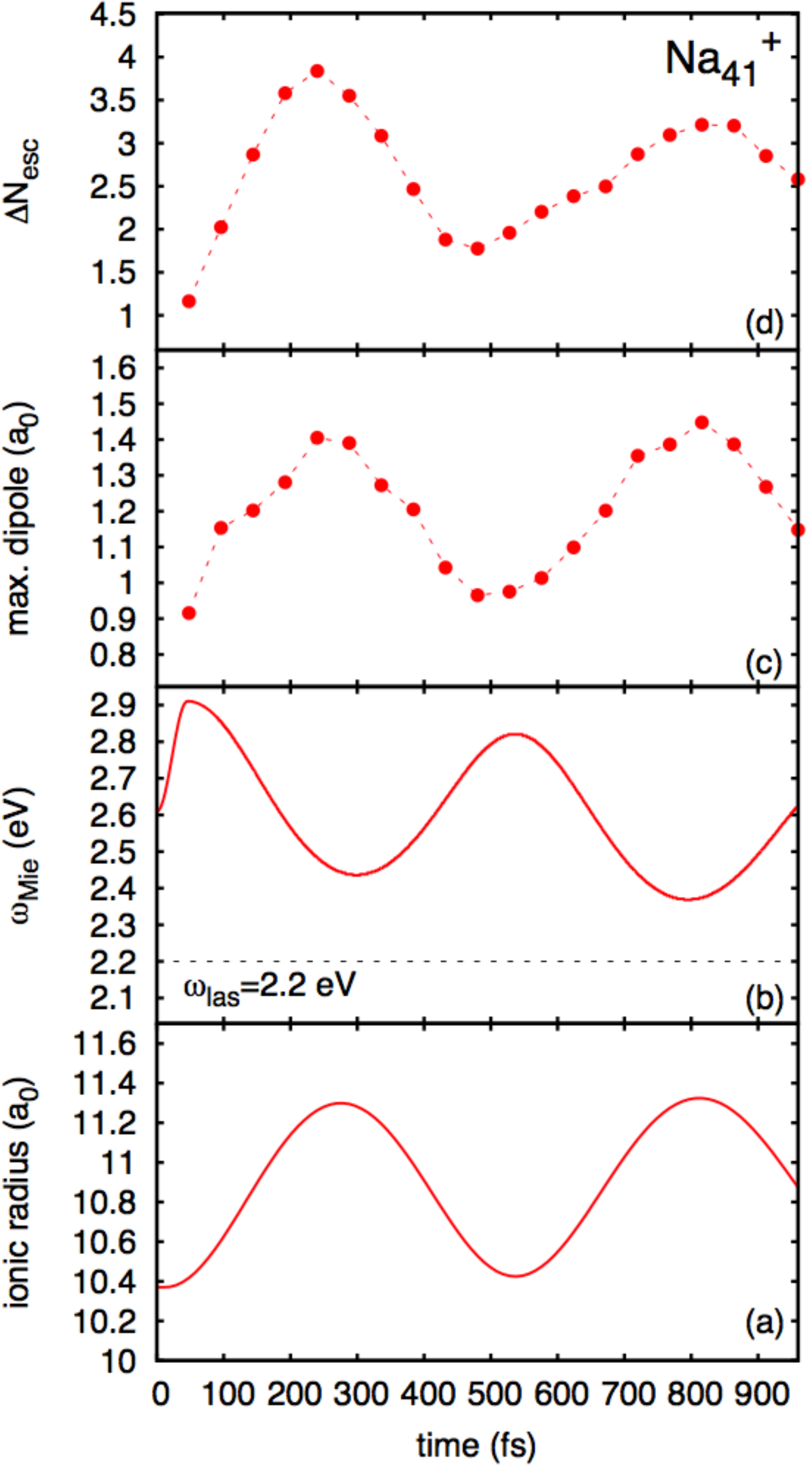}}
\caption{
Pump and probe spectroscopy of ionic breathing vibrations of
${\mathrm{Na}_{41}}^+$ using the Mie plasmon resonance as indicator.
Panel (a): time evolution of the ionic r.m.s. radius,
$\sqrt{\sum_I {\bf R}_I^2}$,  after the pump pulse.
Panel (b): time evolution of the Mie plasmon frequency after the
pump pulse. The laser frequency $\omega_\mathrm{las}=2.2\,\mathrm{eV}$ is
indicated as horizontal
dashed line for comparison.
Panel (c): maximum amplitude of the dipole response 
to the probe pulse as a function of time delay.
Panel (d): additional ionization $\Delta N_\mathrm{esc}$ induced
by the probe pulses as a function of time delay.
Pump and probe pulses have the same properties: photon frequency
$\omega_{\rm las}=2.2\,\mathrm{eV}$,
intensity $I=1.1\!\times\!10^{12}$ $\,\mathrm{W/cm}^2$,
and a sin$^2$ shape with
FWHM$=24\,\mathrm{fs}$.
The pump laser produces very quickly an initial emission of
$N_\mathrm{esc}=3$ electrons, thus {delivering a total} charge state $q=4^+$.
After \cite{And02}.
\label{fig:na41p-PP}
}
\end{figure}

The dominance of the Mie plasmon peak in metal clusters allows a
particular P\&P strategy which does not rely on directly hitting the
resonance but uses just the distance of the Mie plasmon frequency to
the laser frequency to map the underlying ionic dynamics. This
strategy has been studied in detail for the global breathing (radius
oscillations) of Na clusters in \cite{And02} and for the dynamics of
cluster deformation in \cite{And04}. We illustrate the scheme here for
the case of breathing.
Fig.~\ref{fig:na41p-PP} shows the result of a theoretical exploration
for the cluster ${{\rm Na}_{41}}^+$ using TDLDA
for electronic
dynamics coupled to molecular dynamics for the ionic motion
\cite{Cal98d,Cal00,Rei03a}.
The cluster ${{\rm Na}_{41}}^+$ is nearly spherical.  The pump pulse ionizes
it quickly to charge state ${{\rm Na}_{41}}^{4+}$.  This produces a Coulomb
pressure which triggers slow breathing oscillations of the whole
cluster, while deformation is negligible along the whole dynamics.  The
radius oscillations after the mere pump pulse are shown in panel (a)
of the figure.  The Mie plasmon resonance depends on charge state and
cluster radius. An estimate is shown in panel (b). One sees the fast
initial blue-shift due to the fast initial ionization to
$q=4^+$. After that, one finds oscillations which perfectly follow the
radius oscillations according to $\omega(t)\propto R(t)^{-3/2}$
\cite{Rei96c}.  The laser frequency for the probe pulse is also
indicated. It was chosen safely below the Mie resonance such that the
actual Mie frequencies never cross. The electronic response to the
probe pulse is small if the Mie frequency is far from the laser, and
large if it comes close.  This can be seen in panel (c) from the
maximum amplitude of the dipole signal during the probe pulse. The
strong dipole response leads to further ionization shown as additional
number of escaped electrons $\Delta N_\mathrm{esc}$ in panel (d). It
is, of course strongly correlated with the dipole amplitude. Tracing
the chain of correlations back to panel (d), we can conclude that the
extra ionization directly maps the global ionic radius using the
scheme with the remote laser frequency as an ``observer''.  Net
ionization thus provides a direct (time resolved) analysis of ionic
motion, in that case dominated, on the rather short times considered,
by radial oscillations. The actual long term evolution of the system
is, in fact, Coulomb explosion. The interesting aspect is that the
long path to explosion is accompanied by monopole (radial)
oscillations which are directly visible in the ionization signal
(actually via the plasmon peak).

Almost all P\&P studies on clusters have used the ionization yield as an
observable. One expects that one could learn more from more detailed
observables, particularly from time-resolved PES and PAD.  Such
experiments are, of course, much more complex and still more demanding
than the, already intricate, traditional P\&P measurements. Nonetheless, 
first studies in that direction have been published
\cite{Fen07,Pas12a} so far in the regime of hefty excitations. The
field is widely open for further studies in more moderate excitation
regimes.

All the P\&P studies, including the above example, show that the
ionization yield can depend sensitively on the choice of the laser
pulse characteristics. As the temporal profile and all other laser
parameters can be tuned in an extremely flexible manner, the question
naturally arises whether one could tune laser pulses for maximum yield
(or other desired reaction properties). This is the idea of ''optimal
control'' which is of particular importance in chemistry and molecular
physics, see e.g. \cite{Dam02,Kou05a}. Again, the application to large
clusters is more involved and allows more strategies to be tracked. An
interesting study using optimal control, e.g., to trigger the yield of
highly charged Ag atoms from Ag clusters can, nevertheless, be found in
\cite{Tru10a}.

While addressing promising future developments of P\&P studies, we
ought to mention the upcoming availability of attosecond pulses. These
allow P\&P studies which resolve features of electronic dynamics. We
will discuss that in the next section \ref{sec:P&P_ele}.

\subsubsection{Towards P\&P experiments with attosecond pulses}
\label{sec:P&P_ele}

Electron dynamics can also be analyzed at its own pace if one is
  able to handle pulses much shorter than typical electronic time
scales in the fs regime. This is nowadays experimentally accessible
down to some hundreds of attoseconds, at least in the form of a train of
attosecond pulses. This yields access to details of electronic
dynamics subject to electromagnetic perturbations. Early convincing
tests were performed in simple atoms such as He and Ar \cite{Joh07} on
the basis of a P\&P setup involving a UV atto-train on top of an IR
field. It was shown that the total
ionization may exhibit marked oscillations as a function of the delay
between UV train and IR signals, once the repetition rate of seven
attopulses per train is chosen to be half the IR period. The analysis of these experiments
was supported by simple simulations using the Time-Dependent
Schr\"odinger Equation (TDSE) with a single active electron. The
TDSE also served as a basis for further theoretical investigations,
either directly \cite{Ton10a,Ton10b,Mur13} or in perturbation theory
\cite{Riv09}.  These approaches gave convincing clues on the origin of
the modulation of the ionization but so far, no robust many-electron
theory is available to explain the observations.

More recently, experiments were generalized to simple
molecules with qualitatively similar results as in the atomic case,
although with a slightly different combination of IR and UV
pulses~\cite{San10,Kel11,Siu11}. The first fully microscopic calculations were performed on
this occasion and led to results remarkably compatible with
experiments~\cite{Nei13}. 
The actual interpretation of the underlying mechanism is
nevertheless, again, to be understood in more detail. Still, the
remarkable agreement between theory and experiments is worth being
presented and discussed in one test case.

We consider here the N$_2$ molecule as a test case~\cite{Nei13}. The laser pulse consists in an IR component~: 
 \begin{equation}
V_{\rm IR}(t) = E_{\rm IR} f_{\Delta T}(t) \sin(\omega_{\rm IR} t) \quad,
\label{eq:ir}
\end{equation}
with a frequency $\omega_{\rm IR}=1.58$~eV and 
with a sin$^2$ pulse profile $f_{\Delta T}$ of FWHM$=\Delta T/2 \simeq 29$~fs 
(see Eq.(\ref{eq:cos2})). At the same time is superimposed a train of $n$ attopulses, 
each of them labeled by $i$, which reads~: 
\begin{subequations}
\label{eq:atto}
\begin{eqnarray}
  V_{\rm atto} (t) 
  &=& 
  \sum_{i=1}^n E_{\rm atto}(\tilde{t}) \, f_{\delta t}(\tilde{t}-t_i)
  \sin\left[\omega_{\rm UV}(\tilde{t}-t_i) \right] \, e^{-4\tilde{t}^2/\Delta T^2}
  \;,
\\
  t_i 
  &=&
  \Delta t +(i-1) [\delta t + \delta t']
  \;,
\\
  \tilde{t}
  &=&
  t- \Delta t
  \;,
\end{eqnarray}
\end{subequations}
where $\delta t \simeq 0.29$ fs is the actual duration of each
individual attopulse, and $\delta t'$ the time separation between two successive
attopulses. The attopulse train (APT) is delayed by a variable delay $\Delta t$
with respect to the IR pulse (starting at t=0). Both the IR pulse and
the APT are linearly polarized. The individual attopulse shape 
is again a sin$^2$ profile (see Eq. (\ref{eq:cos2})) of FWHM$=\delta t/2$.  
A key point of the setup is to fix the time interval
between two successive attosecond signals. We choose here $\delta t +
\delta t' = T_{IR}/2 \simeq 1.30 fs$, which is exactly half the IR
period.  The amplitude of the APT is further modulated by a
Gaussian envelop of width such that the total APT duration is
about half (29 fs) the total duration of the IR pulse. 
This fixes the number $n$ of attopulses which is, in
this case, 22. The peak intensity of the IR pulse is chosen to
be $I_{\rm IR} = 10^{12}$~W/cm$^2$, while that of the attopulses is, at
maximum of the Gaussian envelope, $I_{\rm atto,0} = 10^{10}$~W/cm$^2$. 
Finally, the frequency of the APT lies in the UV
domain, $\omega_{\rm UV} = 20.4$~eV, so that each individual attopulse
contains about 1.5 UV oscillation.  The laser parameters are such that
the pure IR pulse does not lead to ionization, while a pure UV train does
lead to some ionization through
one-photon processes because the IP of N$_2$ is
around 16.3 eV $< \omega_{\rm UV}$. The remarkable result of
the experiments is that combining the IR and the attopulses leads to
a significant enhancement of ionization (while it only adds 1.58 eV on
top of the UV photons already above the continuum). Moreover, this
enhanced ionization is strongly modulated by the delay $\Delta
t$. Ionization actually exhibits marked oscillations as a function of
delay with a period equal to half the IR period. Maxima of
oscillations are attained for delays such that the attopulses are in
phase with maxima or minima of the IR pulse, which explains the
doubled frequency of ionization maxima as compared to the IR
frequency.

\begin{figure}[htbp]
\centerline{\includegraphics[width=0.7\linewidth]{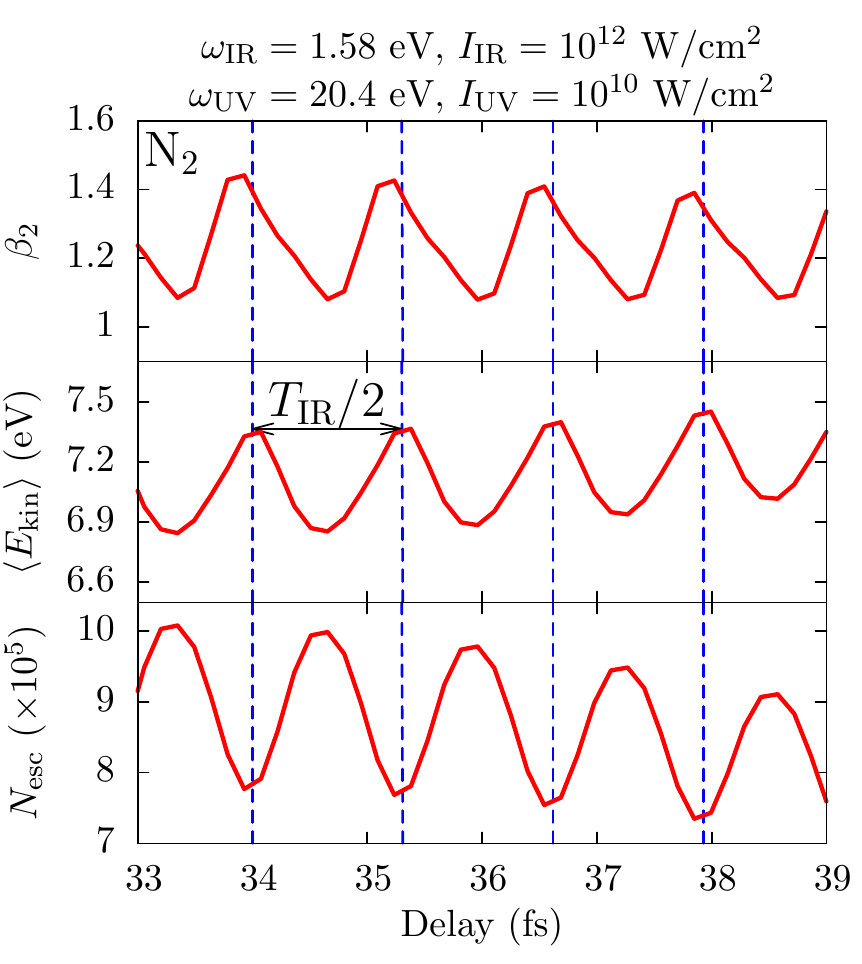}}
\caption{N$_2$ ionization properties irradiated by an IR pump and an
  attopulse train (see text for details) as a function of delay time
  between attopulse train and IR pulse. Bottom~: total ionization. Middle~: average kinetic energy per emitted
  electron, see Eq.~(\ref{eq:avekin1}). Top~: anisotropy
  parameter $\beta_2$, see Eq.~(\ref{eq:anisotropy}).  The faint dashed lines indicate the sequence
  of maxima and minima regularly separated by half the IR period. {Adapted from~\cite{Nei13}.}
\label{fig:atto_n2}
}
\end{figure}

The case is illustrated in Fig.~\ref{fig:atto_n2} where we have
plotted the total ionization, the average kinetic energy of emitted
electrons and the anisotropy $\beta_2$ characterizing the PAD (see
Eq.~(\ref{eq:anisotropy}) and Sec.~\ref{sec:pad_results}). 
As one
UV photon suffices for ionization, $\beta_2$ provides a complete
 characterization of the PAD.  The average kinetic energy is defined as
\begin{equation}
  \langle E_{\rm kin}\rangle
  =  
  \frac{\hbar^2}{2m} \int \textrm d \mathbf r \, \frac{{\bf j}^2({\bf r})}{\rho({\bf r})} \,
\left[ 1-{\cal M}^2({\bf r}) \right] \quad,
\label{eq:avekin1}
\end{equation}
where {$m$ is the electron mass}, ${\bf j}({\bf r})$ is the local current, $\rho({\bf r})$ the
local density and ${\cal M}({\bf r})$ the mask
function used to evaluate emitted electrons  (see Eq.~(\ref{eq:mask})).  It provides a simple
measure of the PES in terms of one number.  All three signals in
Fig.~\ref{fig:atto_n2} display remarkable oscillations as a function
of delay time $\Delta t$ with a period equal to half the IR
period. Both $\langle E_{\rm kin}\rangle$ and $\beta_2$ oscillate in phase
and in opposite phase with the total ionization respectively. Indeed, if the
deposited energy content is about the same, the higher the ionization,
the lower their average kinetic energy. And not surprisingly, the more
energetic the emitted electrons, the more aligned the emission along the
laser polarization axis and the larger the $\beta_2$. These
oscillations of the total ionization perfectly match those observed
experimentally. A comparison with PES and PAD has to wait until
  these quantities are experimentally available.

We finally end this discussion by mentioning recent experimental P\&P dynamics which use
XUV pulses for both the pump and the probe. This is at variance with the example discussed
above, where the pump is an IR pulse and the XUV probe is constructed
from some of its high order harmonics using the so-called RABITT
(Reconstruction of Attosecond Beating by Interference of Two-photon Transitions)
technique~\cite{Pau01,Mul02}. To distinguish the two kinds of P\&P, one sometimes quotes
them as IR-pump-XUV-probe and XUV-pump-XUV-probe experiments respectively. One type of
setup uses a coherent splitting of a XUV light produced by a FEL. For instance, this
technique has been successfully applied to small molecules as N$_2$ and
O$_2$~\cite{Mag12}, or C$_2$H$_2$~\cite{Jia13}. In these latter examples, the XUV
frequency is 38~eV, the photon intensity between $10^{11}$ and $10^{13}$~W/cm$^2$, and the
XUV pulse duration of 30~fs. The delay time resolution is 1 fs, and the whole delay time
can vary over the $\pm 350$~fs range. The advantage of using XUV light for the pump and
the probe is to ionize the species under study by absorption of a few photons, at variance
with an IR pulse. Therefore, this P\&P setup can follow the induced Coulomb explosion at a
time scale of a few fs. Very recently, some experiments went even further by taking
advantage of high harmonic generation from an IR pulse irradiating an atomic gas jet~: the
produced XUV attopulses are then separated from the IR pulse, filtered to keep only one
pulse which is at the end split into two coherent XUV as pulses. This brand new
technology has been applied in the irradiation of Xe atoms~\cite{Tza11} and the H$_2$
dimer~\cite{Car14}. The XUV intensities are about $10^{13-14}$~W/cm$^2$, their duration
about 600~as, their frequency is 14~eV, and the delay time is well below 1 fs. Such an
experimental apparatus thus enables to track the Coulomb explosion dynamics over the whole
reaction path and on a time scale never attained before.
In both XUV-XUV P\&P experiments, fragment or ion yields are measured as a
function of delay time. To the best of our knowledge, no electronic observable has been
measured so far. There are also very few real-time calculations of such a
dynamics~\cite{Car14,Pal14}. They employ a time-dependent Schr\"odinger equation of the
full electronic and nuclear wave function (note however that only vibronic modes -- no
rotational mode -- are considered in these calculations). For light atoms as those in H$_2$,
a quantal description of the nuclei is probably compulsory. The necessity  of such a fully quantal 
treatment is more questionable for heavier atoms, as in N$_2$. Anyway, the computational
cost of such calculations becoming too prohibitive for larger covalent systems, this probably calls for a
classical treatment of the ions, even if H$^+$ nuclei come into play.

\subsection{Dynamical aspects in photoelectron spectra}
\label{sec:pes_dyn}

We have discussed above how (static) s.p. spectra can be
extracted from the peaks observed in PES using Eq.~(\ref{eq:ekin1}),
see for instance Fig.~\ref{fig:sic_pes}. This identification can be
worked out by perturbation theory and requires a laser with moderate
intensity and high frequency resolution.  True dynamical processes
exploit more of the versatility of lasers.  The aim of this section is
thus to discuss the impact of the laser parameters frequency,
intensity, and pulse duration on PES, and find out how dynamical
aspects can be analyzed through PES.

\subsubsection{Impact of pulse duration}
\label{sec:pes_fwhm}

\begin{figure}[htbp]
\centerline{
\includegraphics[width=0.65\linewidth]{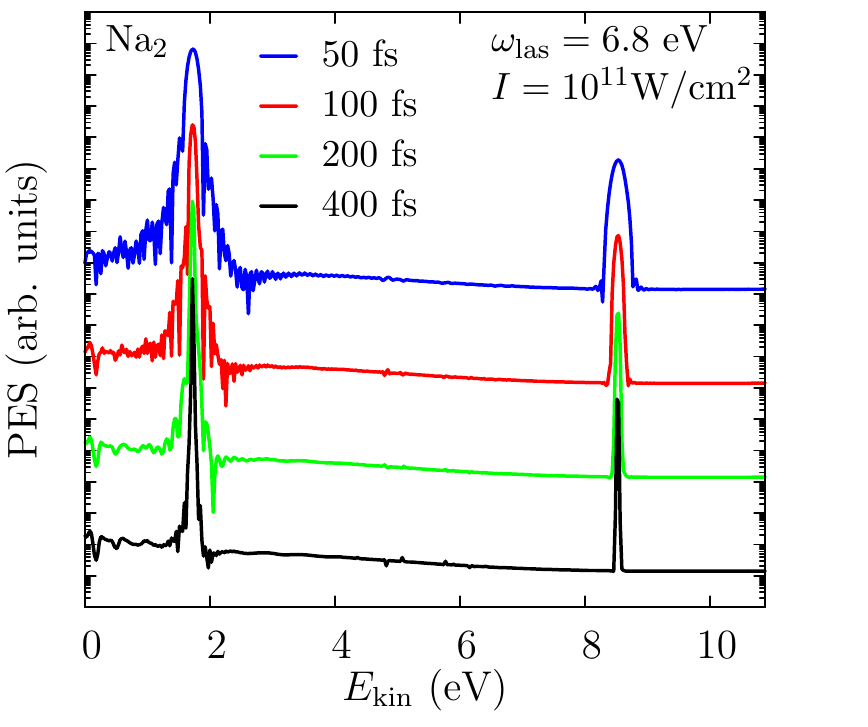}
}
\caption{Photoelectron spectra of Na$_2$ after irradiation by laser
  pulses of indicated characteristics. Some PES have been up-shifted
  for the sake of clarity.}
 \label{fig:na2_fwhm}
\end{figure}

Any laser pulse of finite duration delivers a distribution of frequencies
about its mean frequency $\omega_\mathrm{las}$. The longer the pulse,
the sharper this distribution. One can evaluate the width of this
distribution by calculating $\sigma_\mathrm{las} = \int_0^\infty
(\omega-\omega_\mathrm{las})^2 \, | \tilde{I}(\omega) | \, \textrm
d\omega$ where $\tilde{I}(\omega)$ is the Fourier transform of the
time-dependent laser intensity $I(t)$.  As an illustration, we give in
the following table some widths related to pulse duration at
$\omega_\mathrm{las}=20$~eV\cite{WopPhD}.
\begin{table}[htbp]
\begin{center}
\begin{tabular}{c||c|c|c|c|c|c}
\hline
$T_\mathrm{pulse}$  (fs) & 10 &  30 & 60 & 75 & 200 & 1000\\
\hline
$\sigma_\mathrm{las}$  (eV) & \ \ 0.44 \ \ & \ \ 0.16 \ \  & \ 0.083 \ & \ 0.068 \ & \ 0.027 \  &\  0.0058 \ \\
\hline
\end{tabular}
\label{tab:sigma_las}
\caption{Width of frequency distribution for different laser pulse durations, around a
mean value of $\omega_\mathrm{las}=20$~eV.}
\end{center}
\end{table}
One consequence of the finite frequency width of the laser pulse is
that a PES (in the perturbative regime) does not display a sharp spike
exactly at $E_\mathrm{kin}=\varepsilon_i + n\omega_\mathrm{las}$, but
a more or less soft peak around this $E_\mathrm{kin}$. Finite pulse
duration thus produces a broadening of the PES peaks, the larger the
shorter the pulse. To exemplify this effect, Fig.~\ref{fig:na2_fwhm}
displays PES of Na$_2$ irradiated by laser pulses of intensity
$10^{11}$~W/cm$^2$, frequency $\omega_\mathrm{las}=6.8$~eV, and
a couple of different pulse durations from 50 to 400~fs.
We clearly observe an increasing resolution of the PES peaks with
increasing pulse duration. This holds as long as the intensity of the
laser pulse remains sufficiently low. Higher intensities can blur this
picture because ongoing ionization induces a drift of the peaks due to
a Coulomb shift of the levels. This will be addressed in
{Secs.~\ref{sec:pes_I} and \ref{sec:TRPES}}.

\subsubsection{Impact of laser polarization}
\label{sec:polariz}

\begin{figure}[htbp]
\centerline{\includegraphics[width=0.6\linewidth]{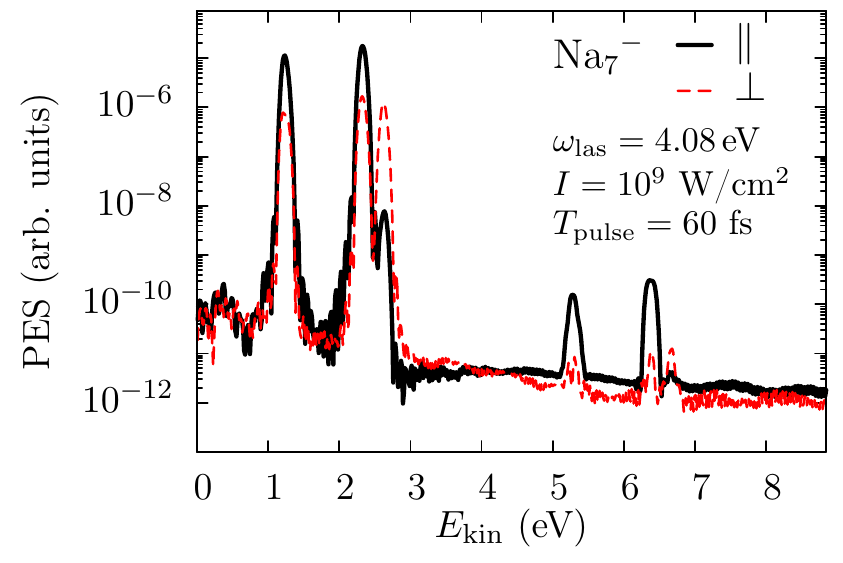}}
\caption{Photoelectron spectra of ${\mathrm{Na}_7}^-$ for a given
  orientation, irradiated by a laser with parameters as
  indicated. Black full curve: PES along the laser polarization
  axis. Red dashes: PES in the direction perpendicular to it.
  Calculations were done in cylindrical approximation for
  the electronic potentials.}
\label{fig:na7m-pes_para_perp}
\end{figure}

In section~\ref{sec:PADetal}, we have seen that a combined PES/PAD,
that is an energy- and angle-resolved photoelectron spectrum, can
deliver a lot of information on the dynamics of the photoemission.
In the perturbative regime, it reveals the angular distribution
  for emission from specific s.p. states. A simplified view
  can be obtained by restricting the analysis to two specific
  directions~: one parallel to the laser polarization axis
  ($\theta=0,180^\circ$) and one perpendicular to it
  ($\theta=90^\circ,270^\circ$).  The point is illustrated in
Fig.~\ref{fig:na7m-pes_para_perp} where calculated parallel and
perpendicular PES of ${{\rm Na}_7}^-$ are plotted.  
The sodium cluster
has a fixed orientation here (orientation averaging will be discussed
in Sec.~\ref{sec:pad_results}).  The calculated s.p.
energies are $\varepsilon_1=-2.82$~eV, $\varepsilon_2 = -1.72$~eV, and
a pair of almost degenerate $\varepsilon_{3,4} = -1.43$~eV. As in the
case of Na$_5$ (see Fig.~\ref{fig:sic_pes}), the peaks of the PES
perfectly fulfill the relation $E_\mathrm{kin}=\varepsilon_i+\nu
\omega_\mathrm{las}$ with $\nu=1,2$. The first group of peaks
  between 1 and 3 eV corresponds to $\nu=1$. States 1 and 2 
  predominantly emit along the laser polarization axis, while states 3 and 4
  show a clear preference of emission in perpendicular direction. The
2-photon process (between 5 and 7 eV) suppresses even more strongly
the perpendicular direction, and thus the parallel photoemission dominates.
This indicates a general feature of multiphoton emission~: the
  higher the photon number $\nu$, the larger the anisotropy $\beta_2$
  corresponding to an increasing dominance of emission parallel to
  the laser polarization axis.

\subsubsection{Impact of laser frequency}
\label{sec:mono&multi}

\begin{figure}[htbp]
\centerline{
\includegraphics[width=0.5\linewidth]{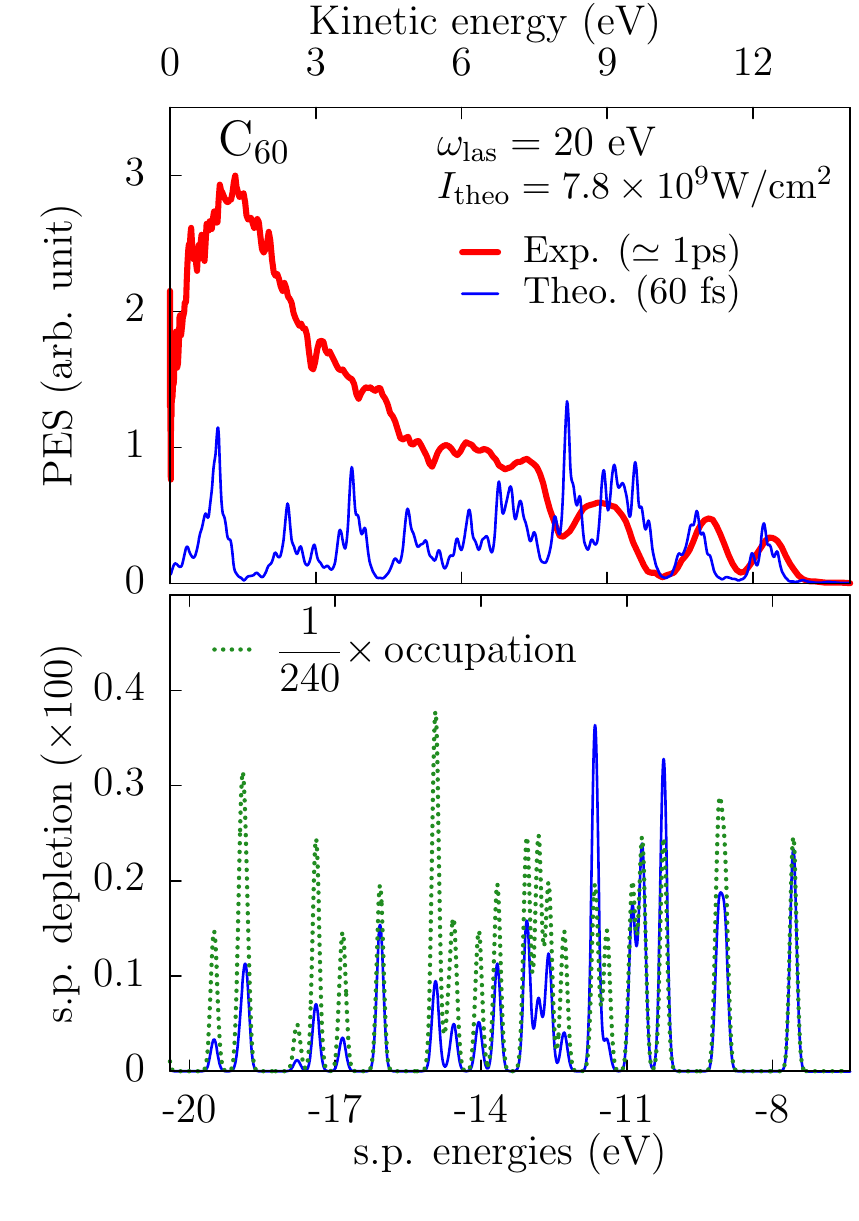}
\includegraphics[width=0.5\linewidth]{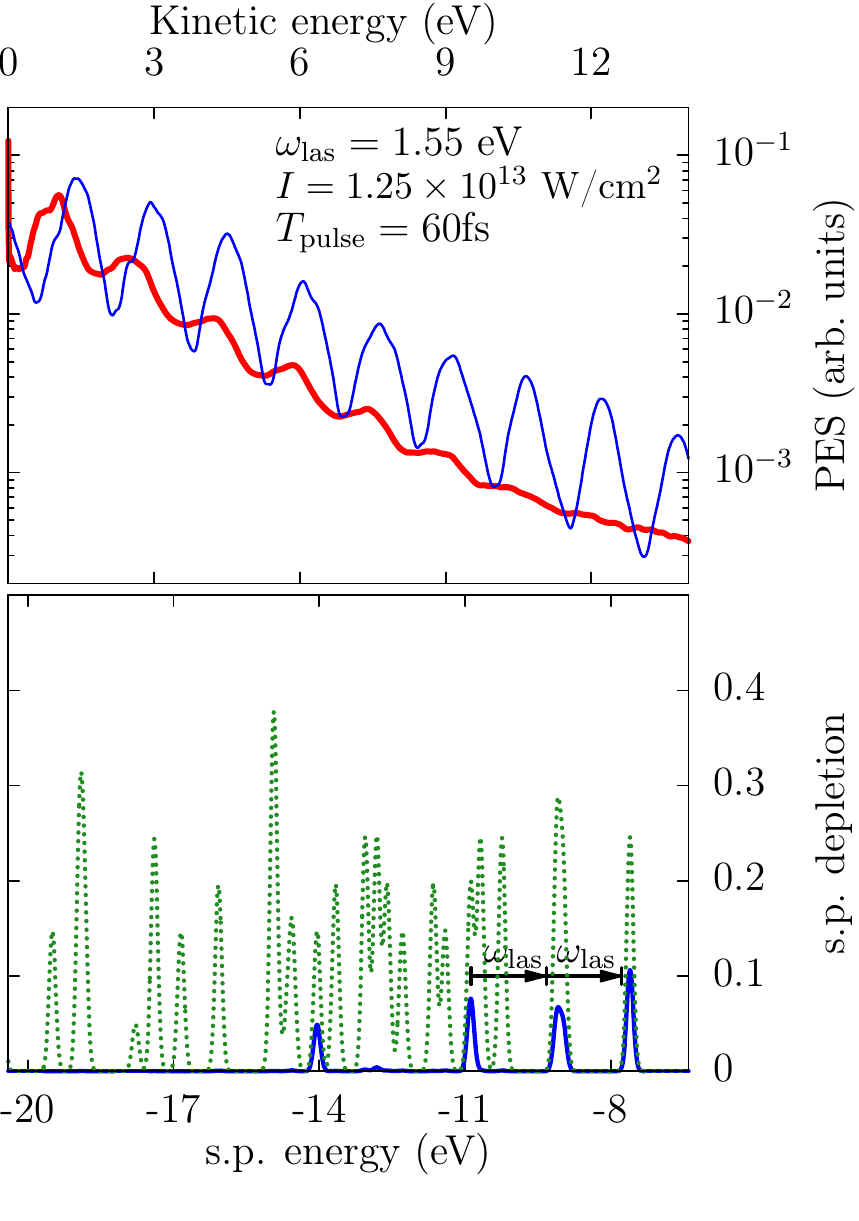}
}
\caption{Top~: Theoretical and experimental photoelectron spectra of
  C$_{60}$ {with radius of 6.763~a$_0$ and orientation averaging}. 
Bottom~: Calculated single particle depletion (blue full lines) compared
  with static occupation numbers (green dots). Left~: one-photon regime
  ($\omega_\mathrm{las}=20$~eV), laser pulse length of 60 fs and
  intensity of $7.8\times 10^9$~W/cm$^2$ (theory) or a synchrotron
  irradiation of duration of about 1 ps. Right~: multi-photon regime
   with $\omega_\mathrm{las}=1.55$~eV, laser pulse length of 60 fs and
  intensity of $1.25\times 10^{13}$~W/cm$^2$, both in theory and
  experiment. In the bottom left panel, the single particle depletions
  are multiplied by 100 for a better comparison.  Adapted from
  \cite{Bar14}.}
 \label{fig:pes-deple_c60}
\end{figure}

The estimate (\ref{eq:ekin1}), i.e.
  $E_\mathrm{kin}=\varepsilon_i + \nu\omega_\mathrm{las}$, of PES peaks
  establishes correctly the relation between peaks at $E_\mathrm{kin}$
  and corresponding s.p. energies $\varepsilon_i$. However,
  it does not tell anything about the strength with which the peaks
  appear. And here, we can have dramatic differences between
  one-photon processes and multiphoton ones.  As an illustration, we show in
Fig.~\ref{fig:pes-deple_c60} ionization pattern of C$_{60}$ in the
one-photon (left panels) and in the multi-photon (right panels) regimes~\cite{Bar14}.
{Orientation averaging presented in Sec.~\ref{sec:orientaver} has 
been applied in the theoretical calculations, to allow a comparison with 
experimental measurements.}
The upper panels compare theoretical and experimental PES, and the
  lower panels show as complementing information the depletion (blue
  full lines) and the occupancy (green dotted lines) of the
  corresponding s.p. levels.  
{The one-photon case was already presented in Fig.~\ref{fig:c60-vmi_xuv} which compares theoretical 
and experimental combined PES/PAD and VMI.}
The laser pulse in this case experimentally stems from synchrotron radiation at
20~eV with a pulse duration of about 1~ps. Theoretical calculations
were done for the same frequency, intensity $I=7.8\times
10^9$~W/cm$^2$, but shorter pulse $T_\mathrm{pulse}=60$~fs for
practical reasons.  In the one-photon regime ($\omega_\mathrm{las}
\gg$ IP), we do not expect that the pulse duration is essential.
Calculations yield a total ionization of about 0.006. In the
multi-photon case, the theoretical parameters are chosen as in the
experiment, that is $T_\mathrm{pulse}=60$~fs, $I=7.8\times
10^9$~W/cm$^2$, and $\omega_\mathrm{las}=1.55$~eV. Here, the
calculated total ionization is about 0.07.
Comparing both cases (one- and multi-photon), PES and depletion
pattern are completely different. The one-photon case also shows a
marked difference between experiment and theory. This has to be
discussed in detail.

The bottom panels of Fig.~\ref{fig:pes-deple_c60} display the
s.p. levels (with occupation weight) and their
depletion. For $\omega_\mathrm{las}=20$~eV (lower left panel), all
states can emit by a one-photon process.
And indeed, we observe that
most states contribute to the total ionization. As expected, the
theoretical PES resembles the s.p.  depletion pattern very
much~\cite{Din12}. 

We now turn to the multi-photon case (right panels of
Fig.~\ref{fig:pes-deple_c60}). With the laser frequency chosen here,
at least 6 photons are needed to bring electrons from the HOMO into
the continuum. Since the probability of ejection decreases with the
number of photons required, only the least bound states can be
significantly depleted. And this is what is observed in the right
bottom panel~: the states which emit the most are the HOMO, HOMO$-1$
and HOMO$-3$, precisely separated by about
$\omega_\mathrm{las}=1.55$~eV. Therefore, we cannot expect that the
PES maps the whole s.p. energy spectrum. Indeed, the
theoretical PES exhibits oscillations consisting in a broad peak
repeated several times, separated by $\omega_\mathrm{las}$ from one
copy to the other. These oscillations constitute the typical multiphoton ionization (MPI)
pattern (see Sec.~\ref{sec:mecha}). Each one of these MPI peaks is
rather well bundled due to the fact that the few emitting states line
up rather well with the photon frequency.

In both cases (one- and multi-photon), the experimental PES differ 
from the theoretical ones. The difference
looks particularly large for the one-photon case (upper left
panel). Here, the PES are still fairly comparable at higher energies
(7--13 eV).  While theoretical calculations were done at the ground
state configuration (zero temperature), the experimental peaks are
broadened due to ionic vibrations of C$_{60}$ which are rather large
at the experimental temperature of 900~K.  A huge discrepancy between
theory and experiment is observed at low electron kinetic energy.  The
experimental PES grows almost an order of magnitude above the
theoretical one. We think that these low-energy electrons stem from
electron-electron collisions which hinder part of the electrons from
being directly emitted, but rather lead to auto-ionization mechanisms
or thermal electron emission. Such dynamical electron-electron
correlations are not included in TDLDA. So we are missing here most
of the low-energy electrons. This could be cured with theories beyond
TDLDA which will be discussed in Sec.~\ref{sec:dissipe}. For the
time being, the comparison between experiment and theory is relevant
only for the highest photoelectron energies dominated by direct
electron emission. And there, the agreement is very satisfying.

A difference between experimental and theoretical PES is also seen in
the multi-photon regime (upper right panel).  
We first note a shift between the position of the theoretical peaks
and that of the experimental ones. This might be due to a slight
uncertainty in the determination of the experimental intensity and/or
pulse duration, which can then produce a different ionization stage.
And we will see that this can cause a sizeable (Coulomb) redshift, see 
Secs.~\ref{sec:pes_I} and \ref{sec:TRPES} below. Moreover,
the amplitude of
oscillations of PES decreases with increasing kinetic energy in the
measurements, while it remains more or less constant in the
theoretical calculations.  Once again, the dynamical correlations,
which are missing in the theory, are most probably the mechanism
responsible for the damping of the oscillations in the experimental
data.

\subsubsection{Impact of laser intensity}
\label{sec:pes_I}

\begin{figure}[htbp]
\centerline{
\includegraphics[width=0.49\linewidth]{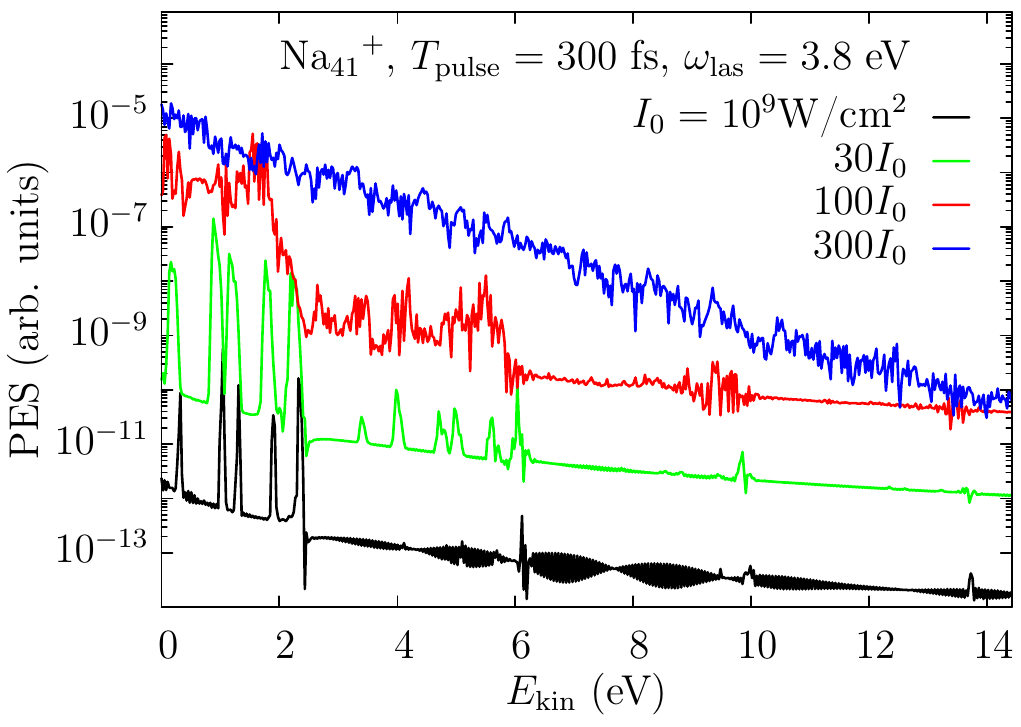}
\includegraphics[width=0.49\linewidth]{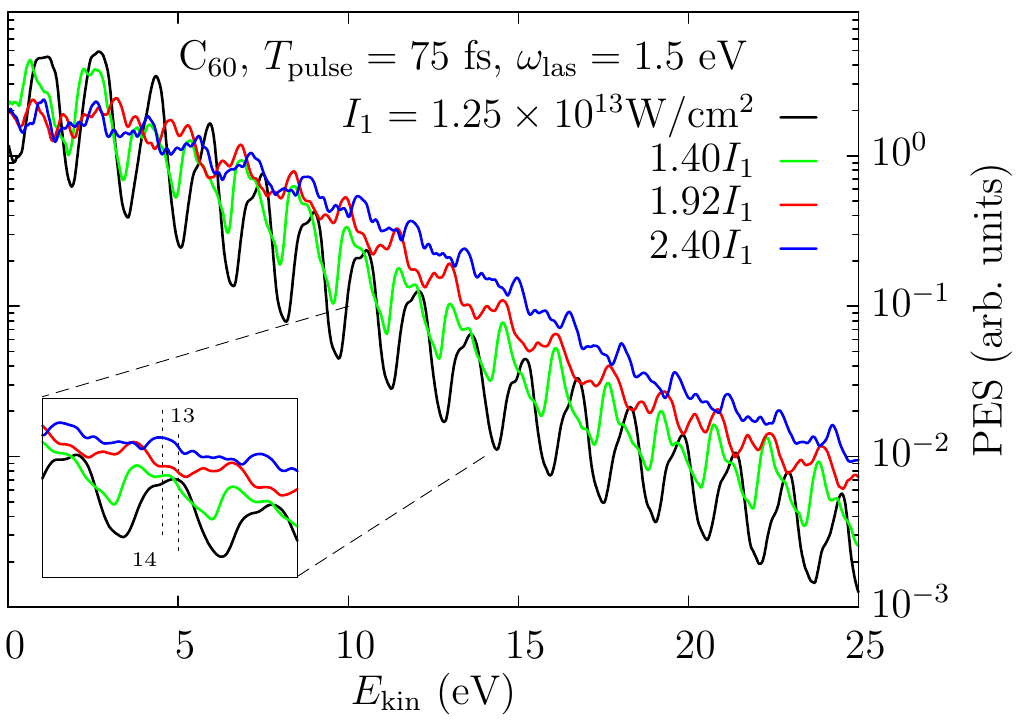}
}
\caption{Calculated photoelectron spectra for various
  intensities. Left~: case of ${{\rm Na}_{41}}^+$ irradiated by laser pulses
  of duration of 300~fs, frequency of 3.8~eV, and $I_0=10^9$~W/cm$^2$ 
{(calculations done with CAPS)}.
  Right~: case of C$_{60}$ {(calculated in full 3D with radius of 6.44~a$_0$ 
and with  orientation averaging)}
  irradiated by laser pulses of duration of 75~fs, frequency of
  1.5~eV, and $I_1=1.25\times 10^{13}$~W/cm$^2$. The inset zooms in
  the 10--14~eV range with the 13-photon ionization of the HOMO and
  the 14-photon one of the HOMO$-1$.}
\label{fig:pes_I}
\end{figure}

A couple of {experimental} PES had already been shown in
  Fig.~\ref{fig:campbell1} for the case of C$_{60}$ irradiated by
  laser pulses of various intensities. The results seemed to be all in
  the same regime marked by smooth, exponentially decreasing PES
  throughout.  Another example was just given above in
  Fig.~\ref{fig:pes-deple_c60} where the top right panel shows
  typical MPI pattern of repeated peaks which have at least an
  exponentially decreasing envelope. For a more systematic survey, we
  now discuss computed PES for two different
  clusters.  As a first example, we discuss a set of PES of
${{\rm Na}_{41}}^+$ (see left panel of Fig.~\ref{fig:pes_I}) obtained with
the same laser frequency (3.8~eV) and pulse duration (300 fs) but with
varying intensities (from bottom to top) starting from
$I_0=10^9$~W/cm$^2$ and up to $300\,  I_0$.
The chosen frequency of 3.8~eV is basically off-resonant, as the
dominant plasmon response of ${{\rm Na}_{41}}^+$ lies in the spectral range
between 2.5 and 3.1~eV. Since the IP of ${{\rm Na}_{41}}^+$ is 5.3~eV, the
first peaks at low kinetic energies in Fig.~\ref{fig:pes_I} stem
from two-photon processes. The pulse duration is 300~fs to allow a
high spectral resolution of the PES peaks, as is observed in the
lowest PES (black curve). The laser intensity is also small enough to
stay in the perturbative regime, since the total ionization
$N_\mathrm{esc}$ is 0.004 in this case. When the laser intensity is
increased by a factor 30 (light green curve), one can spot a slight
broadening of the peaks, although the PES still shows clear
signatures of the underlying s.p. spectrum. The broadening
develops to the side of lower kinetic energies, in total yielding a
weak red-shift of the peaks. It is to be noted that the total ionization
amounts to $N_\mathrm{esc}=0.1$ now. This ionization enhances the
Coulomb binding in the course of the electron emission which, in turn,
leads to a down-shift of the s.p. energies, called henceforth
``Coulomb shift'' \cite{Poh00}~: electrons which are emitted later in
the process see a deeper binding and thus escape with lower kinetic
energy. {A time-resolved PES would allow one to observe more clearly 
how the Coulomb shift builds in the course of time. This will be discussed 
in more detail in Sec.~\ref{sec:TRPES} and Fig.~\ref{fig:na8_trpes}.}
Note also the appearance of multi-photon peaks with this larger
intensity. Two-photon processes grow $\propto I^2$ and so the
absorption of more photons becomes more probable.

At the next stage, $I=100\,I_0$, the PES is already smeared to broad
steps. Here, $N_\mathrm{esc}=0.54$ and the Coulomb shift significantly blurs the PES. 
But still, we can distinguish blocks of one-, two-, and
three-photon processes.
Finally, the highest intensity of $I=300\,I_0$ produces an ionization
of $N_\mathrm{esc}=4.1$ and the PES are fully smoothed to an
exponential decrease with almost no structure left, resembling a
``thermal'' PES. But an exponential PES alone is not a sufficient
indicator of thermalization.  More information contained in PAD can
help in that respect, as was briefly mentioned in
Sec.~\ref{sec:PADetal}. We will address this point in detail in
Secs.~\ref{sec:pad_results} and particularly \ref{sec:dissipe}.

We finally discuss the case of PES in a highly multi-photon regime,
shown in the right panel of Fig.~\ref{fig:pes_I}. The studied system
is here C$_{60}$ and the laser has a pulse duration of 75~fs and a
frequency of 1.5~eV. Five laser intensities $I$ have been considered~:
$I_0=1.25\times 10^{13}$~W/cm$^2$, $1.2\,I_0$, $1.4\,I_0$, $1.92\,I_0$, and
$2.4\,I_0$. The total ionization of course increases with $I$~:
$N_\mathrm{esc}=0.03$, 0.06, 0.12, 0.31, and 0.60. With the chosen
frequency, we need at least 6 photons to extract electrons (the IP is
here 8~eV). At the lowest intensities, we clearly observe the typical
MPI patterns of repeated copies of the peak.  The peaks are gradually
red-shifted when we increase the laser intensity which is, again, the
Coulomb shift. Moreover, the red-shift increases with laser intensity due to
increasing ionization and finally washes out all structures at the
highest $I$. The disappearance of the MPI peaks is illustrated in the
inset zooming into the 13-photon ionization of the HOMO and the
14-photon ionization of the HOMO$-1$. The net conclusions from that
case are the same as those from ${{\rm Na}_{41}}^+$. But here, it becomes even
more obvious that the envelope of MPI follows in any regime an
exponential decrease. Taking this together with the smoothing due to
high ionization delivers then the purely exponential profile
resembling thermal emission.


\subsubsection{More on the role of plasmon}

\paragraph{Competition between laser and plasmon frequencies}
\label{sec:las_vs_plasm}

\begin{figure}[htbp]
 \centering
 \includegraphics[width=0.7\linewidth]{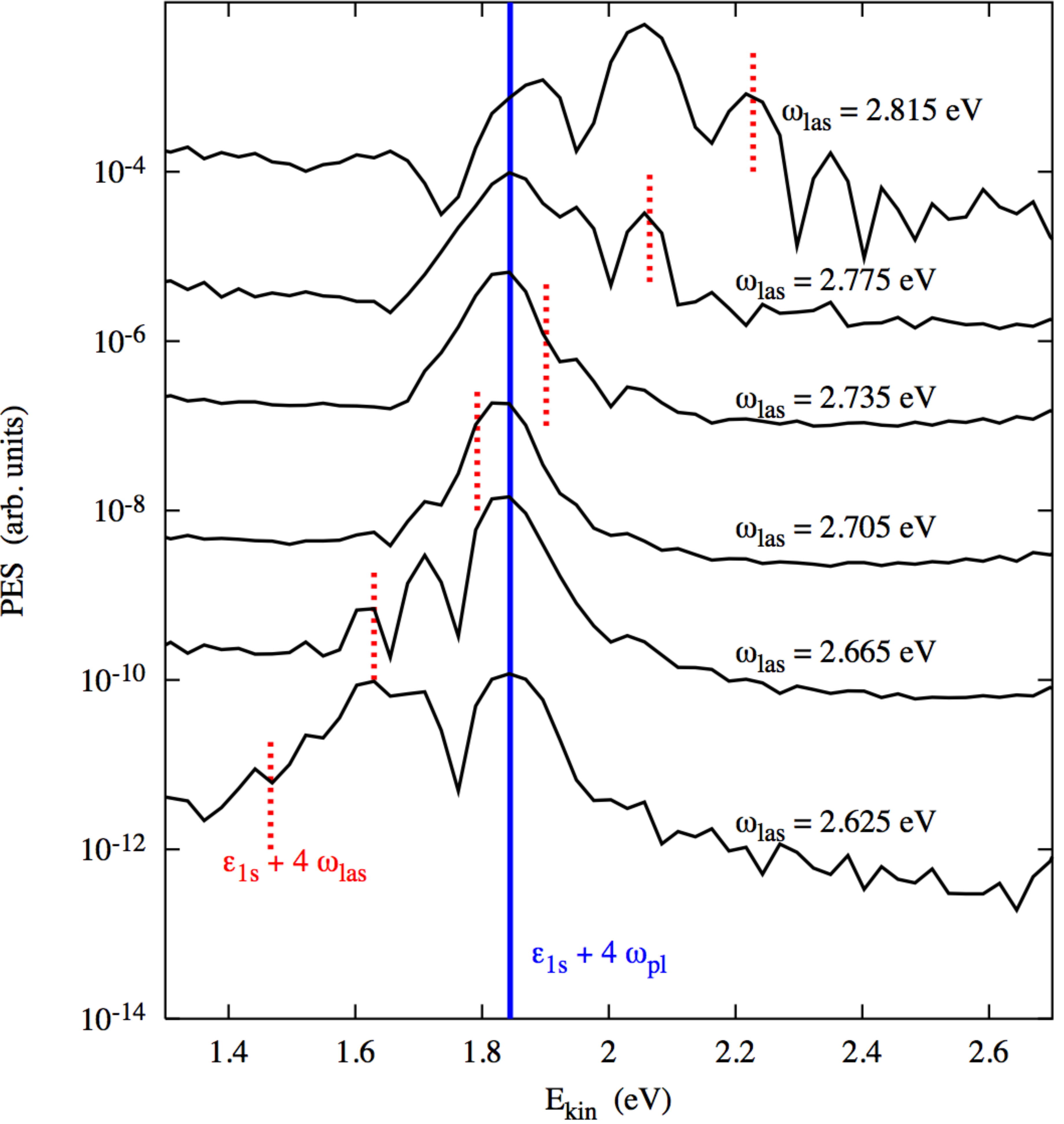}
\caption{Photoelectron spectra in the energy range of the 4-photon
  process (red vertical dots at $\varepsilon_{1s} +4
  \omega_\mathrm{las}$) of the $1s$ state of ${\mathrm{Na}_9}^+$,
  after irradiation by laser pulses with duration of 48~fs, intensity
  of $10^9$~W/cm$^2$, and six different frequencies
  $\omega_\mathrm{las}$ as indicated. The blue solid vertical line
  indicates the the 4-plasmon process located at $\varepsilon_{1s} + 4
  \omega_\mathrm{pl}$, with $\omega_\mathrm{pl}=2.7$~eV.  Adapted
  from~\cite{Poh01}.}
\label{fig:na9p-pes-plasm}
\end{figure}

Thus far, we have discussed PES emerging from the interplay
  between s.p. energies and the pulse frequency. This simple
  view has to be modified in the vicinity of strong excitation modes
  of the system.  In particular, the dominant Mie surface plasmon in
  metal clusters can also have a large impact on the PES. To
illustrate this point, we discuss the irradiation of ${{\rm Na}_9}^+$
by laser pulses with duration of 48~fs, intensity of $10^9$~W/cm$^2$,
and six different laser frequencies $\omega_\mathrm{las}$ in the
vicinity of the Mie plasmon frequency of ${{\rm Na}_9}^+$,
$\omega_\mathrm{pl}=2.7$~eV. The resulting PES are depicted in
Fig.~\ref{fig:na9p-pes-plasm}.
The peak of the pure four-photon process, located at $\varepsilon_{1s}
+4 \omega_{\rm las}$, is indicated for each laser frequency by
vertical dots. The position of this peak moves to the blue with
increasing $\omega_\mathrm{las}$. Additionally, one observes a peak
whose position does not depend on the laser frequency, indicated by
the solid vertical line. Its position matches the energy of a
4-plasmon process, i.e. $\varepsilon_{1s} +4 \omega_\mathrm{pl}$. When
$\omega_\mathrm{las}$ is sufficiently separated from
$\omega_\mathrm{pl}$ (see the two lowest and the two uppermost
curves), one can easily disentangle the four-photon process from the
four-plasmon one. The four-plasmon peak actually dominates the PES in
most of the cases, a further indication of the already discussed
resonant ionization mechanism (see Fig.~\ref{fig:h2o_irrad} and
\ref{fig:nesc_om} in Sec.~\ref{sec:basic_ioniz}).  In most cases,
one can even conceive a coexistence of plasmon and photon
excitations. For instance, the uppermost curve shows three prominent
peaks, that is the four-plasmon peak, the four-photon peak, and in between a
two-plasmon--two-photon peak. For reasons not yet well understood, we
cannot find significant signals of a mix of one-plasmon--three-photon
or three-plasmon--one-photon processes.

\paragraph{Towards time-resolved PES}
\label{sec:TRPES}

\begin{figure}[htbp]
 \centering
 \includegraphics[width=\linewidth]{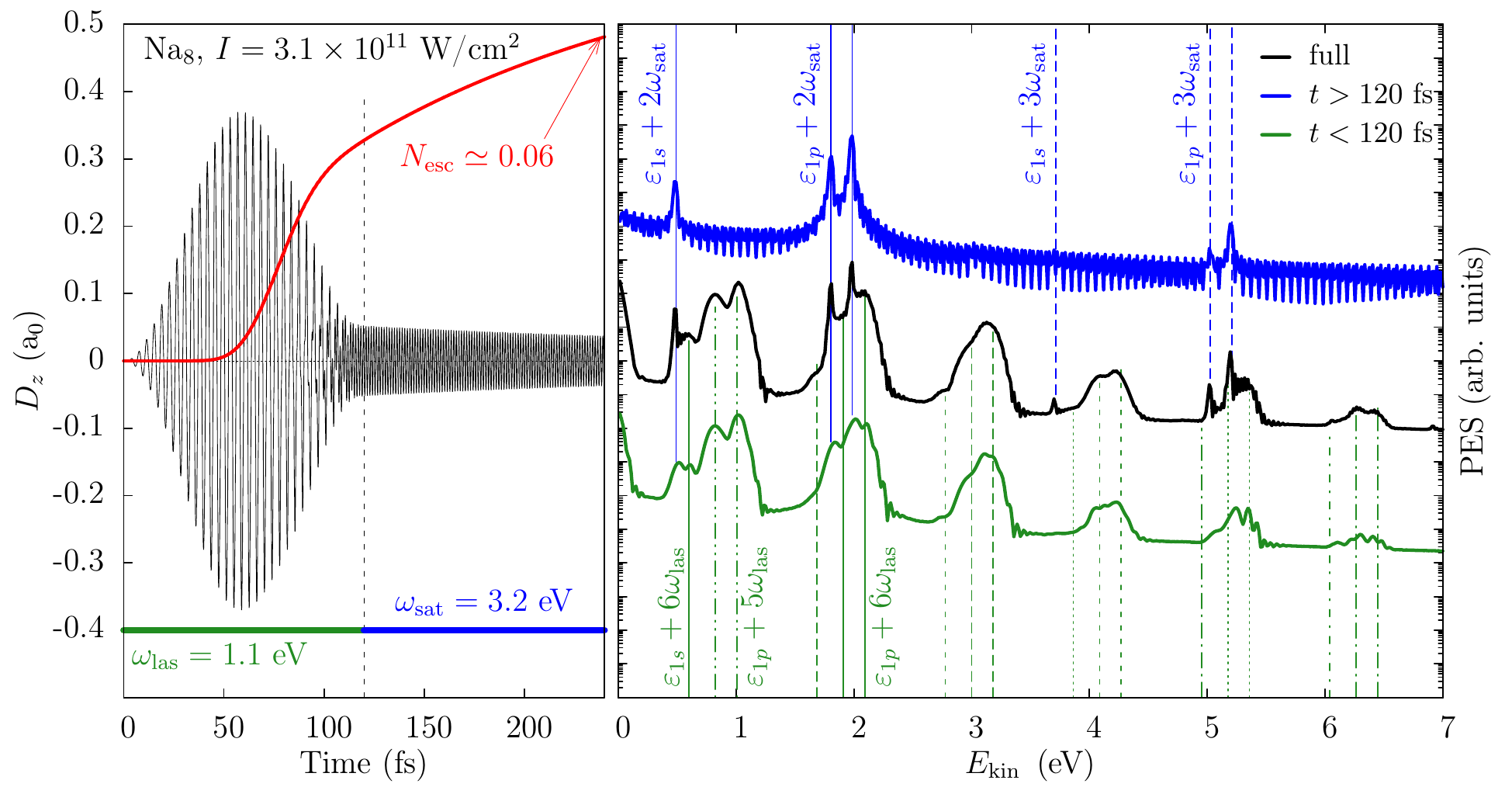}
\caption{Electronic dynamics of Na$_8$ after irradiation by a laser
  with polarization along $z$, duration of 120~fs, frequency of
  $\omega_\mathrm{las}=1.1$~eV and intensity of $3.1\times
  10^{11}$~W/cm$^2$ {(calculations with pseudopotentials and in full 3D)}.  
Left~: Time
  evolution of electronic dipole along the symmetry axis of Na$_8$,
  denoted by $z$, and of ionization $N_\mathrm{esc}$.  The horizontal
  bars emphasize time spans dominated by the indicated frequencies,
  that is by $\omega_\mathrm{las}$ below 90 fs and by a satellite
  frequency $\omega_\mathrm{sat}=3.2$~eV. Right~: corresponding PES
  evaluated in different time windows~: during the first 120 fs (green
  or light curve), after 120 fs (blue or dark curve), and for the full
  time span (black curve).  The various PES have been augmented
    with scale factors to separate them in the plot. The vertical
  lines indicate the $1s$, the degenerate $1p_{xy}$ and the $1p_z$
  energies shifted by multiples of $\omega_\mathrm{las}$ (lower lines)
  or of $\omega_\mathrm{sat}$ (upper lines). All single electron
  energies have been down-shifted by 0.18~eV to account for the Coulomb
  shift due to the total ionization of 0.06 at the end of the
  simulation time. Adapted from~\cite{Wop13}.}
 \label{fig:na8_trpes}
\end{figure}

The above example dealt with the excitation of a resonance mode
  through a close, although not exactly matching, laser
  frequency. Resonant modes may also be excited by laser pulse whose
  frequency $\omega_\mathrm{las}$ is far away from the resonance, but
  where a multiple of $\omega_\mathrm{las}$ coincides with the mode.
  Such a situation should also {leave} traces in the PES. We exemplify that
  for the case of  Na$_8$ irradiated by a laser polarized along the
symmetry axis of the cluster, denoted by $z$, a pulse duration of
120~fs, a frequency of 1.1~eV and an intensity of
$3.1\times 10^{11}$~W/cm$^2$. The Na$_8$ cluster consists in two
squares parallel to a plane (denoted by $x$ and $y$) and which are
twisted by $45^\circ$ around the $z$ axis. It possesses three states of
energies $\varepsilon_{1s}=-5.75$~eV, $\varepsilon_{1p_{xy}}=-4.5$~eV
and $\varepsilon_{1p_z}=-4.2$~eV. The optical response is dominated by
the Mie plasmon at 2.5~eV. But there are also further strong peaks
\MD{in the optical response},
especially one located at $\omega_{\rm sat}=3.2$~eV. With
$\omega_\mathrm{las}=1.1$~eV, the laser pulse is clearly
off-resonant. One thus expects a time evolution of the electronic
dipole in phase with the laser pulse, as it was the case in the
off-resonant irradiation of a water molecule, see
Fig.~\ref{fig:h2o_irrad}. This is indeed the case during almost the
whole pulse duration, as is visible in the left panel of
Fig.~\ref{fig:na8_trpes}.
A higher frequency however appears from 100~fs on, precisely at
$\omega_\mathrm{sat}$. It persists even after 120~fs when the laser is
switched off, since sizable oscillations remain at this higher frequency.
One should moreover notice that
$\omega_\mathrm{sat}\simeq{3}\, \omega_\mathrm{las}$, which indicates that
absorption of three photons from the laser pulse triggers this
excitation. It is a typical example for higher harmonic
generation. The persistence of dipole oscillations provokes a
continuous electronic emission. Thus the total ionization $N_{\rm
  esc}$, also plotted in the left panel of Fig.~\ref{fig:na8_trpes} as a red curve,
does not level off after 120~fs but rather increases with a constant
slope, reaching the value of 0.06 at the end of the simulation time.

It is interesting to observe how the PES builds up in time in
  such a case. To this end, we plot in the right panel of
 the PES calculated for the full time span,
 and compare it to that calculated for the first
120~fs (green lower curve) and that after 120~fs (blue upper
curve). 
One observes a slight down-shift of the peaks from the early to the 
late time
windows. This provides a time-resolved illustration of the Coulomb shift
(discussed in Sec.~\ref{sec:pes_I}).
Due to the final $N_{\rm esc}=0.06$, an
  average Coulomb shift of the s.p. energies of $\delta\epsilon=-0.18$~eV
emerges. Hence the latter energies, indicated by vertical lines, have been shifted by
  $\delta\epsilon=-0.18$~eV to achieve a better matching with the PES peaks. The full
PES (black middle curve) shows patterns repeated with equal spacing
which are, at first glance, MPI peaks, as
seen before for C$_{60}$ (see right panel of
Fig.~\ref{fig:pes-deple_c60}).  The first peak near zero kinetic
energy is related to a four-photon process emitting out of the $1p_z$
state.  Most of the peaks in this full PES can be identified with
$E_\mathrm{kin}=\varepsilon_i+\nu \omega_\mathrm{las}$, as indicated
by the bottom vertical lines. There are, however, further peaks not
explained in terms of photon frequency.  To disentangle the
peaks, we have also evaluated the PES in two time windows, an early
one during the laser pulse, i.e. 0-120~fs, and a late one after the
pulse is over. The PES for the early window (lower green curve) is fully
explained by MPI with the laser frequency.  The PES from the late
window (upper blue curve) shows sharp peaks which can be identified as
multi-resonance peaks located at
$\varepsilon_i+\mu\omega_\mathrm{sat}$, with $\mu=2,3$ (we have also
applied here the same red-shift $\delta\epsilon$). No MPI peak from the
laser shows up in the late window.  Note also that the multi-resonance
peaks already slightly develop during the laser pulse (see vertical
lines from above). 

In this example, we have thus demonstrated that a dynamical competition
between various frequencies, here that from the laser pulse and that from a
higher resonance matching the third harmonics of the laser, can
provide mixed mappings of the PES and can thus give rise to a complex
structure of the PES. A time-resolved PES analysis can be a way
  to disentangle the different contributions. To that end, even a
  coarse time resolution {as performed here} may be sufficient.

\subsection{Photoelectron angular distributions (PAD): a sensitive tool}
\label{sec:pad_results}

This section is devoted to PAD. We are using free clusters as
  examples. Thus we consider orientation averaged PAD throughout, see
  Sec.~\ref{sec:orientaver}. Remind that these can be expanded in
  terms of Legendre polynomials $P_{2k}(\cos\theta)$ according to
  Eq.~(\ref{eq:anisotropy}). The expansion parameters $\beta_{2k}$
  carry all information about the orientation averaged PAD. The
  largest non-vanishing $\beta_{2\nu}$ is related to the number $\nu$
  of photons involved in the process. The most important parameter is
  the anisotropy $\beta_{2}$ which is also the only relevant parameter
  for one-photon processes. Therefore, $\beta_{2}$ will play a key
  role in the following presentations.

\subsubsection{One-photon regime}
\label{sec:pad_mono}

\paragraph{Stationary state picture and Bethe-Cooper-Zare formula}

The study of PAD has a long standing history, especially in
atoms. Early works by Bethe \cite{Bet33aB} and Cooper and Zare
\cite{Coo68a,Coo68b} are still routinely used in today's cluster literature
\cite{Bar09,Mau09}.  The Bethe-Cooper-Zare formula delivers a compact
expression for the evaluation of $\beta_2$ in spherical potentials. At
variance with our standard way of evaluating PAD (see Sec.~\ref{sec:pad}), 
it provides a stationary state picture which thus
requires evaluation of both bound {\it and} continuum electronic
states to describe initial and final electronic states. By
construction, it does not include possible electronic rearrangement
effects following electronic emission, as was demonstrated in
\cite{Wop13}. This limits the applicability to cases where electronic
rearrangement (through Coulomb residual interaction) can be
neglected. Furthermore, being developed for atomic physics, the
Bethe-Cooper-Zare is strictly limited to spherical external
potentials. Nonetheless, it may be useful as zeroth order estimate
and reference.

The Bethe-Cooper-Zare formula was first derived {\it in
    first-order perturbation theory} for one-electron atoms in
  \cite{Bet33aB}. But it can also be applied to many-electron systems in
  an independent-state picture where the many-body wave function is a
  simple anti-symmetrized product of s.p. orbitals
  \cite{Coo68a,Coo68b}.  For a given electronic level $i$, the total cross-section 
$\sigma^{(i)}$ for emission and its anisotropy
  $\beta_2^{(i)}$ are given by \cite{Coo69,Buc70}:
\begin{align}
 \sigma^{(i)}  &= \frac{(4\pi)^2\mathcal{N}}{3}\cdot\frac{L\mathcal{R}_-^2+(L+1)\mathcal{R}_+^2}{(2L+1)}\:,              \label{eq:bczyield}           \\
 \beta_2^{(i)} &= \frac{L(L\!-\!1)\mathcal{R}_-^2+(L\!+\!1)(L\!+\!2)\mathcal{R}_+^2-6L(L\!+\!1)\mathcal{R}_-\mathcal{R}_+\cos\Delta}       
                 {(2L+1)[L\mathcal{R}_-^2+(L+1)\mathcal{R}_+^2]}\:,                                                \label{eq:bczbeta}
\intertext{with}
 \mathcal{R}_\pm &= \int_0^\infty \textrm dr\,r^3 R_{L\pm 1}^{(f)}(r)R_L^{(i)}(r)\quad\text{and}\quad\Delta=\Delta_{L+1}-\Delta_{L-1}\:,  \label{eq:bczrpm}\\
 \mathcal{N}    &= \frac{4\pi^2e^2\omega_\mathrm{las}}{\hbar c}\:.\notag
\end{align}
where $L$ is the angular momentum of the initial state.  Once given
the (spherical) potential, the radial wave functions of bound state
$R_L^{(i)}$ and continuum state
$R_{L\pm 1}^{(f)}$ can be calculated by solving the associated radial
Schr\"odinger equation.  The phases $\Delta_{L\pm 1}$ entering the
continuum states can be obtained in a standard manner from the
asymptotic behavior of the outgoing wave $R_{l}^{(f)}$
\cite{Sch68aB}.

Note that in the particular case of a spherical wave function ($L=0$),
the Bethe-Cooper-Zare formula exactly delivers $\beta_2^{(s)}=2$
(maximum possible value of $\beta_2$ in the one-photon domain).  In this
case, the angular distribution of $s$ states is not influenced by the
radial form of the outgoing wave, so that the potential does not affect
the angular distribution; it only impacts the cross-section
\eqref{eq:bczyield}.
 
In spite (or maybe because) of its simple compact form, the
Bethe-Cooper-Zare formula has thus to be taken with a grain of caution
in realistic cases, because of its strong dependence on the shape of
bound and unbound electronic wave functions (see Fig.~\ref{fig:na8-beta2_om} below). 
 Furthermore, it remains a stationary
state picture thus well inside the perturbative regime, when
applicable (spherical potential).  Its range of application is thus
limited.

\paragraph{On the sensitivity of $\beta_2$ to model assumptions}
\label{sec:modelassum}

There are several approximations around in the description of
  clusters and molecules. The above mentioned Bethe-Cooper-Zare
  formula for instance forces spherical symmetry and neglects electronic
  rearrangement. Fully dynamical calculations often employ reduced
  symmetries as, e.g., in CAPS or by using a jellium model for the
  ionic background. Many of these approximations are validated for
  describing spectra and global emission properties. However, PAD is
  very sensitive {to this kind of theoretical details}, and one has to check carefully the impact of
  approximations.

We test the sensitivity for the simplest case of one-photon
  processes which are fully characterized by the anisotropy
  $\beta_2$. To have a systematic test, we study variations of
  $\beta_2$ as a function of laser frequency. We take as a test case
  Na$_8$ and consider $\beta_2^{(1p)}(\omega_\mathrm{las})$, that is, the
  anisotropy parameter for emission out of the $1p$ state~\cite{Wop12}. To
explore the impact of dynamical rearrangement effects in the PAD, the
left panel of Fig.~\ref{fig:na8-beta2_om} compares results of a
fully dynamical TDLDA-ADSIC calculation (full blue line) with the
result of the Bethe-Cooper-Zare formula {or, alternatively, with}  a TDLDA calculation in which the
electrons are propagated in the frozen ground-state Kohn-Sham
potential (dashed red line).
\begin{figure}[htbp]
\centerline{\includegraphics[width=\linewidth]{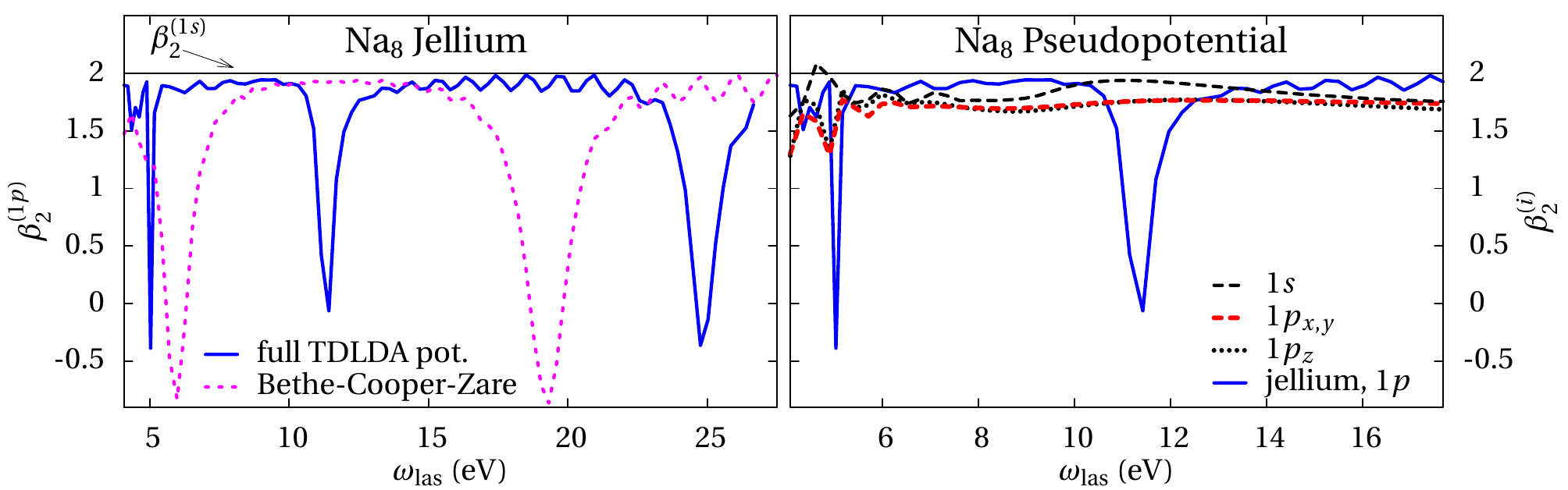}}
\caption{Anisotropy parameters $\beta_2^{(i)}$ of single particle
  states $i$ in Na$_8$, as a function of laser frequency
  $\omega_\mathrm{las}$. Left~: $\beta_2$ of the $1p$ shell in Na$_8$
  described by a spherical jellium ionic background (Wigner-Seitz
  radius $r_s = 3.65$~a$_0$, surface thickness $\sigma = 1$~a$_0$),
  calculated in TDLDA-ADSIC (full blue line), and in  {a Bethe-Cooper-Zare approach, 
see Eq.~(\ref{eq:bczbeta})} (pink dotted
  curve). For the $1s$ level, $\beta_2^{(1s)}=2$ at any
  $\omega_\mathrm{las}$. Right~: $\beta_2$ of the $1s$, $1p_{x,y}$,
  and $1p_z$ shells in Na$_8$ described by explicit ions and
  pseudopotentials, and averaged over six orientations. For a better
  comparison, $\beta_2^{(1p)}$ from the jellium calculation is
  superimposed. Adapted from~\cite{Wop12}.}
\label{fig:na8-beta2_om}
\end{figure}
{Note that the result of the Bethe-Cooper-Zare formula is basically
  identical to a dynamical calculation with the KS potential kept
  fixed at its static form and driven at very low laser intensity to
  stay safely in the one-photon regime.} The laser pulse duration is
of 60~fs and its intensity is scaled with the frequency ($I =
10^{13}$~W/cm$^2$ $\times \omega_\mathrm{las}$) to keep the total
ionization in the range between $10^{-4}$ and 0.1, and thus to stay in
a perturbative regime. The laser frequency is varied between 4.1 and
29~eV, so that only one photon is needed to promote the electron from
the $1p$ state into the continuum ($\varepsilon_{1p} =-4.08$~eV). The
ionic background of Na$_8$ is treated by a spherical jellium
(Wigner-Seitz radius $r_s = 3.65$~a$_0$, surface thickness $\sigma =
1$~a$_0$) which provides exactly the atomic situation for which the
Bethe-Cooper-Zare was developed.
Remind that $\beta_2^{(1p)}$ can vary between $-1$ and 2 for the
one-photon processes considered here. Within both approaches,
$\beta_2^{(1p)}$ is close to 2 for most frequencies. That means that 
the photoelectrons are mostly
emitted along the polarization axis of the laser. There are however frequencies
$\omega_\mathrm{las}$ at which we find pronounced dips down to
negative values.  Qualitatively, the pattern of the
  two cases are similar.  However, the deep dips in
  $\beta_2^{(1p)}(\omega_\mathrm{las})$ occur at very different
  places. This shows that dynamical effects, as the interaction of
the photoelectrons with the residual cluster, strongly impact the PAD.
A purely perturbative formula is therefore dangerous for molecules which 
develop remarked rearrangement effects, as e.g. metal clusters.

Moreover, the anisotropy parameter is very sensitive to the ionic
background itself. This is demonstrated in the right panel of
Fig.~\ref{fig:na8-beta2_om} where the jellium results are compared
to calculations with explicit ions and pseudopotentials (see Sec.~\ref{sec:ions}). 
The jellium model was tuned such that both models for
the ionic background provide about the same IP (4.08~eV for the
jellium and 4.28~eV for the pseudopotentials).  The spherical jellium
  delivers two occupied states, a $1s$ states with two occupancies and
  a degenerate $1p$ state holding six electrons. The non-spherical
ionic structure breaks the degeneracy into a $1p_z$ state and two
still degenerated $1p_{xy}$ states, and delivers a $1s$ which is not
perfectly spherical anymore. To extract a sound $\beta_2$ from the
PAD, we apply orientation averaging with the six reference
orientations appropriate for one-photon processes
(see Sec.~\ref{sec:orientaver}).  The $\beta_2$ of these three states
exhibit only faint dependence on $\omega_\mathrm{las}$, and stay
close to 1.7. The jellium model, on the contrary, systematically
delivers higher values of $\beta_2$, with the exception of a few
marked dips. It seems that the marked structures from the highly
symmetric jellium model are averaged out to a nearly constant
anisotropy. This can be explained by the rescattering of the
photoelectrons on the ionic structure before leaving the cluster.

\paragraph{More on the dependence of $\beta_2$ on laser frequency}

The results discussed in Sec.~\ref{sec:modelassum} suggest that
the ionic structure washes out strong variations in the frequency
dependence of the anisotropy. We will address here two exceptions from
this general observation.

\begin{figure}[htbp]
\centerline{\includegraphics[width=0.9\linewidth]{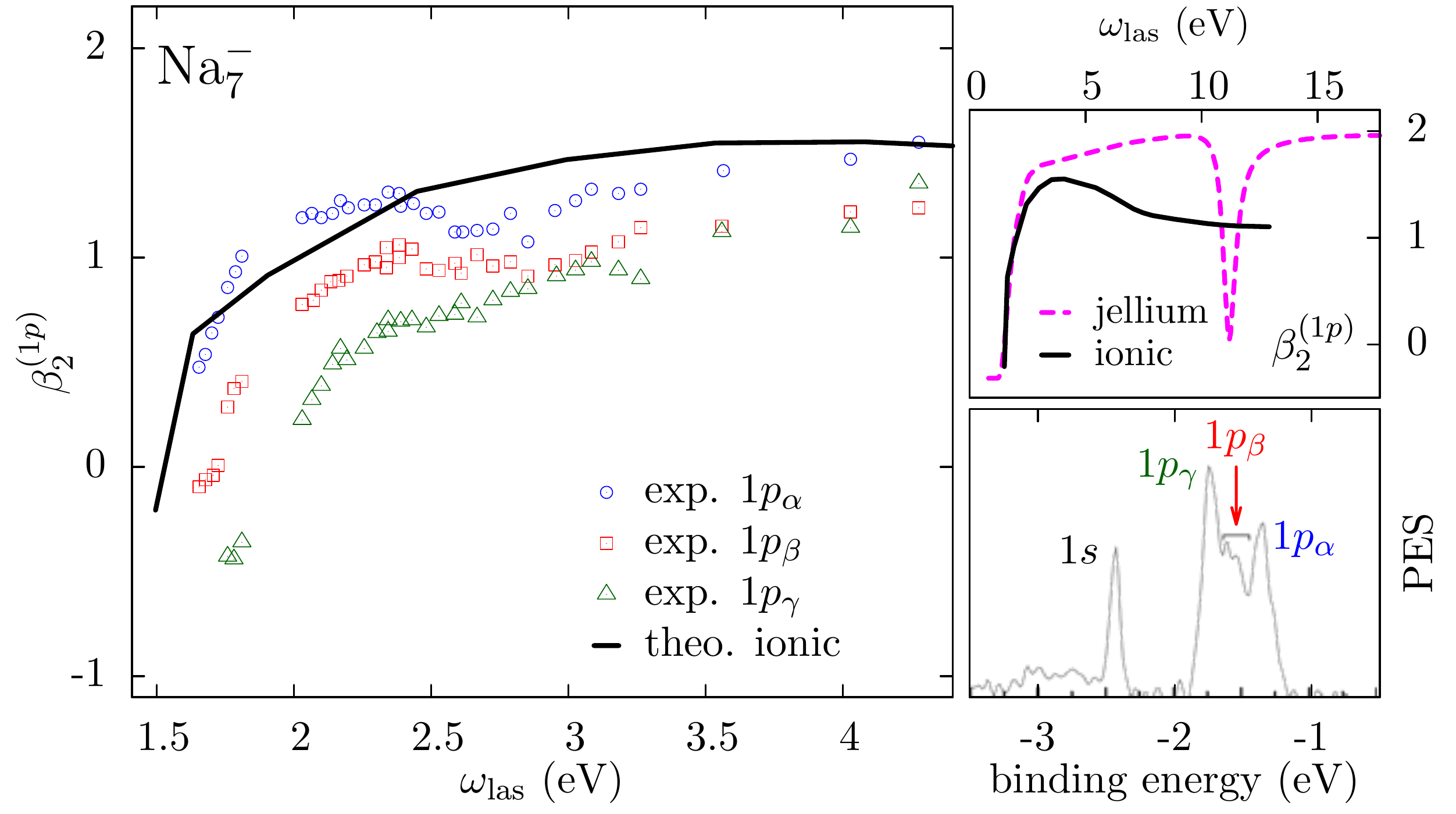}}
\caption{Top right~: comparison of calculated $\beta_2^{(1p)}$ of
  ${\mathrm{Na}_7}^-$ with a jellium background and an explicit ionic
  one, after irradiation by a laser with $I=10^9$~W/cm$^2$ and
  $T_\mathrm{pulse}=60$~fs.  Bottom right~: experimental PES of
  ${\mathrm{Na}_7}^-$ irradiated by a laser of intensity
  $<10^5$~W/cm$^2$ and duration of about 10~ns~\cite{Bar08}.  Left~:
  Experimental (open symbols, \cite{Bar08}) and theoretical (full
  curve, \cite{Wop10a}) anisotropy parameter of the $1p$ states of
  ${\mathrm{Na}_7}^-$, as a function of laser frequency.  $1p_\alpha$,
  $1p_\beta$ and $1p_\gamma$ correspond to state assignments of peaks
  observed in the experimental PES (bottom right).}
\label{fig:na7m-beta2}
\end{figure}

As a first example, we consider ${\mathrm{Na}_7}^-$ for which
experimental PAD exist~\cite{Bar08}. We should mention that this
case is numerically extremely demanding, because of the negative
charge and subsequently low IP (1.43~eV). We had to use a huge numerical box
($160^3$ mesh points and an overall box length of 280 $a_0$).
The left panel of
Fig.~\ref{fig:na7m-beta2} compares the calculated anisotropy for
emission out of the group of $1p$ states,
$\beta_2^{(1p)}(\omega_\mathrm{las})$, with the experimental data. 
All laser frequencies shown in the figure correspond to emission from
the $1p$ states in the one-photon regime.  
The experimental data show three curves. These are associated
  with the three sub-peaks of the $1p$ states found in the experimental
  PES, see lower right panel. The theoretical calculations did not
  disentangle these sub-peaks and show the $\beta_2^{(1p)}$ from the
  PAD averaged over the whole $1p$ group. The theoretical and
  (averaged) experimental curves nicely agree with each other. For
  higher frequencies, we see again the smooth trend as was already
  observed in the example of Na$_8$ in Fig.~\ref{fig:na8-beta2_om}. The
  data stay systematically a bit below the theoretical
  $\beta_2^{(1p)}$. This is probably due to electronic collisions not
  accounted for in TDLDA. The great surprise is the deep dip at low
  frequencies which is not an artifact because it is also clearly seen
  in the experimental results. It is due to a very special situation
  for this loosely bound anion. Indeed, we are near threshold, and the electron
  cloud is thus emitted with near zero momentum.  The KS potential seen by
  the escaping electron is extremely shallow and the outgoing
  electronic wave function has an extremely long wavelength
  throughout. Thus it cannot resolve the ionic structure and the
  rescattering mechanism which wipes out the dips (see discussion of
  Fig.~\ref{fig:na8-beta2_om}) becomes obsolete. This is demonstrated in
  the upper right panel of Fig.~\ref{fig:na7m-beta2} where we
  compare $\beta_2^{(1p)}(\omega_\mathrm{las})$ for jellium and
explicit ionic background over a larger frequency span. As in the case
of Na$_8$, the jellium model produces values near 2 and a pronounced
dip around 11~eV, while ionic structure delivers a generally smoother
curve with a maximum around 1.5. But both, jellium and ionic
background, deliver the same deep dip towards threshold. This
confirms nicely that the extremely long wavelength of the outgoing
electron state reduces the spatial resolution such that the soft
jellium and detailed ions cannot be distinguished anymore. This is
also the reason why an older calculation of $\beta_2$ for
${\mathrm{Na}_7}^-$ at low frequencies using a jellium model and
linear response could provide realistic results
\cite{Sol10}.

\begin{figure}[htbp]
\centerline{\includegraphics[width=0.6\linewidth]{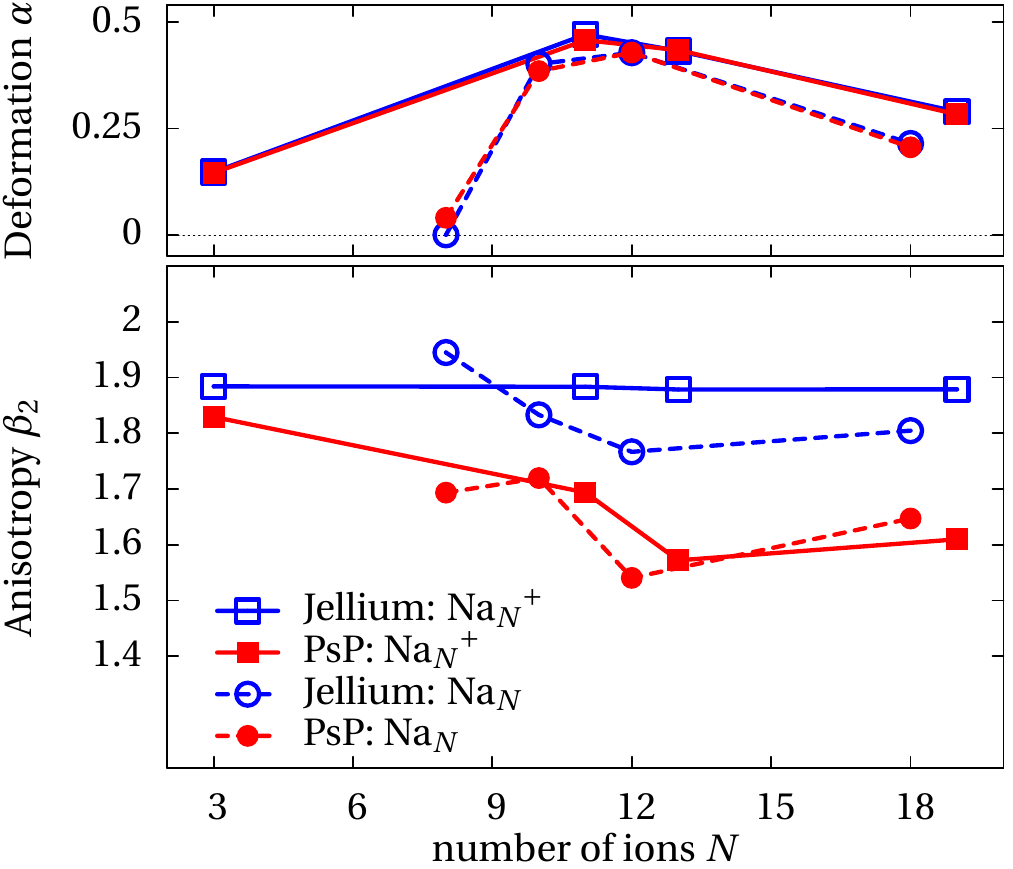}}
\caption{Comparison of quadrupole deformation $\alpha$ (top) and total
  anisotropy parameter $\beta_2$ (bottom) for small neutral (circles)
  and cationic (squares) clusters, obtained in a jellium description
  of the ionic background (open symbols) or with an explicit ionic
  structure (closed symbols). Adapted from~\cite{Wop10a}.
\label{fig:NaN-anisotrop}}
\end{figure}

In contrast to the PES which exhibits a strong dependence on laser
intensity, an orientation-averaged PAD seems to be not very sensitive
to it. Indeed, when irradiated by laser pulses of duration of 30~fs
and frequency of 34~eV but with two different intensities ($10^{10}$
and $10^{12}$~W/cm$^2$), the obtained PAD both delivers the same
$\beta_2=0.38$, while the PES at the highest intensity is blurred and
red shifted~\cite{WopPhD}.   Note that
the anisotropy parameter is here much smaller than for small Na
clusters, see Fig.~\ref{fig:NaN-anisotrop}. This is again due to the
influence of the ionic structure~: for sodium clusters, $\beta_2$
decreases by about 25~\% when going from jellium to ionic
background. This effect should be even stronger in C$_{60}$, since the
number of ions is much higher than for the considered Na$_N$ with $N =
3-19$. Additionally, the coupling of the electrons to the ions in
carbon atoms is stronger than in simple metal clusters.
 
\begin{figure}[htbp]
\centerline{\includegraphics[width=\linewidth]{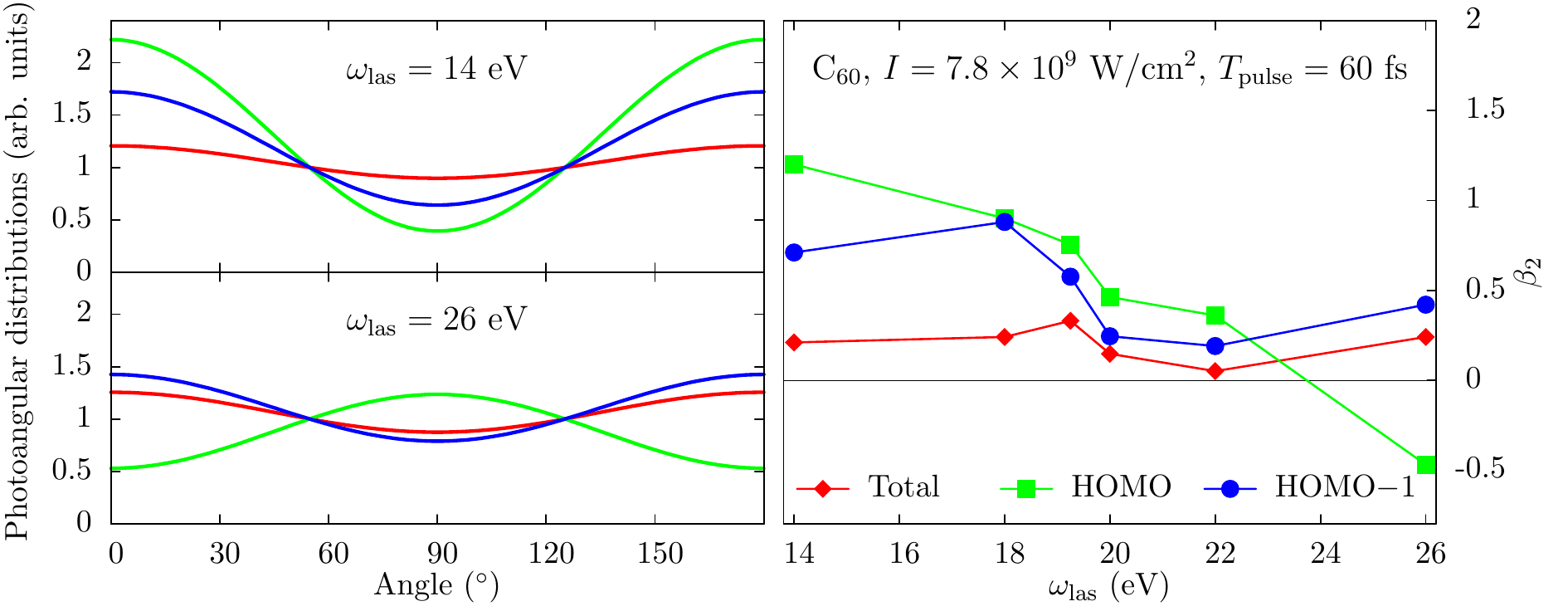}}
\caption{{Left~: Calculated orientation-averaged PAD from C$_{60}$ with radius of 6.763~a$_0$
irradiated by a laser pulse of intensity $I=7.8\times 10^9$~W/cm$^2$, duration of 60 fs, and frequency
$\omega_{\rm las}$ of 14~eV (top) or 26~eV (bottom). Right~: anisotropy parameter
$\beta_2$ as a function of $\omega_{\rm las}$. Red curves: total PAD and $\beta_2$ from
all single particle states~; light green curves~: the same but from the HOMO only~; blue curves~: the
same but from the HOMO$-1$ only.
}}
\label{fig:c60-beta2_om}
\end{figure}

More interesting is the frequency dependence of the PAD and
$\beta_2$. An orientation averaging procedure is applied here
(see Sec.~\ref{sec:direct_oapad}).  We focus on the photo-emission from
HOMO and HOMO$-1$ because experimentally, these are the only states
which clearly emerge above the background{~\cite{Bar14a,Bar14}. 
 Various calculated PAD are presented in the left column of Fig.~\ref{fig:c60-beta2_om}, and the extracted
$\beta_2$ are plotted in the right panel.}
The total $\beta_2$ (black curve) does not depend on
$\omega_\mathrm{las}$ very much. 
The anisotropy parameter of the HOMO and HOMO$-1$ exhibit by contrast
larger variations and both states deliver different behaviors. The
$\beta_2$ of the HOMO$-1$ (blue curve) is always positive
{with a minimum value at 0.18 around 22 eV.
On the contrary, the $\beta_2$ of the HOMO (green curve) steadily decreases 
with $\omega_{\rm las}$ and changes of sign a bit before 24~eV. Below 22~eV,}
the $\beta_2$ of the HOMO is also higher than the one of the
HOMO$-1$. 
The case once more demonstrates the extreme sensitivity of the 
$\beta_2$ as an observable characterizing a dynamical scenario. Mind that, in 
this monophoton domain, the anisotropy parameter is bound 
between $-1$ and $+2$, so that the variations 
in  Fig.~\ref{fig:c60-beta2_om} are quite significant. 
Together with its strong model sensitivity (see for example the 
discussion on Fig.~\ref{fig:na8-beta2_om}), this points out that 
$\beta_2$ is certainly a very rich quantity to be measured and 
computed in a highly refined manner.

\paragraph{Dependence of $\beta_2$ on cluster deformation}

To explore the impact of cluster deformation on the PAD, we now turn
to a series of small neutral and cationic metal clusters which cover
planar (${\mathrm{Na}_3}^+$), prolate (Na$_{10}$,
${\mathrm{Na}_{11}}^+$), oblate (${\mathrm{Na}_{13}}^+$ and
Na$_{18}$), and triaxial (Na$_{12}$, ${\mathrm{Na}_{19}}^+$)
systems. We consider detailed ionic background as well as a
  deformed jellium approach to it. The jellium deformation is tuned in
  each case to reproduce the global deformation of the ionic
  configuration. The shape can be quantified by the quadrupole
deformation $\alpha$ defined by
$\alpha=\sqrt{\sum_{m=-2}^2\alpha_{2m}^2}$ with $\alpha_{2m} =
4\pi\overline{r^2Y_{2m}}/{5N R_\mathrm{rms}^2}$, $R_\mathrm{rms}$ the
ionic root mean square radius, $N$ the total number of ions, and
$Y_{2m}$ the spherical harmonics for $l=2$. In
Fig.~\ref{fig:NaN-anisotrop}, we compare $\alpha$ with the total
anisotropy parameter $\beta_2$.
As in the case of Na$_8$ previously discussed, $\beta_2$ is extracted
from orientation averaged PAD, each PAD obtained after irradiation by
a laser in the mono-photon regime, that is
$\omega_\mathrm{las}=7.5$~eV for neutral species and 10~eV for
cationic ones (due to a stronger binding there) with
$I=10^{11}$~W/cm$^2$. The total ionization always remains between
$10^{-4}$ and $10^{-3}$.  We first note that $\beta_2$ shows only
  small variations, particularly for the jellium model. We
  cannot spot any correlation of the anisotropy $\beta_2$ with the
  deformation $\alpha$. The difference between neutral clusters and
  cations is also very small. However, as observed previously, there
  can be a large sensitivity to the structure of the ionic
  background. The $\beta_2$ is systematically smaller (more isotropic)
  when detailed ions are considered. It is particularly interesting to
  note that the discrepancy grows systematically with increasing
  cluster size $N$. This trend is corroborated by results from larger
  clusters. For example, the total anisotropy for C$_{60}$ comes close
  to zero, see Fig.~\ref{fig:c60-multi_I} for low intensities. The
  trend complies with the interpretation that rescattering with ions
  enhances the isotropic background: the more scatterers, the closer
  to isotropy.

\subsubsection{PAD in the multiphoton regime}
\label{sec:pad_multi}

{The orientation averaged PAD in the multiphoton regime develops
  more detailed angular dependence as indicated by
  Eq.~(\ref{eq:anisotropy}). For clarity, we recall it here~:}
\begin{equation*}
 \frac{\mathrm{d}\sigma}{\mathrm{d}\Omega}
  \propto
  1+\beta_2 P_2(\cos\vartheta)
  +\beta_4 P_4(\cos\vartheta)
  + \ldots
\end{equation*}
In a strictly perturbative regime (low laser intensity), the
  series terminates at $P_{2\nu}(\cos\vartheta)$ where $\nu$ is the
  order of the multiphoton process which is determined by the
  relation of IP to photon frequency.  This changes with increasing
  intensity where always all amounts of photons could be
  possible. Thus the series is, in principle, unterminated and we
  expect that the contributions of higher $\beta_{2n}$ increase with
  increasing intensity $I$. We thus briefly analyze the impact of
  intensity on the PAD in two emblematic cases, C$_{60}$ and
  Na$_8$. As we are going beyond the perturbative regime, we will
  consider higher $\beta_{2n}$ beyond the anisotropy $\beta_2$. It is
  to be noted that orientation averaging in the multiphoton regime
  has to be done explicitly by integration over orientations. Due to
  the high symmetry of the two test cases, only 18 integration points
  need to be computed.

\begin{figure}[htbp]
\centerline{
\includegraphics[width=0.49\linewidth]{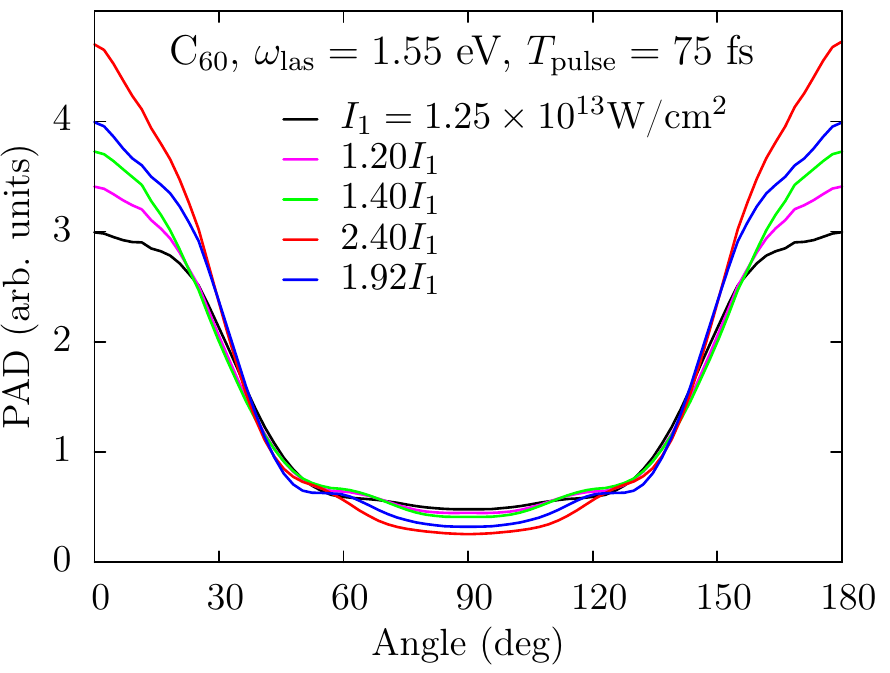}
\includegraphics[width=0.49\linewidth]{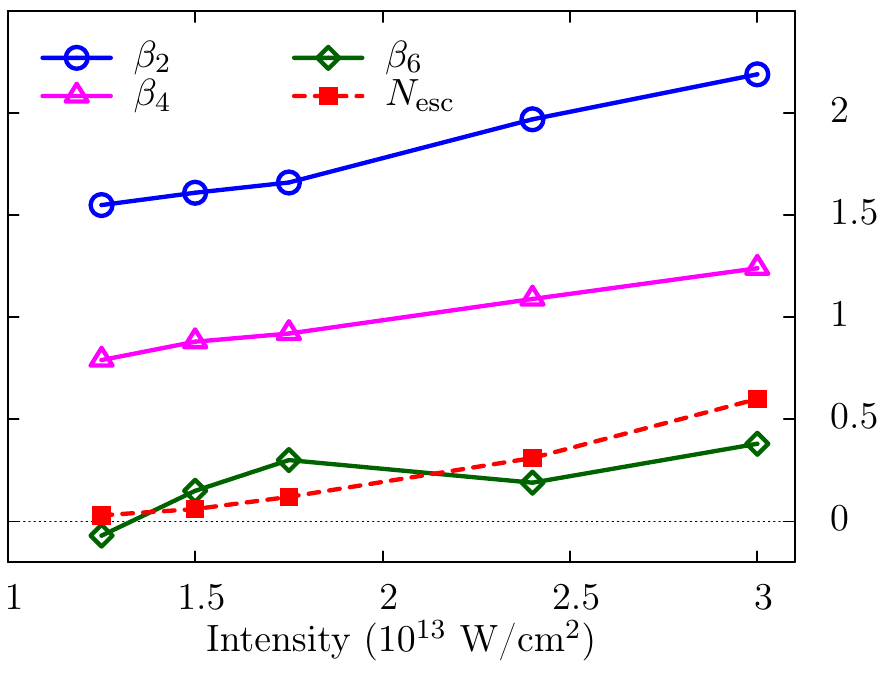}
}
\caption{Anisotropy parameters of high orders (right) extracted from
  PAD (left) of orientation-averaged C$_{60}$ irradiated by laser
  pulses of frequency of 1.55~eV, duration of 75~fs, for different
  laser intensities. For the sake of completeness, the total
  ionization $N_\mathrm{esc}$ is plotted in the right panel. Adapted
  from~\cite{WopPhD}.}
\label{fig:c60-multi_I}
\end{figure}

We start with the case of C$_{60}$ irradiated by a laser of frequency
of 1.55~eV and pulse duration of 75~fs, with the same set of
increasing intensities as in use for the systematics of PES in the right
panel of Fig.~\ref{fig:pes_I}. The IP of C$_{60}$ being 8~eV, it
requires at least 6 photons to extract electrons from occupied
states.
The PAD obtained
in C$_{60}$ with different laser intensities are shown in the left
panel of Fig.~\ref{fig:c60-multi_I}, while the first anisotropy
parameters $\beta_{n}$ extracted from these PAD are plotted in the
right panel.
The total ionization $N_\mathrm{esc}$ (red boxes and dashed line in
the right panel) increases rapidly with $I$ as expected.  The PAD
shown in the left panel become more and more aligned along the laser
polarization with increasing $I$.  Accordingly, all $\beta_{2l}$,
shown in the right panel, increase with $I$. Note that $\beta_2>2$
becomes possible in this non-linear regime. The limitation $-1\leq
\beta_2\leq 2$ applies only to strict one-photon processes.

\begin{figure}[htbp]
\centerline{
\includegraphics[width=\linewidth]{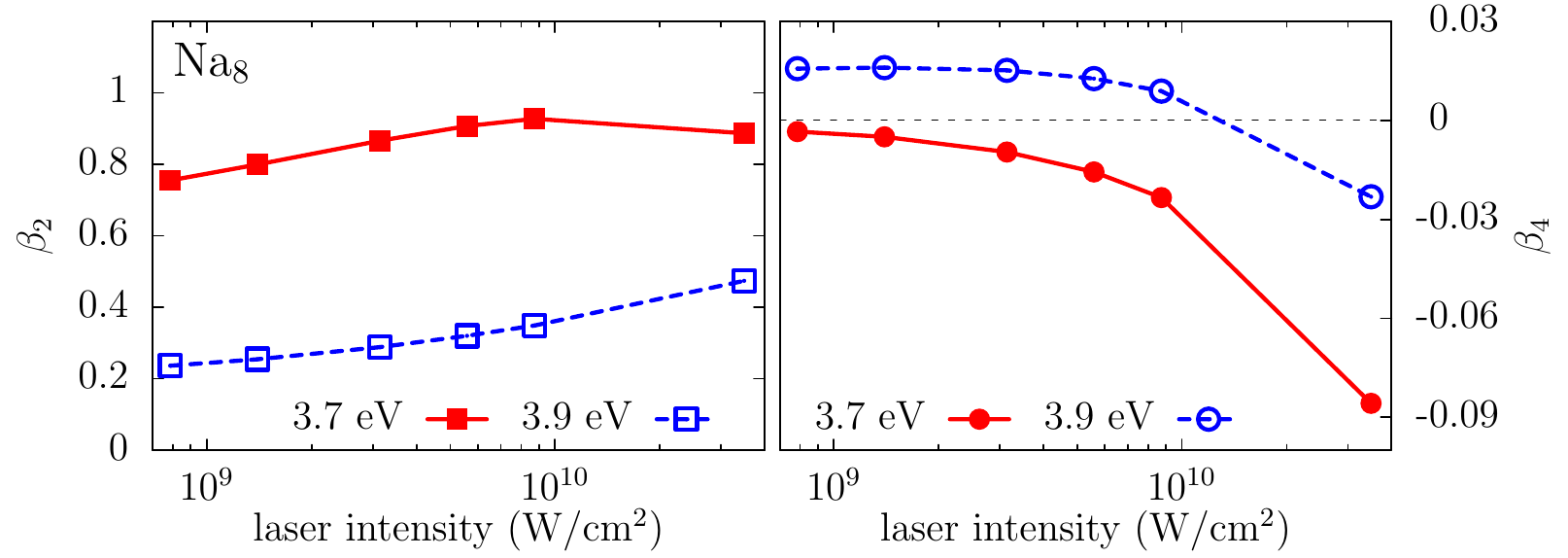}
}
\caption{Anisotropy parameters $\beta_2$ (left) and $\beta_4$ (right)
  of Na$_8$ extracted from PAD with an averaging procedure over 18
  orientations, after irradiation of laser pulses of duration of 60~fs
  and frequencies of 3.7 (red full curves) or 3.9~eV (blue dashes).
  Adapted from~\cite{Wop10a}.}
\label{fig:na8-beta_I}
\end{figure}

The next test case is the Na$_8$ cluster. Differently as in previous
sections, we now run it for laser frequencies below ionization
threshold, namely $\omega_\mathrm{las}=3.7$ and 3.9~eV while the IP is
4.4~eV. At least two-photon processes are required for ionization,
probably higher ones with increasing intensity
$I$. Fig.~\ref{fig:na8-beta_I} shows the intensity dependence of
$\beta_2$ (left panel) and $\beta_4$ (right panel).
The total ionization ranges from 0.001 to 0.1, which means that all
cases constitute rather moderate excitation dynamics. We see again
generally an increase of total anisotropy $\beta_2$ with intensity.
Still, the growth of $\beta_2$ is slower than in the previous example
C$_{60}$ because we deal here only with two-photon processes. The next
coefficient $\beta_4$ also shows a marked trend to larger negative
values with increasing intensity. It is interesting to note that both
coefficients, $\beta_2$ and $\beta_4$, are very sensitive to the laser
frequency. This is also a feature of MPI while frequency dependences 
are more moderate for one-photon processes, see e.g. Sec.~\ref{sec:modelassum}.

\subsection{Impact of temperature in PES and PAD}
\label{sec:temper}

\subsubsection{Effect of ionic motion on PES and PAD}
\label{sec:ionicT}

So far, the presented calculations of PES and PAD were performed
  at ionic ground-state configuration, i.e. at a temperature of
  0~K. This ideal situation is hardly ever feasible in an
  experiment. Depending on the production conditions, cluster beams
  have temperatures in the range of several 100~K. This means that we
  encounter usually an ensemble of ionic configurations fluctuating
  around the ground-state configuration. In the following, we will
  discuss the impact of thermal shape fluctuations on PES/PAD.

\begin{figure}[htbp]
\centerline{\includegraphics[width=\linewidth]{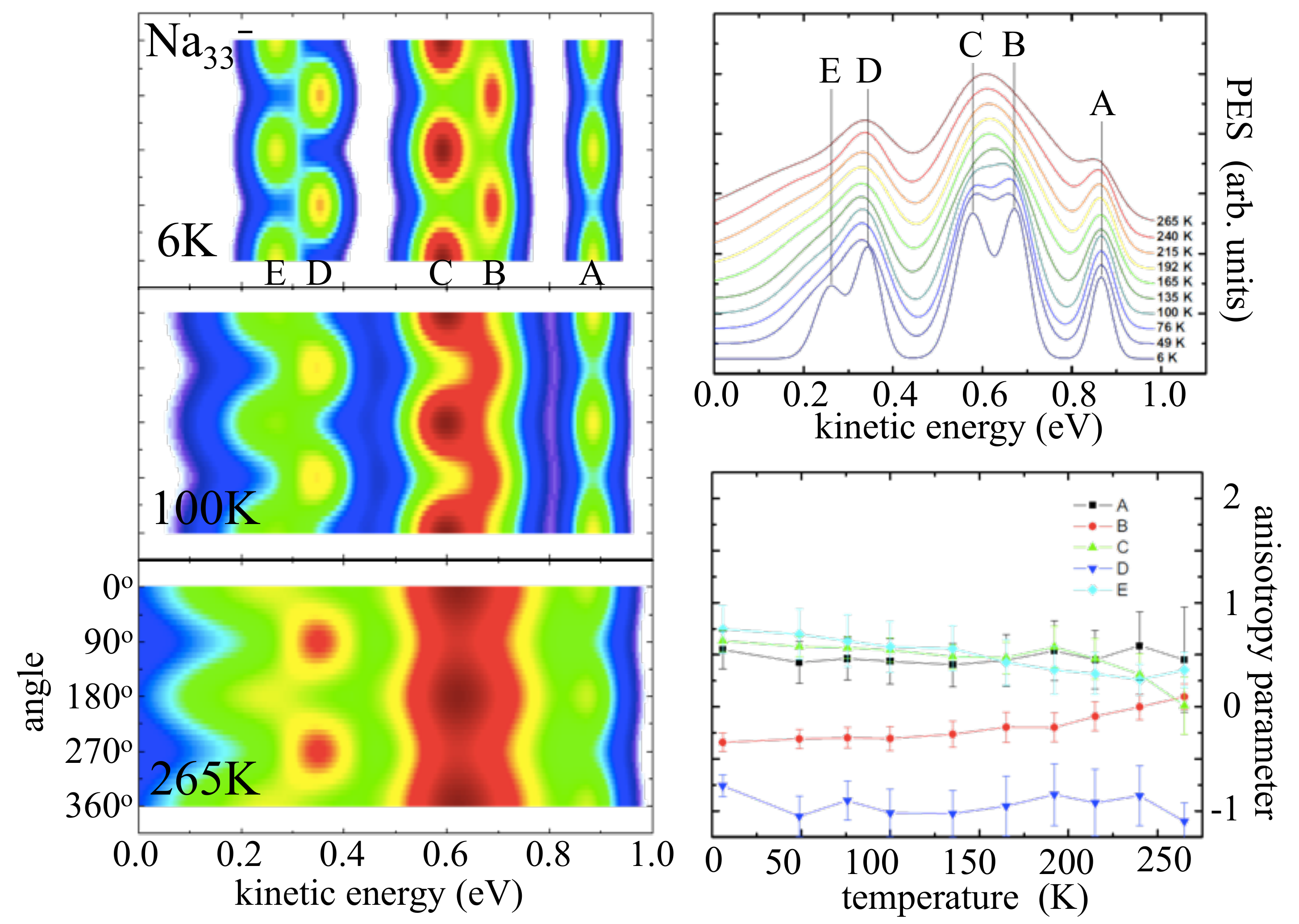}}
\caption{Left~: Combined PES/PAD of ${\mathrm{Na}_{33}}^-$ obtained by
  fit of five Gaussian peaks to the raw distributions at three
  different ionic temperatures as indicated. Right top~: extracted PES
  for temperatures ranging from 6~K to 265~K. Right bottom~: extracted
  anisotropy parameter $\beta_2$ as a function of ionic
  temperature. Adapted from~\cite{Bar08}.}
\label{fig:na33m_pespad}
\end{figure}

Experimentalists are well aware of the temperature problem and
  have developed several techniques for dedicated cooling of cluster
  beams.  Ion traps are particularly powerful devices for a clean
  handling of cluster beams \cite{Sch99c}. We show here results from
  recent experiments which used a trap and cooling with a He buffer
  gas to produce beams of Na anionic clusters with well defined
  temperatures between 6 and 265~K~\cite{Bar08}.  The upper
  temperature is above the melting point of $\approx 250$~K for small
  Na clusters~\cite{Hab05}. The test case ${\mathrm{Na}_{33}}^-$ is a
  cluster anion which has naturally a low IP. Thus one can easily
  realize one-photon processes with standard laser pulses.   A
selection of combined PES/PAD for ${\mathrm{Na}_{33}}^-$ is shown in
the left panel of Fig.~\ref{fig:na33m_pespad}. The upper right panel
shows the PES at different temperatures $T$ and the lower right panel
the anisotropy $\beta_2$ as a function of $T$.
The PES/PAD in the left panels and the PES in the upper right panel
(both analyzed at low temperature) allow one to identify five emitting
states, labeled A, B, C, D, and E.  The structures produced by these
states in the PES (and PES/PAD) are gradually blurred with increasing
temperature. This is easily understood from the fact that
  single-electron energies can be very sensitive to changes in the
  cluster shape. The thermal ensemble thus represents a more or less
  broadened distribution of s.p. energies which, in turn,
  is mapped into PES and PES/PAD.  On the other hand, the
anisotropies $\beta_2$ exhibit only a weak dependence on $T$. This
probably reflects the fact that the angular momentum characteristics
(stemming from their wave functions) of the s.p. states are more robust than
their energies.

\begin{figure}[htbp]
\centerline{\includegraphics[width=0.85\linewidth]{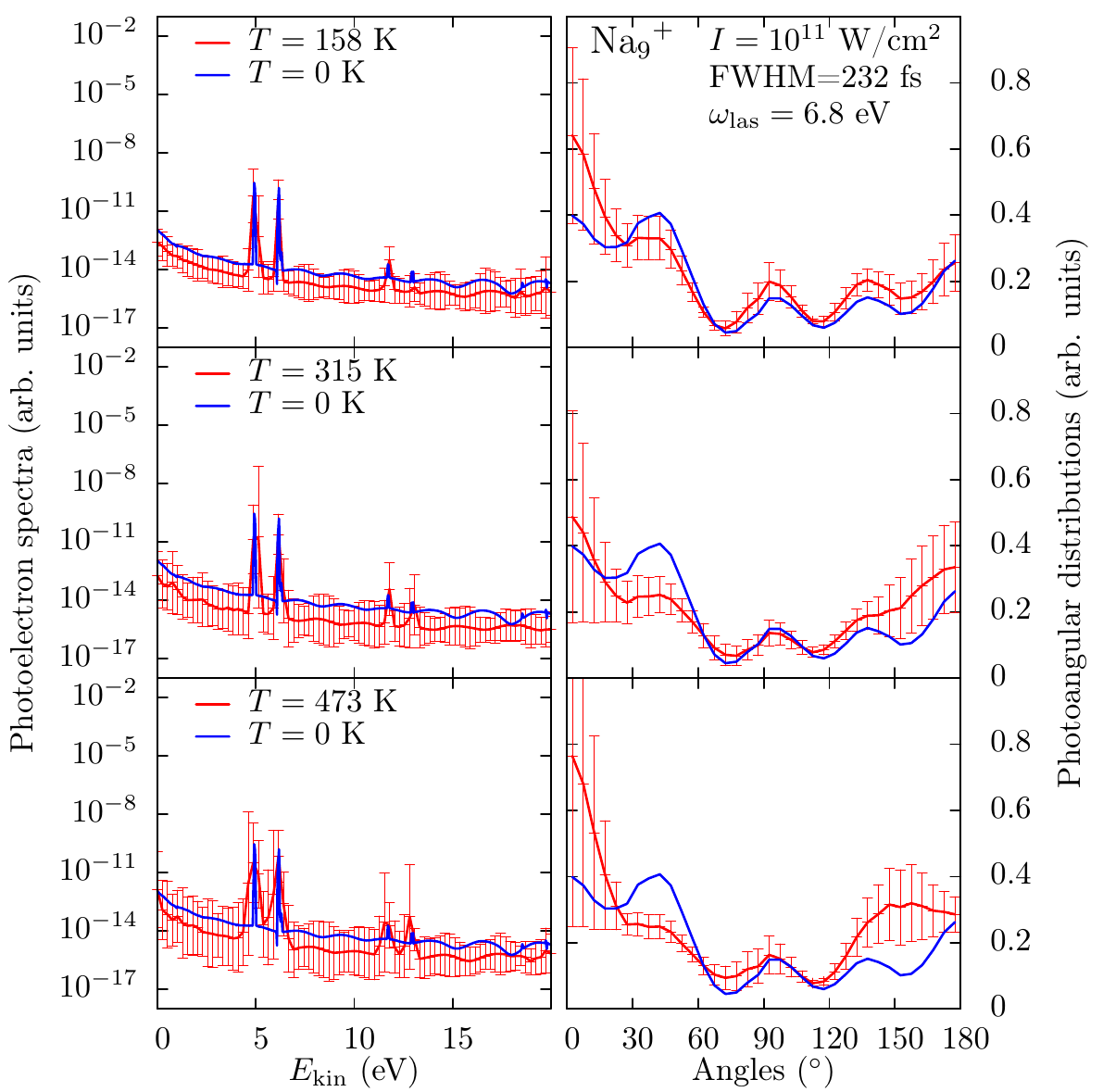}}
\caption{Left column~: Photoelectron spectra of ${\mathrm{Na}_9}^+$ of
  fixed orientation, irradiated by laser pulses polarized along the
  symmetry axis of the cluster, with FWHM of 232 fs, intensity of
  $10^{11}$~W/cm$^2$, and frequency of 6.8~eV, at an ionic temperature
  $T$ of 158~K (top), 315~K (middle) or 473~K (bottom). The blue curve
  shows the PES at $T=0$~K. Right column~: corresponding PAD.}
\label{fig:na9p_w6.8_tempion}
\end{figure}

The numerical simulation of a thermal ensemble is conceptually
  straightforward, although somewhat cumbersome.  One starts an ionic
  dynamics from the ground-state configuration by initializing ionic
  velocities stochastically according to a Maxwellian distribution for
  the given temperature $T$. This state is then propagated by TDLDA-MD
  for a few ps. About each 100~fs, a snapshot of the actual
  configuration is taken. The set of all snapshots constitutes the
  thermal ensemble of cluster configurations. Now, in a laser-induced
  dynamics propagated for each sample, the wanted observables
  (e.g., PES and PAD) are evaluated and incoherently superimposed.
  This altogether yields the observable for the ensemble. We have
performed such a study for ${\mathrm{Na}_9}^+$ irradiated by laser
pulses of intensity of $10^{11}$~W/cm$^2$, and FWHM of 232 fs. This
pulse duration allows a high resolution of the PES peaks (see the
effect of pulse duration in Fig.~\ref{fig:na2_fwhm}) such that line
broadening comes predominantly from thermal effects. The pulse
duration lies within the time scale of ionic motion. Thus the
dynamical propagation is done at the level of TDLDA-MD to include
properly ionic motion. For reasons of simplicity, we are not
performing orientation averaging such that we see exclusively the
thermal effects. The laser polarization is chosen along the symmetry
axis of the $T=0$ configuration.  For a first test, we use a laser
frequency at 6.8~eV just below the IP. The $1s$ state
($\varepsilon_{1s}=-8.5$~eV) and degenerate $1p$ states
($\varepsilon_{1p}=-7.2$~eV) then emit through a two-photon
process.  We compare in Fig.~\ref{fig:na9p_w6.8_tempion} the PES and
the PAD calculated at $T=0$ and at three increasing temperatures
$T=158$, 315 and 473~K. 
As expected, the higher $T$, the broader the PES peaks.  There is also
a faint red shift of the peaks with increasing temperature.
Nonetheless, it is surprising how well the structures survive in the
PES even high above melting temperature (about 250~K).  What the PAD
is concerned, remind that it is computed without orientation
averaging. It thus shows more structure and is asymmetric (reflecting
the asymmetry of ${\mathrm{Na}_9}^+$). The thermal effects on the PAD
are a bit larger than for the PES, but remain still small, in
accordance with the experimental results for ${\mathrm{Na}_{33}}^-$ in
Fig.~\ref{fig:na33m_pespad}.
Fig.~\ref{fig:na9p_w6.8_tempion} also shows through the error bars the
uncertainty associated with thermal fluctuations. These are computed
in standard manner as the variance of PAD and of the logarithm of the
PES yield from the statistical ensemble. The uncertainties grow with
temperature. They stay rather small for the PES, showing once more that
these structures are rather robust. The error bars are larger for the PAD
in forward (0$^\circ$) and backward (180$^\circ$) direction. Fortunately, these
forward and backwards cones have a small integration weight, such that
global measures as, e.g., the anisotropy $\beta_2$ are again robust.

\begin{figure}[htbp]
\includegraphics[width=\linewidth]{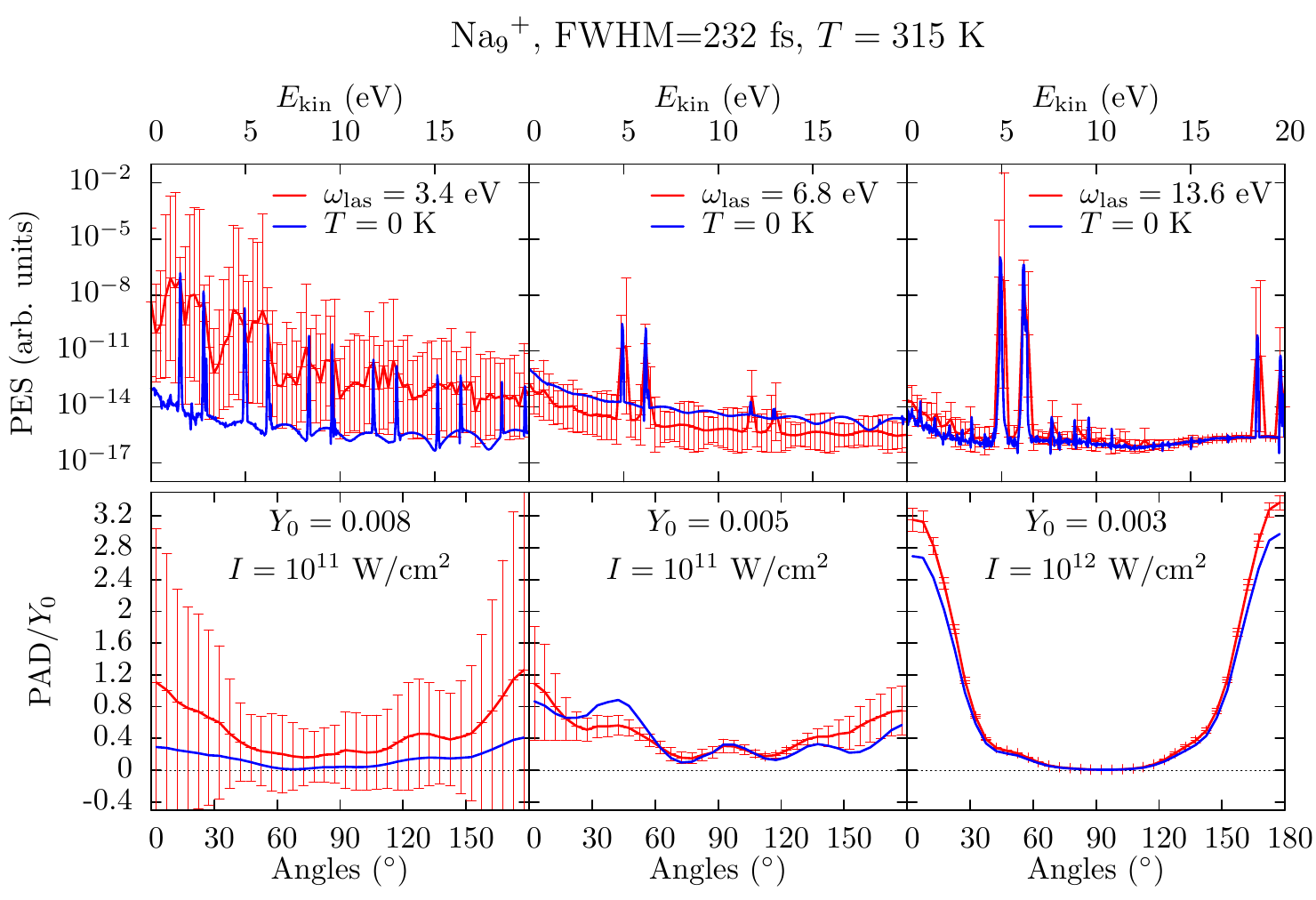}
\caption{Top row~: Photoelectron spectra of ${\mathrm{Na}_9}^+$ with
  initial ionic temperature of 315~K, irradiated by laser pulses of
  FWHM of 232 fs, intensity of $10^{11}$~W/cm$^2$, and frequency of
  3.4~eV (left), 6.8~eV (middle), and 13.6~eV (right).  The blue curve
  shows the PES at $T=0$~K. Bottom row~: corresponding PAD.}
\label{fig:na9p_tempion_om}
\end{figure}

We now concentrate on the case of $T=315$~K and complement the
$\omega_\mathrm{las}=6.8$~eV by two other frequencies~: one smaller
with 3.4~eV well below IP and one larger with 13.6~eV well above
IP. The laser polarization and duration are the same as before. The
intensity is $10^{11}$~W/cm$^2$ for the two lower
$\omega_\mathrm{las}$ and $10^{12}$~W/cm$^2$ for
$\omega_\mathrm{las}=13.6$~eV to deliver comparable ionization for all
cases.  Fig.~\ref{fig:na9p_tempion_om} shows the resulting PES and
PAD (again compared with the $T=0$ case).
For the highest frequency (right panels), the impact of temperature is
quite weak: we still observe a slight broadening of the peaks in the
PES. However, the effect on the PAD is negligible. This result is
intuitive because for such a high laser frequency, the photoelectrons
are extracted by absorption of a single photon and basically follow
the laser field, as is visible from the fact that the PAD is peaked
along $\vartheta=0^\circ$ and $180^\circ$. On the contrary, the lowest
frequency (left panels) lies deeply in the multi-photon regime. And
the uncertainties produced by the ionic motion at this temperature are
extremely large. Apart from that, MPI peaks are still visible in the
PES, although much broadened.  At the side of the PAD, the
uncertainties are larger than the signal.  This means that in the
multi-photon regime, the PAD is very sensitive to ionic temperature.
The intermediate frequency (middle panels) lies in between in all
respects. Temperature effects are already well visible, but not as
disastrous as for low frequency. This example shows that the {ionic}
temperature should better be well controlled and kept at a
sufficiently low value to allow a quantitative analysis of PAD.

\subsubsection{Impact of electronic dissipation on PES and PAD}
\label{sec:dissipe}
\definecolor{gray}{gray}{0.5}

In Sec.~\ref{sec:pes_I}, we found that a smooth exponential PES
develops for sufficiently high intensities, see
Fig.~\ref{fig:pes_I}. Figure~\ref{fig:campbell1} did also show a
series of exponential PES from a measurement with high intensities
(fluences) and a long pulse. The question is to what extent this
could be a signature of thermal emission after full thermalization of
the cluster. 

We start the discussion with looking back at Fig.~\ref{fig:pes_I}.
From the inverse slope of the exponential, we would read off an apparent
temperature of $T_{\rm app}=1.4$~eV. From the TDLDA calculations,
we also find that the energy deposited by
the laser pulse in the cluster is about $E^*\simeq{1.8}$~eV. 
Assuming that all this energy is converted into thermal energy
would lead to an intrinsic temperature of about
$E^*/160\simeq{0.01}$~eV which is two orders of magnitude smaller than
$T_\mathrm{app}$. This, together with the fact that TDLDA does not
account for electronic thermal effects, clearly rules out the thermal
origin of the observed exponential slope.

The experimental results from \cite{Kje10} in
Fig.~\ref{fig:campbell1} (see Sec.~\ref{sec:PES}) did show also a
remarkable series of smooth exponential PES.
\begin{figure}[htbp]
\centerline{\includegraphics[width=0.75\linewidth]{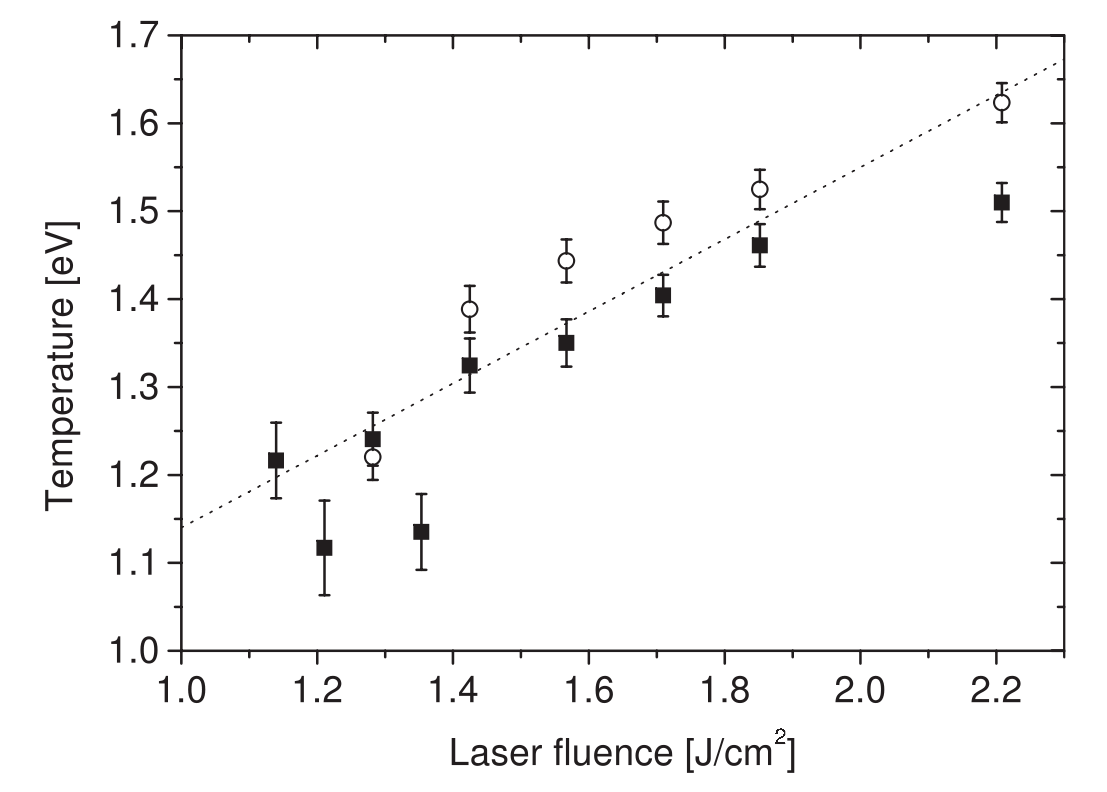}}
\caption{
Experimental apparent electronic temperature as a function of laser
fluence as extracted from PES for C$_{70}$ (squares) and  C$_{60}$ (circles).
The dotted line is a fit to the 
temperatures vs laser fluence obtained in the case of C$_{60}$ in \cite{Han03a}.
Note that the laser fluence has
an overall systematic uncertainty of 10\%, 
which is not included in the present error bars. From \cite{Kje10}.
\label{fig:campbellT}
}
\end{figure}
Fig.~\ref{fig:campbellT} gathers the apparent temperatures (inverse
slopes) extracted from these PES. {Again we find rather large
  values in the range from 1 up to 1.6 eV, near the 1.8~eV of the
  previous example. The remaining excitation energy is not available
  for these experiments. Nevertheless, it is questionable that these
  exponential slopes should correspond to real temperatures of the
  system in thermal equilibrium.  This is why one wisely has coined
  the notion ``apparent temperature'' for the slope of the PES, see
  Fig.~\ref{fig:campbell1} from \cite{Kje10}. The example of
  Fig.~\ref{fig:pes_I} has shown that a smooth exponential pattern can
  also be explained by TDLDA calculations.  In fact, the exponential
  profile can be nicely fully explained in terms of multiphoton
  perturbation theory \cite{Poh04a}. Any MPI yields an exponentially
  decreasing slope. The Coulomb shift increasing with ionization stage
  increasingly washes out the MPI peaks to yield eventually a purely
  exponential PES.}

\begin{figure}[htbp]
\centerline{\includegraphics[width=.65\linewidth]{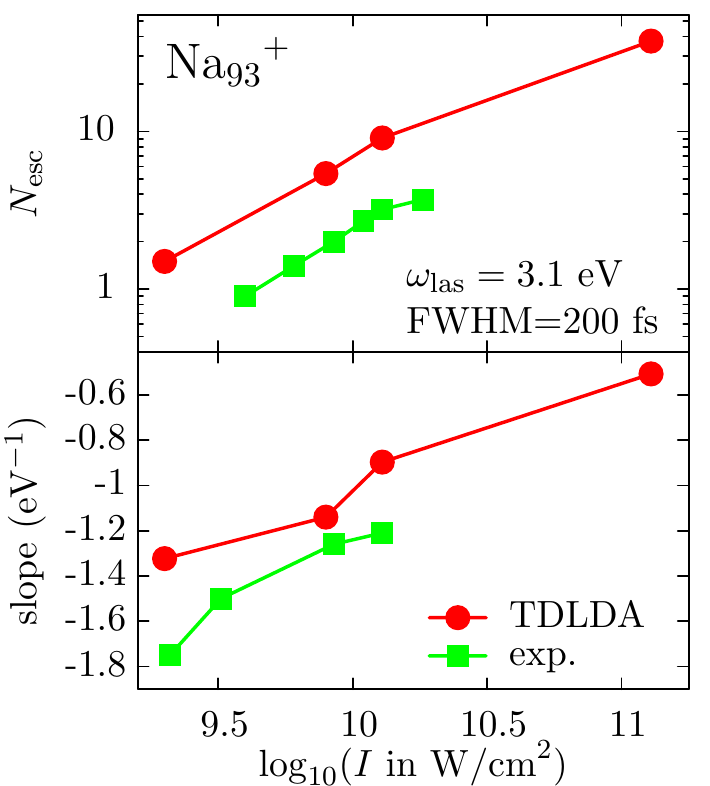}}
\caption{Ionization characteristics of ${{\rm Na}_{93}}^+$ irradiated by 
lasers of increasing intensities (in W/cm$^2$)
 but fixed photon frequency at 3.1 eV 
and pulse FWHM at 200 fs. Calculations (red circles) are compared to experimental data (green squares)
from \cite{Sch01}. Ionic dynamics has been included with a  Maxwellian distribution of velocities according to a temperature of 100 K. 
Upper panel: total ionization in logarithmic scale.
Lower panel: slope of the PES. Adapted from \cite{Poh04a}.
\label{fig:na93p-slope}
}
\end{figure}

{In order to check how far pure TDLDA (free of any thermalization) can
describe exponential PES, we consider ${{\rm Na}_{93}}^+$ for which
calculations can be compared with experimental data.  In this
experiment both total ionization and PES have been measured. The
exponential shape of the PES was attributed to thermal effects
\cite{Sch01}.  The resulting apparent temperature (inverse slope) and
total ionization are plotted as a function of laser intensity in 
Fig.~\ref{fig:na93p-slope}.} 
{The experimental results are compared to
standard TDLDA calculations performed under the same laser conditions.
The theoretical results are surprisingly close to the data, which indicates
that a purely thermal interpretation is not compelling.  Consider the
apparent temperatures (inverse of the slopes). They are in range
0.7--2~eV for TDLDA and 0.6--0.8~eV for the experiment.  This would
amount to about 300~eV intrinsic energy in case of full thermal
equilibrium ($\approx$150~eV if only electrons were thermalized)~: this is too
much as compared to the typical cluster binding in ${{\rm Na}_{93}}^+$. Thus
we are surely far from full thermalization in this case.  On the other
hand, there are small, but systematic, differences which may be a trace
of thermal effects. The apparent temperature (lower panel) and the
ionization (upper panel) is somewhat lower in the data than in TDLDA.
This indicates that the data represent, in fact, a mixed situation,
not fully thermalized yet, but somewhere on the way.
}

There still remains the task to distinguish direct from thermal
  electron emission. We have argued above that exponential PES are
  only a necessary condition, not a sufficient proof. Additional
  information for a better discrimination is delivered by the
  PAD. Isotropic PAD are another necessary, usually much more
  conclusive, condition for thermal emission. For example, the
  experimental PES/PAD in Fig.~\ref{fig:c60-vmi_xuv} shows a larger
  inner spot of isotropic emission at low energies which can be
  associated with thermal electrons. At the theoretical side, a
  relevant description of thermalization requires dynamical
  correlations beyond TDLDA. This has been achieved at the
  semi-classical level, see Sec.~\ref{sec:semiclassicalroute}.  It
  is still a great challenge for a fully quantum mechanical modeling,
  see Secs.~\ref{sec:relax_ansatz} and \ref{sec:stdhf}. In the
  following, we will briefly present a simple estimate for the
  contribution of thermal electrons which provides at least a first
  impression of the impact of thermal electrons on PAD.
 
\begin{figure}[htbp]
\centerline{\includegraphics[width=0.65\linewidth]{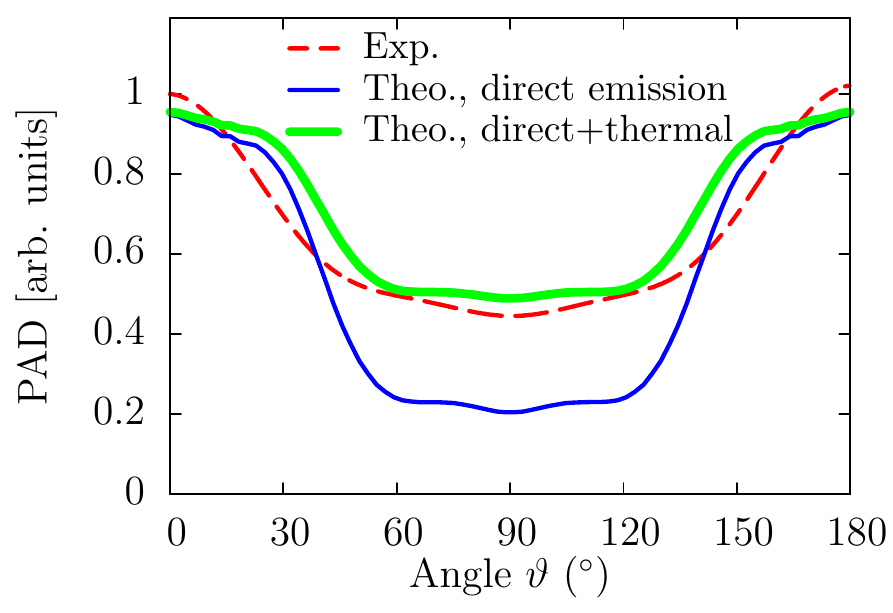}}
\caption{Comparison of 
experimental photoangular distribution from C$_{60}$ 
HOMO and HOMO$-1$ states (red dashed curve)  in the multiphoton regime   with total
theoretical PAD. One calculation takes into account only direct electronic
emission (blue thin line) and the other one includes an estimate of the  
additional  thermal 
component  due to electronic temperature
(green thick line), see text for details. From \cite{Bar14}.
\label{fig:pad_ati}
}
\end{figure}

The case of C$_{60}$, illustrated in Fig.~\ref{fig:pad_ati},
addresses the complementing PAD signal, which is expected to contain a
significant isotropic component if strong thermal effects are present.
We consider here the MPI regime, irradiated by a laser of frequency
1.55 eV, pulse FWHM of 20 fs and intensity of $1.25\times 10^{13}$~W/cm$^2$. 
The experiments deliver a
PAD energy-integrated over the HOMO, HOMO$-1$, and HOMO$-2$ and the
theoretical results are integrated over the same interval~\cite{Bar14}. Although
both results have similar pattern, the TDLDA distribution is much more
anisotropic than the experimental one. This reflects once again the fact that
TDLDA underestimates electron-electron collisions which are the
doorway to thermal effects. It is thus interesting to test whether
thermal effects might explain the observed discrepancy.  For a simple
estimate, we proceed as follows. We assume that the residual
electronic excitation energy which is found to be 1.8 eV is fully
thermalized. Thermal energy is later on converted into a thermal
electronic emission. The IP of C$_{60}$ is 7.8 eV. The 1.8 eV
excitation energy thus suffices to emit about 0.22 electrons (when
neglecting the possible C$_2$ dissociation channel which requires
larger energy). We then add up the contribution of the extra 0.22
emitted electrons as a thermal, isotropic background to the PAD.
This leads to the curve labeled ``direct$+$thermal" in Fig.~\ref{fig:pad_ati},
 which now agrees fairly well with the experimental
curve. Of course, the reasoning basically provides an argument and does
not constitute a theory by itself, but it once again confirms that
bare TDLDA underestimates electronic collisions, while they are
obviously non negligible in the MPI regime.

\subsubsection{The semi-classical route}
\label{sec:semiclassicalroute}

\begin{figure}[htbp]
\centerline{\includegraphics[width=\linewidth]{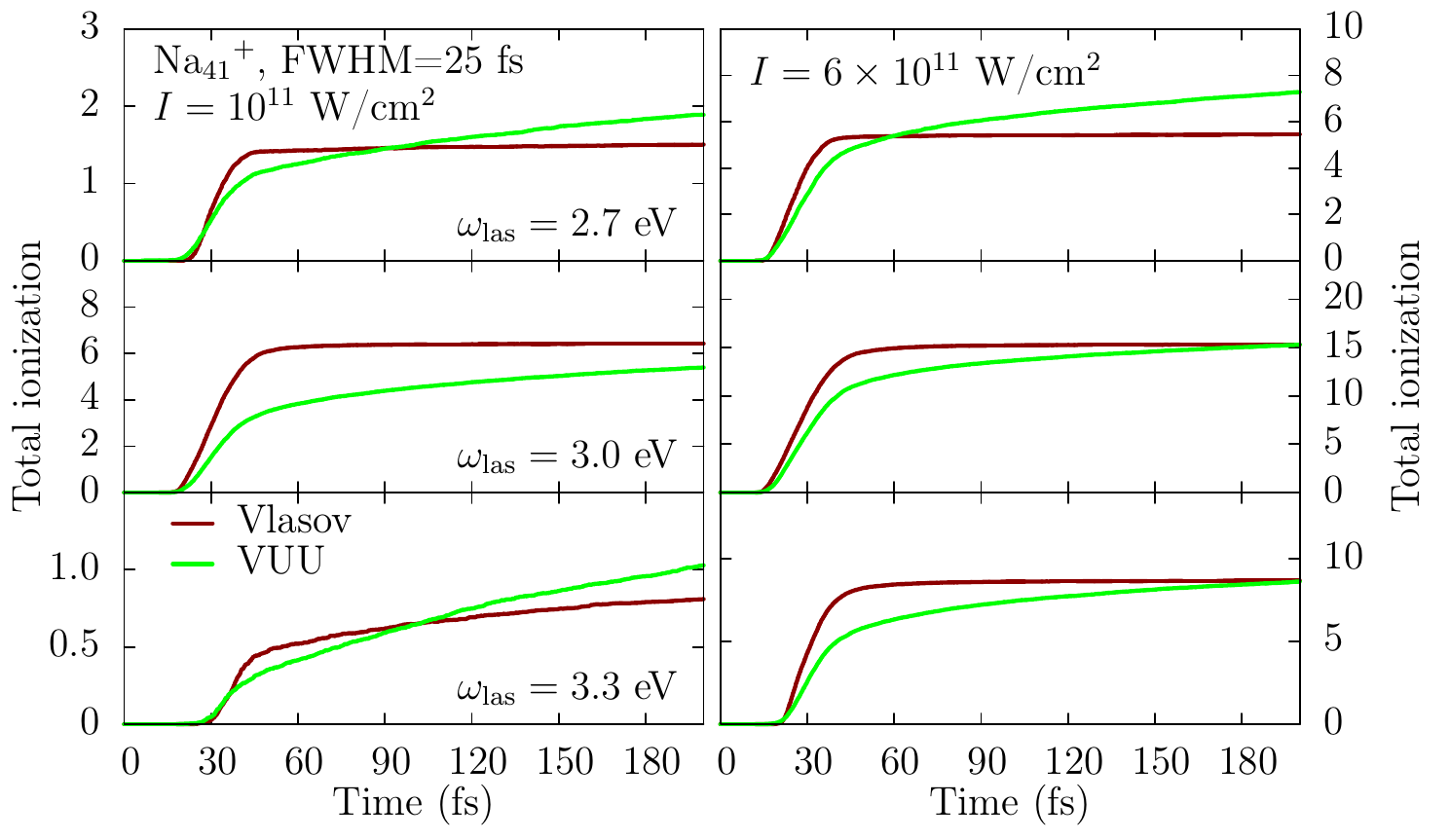}}
\caption{Comparison of the time evolution of the total ionization in
  Vlasov (brown lines) and VUU (light green lines) approaches. The
  system is ${{\rm Na}_{41}}^+$ irradiated by lasers of cos$^2$ profiles
  with FWHM of 25 fs. Intensities are 10$^{11}$ (left panels) and
  $6\times 10^{11}$ W/cm$^2$ (left panels). Frequencies are varied
  between 2.7 and 3.3 eV, which covers the dominant optical response
  peak located around 3 eV for this system.
\label{fig:na41p_vuu_example}
}
\end{figure}

As already discussed in Sec.~\ref{sec:temperature_theo}, thermal
effects in finite systems can presently only be attained via a
semi-classical approximation, leading to semi-classical kinetic
equation such as VUU. The underlying Vlasov equation is the
semi-classical limit of TDLDA. VUU employs additionally the
Uehling-Uhlenbeck collision term which accounts for the dynamical
electron-electron correlations.  Ionization, as a basic dynamical mechanism,
has been discussed within VUU in several papers
\cite{Gig02,Gig03,Fen04,Fen07,Koe08}. The interesting point for the
present discussion is to analyze the impact of VUU as compared to
Vlasov. We shall discuss the point on the example of ${{\rm Na}_{41}}^+$. We
fix the pulse duration at a FWHM of 25 fs. We consider two laser
intensities, $10^{11}$ W/cm$^2$ and $6\times10^{11}$ W/cm$^2$, and
three frequencies, $\omega_\mathrm{las}=2.7$, 3.0, and 3.3~eV, around
the plasmon frequency of the system (3 eV). The time evolution of the
ionization is shown in Fig.~\ref{fig:na41p_vuu_example}.
It strongly depends on $\omega_\mathrm{las}$, as expected at the
passage of a resonance \cite{Gig01a}.  
The most interesting feature is
the shape of the emission profile. In the Vlasov calculations,
ionization grows very quickly in the early stages and levels off once
the laser pulse is switched off. Note that the huge ionization
blue-shifts the plasmon resonance such that the case $\omega_{\rm las}=3$~eV 
becomes off-resonant
and subsequently emission is terminated after the pulse (see Sec.~\ref{sec:basic_ioniz}).  
The VUU results look much different. Emission
is suppressed in early stages. This is overcompensated by a steadily
continuing emission later on. The pattern are very similar for all
three laser frequencies, while the amplitude of the effect (and the
relative values of Vlasov and VUU ionization at early times) depends
sensitively on frequency. This is a long known effect that field
amplification by the plasmon resonance naturally leads to enhanced
emission \cite{Rei98b} in TDLDA and correspondingly in Vlasov
\cite{Gig01a,Gig03}. 
VUU shows the same resonant behavior, but
modulates the time profile of emission. It reduces ionization in early
stages because electron-electron collisions remove energy from the
direct emission channel, and it enhances emission in later stages by
releasing gently the stored energy.

\begin{figure}[htbp]
\centerline{\includegraphics[width=\linewidth]{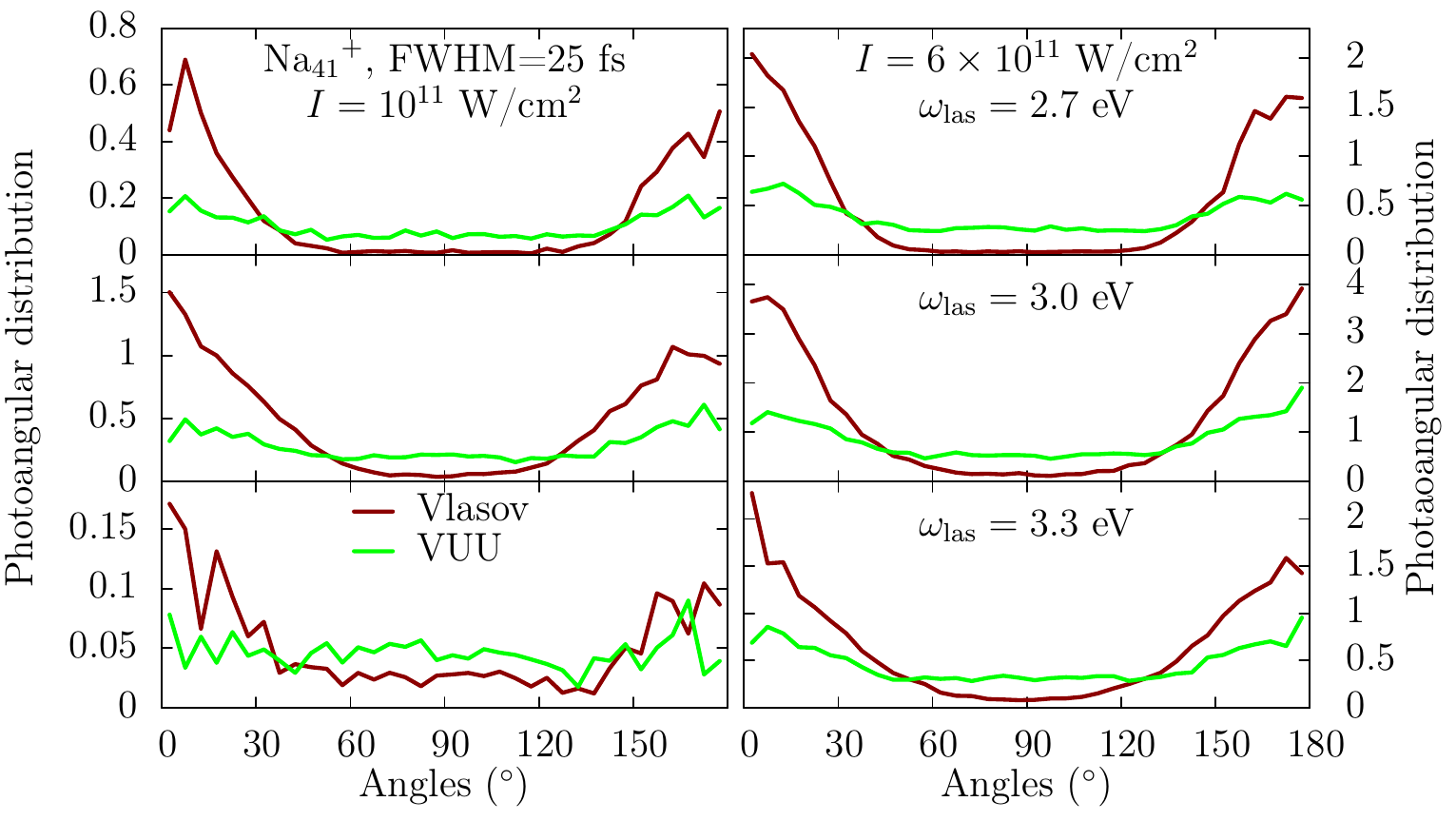}}
\caption{Comparison of angular distributions in Vlasov (brown lines)
  and VUU (light green lines) approaches.  Same system and laser
  conditions as in Fig.~\ref{fig:na41p_vuu_example}. 
\label{fig:pad_vuu} }
\end{figure}

The analysis of PES is of limited interest in Vlasov and VUU as at low
energy neither can identify electron single particle energies because
of their semi-classical nature.  The obtained PES are thus always more
or less exponentially decreasing, whatever the laser conditions. More
interesting is the PAD which can be easily evaluated in Vlasov and VUU
and which does not suffer so much from the semi-classical
approximation. Furthermore, it is an ideal observable to identify
thermal effects in terms of isotropy of the PAD, as we have seen in
Sec.~\ref{sec:dissipe}. The point is illustrated in Fig.~\ref{fig:pad_vuu} for 
the same test case as in Fig.~\ref{fig:na41p_vuu_example}, and for a few laser frequencies, 
again around plasmon frequency of ${{\rm Na}_{41}}^+$. 
Orientation averaging has not
been performed, which is an acceptable approximation here because
${{\rm Na}_{41}}^+$ is large and close to sphericity.  For all laser
frequencies, the PAD from Vlasov exhibit strongly oriented emission
along the laser polarization. The effect becomes larger when $\omega_\mathrm{las}$ 
comes closer and closer to the plasmon frequency. This is again a
consequence of field amplification near the resonance.  A similar
trend with frequency is observed in the VUU calculations, although
less pronounced.  More striking is the fact that the PAD from VUU are
much less peaked than the Vlasov ones. Very clearly, the VUU results
have a strong isotropic component.  The effect is especially clear far
from the plasmon resonance but remains very visible close to it. We
observe then a competition between field amplification (from
resonance) and thermalization \cite{Gig01a}. 
In any case, we see a
significant enhancement of ionization perpendicular to the laser
polarization, as compared to Vlasov, a clear signature of isotropy.


\section{Future directions}
\label{sec:future}

We have seen in the previous sections the richness and variety of
observables today accessible in experiments and theory. Of course,
there remain many open questions for future research.  We want to
illustrate in this section two lines of development which we consider
as especially promising and which require dedicated efforts both from
the experimental and theoretical side.

The first aspect again focuses on the analysis of dynamics in terms of
PES and PAD from which we have seen that it is a powerful tool.  We
will now consider the analysis in terms of PES and PAD in connection
with the short pulses delivered by a bypassing ionic projectile. As we
shall see, PES and PAD can again provide useful insights into the
underlying dynamics.

The second topic concerns electronic thermalization which was already
addressed to some extent in Sec.~\ref{sec:dissipe}.  The excitation
energy deposited originally by the laser pulse is released in the
first stages of the dynamics by direct electron emission. However,
part of the deposited energy is progressively converted into
incoherent electronic excitation of "thermal" nature. This takes place
on a moderate time scale of some tens of fs.
 Analysis of such effects is difficult and requires
detailed experiments (see Figs.~\ref{fig:campbell1} and
\ref{fig:campbellT}). From the theoretical point of view, the situation
is even worse as it requires the development of deep extensions of
available current theories such as real-time TDDFT. In the following,
we address both these directions in more detail.

\subsection{An excursion into irradiation by charged projectiles}
\label{sec:projectile}

We briefly discuss in this section PES and PAD of "photo"-electrons
emitted after collision with a fast charged projectile.  We put the
word "photo" in quotation marks because the electromagnetic pulse has
not a well defined frequency here. Although rare, there exist data
measured on atoms and mono-atomic dimers with a special focus at very
high kinetic energies ($E_{\rm kin}>40$~eV) of the projectiles. This
was first motivated by the observation of non-monotonous patterns in
the PES after irradiation of O$_2$ and N$_2$ with photons in the
30--60~eV range~\cite{Sam65}, which were explained theoretically one
year later~\cite{Coh66} by Young-type interferences between electronic
wave functions of electrons coherently emitted from identical atomic
centers. Various experiments have been performed on H$_2$ bombarded by
He$^+$ and He$^{2+}$ of 20 and 40 keV~\cite{Fre05} or by 8 keV
electrons~\cite{Cha08,Cha10}, on N$_2$ colliding with 1--5 MeV
H$^+$~\cite{Bar08b}, and on O$_2$ colliding with 3.5 MeV/$u$ C$^{6+}$
ions~\cite{Nan12} or 30 MeV O$^{5+}$ and O$^{8+}$
ions~\cite{Win09}. Very recently, collisions of 3.5 MeV/$u$ C$^{6+}$
ions on uracile~\cite{Agn13} and of 4.5 MeV/$u$ O$^{8+}$ ions on
H$_2$O~\cite{Nan13} have been reported. The kinetic energy of the
ejected electrons ranges from a few eV up to 600~eV, and the emission
is measured between 20$^\circ$ and 150$^\circ$.

\begin{figure}[htbp]
 \centerline{
\includegraphics[width=\linewidth]{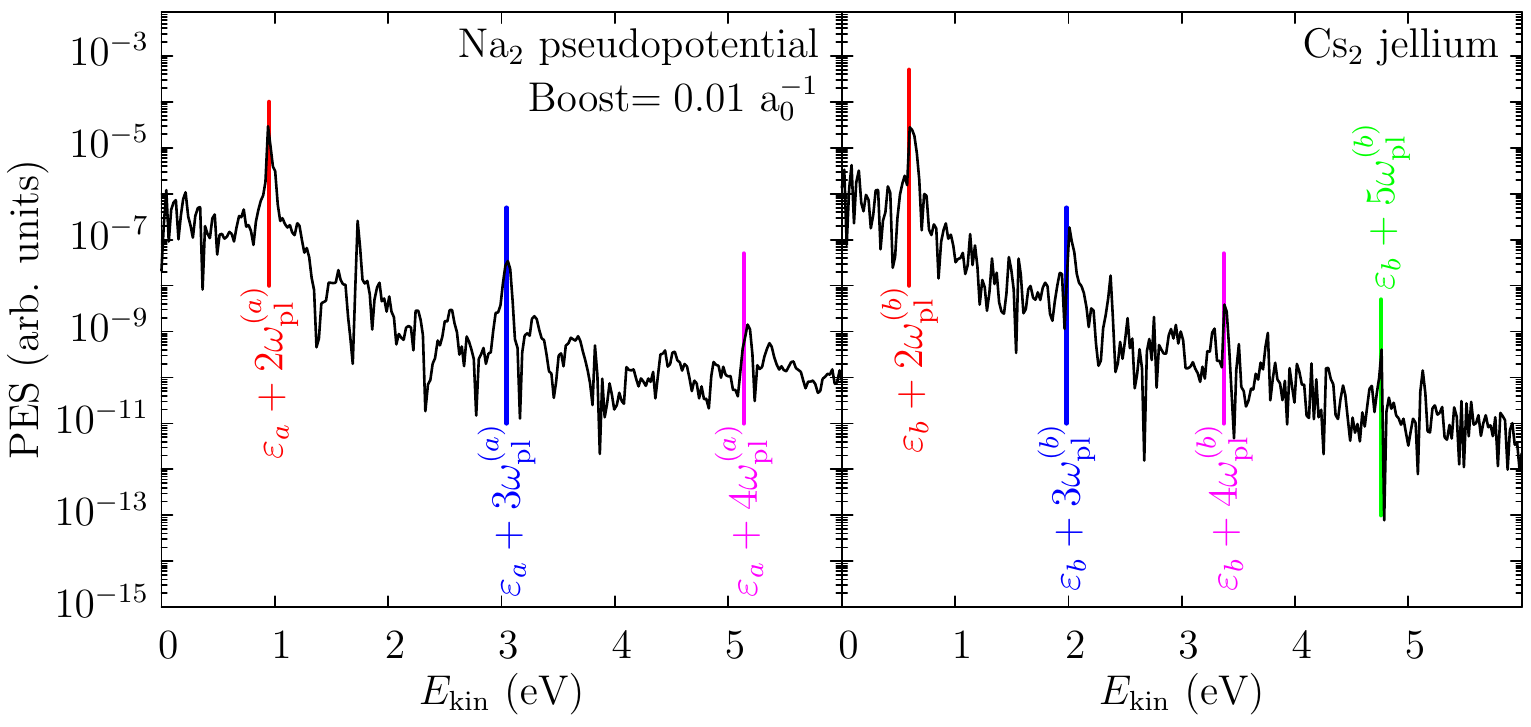}
}
\caption{Photoelectron spectra of Na$_2$ described by pseudopotentials
  (left) and Cs$_2$ described by a jellium background, excited by an
  instantaneous boost of 0.01 a$_0^{-1}$ applied to the wave
  functions at $t=0$. The vertical lines are positioned at energies
  corresponding to the single particle energy of Na$_2$ (of Cs$_2$),
  $\varepsilon_a=-3.23$~eV ($\varepsilon_b=-2.18$~eV), blue-shifted by
  multiples of the plasmon frequency $\omega_{\rm pl}^{(a)}=2.1$~eV
  ($\omega_{\rm pl}^{(b)}=1.4$~eV).}
\label{fig:pes_coll_pedago}
\end{figure}

The pulses from fast projectiles are extremely short and cover a
  very broad band of frequencies. At first glance, this looks like a
  disadvantage as there is thus no specific frequency information in
  the pulse. However, it has the advantage that it enables to extract
  unambiguously effects from the system's modes. To illustrate this
point, we start with two pedagogical examples which have a dominant
dipole mode, namely the case of Na$_2$ and Cs$_2$. The Na$_2$ is
described by explicit ions and pseudopotentials, while we use a
deformed jellium background for the description of Cs$_2$. The
interaction of the irradiated system with a very fast charged
projectile can be modeled by an instantaneous boost of the electronic
wave functions at $t=0$ \cite{Cal97b}.  Note that this procedure is
the same as the one we use for the calculation of an optical response,
see Sec.~\ref{sec:sic_optresp}. 
\MD{
More precisely, we apply a boost $\mathbf p$ to each occupied s.p.
wave function $\varphi_{j,{\rm gs}}$ of the ground state, and take the obtained wave 
functions as the initial states~:
\begin{equation}
\varphi_j(\mathbf{r},t\!=\!0)=\exp(\mathrm{i} \mathbf p \cdot \mathbf r) 
\varphi_{j,\mathrm{gs}}(\mathbf{r})
\label{eq:boost}
\end{equation}
This mimics the effect of the Coulomb field caused by a fast by-passing charged projectile. 
For simplicity, let us consider a boost in the $z$ direction only. If the projectile is fast enough,
we can assume that it travels on a straight line with constant velocity $v_{\rm proj}$.
Therefore, one can evaluate the net force integrated over the collision, and the latter 
exhibits a component only in direction to 
the point (here the $z$ axis) of closest impact \cite{Bae06a}. The induced boost for a projectile
of charge $Z$ and an impact parameter $b$ then reads 
\begin{equation}
p= \frac{4Ze^2}{bv_{\rm proj}} \quad.
\label{eq:boost_p}
\end{equation} 
One can see that, for a given value of the boost, larger $b$ and/or $v_{\rm proj}$ can be
compensated by increasing the charge $Z$ of the projectile. To derive
Eq.~(\ref{eq:boost_p}), the passage time of the projectile, which reads $b/v_{\rm proj}$,
should be much smaller than a typical electron reaction time $\omega_\mathrm{el}^{-1}$. If
one uses the value $\omega_\mathrm{el}=1$~Ry and an impact parameter $b=10$~a$_0$ for a
rough estimate, we have the constraint $v_{\rm proj} \gg b \omega_{\rm el}=200$~a$_0$/fs.
This lower value corresponds to a kinetic energy of 700~keV for a colliding proton.
Inserting this value of $v_{\rm proj}$ in Eq.~(\ref{eq:boost_p}) with $b=10$~a$_0$ and
$Z=1$ yields a maximal value $_{\rm max}=0.08$/a$_0$, which is in the range of the boosts
used in the following.

We now come back to the first two test cases that we have studied, namely Na$_2$ and Cs$_2$.
Both systems exhibit a very clean plasmon peak at 2.1~eV and 1.4~eV respectively.} 
Fig.~\ref{fig:pes_coll_pedago} shows the obtained PES.
On top of the exponential decrease emerge some peaks. One can identify
the dominant ones as multi-plasmon excitations, similar to those
already discussed in Sec.~\ref{sec:las_vs_plasm}~: the energy given
by the boost to the system is mainly stored in the dominant dipole
mode, let us call it for simplicity the ``plasmon''. Since the plasmon
frequency is below the ionization threshold for each case, two or more
plasmons are needed in order to ionize the system. And indeed, the
double and triple plasmon processes are clearly visible in the PES. It
becomes more difficult to disentangle higher orders from the
background.

\begin{figure}[htbp]
 \centerline{
\includegraphics[width=\linewidth]{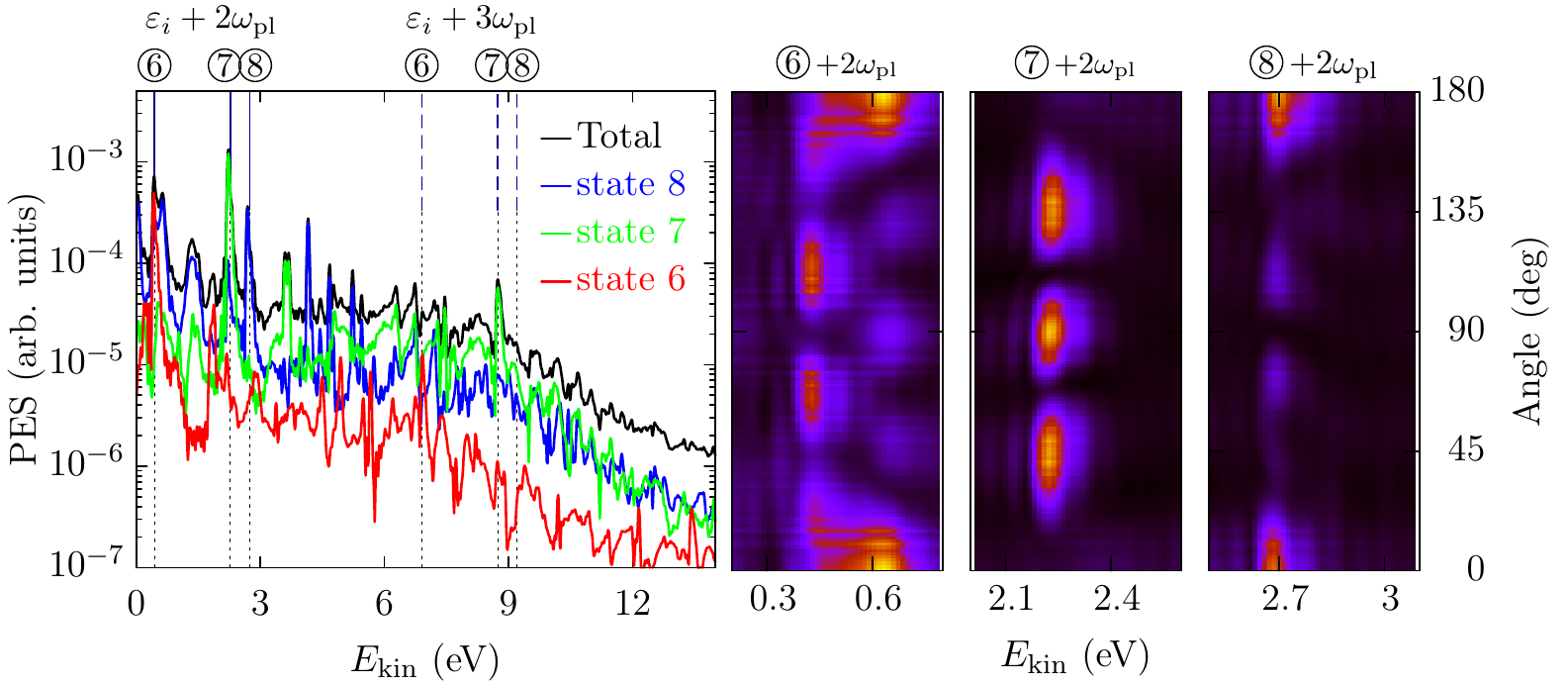}
}
\caption{Electron emission from C$_5$ chain after excitation by an
  instantaneous boost of 0.07 a$_0^{-1}$ along the chain direction of the
  electronic wave functions at $t=0$. Left panel~: PES
  (in the longitudinal direction) of all states (black), of
  the HOMO (state 8, blue), HOMO$-1$ (state 7, green), and HOMO$-2$
  (state 6, red). The vertical lines indicate the 2- and the 3-plasmon
  processes from these three least bound states,
  with a Coulomb shift (see Sec.~\ref{sec:pes_I}) of 0.3~eV applied
  to account for the total ionization of 0.066 at the end of the
  simulation time. Right~: Density maps of combined PES/PAD focused in
  the 2-plasmon excitation energy window of states 6, 7, and 8
  respectively. The angle is measured with respect to the chain
  direction.}
 \label{fig:c5boost}
\end{figure}

We now turn to a more involved case, that is the C$_5$ chain. It is
described by an explicit ionic structure and the 20 valence electrons
are shared among 8 different electronic levels. Note that the HOMO$-2$
and HOMO$-3$ are doubly degenerated. We distinguish longitudinal modes
along the symmetry axis (elongated direction) and transversal modes
perpendicular to it. The transversal optical response (not shown here)
is suppressed by more than one order of magnitude with respect to the
longitudinal response and it is substantially fragmented.  Therefore,
we do not expect that the peaks in transversal modes can significantly
contribute to the PES. We concentrate the following discussion only
longitudinal modes.  The photoabsorption spectrum of C$_5$ in this
direction is dominated by one single, strong and sharp resonance at
$\omega_\mathrm{pl}=6.5$~eV. The notion plasmon is justified here
because this is a truly collective oscillation in the sense of a Mie
surface plasmon.  The left panel of Fig.~\ref{fig:c5boost} shows the
total PES stemming from all states, and the state
specific PES of the HOMO (state 8), HOMO$-1$ (state 7) and HOMO$-2$
(state 6).
 The dominant peaks at low energies can be
clearly identified as 2-plasmon excitations from states 6, 7, and 8,
which are emphasized by vertical lines. The state-resolved PES confirm
this interpretation. For instance, the peak in the total PES at 2.3~eV
comes from the PES of state 7 (green curve). The same occurs in the
two other peaks for states 6 and 8 excited by 2-plasmon processes. As
for the 3-plasmon peaks, one can catch some of the peaks, especially
that of state 7. We also observe other peaks which are most likely
images of the s.p. spectrum but for frequencies different
from the plasmon one.  In contrast to a frequency-selective laser
pulse, the boost excites here all possible modes. Due to the strength
of the excitation, there might also be some cross-talk to the
transverse modes. It is therefore expected that the present phenomenon
can only be seen in systems with a rather "clean" dipole response
characterized by sharp plasmon resonances in all directions and not
too much a fragmented spectrum.

The right panels in Fig.~\ref{fig:c5boost} display the full combined
PES/PAD zoomed onto the features corresponding to the doubly and
triply excitations from states 6, 7, and 8. The angle is measured with respect
to the longitudinal direction of C$_5$. The striking feature is that,
although the boost is performed along the chain, the electrons are not
exclusively emitted in this direction. For instance, states 6 and 8
exhibit in addition a sizable emission at 60$^\circ$ and
120$^\circ$. On the contrary, state 7 preferentially emits at
45$^\circ$, 90$^\circ$ and 135$^\circ$. Clearly, these combined
PES/PAD allow one to relate the emission behavior to the symmetry of
the depleted wave functions as in cases with lasers, see e.g.  Fig.~\ref{fig:na33m_pespad}.

\subsection{Towards quantum dissipative electron dynamics}
\label{sec:qdissip}

\subsubsection{From VUU to quantum world}
\label{sec:vuu_2_q}

Although VUU (see Sec.~\ref{sec:semiclassicalroute}) provides a way to
describe dissipation in dynamical scenarios, it is limited to large
excitation energies and to simple materials as, e.g., alkaline
clusters. Both limitations are direct consequences of the
semi-classical nature of VUU.  Simple metals can be described because
their electron cloud comes close to a Fermi gas.  Reducing the
excitation energy or considering other systems (as e.g.  C$_{60}$)
requires to account for quantum effects which renders VUU inapplicable.

This calls for a quantum kinetic theory. This can be seen two ways,
either as quantum counterparts of the VUU equation (\ref{eq:vuu})
or as TDLDA complemented by a quantum generalization of the UU
collision term (\ref{eq:UU}).  Anyway, such a theory deals with
impure quantum states which are described at the level of the one-body
density matrix .  The corresponding dynamical equation for
$\hat{\rho}$ reads
\begin{equation}
 \mathrm{i}\frac{\partial \hat{\rho}}{\partial t} 
 =  
 [\hat{h},\hat{\rho}] 
 + 
 I_\mathrm{coll}[\hat{\rho}]
 \quad.
\label{eq:kineteq}
\end{equation}
The commutator with the mean-field Hamiltonian $\hat{h}$ describes the
mean-field evolution according to TDLDA.  It is complemented by a
collision term $I_\mathrm{coll}[\hat{\rho}]$ which, however,
becomes awfully involved in the quantum case \cite{Goe86a}.
There does not yet exist any routine solution to the problem in finite
systems, in spite of the many investigations, particularly in nuclear
physics \cite{Ber88,Dur00,Abe96}. So far, most practical solutions
rely on a (partial or full) semi-classical treatment \cite{Ber88}.
The many detailed experiments on cluster dynamics discussed in the
previous sections revive the call for a manageable quantum kinetic
theory. We discuss in this section two promising directions of
research along that line.


\subsubsection{A relaxation time ansatz}
\label{sec:relax_ansatz}


VUU in the semi-classical domain and stochastic TDHF/TDLDA (see
Sec.~\ref{sec:stdhf}) describe dissipation in a very detailed, thus
expensive, manner.  In cases of moderate fluctuations, the system as such
remains intact and the outcome is rather obvious~: the dissipative
dynamics drives steadily towards thermal equilibrium of a still compact
system. This suggests a simplification in terms of the relaxation-time
approximation which had been used since long in the homogeneous
electron gas \cite{Ash76}.  An implementation for finite fermion
systems had been proposed in the nuclear context in \cite{Won83a}. But
the computational limitations at that time did not allow realistic
applications. Just recently, we have taken up this old idea of a
relaxation time approximation and started to implement for cluster
dynamics. We give here a brief preview of this ongoing work.

The relaxation-time approximation starts from Eq.~(\ref{eq:kineteq}).
The collision term $I[\hat{\varrho}]$ is approximated by
\begin{subequations}
\label{eq:dissipTDLDA}
\begin{equation}
  \mathrm{i}\partial_t\hat{\varrho}
  -
  \left[\hat{h},\hat{\varrho}\right]
  =
  \frac{1}{\tau_\mathrm{relax}}
  \left(
  \hat{\varrho}-\hat{\varrho}_\mathrm{equil} \left[\rho(\mathbf{r}),\mathbf{j}(\mathbf{r}) \right]
  \right)
  \quad.
\label{eq:relaxtime1}
\end{equation}
The right-hand-side is the effective collision term which forces the
system to converge towards the equilibrium. Note that this employs the
{\em local} equilibrium
$\hat{\varrho}_\mathrm{equil}[\rho(\mathbf{r}),\mathbf{j}(\mathbf{r}),E]$
which depends on the instantaneous local density, current and energy
$E$ from the given $\hat{\varrho}(t)$. It reads
\begin{equation}
  \hat{\varrho}_\mathrm{equil}
  =
  \sum_\alpha
  |\varphi_\alpha\rangle n_\alpha^\mathrm{(equil)}\langle\varphi_\alpha|
  \quad,\quad
  n_\alpha^\mathrm{(equil)}
  =
  \frac{1}{1+\exp((\varepsilon_\alpha-\epsilon_\mathrm{F})/T)}
\end{equation}
and can be computed with density- and current-constrained TDLDA
\cite{Cus85b,Uma06a}.  The temperature $T$ is tuned iteratively such
that the total energy matches the wanted value $E$.
The key parameter is the local relaxation time $\tau_\mathrm{relax}$
for which we need a reliable choice. To that end, we recur to a
semi-classical estimate of relaxation time
\cite{Ber78a,Dan84a,Dan84b,Dan84c}.  It is based on the Fermi gas
model in which the relaxation time becomes simply
\begin{equation}
  \frac{\hbar}{\tau_{\rm relax}}
  =
  \frac{16}{15} \frac{m}{\hbar}\sigma_\mathrm{ee} T^2
  \quad,
\label{eq:rateFgas}
\end{equation}
\end{subequations}
where $k_B=1$ and $\sigma_\mathrm{ee}$ is the effective in-medium
cross section for electron-electron collisions. For metal clusters,
we find $\sigma_{ee}\approx 4\pi r_s^2$ \cite{Gig02}.
Eqs. (\ref{eq:dissipTDLDA}) together constitute the dissipative TDLDA
in relaxation-time approach.
 
\begin{figure}[htbp]
\centerline{\includegraphics[width=\linewidth]{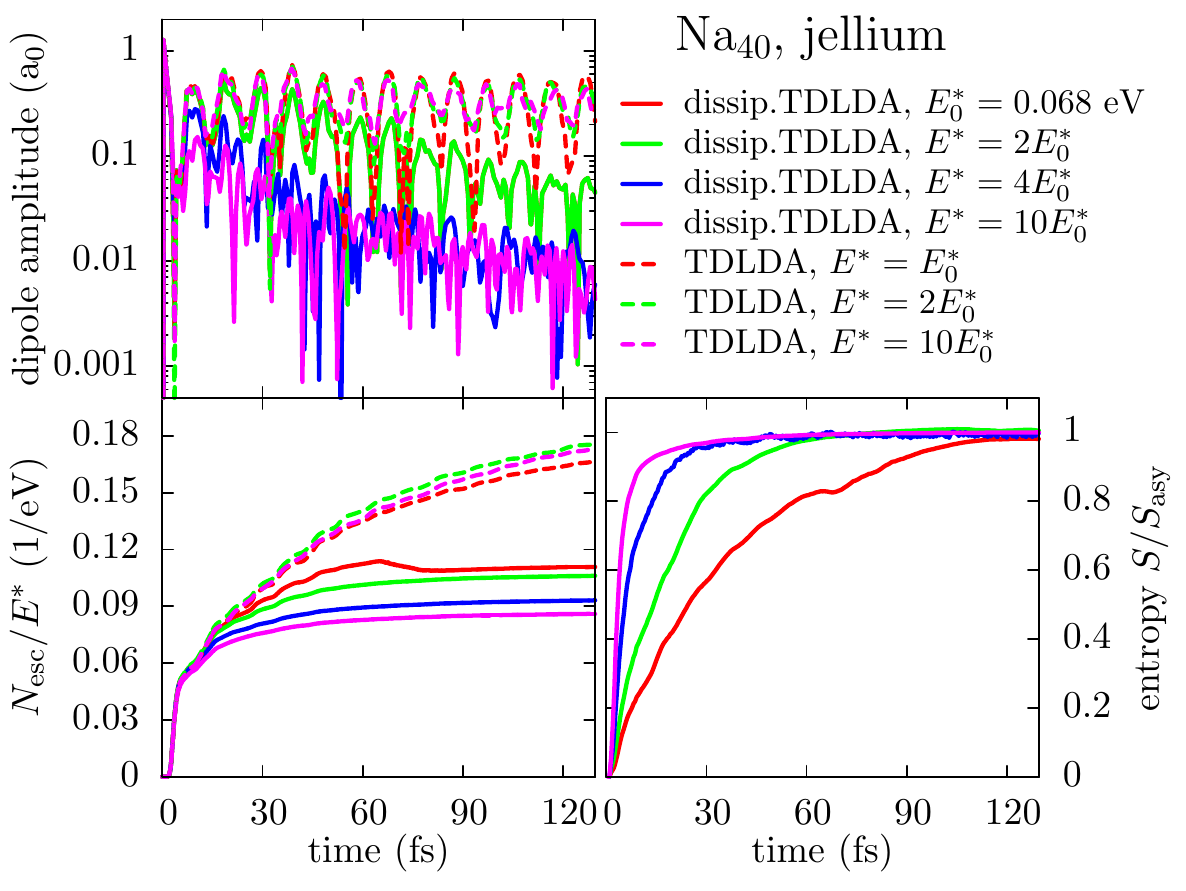}}
\caption{Time evolution of basic observables in Na$_{40}$ with soft
  spherical jellium background (\ref{eq:softJ}) using $r_s=3.65$ a$_0$
  and $\sigma_\mathrm{jel}=1$ a$_0$, after instantaneous boost
  with corresponding excitation energy $E^*$ as indicated~: Ionization 
$N_\mathrm{esc}$ normalized to $E^*$ (bottom left), envelope of the 
dipole signal (top left), and single particle
  entropy $S$ normalized to the asymptotic entropy $S_\mathrm{asy}$ (bottom right).
  Compared are results from pure TDLDA (dashes) with those from TDLDA
  with dissipation in relaxation-time approximation (full curves).
\label{fig:Na40-boost-short}
}
\end{figure}

Fig.~\ref{fig:Na40-boost-short} shows a first result from the newly
developed dissipative TDLDA scheme. 
We consider as a test case Na$_{40}$
after instantaneous boost of its electron cloud at various boost
energies. The s.p. entropy
$S=\sum_\alpha\left(n_\alpha\log{n}_\alpha+(1-n_\alpha)\log(1-n_\alpha)\right)$
is shown on the right panel. It demonstrates most clearly the
evolution towards thermal equilibrium. The global relaxation time
shrinks visibly with increasing excitation. This is a general feature
already well known from VUU. It is related to the fact that the phase
space for transition opens up with increasing energy.  The relaxation
times deduced from this figure range from about 40 fs for low
excitation down to few fs for very energetic cases. This is in range
of measured values \cite{Kle97,Leh00}.
The time evolutions of the dipole envelope (top left) from pure TDLDA (dashed lines) 
are only very slowly decaying
and look at the scale of this figure nearly
constant. Activating dissipation leads to a clear decay of the
signal~: the higher the excitation, the stronger the decay (in accordance
with the right panel). But note that this decay starts only after some
delay while the evolution in the early stages is very similar to
TDLDA. Finally, the left lower panel shows the time evolution of
ionization. The ongoing dipole oscillations in case of TDLDA leads to
ongoing electron emission. The case with dissipation shows a leveling
off for the ionization. The dipole signals has been damped away and
the excitation energy is converted to intrinsic, thermal energy. This
energy later on leads to a thermal electronix emission at a much
slower time scale, thus not visible here. After all, we see that
dissipative TDLDA can provide a pertinent of the thermalization of an
excited electron cloud.

\subsubsection{Stochastic Time-Dependent Hartree-Fock}
\label{sec:stdhf}

An alternative route to kinetic theory is provided by stochastic
methods describing the system as an ensemble of (pure) mean-field
states. This leads to Stochastic Time-Dependent Hartree-Fock (STDHF),
or Stochastic TDLDA (STDLDA) when combined with density functionals.
It was originally formulated in the context of nuclear collisions
\cite{Rei92c}, whence the acronym TDHF, but it can formulated for
whatever system in which a quantum mean field provides a sound
description of the ground state and to low energy properties. In the case of
clusters and molecules TDLDA provides the obvious effective mean field
theory as a starting point, thus coming to STDLDA. For simplicity, we
will use the notion STDHF further on.  STDHF contains all the
ingredients of a standard kinetic equation, complemented by proper
statistical fluctuations.  It accounts for collisional correlations
(from electron-electron collisions).  They are treated in incoherent
manner and should not be mixed with coherent correlations as they
typically dominate in low-energy processes.

The original formulation of STDHF started from the quantum Liouville
equation for density matrices.  The early studies could show that the
ensemble description of STDHF can be reduced to a quantum Boltzmann
equation complemented by a to the quantum Boltzmann Langevin equation
\cite{Rei92c}.  The latter was introduced in \cite{Bix69,Zwa73} and
has been particularly studied in the nuclear context
\cite{Ayi88,Ran92a,Abe96,Nap13}.  STDHF is thus a well founded theory
containing all the ingredients necessary for a description of
dissipative electronic features. It has unfortunately never been
explored at full quantum level because of its complexity. It is
only recently that the first calculations were performed in model
systems \cite{Sla14}
with a proper reformulation of the theory. The
first results are quite promising and we will thus discuss here
briefly the formalism and typical results obtained in a simple system.

The STDHF describes the system by an ensemble of $\mathcal{N}$ Slater
states $\{|\Phi^\alpha \rangle, \alpha = 1, \ldots,\mathcal{N}\}$. Each
state $|\Phi^\alpha\rangle$ is associated with a set of
single-particle (s.p.)  states
$\{\varphi^\alpha_i,i=1,\ldots,\Omega\}$. The labels $i=1, \ldots, N$ with $N<\Omega$ stand
for the occupied (hole) states. We also include in the description a
sufficient amount of unoccupied (particle) states $i=N\!+\!1,
,\ldots,\Omega$ which will serve as a "reservoir" of levels for
transitions to come.  With this ensemble of s.p.  states we can now
unfold hierarchy of $n$-particle-$n$-hole ($nph$) excitations.  The
first ones to be appear are $2ph$ excitations because $1ph$
excitations are already accounted for in the mean-field propagation
(TDHF or TDLDA).  The correlated wave function (starting from an
uncorrelated situation) can then be expanded as
\begin{equation}
|\Psi^\alpha(t)\rangle =
|\Phi^\alpha(t)\rangle+\sum_{pp'hh'}c^\alpha_{pp'hh'}(t)|\Phi^\alpha_{pp'hh'}(t)\rangle \quad,\\
\end{equation}
with
\begin{equation}
|\Phi^\alpha_{pp'hh'}\rangle
 =
 \hat{a}_p^\dagger\hat{a}_{p'}^\dagger
 \hat{a}_{h'}^{\mbox{}}\hat{a}_{h}^{\mbox{}}|\Phi^\alpha\rangle.
\end{equation}
Note that the $2ph$ states $|\Phi^\alpha_{pp'hh'}\rangle$ are
also Slater states.  Starting from an uncorrelated situation, one
then propagates a correlated state $|\Psi^\alpha(t)\rangle$ up to a
certain time $\tau$ at which it is sampled in terms of an ensemble
$\{|\Phi^\alpha_\kappa\rangle,w^\alpha_\kappa\}$, where
$\kappa\in\{0,pp'hh'\}$.  The weight
$w^\alpha_\kappa=|c^\alpha_{pp'hh'}|^2$ is the probability with which
$|\Phi^\alpha_\kappa\rangle$ appear. It is evaluated by means of
time-dependent many-body perturbation theory which finally
leads to a transition probability following Fermi's golden rule as
\cite{Rei92c}
\begin{equation}
  w^\alpha_{pp'hh'}
  =
 \tau  \left|  \langle\Phi^\alpha_{\kappa}|\hat{W}|\Phi^\alpha\rangle  \right|^2
  \delta(\varepsilon^\alpha_{p}\!+\!\varepsilon^\alpha_{p'}\!
               -\!\varepsilon^\alpha_{h}\!-\!\varepsilon^\alpha_{h'}) \quad,
\label{eq:jumprate}
\end{equation}
where we have introduced the residual interaction $\hat{W}$
complementing the mean-field hamiltonian $\hat{h}$.  The original
state $|\Phi^\alpha\rangle$ itself has a weight $w^\alpha_0=1-\sum
w^\alpha_{pp'hh'}$ (attributing $w^{\alpha}_0$ to the "no transition"
case) which is the complement of all the other transition
probabilities.  The Dirac $\delta$-function has to be taken with a
word of caution.  The full expression involves an operator $\delta$
function of the mean-field Liouvillean \cite{Rei92c}. The
approximation (\ref{eq:jumprate}) involves s.p. energies taken as
expectation values over the s.p. states.  These, however, are
ambiguous to the extent that one has always the freedom of a unitary
transformation amongst the occupied states. We define the
$\varepsilon^\alpha_i$ uniquely by diagonalizing the actual mean-field
hamiltonian $\hat{h}^\alpha$ separately amongst occupied states and
unoccupied states.  In practice, the Dirac $\delta$-function has to be
augmented by a finite width to account for the discrete nature of
spectra in finite systems \cite{Sla14}. The choice of $\tau$ and
$\hat{W}$ also requires some caution as the scheme, being based on
time-dependent perturbation theory, has to remain in the weak coupling
limit as typical of standard kinetic theory.  In particular the
sampling interval $\tau$ should be long enough to allow a sufficient
number of "collisions" to take place, which then justifies stochastic
reductions and loss of coherence, but short enough to remain
perturbative, namely with $w^\alpha_0\ll 1$ \cite{Hov55,Rei92c}.

The STDHF/STDLDA ensemble propagation can thus be summarized as
follows. We define an initial state $|\Phi_0\rangle$ and each member of
the ensemble is initially set to
$|\Phi^\alpha(0)\rangle=|\Phi_0\rangle$.  We then propagate each
$|\Phi^\alpha(t)\rangle$ individually, first from $t=0$ to $\tau$ by
TDHF/TDLDA.  At time $\tau$, all $2ph$ states about
$|\Phi^\alpha(\tau)\rangle$ are evaluated as well as the associated
jump probabilities $w^\alpha_\kappa$ following Eq.(\ref{eq:jumprate}).
One state $|\Phi^\alpha_\kappa\rangle$ is then randomly selected
according to its weight $w^\alpha_\kappa$.
$|\Phi^\alpha_\kappa\rangle$ is then again propagated according to
TDHF/TDLDA from $\tau$ to $2\tau$ up to $2\tau$ at which a similar
sampling takes place, and so on.  The above procedure is then
restarted from initial time for each member of the ensemble
separately. This altogether provides the STDHF ensemble
$\{|\Phi^\alpha(t)\rangle, \alpha=1,\ldots,\mathcal{N}\}$ :
\begin{equation*}
\left\{
\begin{array}{cccccc}
|\Phi^\alpha \rangle &\stackrel{\rm TDHF}{\longrightarrow} &\underbrace{ \{ |\Phi^\alpha_\kappa \rangle, w^\alpha_{\kappa} \} }&  & &\\
&&  \mathrm{Sampling} &&&\\
 &  & |\Phi^\alpha_{\kappa_0} \rangle & \stackrel{\rm TDHF}{\longrightarrow} & 
 \underbrace{\{ |\Phi^\alpha_{\kappa'} \rangle,w^\alpha_{\kappa'}\}} &\\
&&&&...&\\
t\!=\!0 && \tau_{} && 2 \tau_{}  &...\\
\end{array}
\right\}
\rotatebox[origin=c]{90}{$\alpha=1,\ldots,\mathcal{N}$}
\end{equation*}

The ensemble $\{|\Phi^\alpha \rangle,\alpha=1,\ldots,\mathcal{N}\}$
allows one to compute any observable by standard statistical averages.  In
particular, one-body or two-body operators (both correlated) can be
directly constructed from the ensemble. The one-body density matrix
reads
\begin{equation}
 \hat{\rho}
 = 
 \frac{1}{\mathcal{N}}
 \sum_{\alpha = 1}^\mathcal{N} \hat{\rho}^\alpha = \ \sum_{\alpha = 1}^\mathcal{N} \sum_{i=1}^{N}
 |\varphi_{i}^\alpha\rangle \langle\varphi_{i}^\alpha|
 \equiv
 \sum_{\nu=1}^\Omega 
 |\varphi_{\nu}^{(nat)}\rangle n_\nu\langle\varphi_{i}^{(nat)}|
 \label{eq:rho_mat}
\end{equation}
where the second representation employs the natural s.p. orbitals
$|\varphi_\nu^{(nat)}\rangle$ diagonalizing $\hat{\rho}$ and
immediately delivers the associated (fractional) occupation numbers
$n_\nu$. The latter quantities provide a natural tool for analyzing
thermal effects.  

For a first test of STDHF, we use a simple 1D model simulating a dimer
molecule.  The mean-field hamiltonian reads (in $x$
representation and taking $\hbar$=1)~:
\begin{equation}
  \hat{h}^\alpha
  = 
  -\frac{\Delta}{2m} + V_{ext}(x) + \lambda \left(\varrho^\alpha(x)\right)^2.
\label{eq:hn}
\end{equation}
It contains a self consistent term $\lambda
\left(\varrho^\alpha(x)\right)^2$ (with $\lambda=27.2$~eV\,a$_0^2$)
involving the local one-body density
$\varrho^\alpha(x)\varrho(x)=\sum_1^{N} |\varphi_i^\alpha(x)|^2$.
This terms stands for the effect of a simple density functional.  The
external potential $V_{\rm ext}(x)$ has a Woods-Saxon shape~:
$V_{\rm ext}(x)=V_0/(1+\exp((x-x_0)/a))$ with $V_0=-68$~eV, $x_0=15$~a$_0$,
$a = 2$~a$_0$. It is complemented, outside the well, by a confining
harmonic oscillator
ensuring soft reflecting boundary conditions. These
boundaries allow one to avoid direct emission and to focus the analysis on
the building up of thermal effects. Altogether, the model mimics a
typical situation in clusters and molecules with a fixed external
potential delivered by the ions and an energy scale typical of organic
systems.  The residual interaction is, in the present study, chosen
schematically as a simple zero-range force $W(x,x') = W_0\delta(x-x')$
with $W_0= 40.8$~eV which delivers realistic relaxation times
\cite{Sla14}.  Actually, we use 9 physical particles (9 hole states)
complemented by a reservoir of 8 (16 or 24 give similar results)
particle states. The initial excitation is done by a random particle
hole excitation delivering an excitation energy of about 25.8~eV. The
sampling time is time $\tau$ = 1 fs (0.5 and 1.5 fs give similar
results) and the dynamics is followed over 100 fs, which is much larger
than the optical period (1.15 fs) and long enough to study thermal
relaxation. We propagate an ensemble of ${\cal N}$=100 events.

\begin{figure}[htbp]
\centerline{\includegraphics[width=0.6\linewidth]{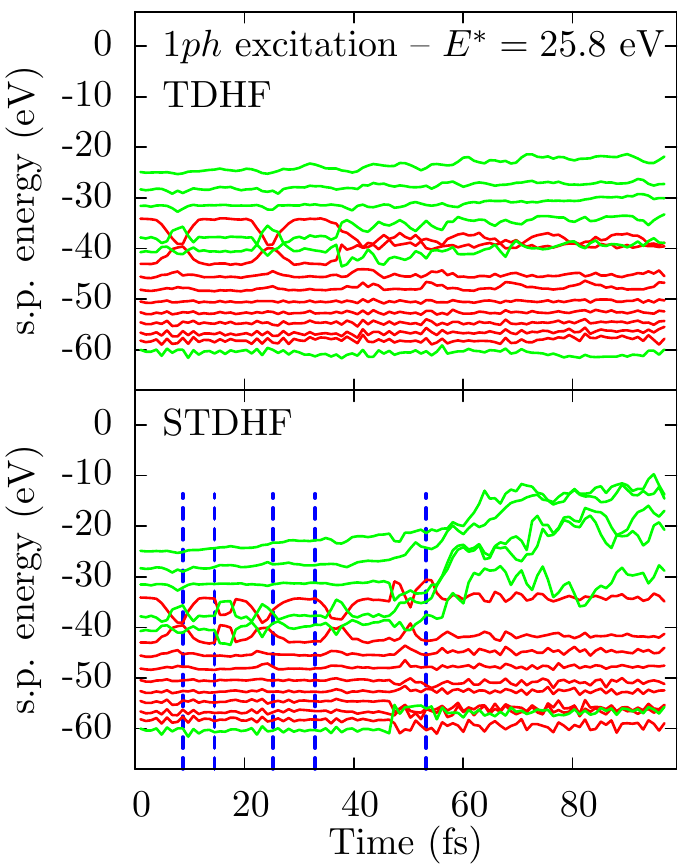}}
\caption{Time evolution of single particle energies in a TDHF
  calculation (upper panel) and in one STDHF event (lower panel) for
  an initial excitation energy of 25.8~eV. Red lines correspond 
to occupied states, green ones to unoccupied ones (see text for details).
In the STDHF case, five $2ph$
  transitions actually occurred (were actually sampled) for this
  event, transitions which are indicated by faint vertical dashed
  lines. }
\label{fig:spe19bis}
\end{figure}

Fig.~\ref{fig:spe19bis} shows the time evolution of
s.p. energies for a typical STDHF event (bottom) and compares it
to the corresponding pure TDHF evolution (top).  
Each member
of the ensemble will actually deliver a different sequence of
transitions which will finally lead to the mixed state representing
the correlated system.  One can identify five $2ph$ transitions
(indicated by faint dashed lines). It is also interesting to note
that, up to minor mean field rearrangements, the TDHF evolution does
preserve the arrangement of particle and hole states in the course of
time evolution. Pure TDHF evolution more or less preserves initial
occupations in time, hindering possible relaxation to a thermal state.
STDHF overcomes overcomes this limitation as can be seen from the
rearrangements in the lower panel.

\begin{figure}[htbp]
\centerline{\includegraphics[width=0.65\linewidth]{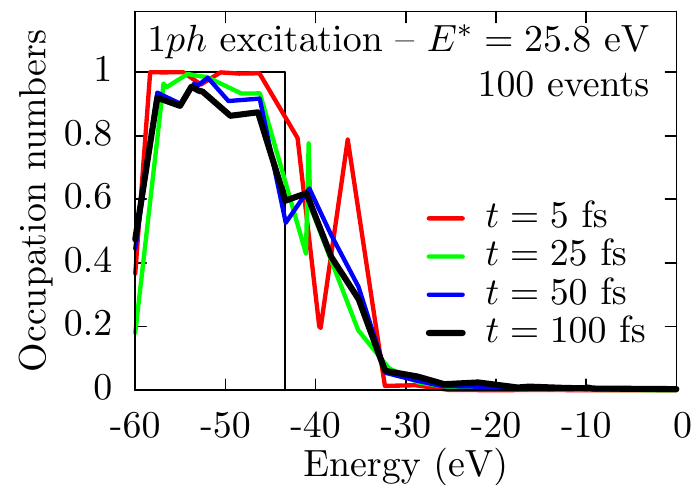}}
\caption{Snapshots of occupation numbers as a function of time in an example of STDHF
calculation, with initial excitation $E^*$ of 25.8~eV and for a 100 event ensemble. 
}
\label{fig:occupoft19}
\end{figure}

Fig.~\ref{fig:occupoft19} displays snapshots of occupation numbers,
extracted from the one body density matrix, see Eq.(\ref{eq:rho_mat}), at
several times along a STDHF propagation. 
The occupation numbers keep for some time a trace of the original excitation, in
particular the initial hole around $-37$~eV (see red curve). This hole is gradually filled
and the occupation numbers are soon washed out, leading asymptotically to an energy
profile typical of thermal equilibrium. Note some unavoidable statistical fluctuations,
still visible at very low energy. This figure therefore demonstrates the capability of
STDHF to account for relaxation effects at the side of electrons in a purely quantum
mechanical manner.
One should also stress that STDHF enables to estimate fluctuations
around average values as computed from one or two-body density matrix,
which again represents a remarkable step forward. The major obstacle
of STDHF lies in the high cost of handling large ensembles which
becomes particularly demanding for low excitation energies where
smaller transition probabilities require better statistics.  This
still more development work is needed to put STDHF fully into action.


\section{Conclusions}
\label{sec:conc}

In this paper we have reviewed the analysis of electron emission
following irradiation of clusters and molecules by light
pulses. Observables from electron emission give detailed insight into
the dynamical response of the irradiated species.  Understanding the
irradiation and emission process is also essential in view of the many
applications in materials science, biology, and medicine. 
High-resolution studies of electron emission have made tremendous
progress over the past few years, both experimentally and
theoretically.  In experiments, new developments in light sources now
provide a broad choice of electromagnetic pulses with widely variable
frequency, intensity, and time profile down to attosecond resolution
in the range of electronic time scales. There is also great progress
at the side of the measurement giving access to increasingly detailed
properties of emitted electrons, high-resolution photo-electron
spectra (PES) or angular distributions (PAD), often combined to
velocity map imaging. As latest achievements, time-resolved PES/PAD are
waiting in the wings.
This remarkable experimental progress calls for elaborate theoretical
treatments at the most microscopic level of description. In this
respect, TDDFT, especially when solved in real time, constitutes an
invaluable tool to simulate the various dynamical scenarios of
irradiation of clusters and molecules. Therefore, many groups all over the
world are heavily working on such approaches. 
In this review, we have presented a series of developments and
results, mostly from the last decade, on irradiation of clusters and
molecules by light pulses and subsequent detailed analysis of electron
emission.  Concerning the observables from emission, we consider total
ionization, PES and PAD, also in connection with time-resolved
measurements. At the theory side, we focus on a microscopic
description in terms of time-dependent density functional theory
(TDDFT). This is practically handled at the level of the
time-dependent local density approximation (TDLDA) augmented by a
self-interaction correction (SIC).
%
We have illustrated the capabilities of our approach on several systems,
ranging from simple molecules like N$_2$ to fashionable nano-clusters
such as C$_{60}$, and also studying archetypal metal systems such as
Na clusters.  We list below a few results that we consider as being
emblematic of these studies.

The theoretical modeling is based on TDDFT which is known to provide a
robust microscopic description of the system dynamics. It enables to
include ionic motion at a classical level. The fully coupled dynamics
is needed in cases of long laser pulses and for thermal ensembles. In
most cases, we consider short pulses and keep ions frozen. 
Starting level for TDDFT is the time-dependent local-density
approximation (TDLDA). However, in order to describe ionization
dynamics properly, one needs a theory fulfilling Koopmans' theorem
which states that the ionization potential (IP) has to be identical
with the single particle (s.p.) energy of the least bound state. This is violated by the
self-interaction error in LDA. It can be cured by augmenting LDA with
a SIC. The latter has been turned manageable by a handling in terms of
two sets of occupied s.p. states, the 2setSIC scheme. A less expensive
alternative is offered by averaged density SIC (ADSIC) which performs
surprisingly well as long as the dynamics stays off the regime of
fragmentation and/or huge ionization. ADSIC has thus been used here in
most cases and it allowed us to obtain remarkably accurate results in
good agreement with experiments.

The numerical handling of ionization dynamics is most efficient in a
coordinate-space representation. To describe ionization, we augment
the coordinate-space grid by absorbing boundary conditions. They
allow one to trivially compute the observables of total ionization and
ionization out of each s.p. state separately (level depletion). PAD
are computed by collecting the electron loss in angular segments on
the grid. A complication arises when comparing PAD with
measurements, since clusters or molecules
in gas phase have an undefined orientation. They
represent, in fact, an isotropic ensemble of orientations. Theoretical
calculations need thus to be complemented by orientation
averaging. For the one-photon domain, we have worked out a compact
formula which can live with only six reference orientations to be
computed. Multi-photon processes require direct integration where we
find that one can obtain reliable results with typically 18--36
integration points, depending on the symmetry of the cluster. PES are
computed by recording the phase oscillations of outgoing wave functions
close to the onset of the absorbing bounds, and by finally Fourier
transforming the temporal oscillations to the energy domain. Special
care has to be taken to cope with strong laser pulses. They may modify
the phase of the wave functions at the sampling point. Fortunately,
this effect can be evaluated analytically and allows us to derive a phase
correction, thus rendering the scheme for computing PES reliable up to
rather large laser intensities (typically
$10^{14}$--$10^{15}$W/cm$^2$.
Altogether, we have thus at hand powerful and versatile tools to
simulate ionization dynamics and to evaluate the observables deduced
thereof.  In the following, we will briefly summarize the results for
each observable separately.

The simple signal of total ionization is already useful when combined
with systematics. For example, the frequency dependence of ionization
maps the underlying dipole response. Ionization becomes the key signal
in pump and probe (P\&P) scenarios which constitute a well established
tool for a time-resolved measurement of ionic motion.  Clusters
are rather complex systems where the motion of single ions is hard to
track. P\&P measurements at least enable to identify global properties
of the ionic configuration, as radius and quadrupole deformation, which
is already very useful information in studies of Coulomb explosion of
clusters.  On a first example, we could show how the now upcoming
attosecond pulses allow time-resolved analysis at electronic pace.

The main body of the paper dealt with results on PES and PAD.
Combined PES/PAD as obtained from velocity map imaging (VMI) contain a
very rich amount of information, but are usually hard to appreciate as
such. The energy- or angular-integrated versions thereof, delivering
PAD and PES, are better suited for detailed analysis and comparisons
between experiments and theory.
A PES in a strictly one-photon domain delivers a map of the clusters
s.p. spectrum. {In experiments,} this was limited previously to negatively charge
cluster anions {due to a low ionization potential}. 
The availability of coherent high-frequency light
sources now allows one to employ the one-photon analysis for neutral
clusters and even cations.  Multi-photon processes make PES more
involved and richer.  In the low-intensity regime, one can identify
multiple copies of the s.p. spectra separated by the photon frequency.
But PES goes beyond just mapping s.p. spectra. It delivers a picture
of the whole dynamical processes. We have illustrated that by working
out the impact of plasmon resonances which can directly drop its
signatures in the PES themselves.
Increasing intensity produces more ionization which Coulomb shifts the
s.p. states gradually downwards, thus broadening the peaks in the PES.
This eventually leads to totally smoothed PES with straightforward
exponential decrease. {This kind of pattern suggests} at first glance an
interpretation as purely thermal electron emission. A closer
inspection from energetic considerations and TDLDA simulations reveals
that it cannot be fully a thermal process. We probably encounter a
mixed situation. Direct emission still prevails and electronic
re-collisions add first thermal effects to the picture. We have also
seen that PES alone cannot distinguish unambiguously between direct
and thermal emission.

The unavoidable orientation averaging of PAD wipes out many details
which were contained in before averaging. Fortunately, there remains a
lot of useful information. Orientation averaged PAD in the one-photon
regime can be characterized by one single parameter, the anisotropy
$\beta_2$. A systematic survey of $\beta_2$ and its frequency
dependence revealed that PAD are extremely sensitive to every detail
of the modeling. This holds even more so for state-resolved
anisotropies $\beta_2^{(i)}$ (where $i$ stands for a s.p. state or
degenerated group thereof).  No compromises are allowed. One must
invoke the full machinery of TDLDA+SIC and a careful description of
the detailed ionic structure to have a chance for a relevant
description.  For example, the $\beta_2(\omega_{\rm las})$ computed with smooth
jellium background shows marked fluctuations which disappear when
ionic background is used. However, there remains one remarkable
exception in the low-frequency tail of $\beta_2(\omega_{\rm las})$ for the
loosely bound anion ${{\rm Na}_7}^-$. This shows a deep dip towards one-photon
emission threshold and it does so independent of the model for the
ionic background.
In the multi-photon domain, a PAD provides crucial information which
helps to distinguish direct from thermal emission. This was nicely visible
in the combined PES/PAD of C$_{60}$ where one could associate
uniquely the region of low kinetic energy with thermal electrons while
higher kinetic energies show clear sign of direct emission with the
PAD being forward/backward dominated.

The collection of results presented in this review has several open
ends which call for further development and investigation. We briefly 
quote a few of them which we consider to be important next steps.
It was already mentioned above that it becomes increasingly possible
to extend time-resolved analysis to the attosecond
domain. Theoretical studies are required to explore the huge space of
new possibilities and to find out the most promising experimental
conditions. 
A further interesting perspective emerges if combining time-resolved
analysis with PES and PAD. This enables, e.g., to track the branching
between direct and thermal emission in the course of time. First
studies in this direction are very promising.
The discussion of PES and PAD left as a yet unsolved problem the
distinction between direct and thermal processes in electron emission.
This calls for a proper theoretical modeling of electron-electron
collisions (dynamical electron correlations) and subsequent
dissipation effects.  Two promising development lines for such an
extended TDLDA have been presented, that is the relaxation time approach which
models dissipation phenomenologically and a stochastic mean field
model which describes the system as an ensemble of mean-field states
incorporating the electron-electron collisions as stochastic jumps
between these states.

The results summarized above prove that observables from electron
emission are an extremely powerful tool to analyze irradiation
processes on clusters and molecules. They allowed one in the past to
reveal many interesting aspects of structure and dynamics of these
systems.  The future directions sketched in the previous paragraph
show that we are not nearly at an end of the investigations. The
field remains lively and highly interesting and challenging for the  future.


\section*{Acknowledgments}
The authors gratefully thank K. Andrae, C. Bordas, F. Calvayrac, J.-M. Escartin, B. and M.
Farizon, B. Faber, T. Fennel, C. Z. Gao, M. Ivanov, P. Kl\"upfel, F. L\'epine, K.-H. Meiwes-Broer, J.
Messud, A. Pohl, P. Romaniello, J.-M. Rost, N. Slama, O. Smirnova, J. Tigges\-ba\"um\-ker, M.
Vincendon, B. von Issendorff, Z. P. Wang, and F.-S. Zhang, for fruitful discussions during the realization of this work.
The authors would like to acknowledge financial support
from ANR-10-BLAN-0428, 
ANR-10-BLAN-0411, 
ANR-11-IS04-0003, 
ITN-CORINF, 
and the Institut Universitaire de France. 
The theoretical work was granted access to the HPC resources of IDRIS 
under the allocation 2013--095115 made by GENCI (Grand Equipement 
National de Calcul Intensif), of CalMiP (Calcul en Midi-Pyr\'en\'ees) 
under the allocation P1238, and of RRZE (Regionales Rechenzentrum Erlangen).

\bibliographystyle{elsart-num}

\end{document}